%% file: main.tex
\PassOptionsToPackage{usenames,dvipsnames}{xcolor}

\documentclass[twocolumn]{aastex63}

\usepackage[usenames,dvipsnames]{xcolor}
\usepackage{lineno}
\usepackage{xspace}
\usepackage{scrextend}
\usepackage{dcolumn,booktabs}
\usepackage{xstring}
\usepackage{comment}
\usepackage{graphicx}
\usepackage{amsmath}
\usepackage[utf8]{inputenc}
\usepackage{textcomp}
\usepackage[normalem]{ulem}
\usepackage{multirow}
\usepackage{tabularx}

\usepackage{fp}
\usepackage{pgf}

\graphicspath{ {figures/} }

\include{definitions}

\begin{document}

\include{PE_macros}
\include{nature_secondary_macros}

\title{
Observation of gravitational waves from two neutron star–black hole coalescences
}

\input{authors.tex}
\collaboration{0}{The LIGO Scientific Collaboration, the Virgo Collaboration, and the KAGRA Collaboration}

\begin{abstract}

We report the observation of gravitational waves from two compact
binary coalescences in LIGO's and Virgo's third observing run with properties consistent with neutron star--black hole (NSBH)
binaries.
The two events
are named \eventoneFullName and \eventtwoFullName, abbreviated as \eventone and \eventtwo; the first was observed by LIGO Livingston and Virgo, and the second by all three LIGO--Virgo detectors.
The source of \eventone has component masses
$\massonesourcemed{GW200105_combined_highspin}^{+\massonesourceplus{GW200105_combined_highspin}}_{-\massonesourceminus{GW200105_combined_highspin}}\,
\msun$ and $\masstwosourcemed{GW200105_combined_highspin}^{+\masstwosourceplus{GW200105_combined_highspin}}_{-\masstwosourceminus{GW200105_combined_highspin}}\,
\msun$, whereas the source of \eventtwo
has component masses $\massonesourcemed{GW200115_combined_highspin}^{+\massonesourceplus{GW200115_combined_highspin}}_{-\massonesourceminus{GW200115_combined_highspin}}\,
\msun$ and $\masstwosourcemed{GW200115_combined_highspin}^{+\masstwosourceplus{GW200115_combined_highspin}}_{-\masstwosourceminus{GW200115_combined_highspin}}\,
\msun$ (all measurements
quoted at the 90\% credible level).
The probability that the
secondary's mass is below the maximal mass of a neutron star is
$\lowestPnsEventOne\%$--$\highestPnsEventOne\%$ and
$\lowestPnsEventTwo\%$--$\highestPnsEventTwo\%$, respectively, for \eventone and \eventtwo,
with the ranges arising from  different astrophysical assumptions.
The
source luminosity distances are
$\luminositydistancemed{GW200105_combined_highspin}^{+\luminositydistanceplus{GW200105_combined_highspin}}_{-\luminositydistanceminus{GW200105_combined_highspin}}~\mpc$ and
$\luminositydistancemed{GW200115_combined_highspin}^{+\luminositydistanceplus{GW200115_combined_highspin}}_{-\luminositydistanceminus{GW200115_combined_highspin}}~\mpc$, respectively.
The magnitude of the primary spin of \eventone is less than  $\spinoneninetypercent{GW200105_combined_highspin}$ at the 90$\%$ credible level, and its orientation is
unconstrained. For \eventtwo, the primary spin
has a negative spin projection onto the orbital angular momentum at $\PEpercentchionenegative{GW200115_combined_highspin}\%$ probability.
We are unable to constrain
the spin or tidal deformation of the secondary component for either event.
We infer an NSBH merger rate density of $\KKLRateGSTLAL{}$~\VolRateUnit when
assuming that \eventone and \eventtwo are representative of the NSBH population, or
$\FGMCRate{}$~\VolRateUnit under the assumption of a broader distribution of component masses.
\\
\end{abstract}

\date{\today}

\section{Introduction}
\label{sec:Intro}
\input{intro}

\input{detector_split_sections}

\section{Source properties}
\label{sec:SourceProperties}
\input{pe}

\section{Nature of the secondary components}
\label{sec:NatureSecondary}
\input{nature_secondary}

\section{Astrophysical Implications}
\label{sec:AstroImplications}
\input{astro}

\section{Conclusions}
\label{sec:Conclusions}
\input{conclusions}

$ $

\input{LVK_acknowledgements_P2000488_v5}

Some of the parameter estimation analyses presented in this paper were
performed using supercomputer clusters at Swinburne University
of Technology (OzSTAR and SSTAR), Cardiff University (ARCCA), the
Max Planck Institute for Gravitational Physics (Hypatia), and the Barcelona Supercomputing Center---Centro Nacional de Supercomputaci{\'o}n.

\software{The detection of the signal and subsequent significance evaluation were performed
with the \textsc{GstLAL}-based inspiral software
pipeline \citep{Cannon:2011vi, Privitera:2013xza, Messick:2016aqy,
Sachdev:2019vvd, Hanna:2019ezx}, built on the \textsc{LALSuite}
software library \citep{LALSuite}, and with
the \textsc{PyCBC} \citep{Usman:2015kfa,Nitz:2018rgo, pycbc-software}
and \textsc{MBTAOnline}~\citep{2016CQGra..33q5012A}
packages. Parameter estimation was performed with
the \lalinference \citep{PhysRevD.91.042003}
and \textsc{LALSimulation} libraries
within \textsc{LALSuite} \citep{LALSuite}, as well as
the \bilby and \pbilby
Libraries \citep{2019ApJS..241...27A, Smith_20} and
the \textsc{dynesty} nested sampling
package \citep{2019arXiv190402180S}.  Estimates of the noise spectra
were obtained using \textsc{BayesWave} \citep{Cornish:2014kda,
Littenberg:2014oda}.  Parameter estimation results were visualized and
shared with the \pesummary software library \citep{Hoy:2020vys}.  Plots were
prepared with Matplotlib \citep{2007CSE.....9...90H}.
Figure \ref{fig:omegascans} was generated using GWpy~\citep{gwpy-software}.
The sky map plot also used Astropy (\href{http://www.astropy.org}{http://www.astropy.org}) a
community-developed core Python package for
Astronomy \citep{astropy:2013, astropy:2018} and \texttt{ligo.skymap}
(\href{https://lscsoft.docs.ligo.org/ligo.skymap}{https://lscsoft.docs.ligo.org/ligo.skymap}).}

We would like to thank all of the essential workers who put their
health at risk during the COVID-19 pandemic, without whom we would not
have been able to complete this work.

\bibliographystyle{aasjournal}
\bibliography{references}

\allauthors
\end{document}

%% file: definitions.tex
\newcommand{\marginal}{GW190426\_152155\xspace}

\newcommand{\eventone}{GW200105\xspace}
\newcommand{\gidone}{S200105ae\xspace}

\newcommand{\utcone}{January 05, 2020 at 16:24:26 UTC\xspace}
\newcommand{\eventoneFullName}{GW200105\_162426\xspace}

\newcommand{\eventtwoFullName}{GW200115\_042309\xspace}
\newcommand{\eventtwo}{GW200115\xspace}
\newcommand{\gidtwo}{S200115j\xspace}

\newcommand{\utctwo}{January 15, 2020 at 04:23:10 UTC\xspace}

\newcommand{\llo}{{LIGO Livingston}\xspace}
\newcommand{\LHO}{{LIGO Hanford}\xspace}
\newcommand{\LLO}{{LIGO Livingston}\xspace}
\newcommand{\virgo}{{Virgo}\xspace}
\newcommand{\cbc}{{CBC}\xspace}

\newcommand{\gstlal}{{\sc GstLAL}\xspace}
\newcommand{\pycbclive}{{\sc PyCBC Live}\xspace}
\newcommand{\pycbc}{{\sc PyCBC}\xspace}
\newcommand{\spiir}{{\sc SPIIR}\xspace}
\newcommand{\mbtaol}{{\sc MBTAOnline}\xspace}
\newcommand{\mbta}{{\sc MBTA}\xspace}
\newcommand{\gdb}{{\sc GraceDB}\xspace}

\newcommand{\lalinference}{{\sc LALInference}\xspace}

\newcommand{\bilby}{{\sc bilby}\xspace}
\newcommand{\pbilby}{{\sc pbilby}\xspace}

\newcommand{\pesummary}{{\sc PESummary}\xspace}

\newcommand{\RangeLLOone}{$\sim$137 Mpc\xspace}
\newcommand{\RangeVirgoone}{$\sim$45 Mpc\xspace}

\newcommand{\RangeLHOtwo}{$\sim$115 Mpc\xspace}
\newcommand{\RangeLLOtwo}{$\sim$133 Mpc\xspace}
\newcommand{\RangeVirgotwo}{$\sim$50 Mpc\xspace}

\newcommand{\MaxUncMagLLOone}{$8.6\%$\xspace}
\newcommand{\MaxUncMagVirgoone}{$5.0\%$\xspace}

\newcommand{\MaxUncPhaLLOone}{$5.9^\circ$\xspace}
\newcommand{\MaxUncPhaVirgoone}{$4.1^\circ$\xspace}

\newcommand{\MaxUncMagLLOtwo}{$8.0\%$\xspace}

\newcommand{\MaxUncPhaLLOtwo}{$5.5^\circ$\xspace}

\newcommand{\GstLALOnlineNetSNRone}{$13.9$\xspace}
\newcommand{\GstLALOnlineFARone}{$\approx24$ per year\xspace}

\newcommand{\OnlineSkymapAreaone}{7700 deg$^{2}$\xspace}

\newcommand{\GstLALOnlineNetSNRtwo}{$11.4$\xspace}

\newcommand{\OnlineSkymapAreatwo}{900 deg$^{2}$\xspace}

\newcommand{\gstlalsinglethreshold}{$4.0$\xspace}

\newcommand{\GstLALOfflineNetSNRone}{$13.9$\xspace}
\newcommand{\GstLALOfflineFARone}{1/($2.8$~yr)\xspace}

\newcommand{\GstLALOfflineNetSNRtwo}{$11.6$\xspace}
\newcommand{\GstLALOfflineFARtwo}{$<\!1$/($1 \times 10^{5}$~yr)\xspace}

\newcommand{\PyCBCLiveNetSNRone}{$13.2$\xspace}

\newcommand{\PyCBCLiveNetSNRtwo}{$11.3$\xspace}

\newcommand{\PyCBCOfflineNetSNRone}{$13.1$\xspace}

\newcommand{\PyCBCOfflineNetSNRtwo}{$10.8$\xspace}
\newcommand{\PyCBCOfflineFARtwo}{$<\!1$/($5.6 \times 10^{4}$~yr)\xspace}

\newcommand{\MBTAOnlineNetSNRone}{$13.3$\xspace}
\newcommand{\MBTAOfflineNetSNRone}{$13.4$\xspace}

\newcommand{\MBTAOnlineNetSNRtwo}{$11.4$\xspace}

\newcommand{\MBTAOfflineNetSNRtwo}{$11.2$\xspace}
\newcommand{\MBTAOfflineFARtricointwo}{1/($182$~yr)\xspace}

\newcommand{\SpiirNetSNRone}{$13.2$\xspace}

\newcommand{\SpiirNetSNRtwo}{$11.0$\xspace}

\newcommand{\VolRateUnit}{Gpc$^{-3}$\,{yr}$^{-1}$\xspace}
\def\SingleRateOhFiveGSTLALMed{\ensuremath{16}}
\def\SingleRateOhFiveGSTLALPlus{\ensuremath{38}}
\def\SingleRateOhFiveGSTLALMinus{\ensuremath{14}}
\def\SingleRateOneFiveGSTLALMed{\ensuremath{36}}
\def\SingleRateOneFiveGSTLALPlus{\ensuremath{82}}
\def\SingleRateOneFiveGSTLALMinus{\ensuremath{30}}

\def\KKLRateGSTLALMed{\ensuremath{45}}
\def\KKLRateGSTLALPlus{\ensuremath{75}}
\def\KKLRateGSTLALMinus{\ensuremath{33}}

\def\SingleRateOneFivePYCBCMed{\ensuremath{40}}
\def\SingleRateOneFivePYCBCPlus{\ensuremath{92}}
\def\SingleRateOneFivePYCBCMinus{\ensuremath{34}}

\def\FGMCRateMed{\ensuremath{130}}
\def\FGMCRatePlus{\ensuremath{112}}
\def\FGMCRateMinus{\ensuremath{69}}

\def\SingleRateOhFiveGSTLAL{\ensuremath{\SingleRateOhFiveGSTLALMed^{+\SingleRateOhFiveGSTLALPlus}_{-\SingleRateOhFiveGSTLALMinus}}}

\def\SingleRateOneFiveGSTLAL{\ensuremath{\SingleRateOneFiveGSTLALMed^{+\SingleRateOneFiveGSTLALPlus}_{-\SingleRateOneFiveGSTLALMinus}}}
\def\SingleRateOneFivePYCBC{\ensuremath{\SingleRateOneFivePYCBCMed^{+\SingleRateOneFivePYCBCPlus}_{-\SingleRateOneFivePYCBCMinus}}}

\def\KKLRateGSTLAL{\ensuremath{\KKLRateGSTLALMed^{+\KKLRateGSTLALPlus}_{-\KKLRateGSTLALMinus}}}

\def\FGMCRate{\ensuremath{\FGMCRateMed^{+\FGMCRatePlus}_{-\FGMCRateMinus}}}

\newcommand{\massonesource}{m_1\xspace}
\newcommand{\masstwosource}{m_2\xspace}

\newcommand{\massratio}{q\xspace}

\newcommand{\spintwomag}{\chi_2\xspace}

\newcommand{\spineffective}{\chi_\mathrm{eff}\xspace}
\newcommand{\spinprecession}{\chi_\mathrm{p}\xspace}

\newcommand{\inclination}{\theta_{JN}\xspace}
\newcommand{\luminositydistance}{D_\mathrm{L}\xspace}

\newcommand{\tidaltwo}{\Lambda_2\xspace}
\newcommand{\mergerratedensity}{\mathcal{R}\xspace}

\newcommand{\msun}{M_{\odot}\xspace}
\newcommand{\mpc}{\text{Mpc}\xspace}

\newcommand{\km}{\text{km}\xspace}
\newcommand{\s}{\text{s}\xspace}
\newcommand{\logB}{\log_{10}\mathcal{B}}



%% file: PE_macros.tex
\newcommand{\luminositydistanceminus}[1]{\IfEqCase{#1}{{GW200105_XPHM_lowspin}{110}{GW200105_v4PHM_lowspin}{120}{GW200105_combined_lowspin}{110}{GW200105_XPHM_highspin}{110}{GW200105_v4PHM_highspin}{120}{GW200105_combined_highspin}{110}{GW200105_NSBH_lowspin}{120}{GW200115_XPHM_lowspin}{90}{GW200115_v4PHM_lowspin}{120}{GW200115_combined_lowspin}{110}{GW200115_XPHM_highspin}{80}{GW200115_v4PHM_highspin}{120}{GW200115_combined_highspin}{100}{GW200115_NSBH_lowspin}{120}}}
\newcommand{\luminositydistancemed}[1]{\IfEqCase{#1}{{GW200105_XPHM_lowspin}{280}{GW200105_v4PHM_lowspin}{270}{GW200105_combined_lowspin}{280}{GW200105_XPHM_highspin}{280}{GW200105_v4PHM_highspin}{270}{GW200105_combined_highspin}{280}{GW200105_NSBH_lowspin}{280}{GW200115_XPHM_lowspin}{300}{GW200115_v4PHM_lowspin}{310}{GW200115_combined_lowspin}{310}{GW200115_XPHM_highspin}{300}{GW200115_v4PHM_highspin}{320}{GW200115_combined_highspin}{300}{GW200115_NSBH_lowspin}{290}}}
\newcommand{\luminositydistanceplus}[1]{\IfEqCase{#1}{{GW200105_XPHM_lowspin}{110}{GW200105_v4PHM_lowspin}{120}{GW200105_combined_lowspin}{110}{GW200105_XPHM_highspin}{110}{GW200105_v4PHM_highspin}{120}{GW200105_combined_highspin}{110}{GW200105_NSBH_lowspin}{110}{GW200115_XPHM_lowspin}{130}{GW200115_v4PHM_lowspin}{160}{GW200115_combined_lowspin}{150}{GW200115_XPHM_highspin}{120}{GW200115_v4PHM_highspin}{160}{GW200115_combined_highspin}{150}{GW200115_NSBH_lowspin}{160}}}
\newcommand{\luminositydistanceonepercent}[1]{\IfEqCase{#1}{{GW200105_XPHM_lowspin}{140}{GW200105_v4PHM_lowspin}{120}{GW200105_combined_lowspin}{130}{GW200105_XPHM_highspin}{140}{GW200105_v4PHM_highspin}{120}{GW200105_combined_highspin}{130}{GW200105_NSBH_lowspin}{120}{GW200115_XPHM_lowspin}{170}{GW200115_v4PHM_lowspin}{150}{GW200115_combined_lowspin}{160}{GW200115_XPHM_highspin}{180}{GW200115_v4PHM_highspin}{150}{GW200115_combined_highspin}{160}{GW200115_NSBH_lowspin}{140}}}
\newcommand{\luminositydistanceninetyninepercent}[1]{\IfEqCase{#1}{{GW200105_XPHM_lowspin}{430}{GW200105_v4PHM_lowspin}{430}{GW200105_combined_lowspin}{430}{GW200105_XPHM_highspin}{430}{GW200105_v4PHM_highspin}{430}{GW200105_combined_highspin}{430}{GW200105_NSBH_lowspin}{430}{GW200115_XPHM_lowspin}{500}{GW200115_v4PHM_lowspin}{540}{GW200115_combined_lowspin}{530}{GW200115_XPHM_highspin}{490}{GW200115_v4PHM_highspin}{550}{GW200115_combined_highspin}{530}{GW200115_NSBH_lowspin}{520}}}
\newcommand{\luminositydistancefivepercent}[1]{\IfEqCase{#1}{{GW200105_XPHM_lowspin}{170}{GW200105_v4PHM_lowspin}{160}{GW200105_combined_lowspin}{160}{GW200105_XPHM_highspin}{170}{GW200105_v4PHM_highspin}{160}{GW200105_combined_highspin}{160}{GW200105_NSBH_lowspin}{160}{GW200115_XPHM_lowspin}{210}{GW200115_v4PHM_lowspin}{190}{GW200115_combined_lowspin}{200}{GW200115_XPHM_highspin}{210}{GW200115_v4PHM_highspin}{190}{GW200115_combined_highspin}{200}{GW200115_NSBH_lowspin}{170}}}
\newcommand{\luminositydistanceninetyfivepercent}[1]{\IfEqCase{#1}{{GW200105_XPHM_lowspin}{390}{GW200105_v4PHM_lowspin}{390}{GW200105_combined_lowspin}{390}{GW200105_XPHM_highspin}{390}{GW200105_v4PHM_highspin}{390}{GW200105_combined_highspin}{390}{GW200105_NSBH_lowspin}{390}{GW200115_XPHM_lowspin}{430}{GW200115_v4PHM_lowspin}{480}{GW200115_combined_lowspin}{460}{GW200115_XPHM_highspin}{420}{GW200115_v4PHM_highspin}{480}{GW200115_combined_highspin}{460}{GW200115_NSBH_lowspin}{450}}}
\newcommand{\luminositydistanceninetypercent}[1]{\IfEqCase{#1}{{GW200105_XPHM_lowspin}{360}{GW200105_v4PHM_lowspin}{370}{GW200105_combined_lowspin}{370}{GW200105_XPHM_highspin}{370}{GW200105_v4PHM_highspin}{370}{GW200105_combined_highspin}{370}{GW200105_NSBH_lowspin}{370}{GW200115_XPHM_lowspin}{390}{GW200115_v4PHM_lowspin}{440}{GW200115_combined_lowspin}{420}{GW200115_XPHM_highspin}{390}{GW200115_v4PHM_highspin}{440}{GW200115_combined_highspin}{420}{GW200115_NSBH_lowspin}{410}}}
\newcommand{\massratiominus}[1]{\IfEqCase{#1}{{GW200105_XPHM_lowspin}{0.04}{GW200105_v4PHM_lowspin}{0.03}{GW200105_combined_lowspin}{0.04}{GW200105_XPHM_highspin}{0.05}{GW200105_v4PHM_highspin}{0.03}{GW200105_combined_highspin}{0.04}{GW200105_NSBH_lowspin}{0.09}{GW200115_XPHM_lowspin}{0.11}{GW200115_v4PHM_lowspin}{0.05}{GW200115_combined_lowspin}{0.08}{GW200115_XPHM_highspin}{0.10}{GW200115_v4PHM_highspin}{0.10}{GW200115_combined_highspin}{0.10}{GW200115_NSBH_lowspin}{0.07}}}
\newcommand{\massratiomed}[1]{\IfEqCase{#1}{{GW200105_XPHM_lowspin}{0.22}{GW200105_v4PHM_lowspin}{0.21}{GW200105_combined_lowspin}{0.21}{GW200105_XPHM_highspin}{0.21}{GW200105_v4PHM_highspin}{0.22}{GW200105_combined_highspin}{0.22}{GW200105_NSBH_lowspin}{0.22}{GW200115_XPHM_lowspin}{0.27}{GW200115_v4PHM_lowspin}{0.22}{GW200115_combined_lowspin}{0.24}{GW200115_XPHM_highspin}{0.27}{GW200115_v4PHM_highspin}{0.25}{GW200115_combined_highspin}{0.26}{GW200115_NSBH_lowspin}{0.20}}}
\newcommand{\massratioplus}[1]{\IfEqCase{#1}{{GW200105_XPHM_lowspin}{0.08}{GW200105_v4PHM_lowspin}{0.05}{GW200105_combined_lowspin}{0.06}{GW200105_XPHM_highspin}{0.09}{GW200105_v4PHM_highspin}{0.06}{GW200105_combined_highspin}{0.08}{GW200105_NSBH_lowspin}{0.15}{GW200115_XPHM_lowspin}{0.25}{GW200115_v4PHM_lowspin}{0.37}{GW200115_combined_lowspin}{0.31}{GW200115_XPHM_highspin}{0.29}{GW200115_v4PHM_highspin}{0.40}{GW200115_combined_highspin}{0.35}{GW200115_NSBH_lowspin}{0.15}}}
\newcommand{\massratioonepercent}[1]{\IfEqCase{#1}{{GW200105_XPHM_lowspin}{0.15}{GW200105_v4PHM_lowspin}{0.15}{GW200105_combined_lowspin}{0.15}{GW200105_XPHM_highspin}{0.14}{GW200105_v4PHM_highspin}{0.15}{GW200105_combined_highspin}{0.14}{GW200105_NSBH_lowspin}{0.10}{GW200115_XPHM_lowspin}{0.13}{GW200115_v4PHM_lowspin}{0.14}{GW200115_combined_lowspin}{0.13}{GW200115_XPHM_highspin}{0.14}{GW200115_v4PHM_highspin}{0.13}{GW200115_combined_highspin}{0.14}{GW200115_NSBH_lowspin}{0.11}}}
\newcommand{\massrationinetyninepercent}[1]{\IfEqCase{#1}{{GW200105_XPHM_lowspin}{0.41}{GW200105_v4PHM_lowspin}{0.30}{GW200105_combined_lowspin}{0.36}{GW200105_XPHM_highspin}{0.45}{GW200105_v4PHM_highspin}{0.47}{GW200105_combined_highspin}{0.46}{GW200105_NSBH_lowspin}{0.45}{GW200115_XPHM_lowspin}{0.64}{GW200115_v4PHM_lowspin}{0.67}{GW200115_combined_lowspin}{0.65}{GW200115_XPHM_highspin}{0.72}{GW200115_v4PHM_highspin}{0.84}{GW200115_combined_highspin}{0.80}{GW200115_NSBH_lowspin}{0.44}}}
\newcommand{\massratiofivepercent}[1]{\IfEqCase{#1}{{GW200105_XPHM_lowspin}{0.17}{GW200105_v4PHM_lowspin}{0.18}{GW200105_combined_lowspin}{0.18}{GW200105_XPHM_highspin}{0.16}{GW200105_v4PHM_highspin}{0.19}{GW200105_combined_highspin}{0.17}{GW200105_NSBH_lowspin}{0.13}{GW200115_XPHM_lowspin}{0.17}{GW200115_v4PHM_lowspin}{0.17}{GW200115_combined_lowspin}{0.17}{GW200115_XPHM_highspin}{0.17}{GW200115_v4PHM_highspin}{0.16}{GW200115_combined_highspin}{0.16}{GW200115_NSBH_lowspin}{0.13}}}
\newcommand{\massrationinetyfivepercent}[1]{\IfEqCase{#1}{{GW200105_XPHM_lowspin}{0.30}{GW200105_v4PHM_lowspin}{0.26}{GW200105_combined_lowspin}{0.28}{GW200105_XPHM_highspin}{0.31}{GW200105_v4PHM_highspin}{0.27}{GW200105_combined_highspin}{0.29}{GW200105_NSBH_lowspin}{0.37}{GW200115_XPHM_lowspin}{0.52}{GW200115_v4PHM_lowspin}{0.59}{GW200115_combined_lowspin}{0.55}{GW200115_XPHM_highspin}{0.56}{GW200115_v4PHM_highspin}{0.65}{GW200115_combined_highspin}{0.62}{GW200115_NSBH_lowspin}{0.35}}}
\newcommand{\massrationinetypercent}[1]{\IfEqCase{#1}{{GW200105_XPHM_lowspin}{0.27}{GW200105_v4PHM_lowspin}{0.25}{GW200105_combined_lowspin}{0.26}{GW200105_XPHM_highspin}{0.27}{GW200105_v4PHM_highspin}{0.24}{GW200105_combined_highspin}{0.26}{GW200105_NSBH_lowspin}{0.33}{GW200115_XPHM_lowspin}{0.45}{GW200115_v4PHM_lowspin}{0.49}{GW200115_combined_lowspin}{0.46}{GW200115_XPHM_highspin}{0.46}{GW200115_v4PHM_highspin}{0.57}{GW200115_combined_highspin}{0.52}{GW200115_NSBH_lowspin}{0.31}}}
\newcommand{\chirpmassdetminus}[1]{\IfEqCase{#1}{{GW200105_XPHM_lowspin}{0.007}{GW200105_v4PHM_lowspin}{0.005}{GW200105_combined_lowspin}{0.006}{GW200105_XPHM_highspin}{0.009}{GW200105_v4PHM_highspin}{0.006}{GW200105_combined_highspin}{0.008}{GW200105_NSBH_lowspin}{0.009}{GW200115_XPHM_lowspin}{0.006}{GW200115_v4PHM_lowspin}{0.008}{GW200115_combined_lowspin}{0.007}{GW200115_XPHM_highspin}{0.006}{GW200115_v4PHM_highspin}{0.008}{GW200115_combined_highspin}{0.007}{GW200115_NSBH_lowspin}{0.006}}}
\newcommand{\chirpmassdetmed}[1]{\IfEqCase{#1}{{GW200105_XPHM_lowspin}{3.618}{GW200105_v4PHM_lowspin}{3.619}{GW200105_combined_lowspin}{3.619}{GW200105_XPHM_highspin}{3.618}{GW200105_v4PHM_highspin}{3.619}{GW200105_combined_highspin}{3.619}{GW200105_NSBH_lowspin}{3.620}{GW200115_XPHM_lowspin}{2.579}{GW200115_v4PHM_lowspin}{2.581}{GW200115_combined_lowspin}{2.580}{GW200115_XPHM_highspin}{2.579}{GW200115_v4PHM_highspin}{2.579}{GW200115_combined_highspin}{2.579}{GW200115_NSBH_lowspin}{2.582}}}
\newcommand{\chirpmassdetplus}[1]{\IfEqCase{#1}{{GW200105_XPHM_lowspin}{0.006}{GW200105_v4PHM_lowspin}{0.005}{GW200105_combined_lowspin}{0.006}{GW200105_XPHM_highspin}{0.008}{GW200105_v4PHM_highspin}{0.005}{GW200105_combined_highspin}{0.007}{GW200105_NSBH_lowspin}{0.014}{GW200115_XPHM_lowspin}{0.007}{GW200115_v4PHM_lowspin}{0.005}{GW200115_combined_lowspin}{0.006}{GW200115_XPHM_highspin}{0.006}{GW200115_v4PHM_highspin}{0.007}{GW200115_combined_highspin}{0.007}{GW200115_NSBH_lowspin}{0.007}}}
\newcommand{\chirpmassdetonepercent}[1]{\IfEqCase{#1}{{GW200105_XPHM_lowspin}{3.606}{GW200105_v4PHM_lowspin}{3.610}{GW200105_combined_lowspin}{3.607}{GW200105_XPHM_highspin}{3.604}{GW200105_v4PHM_highspin}{3.602}{GW200105_combined_highspin}{3.603}{GW200105_NSBH_lowspin}{3.607}{GW200115_XPHM_lowspin}{2.571}{GW200115_v4PHM_lowspin}{2.571}{GW200115_combined_lowspin}{2.571}{GW200115_XPHM_highspin}{2.571}{GW200115_v4PHM_highspin}{2.570}{GW200115_combined_highspin}{2.570}{GW200115_NSBH_lowspin}{2.574}}}
\newcommand{\chirpmassdetninetyninepercent}[1]{\IfEqCase{#1}{{GW200105_XPHM_lowspin}{3.629}{GW200105_v4PHM_lowspin}{3.629}{GW200105_combined_lowspin}{3.629}{GW200105_XPHM_highspin}{3.631}{GW200105_v4PHM_highspin}{3.629}{GW200105_combined_highspin}{3.630}{GW200105_NSBH_lowspin}{3.641}{GW200115_XPHM_lowspin}{2.589}{GW200115_v4PHM_lowspin}{2.588}{GW200115_combined_lowspin}{2.589}{GW200115_XPHM_highspin}{2.588}{GW200115_v4PHM_highspin}{2.589}{GW200115_combined_highspin}{2.589}{GW200115_NSBH_lowspin}{2.593}}}
\newcommand{\chirpmassdetfivepercent}[1]{\IfEqCase{#1}{{GW200105_XPHM_lowspin}{3.611}{GW200105_v4PHM_lowspin}{3.614}{GW200105_combined_lowspin}{3.612}{GW200105_XPHM_highspin}{3.610}{GW200105_v4PHM_highspin}{3.613}{GW200105_combined_highspin}{3.611}{GW200105_NSBH_lowspin}{3.611}{GW200115_XPHM_lowspin}{2.573}{GW200115_v4PHM_lowspin}{2.573}{GW200115_combined_lowspin}{2.573}{GW200115_XPHM_highspin}{2.573}{GW200115_v4PHM_highspin}{2.572}{GW200115_combined_highspin}{2.572}{GW200115_NSBH_lowspin}{2.576}}}
\newcommand{\chirpmassdetninetyfivepercent}[1]{\IfEqCase{#1}{{GW200105_XPHM_lowspin}{3.624}{GW200105_v4PHM_lowspin}{3.624}{GW200105_combined_lowspin}{3.624}{GW200105_XPHM_highspin}{3.626}{GW200105_v4PHM_highspin}{3.624}{GW200105_combined_highspin}{3.625}{GW200105_NSBH_lowspin}{3.634}{GW200115_XPHM_lowspin}{2.585}{GW200115_v4PHM_lowspin}{2.586}{GW200115_combined_lowspin}{2.586}{GW200115_XPHM_highspin}{2.585}{GW200115_v4PHM_highspin}{2.586}{GW200115_combined_highspin}{2.585}{GW200115_NSBH_lowspin}{2.589}}}
\newcommand{\chirpmassdetninetypercent}[1]{\IfEqCase{#1}{{GW200105_XPHM_lowspin}{3.622}{GW200105_v4PHM_lowspin}{3.623}{GW200105_combined_lowspin}{3.623}{GW200105_XPHM_highspin}{3.624}{GW200105_v4PHM_highspin}{3.622}{GW200105_combined_highspin}{3.623}{GW200105_NSBH_lowspin}{3.629}{GW200115_XPHM_lowspin}{2.584}{GW200115_v4PHM_lowspin}{2.585}{GW200115_combined_lowspin}{2.584}{GW200115_XPHM_highspin}{2.584}{GW200115_v4PHM_highspin}{2.585}{GW200115_combined_highspin}{2.584}{GW200115_NSBH_lowspin}{2.587}}}
\newcommand{\spinoneminus}[1]{\IfEqCase{#1}{{GW200105_XPHM_lowspin}{0.10}{GW200105_v4PHM_lowspin}{0.07}{GW200105_combined_lowspin}{0.08}{GW200105_XPHM_highspin}{0.10}{GW200105_v4PHM_highspin}{0.06}{GW200105_combined_highspin}{0.08}{GW200105_NSBH_lowspin}{0.09}{GW200115_XPHM_lowspin}{0.31}{GW200115_v4PHM_lowspin}{0.21}{GW200115_combined_lowspin}{0.29}{GW200115_XPHM_highspin}{0.30}{GW200115_v4PHM_highspin}{0.26}{GW200115_combined_highspin}{0.29}{GW200115_NSBH_lowspin}{0.09}}}
\newcommand{\spinonemed}[1]{\IfEqCase{#1}{{GW200105_XPHM_lowspin}{0.11}{GW200105_v4PHM_lowspin}{0.07}{GW200105_combined_lowspin}{0.09}{GW200105_XPHM_highspin}{0.11}{GW200105_v4PHM_highspin}{0.06}{GW200105_combined_highspin}{0.08}{GW200105_NSBH_lowspin}{0.09}{GW200115_XPHM_lowspin}{0.38}{GW200115_v4PHM_lowspin}{0.22}{GW200115_combined_lowspin}{0.31}{GW200115_XPHM_highspin}{0.36}{GW200115_v4PHM_highspin}{0.28}{GW200115_combined_highspin}{0.33}{GW200115_NSBH_lowspin}{0.10}}}
\newcommand{\spinoneplus}[1]{\IfEqCase{#1}{{GW200105_XPHM_lowspin}{0.20}{GW200105_v4PHM_lowspin}{0.15}{GW200105_combined_lowspin}{0.18}{GW200105_XPHM_highspin}{0.22}{GW200105_v4PHM_highspin}{0.19}{GW200105_combined_highspin}{0.22}{GW200105_NSBH_lowspin}{0.24}{GW200115_XPHM_lowspin}{0.43}{GW200115_v4PHM_lowspin}{0.62}{GW200115_combined_lowspin}{0.52}{GW200115_XPHM_highspin}{0.45}{GW200115_v4PHM_highspin}{0.51}{GW200115_combined_highspin}{0.48}{GW200115_NSBH_lowspin}{0.29}}}
\newcommand{\spinoneonepercent}[1]{\IfEqCase{#1}{{GW200105_XPHM_lowspin}{0.00}{GW200105_v4PHM_lowspin}{0.00}{GW200105_combined_lowspin}{0.00}{GW200105_XPHM_highspin}{0.00}{GW200105_v4PHM_highspin}{0.00}{GW200105_combined_highspin}{0.00}{GW200105_NSBH_lowspin}{0.00}{GW200115_XPHM_lowspin}{0.02}{GW200115_v4PHM_lowspin}{0.00}{GW200115_combined_lowspin}{0.00}{GW200115_XPHM_highspin}{0.01}{GW200115_v4PHM_highspin}{0.00}{GW200115_combined_highspin}{0.01}{GW200115_NSBH_lowspin}{0.00}}}
\newcommand{\spinoneninetyninepercent}[1]{\IfEqCase{#1}{{GW200105_XPHM_lowspin}{0.51}{GW200105_v4PHM_lowspin}{0.30}{GW200105_combined_lowspin}{0.43}{GW200105_XPHM_highspin}{0.56}{GW200105_v4PHM_highspin}{0.48}{GW200105_combined_highspin}{0.51}{GW200105_NSBH_lowspin}{0.43}{GW200115_XPHM_lowspin}{0.94}{GW200115_v4PHM_lowspin}{0.95}{GW200115_combined_lowspin}{0.95}{GW200115_XPHM_highspin}{0.94}{GW200115_v4PHM_highspin}{0.89}{GW200115_combined_highspin}{0.92}{GW200115_NSBH_lowspin}{0.55}}}
\newcommand{\spinonefivepercent}[1]{\IfEqCase{#1}{{GW200105_XPHM_lowspin}{0.01}{GW200105_v4PHM_lowspin}{0.01}{GW200105_combined_lowspin}{0.01}{GW200105_XPHM_highspin}{0.01}{GW200105_v4PHM_highspin}{0.01}{GW200105_combined_highspin}{0.01}{GW200105_NSBH_lowspin}{0.00}{GW200115_XPHM_lowspin}{0.07}{GW200115_v4PHM_lowspin}{0.01}{GW200115_combined_lowspin}{0.02}{GW200115_XPHM_highspin}{0.06}{GW200115_v4PHM_highspin}{0.02}{GW200115_combined_highspin}{0.03}{GW200115_NSBH_lowspin}{0.00}}}
\newcommand{\spinoneninetyfivepercent}[1]{\IfEqCase{#1}{{GW200105_XPHM_lowspin}{0.31}{GW200105_v4PHM_lowspin}{0.22}{GW200105_combined_lowspin}{0.27}{GW200105_XPHM_highspin}{0.34}{GW200105_v4PHM_highspin}{0.25}{GW200105_combined_highspin}{0.30}{GW200105_NSBH_lowspin}{0.34}{GW200115_XPHM_lowspin}{0.81}{GW200115_v4PHM_lowspin}{0.85}{GW200115_combined_lowspin}{0.83}{GW200115_XPHM_highspin}{0.81}{GW200115_v4PHM_highspin}{0.80}{GW200115_combined_highspin}{0.80}{GW200115_NSBH_lowspin}{0.39}}}
\newcommand{\spinoneninetypercent}[1]{\IfEqCase{#1}{{GW200105_XPHM_lowspin}{0.25}{GW200105_v4PHM_lowspin}{0.19}{GW200105_combined_lowspin}{0.22}{GW200105_XPHM_highspin}{0.26}{GW200105_v4PHM_highspin}{0.18}{GW200105_combined_highspin}{0.23}{GW200105_NSBH_lowspin}{0.28}{GW200115_XPHM_lowspin}{0.71}{GW200115_v4PHM_lowspin}{0.71}{GW200115_combined_lowspin}{0.71}{GW200115_XPHM_highspin}{0.71}{GW200115_v4PHM_highspin}{0.73}{GW200115_combined_highspin}{0.72}{GW200115_NSBH_lowspin}{0.32}}}
\newcommand{\spintwominus}[1]{\IfEqCase{#1}{{GW200105_XPHM_lowspin}{0.02}{GW200105_v4PHM_lowspin}{0.02}{GW200105_combined_lowspin}{0.02}{GW200105_XPHM_highspin}{0.41}{GW200105_v4PHM_highspin}{0.20}{GW200105_combined_highspin}{0.29}{GW200105_NSBH_lowspin}{0.01}{GW200115_XPHM_lowspin}{0.02}{GW200115_v4PHM_lowspin}{0.02}{GW200115_combined_lowspin}{0.02}{GW200115_XPHM_highspin}{0.43}{GW200115_v4PHM_highspin}{0.33}{GW200115_combined_highspin}{0.38}{GW200115_NSBH_lowspin}{0.01}}}
\newcommand{\spintwomed}[1]{\IfEqCase{#1}{{GW200105_XPHM_lowspin}{0.03}{GW200105_v4PHM_lowspin}{0.02}{GW200105_combined_lowspin}{0.02}{GW200105_XPHM_highspin}{0.47}{GW200105_v4PHM_highspin}{0.23}{GW200105_combined_highspin}{0.32}{GW200105_NSBH_lowspin}{0.01}{GW200115_XPHM_lowspin}{0.03}{GW200115_v4PHM_lowspin}{0.02}{GW200115_combined_lowspin}{0.03}{GW200115_XPHM_highspin}{0.50}{GW200115_v4PHM_highspin}{0.37}{GW200115_combined_highspin}{0.43}{GW200115_NSBH_lowspin}{0.01}}}
\newcommand{\spintwoplus}[1]{\IfEqCase{#1}{{GW200105_XPHM_lowspin}{0.02}{GW200105_v4PHM_lowspin}{0.02}{GW200105_combined_lowspin}{0.02}{GW200105_XPHM_highspin}{0.46}{GW200105_v4PHM_highspin}{0.41}{GW200105_combined_highspin}{0.55}{GW200105_NSBH_lowspin}{0.03}{GW200115_XPHM_lowspin}{0.02}{GW200115_v4PHM_lowspin}{0.02}{GW200115_combined_lowspin}{0.02}{GW200115_XPHM_highspin}{0.42}{GW200115_v4PHM_highspin}{0.48}{GW200115_combined_highspin}{0.47}{GW200115_NSBH_lowspin}{0.03}}}
\newcommand{\spintwoonepercent}[1]{\IfEqCase{#1}{{GW200105_XPHM_lowspin}{0.00}{GW200105_v4PHM_lowspin}{0.00}{GW200105_combined_lowspin}{0.00}{GW200105_XPHM_highspin}{0.01}{GW200105_v4PHM_highspin}{0.01}{GW200105_combined_highspin}{0.01}{GW200105_NSBH_lowspin}{0.00}{GW200115_XPHM_lowspin}{0.00}{GW200115_v4PHM_lowspin}{0.00}{GW200115_combined_lowspin}{0.00}{GW200115_XPHM_highspin}{0.02}{GW200115_v4PHM_highspin}{0.01}{GW200115_combined_highspin}{0.01}{GW200115_NSBH_lowspin}{0.00}}}
\newcommand{\spintwoninetyninepercent}[1]{\IfEqCase{#1}{{GW200105_XPHM_lowspin}{0.05}{GW200105_v4PHM_lowspin}{0.05}{GW200105_combined_lowspin}{0.05}{GW200105_XPHM_highspin}{0.98}{GW200105_v4PHM_highspin}{0.78}{GW200105_combined_highspin}{0.96}{GW200105_NSBH_lowspin}{0.04}{GW200115_XPHM_lowspin}{0.05}{GW200115_v4PHM_lowspin}{0.05}{GW200115_combined_lowspin}{0.05}{GW200115_XPHM_highspin}{0.97}{GW200115_v4PHM_highspin}{0.96}{GW200115_combined_highspin}{0.97}{GW200115_NSBH_lowspin}{0.04}}}
\newcommand{\spintwofivepercent}[1]{\IfEqCase{#1}{{GW200105_XPHM_lowspin}{0.00}{GW200105_v4PHM_lowspin}{0.00}{GW200105_combined_lowspin}{0.00}{GW200105_XPHM_highspin}{0.05}{GW200105_v4PHM_highspin}{0.02}{GW200105_combined_highspin}{0.03}{GW200105_NSBH_lowspin}{0.00}{GW200115_XPHM_lowspin}{0.00}{GW200115_v4PHM_lowspin}{0.00}{GW200115_combined_lowspin}{0.00}{GW200115_XPHM_highspin}{0.07}{GW200115_v4PHM_highspin}{0.04}{GW200115_combined_highspin}{0.05}{GW200115_NSBH_lowspin}{0.00}}}
\newcommand{\spintwoninetyfivepercent}[1]{\IfEqCase{#1}{{GW200105_XPHM_lowspin}{0.05}{GW200105_v4PHM_lowspin}{0.05}{GW200105_combined_lowspin}{0.05}{GW200105_XPHM_highspin}{0.92}{GW200105_v4PHM_highspin}{0.63}{GW200105_combined_highspin}{0.87}{GW200105_NSBH_lowspin}{0.04}{GW200115_XPHM_lowspin}{0.05}{GW200115_v4PHM_lowspin}{0.05}{GW200115_combined_lowspin}{0.05}{GW200115_XPHM_highspin}{0.92}{GW200115_v4PHM_highspin}{0.85}{GW200115_combined_highspin}{0.90}{GW200115_NSBH_lowspin}{0.04}}}
\newcommand{\spintwoninetypercent}[1]{\IfEqCase{#1}{{GW200105_XPHM_lowspin}{0.04}{GW200105_v4PHM_lowspin}{0.04}{GW200105_combined_lowspin}{0.04}{GW200105_XPHM_highspin}{0.87}{GW200105_v4PHM_highspin}{0.55}{GW200105_combined_highspin}{0.77}{GW200105_NSBH_lowspin}{0.03}{GW200115_XPHM_lowspin}{0.04}{GW200115_v4PHM_lowspin}{0.05}{GW200115_combined_lowspin}{0.04}{GW200115_XPHM_highspin}{0.87}{GW200115_v4PHM_highspin}{0.77}{GW200115_combined_highspin}{0.82}{GW200115_NSBH_lowspin}{0.03}}}
\newcommand{\tiltoneminus}[1]{\IfEqCase{#1}{{GW200105_XPHM_lowspin}{1.12}{GW200105_v4PHM_lowspin}{1.15}{GW200105_combined_lowspin}{1.14}{GW200105_XPHM_highspin}{1.15}{GW200105_v4PHM_highspin}{1.15}{GW200105_combined_highspin}{1.15}{GW200105_NSBH_lowspin}{3.14}{GW200115_XPHM_lowspin}{0.93}{GW200115_v4PHM_lowspin}{1.32}{GW200115_combined_lowspin}{1.16}{GW200115_XPHM_highspin}{0.83}{GW200115_v4PHM_highspin}{1.46}{GW200115_combined_highspin}{1.18}{GW200115_NSBH_lowspin}{3.14}}}
\newcommand{\tiltonemed}[1]{\IfEqCase{#1}{{GW200105_XPHM_lowspin}{1.77}{GW200105_v4PHM_lowspin}{1.65}{GW200105_combined_lowspin}{1.71}{GW200105_XPHM_highspin}{1.66}{GW200105_v4PHM_highspin}{1.66}{GW200105_combined_highspin}{1.66}{GW200105_NSBH_lowspin}{3.14}{GW200115_XPHM_lowspin}{2.31}{GW200115_v4PHM_lowspin}{2.27}{GW200115_combined_lowspin}{2.30}{GW200115_XPHM_highspin}{2.26}{GW200115_v4PHM_highspin}{2.36}{GW200115_combined_highspin}{2.30}{GW200115_NSBH_lowspin}{3.14}}}
\newcommand{\tiltoneplus}[1]{\IfEqCase{#1}{{GW200105_XPHM_lowspin}{0.94}{GW200105_v4PHM_lowspin}{1.06}{GW200105_combined_lowspin}{1.00}{GW200105_XPHM_highspin}{1.04}{GW200105_v4PHM_highspin}{1.13}{GW200105_combined_highspin}{1.08}{GW200105_NSBH_lowspin}{0.00}{GW200115_XPHM_lowspin}{0.57}{GW200115_v4PHM_lowspin}{0.68}{GW200115_combined_lowspin}{0.63}{GW200115_XPHM_highspin}{0.59}{GW200115_v4PHM_highspin}{0.56}{GW200115_combined_highspin}{0.59}{GW200115_NSBH_lowspin}{0.00}}}
\newcommand{\tiltoneonepercent}[1]{\IfEqCase{#1}{{GW200105_XPHM_lowspin}{0.32}{GW200105_v4PHM_lowspin}{0.21}{GW200105_combined_lowspin}{0.26}{GW200105_XPHM_highspin}{0.24}{GW200105_v4PHM_highspin}{0.21}{GW200105_combined_highspin}{0.23}{GW200105_NSBH_lowspin}{0.00}{GW200115_XPHM_lowspin}{0.84}{GW200115_v4PHM_lowspin}{0.45}{GW200115_combined_lowspin}{0.58}{GW200115_XPHM_highspin}{0.93}{GW200115_v4PHM_highspin}{0.42}{GW200115_combined_highspin}{0.57}{GW200115_NSBH_lowspin}{0.00}}}
\newcommand{\tiltoneninetyninepercent}[1]{\IfEqCase{#1}{{GW200105_XPHM_lowspin}{2.95}{GW200105_v4PHM_lowspin}{2.96}{GW200105_combined_lowspin}{2.95}{GW200105_XPHM_highspin}{2.95}{GW200105_v4PHM_highspin}{3.04}{GW200105_combined_highspin}{3.01}{GW200105_NSBH_lowspin}{3.14}{GW200115_XPHM_lowspin}{3.02}{GW200115_v4PHM_lowspin}{3.05}{GW200115_combined_lowspin}{3.04}{GW200115_XPHM_highspin}{3.00}{GW200115_v4PHM_highspin}{3.03}{GW200115_combined_highspin}{3.02}{GW200115_NSBH_lowspin}{3.14}}}
\newcommand{\tiltonefivepercent}[1]{\IfEqCase{#1}{{GW200105_XPHM_lowspin}{0.65}{GW200105_v4PHM_lowspin}{0.50}{GW200105_combined_lowspin}{0.57}{GW200105_XPHM_highspin}{0.51}{GW200105_v4PHM_highspin}{0.51}{GW200105_combined_highspin}{0.51}{GW200105_NSBH_lowspin}{0.00}{GW200115_XPHM_lowspin}{1.39}{GW200115_v4PHM_lowspin}{0.96}{GW200115_combined_lowspin}{1.13}{GW200115_XPHM_highspin}{1.43}{GW200115_v4PHM_highspin}{0.90}{GW200115_combined_highspin}{1.12}{GW200115_NSBH_lowspin}{0.00}}}
\newcommand{\tiltoneninetyfivepercent}[1]{\IfEqCase{#1}{{GW200105_XPHM_lowspin}{2.72}{GW200105_v4PHM_lowspin}{2.71}{GW200105_combined_lowspin}{2.71}{GW200105_XPHM_highspin}{2.71}{GW200105_v4PHM_highspin}{2.79}{GW200105_combined_highspin}{2.74}{GW200105_NSBH_lowspin}{3.14}{GW200115_XPHM_lowspin}{2.88}{GW200115_v4PHM_lowspin}{2.95}{GW200115_combined_lowspin}{2.92}{GW200115_XPHM_highspin}{2.84}{GW200115_v4PHM_highspin}{2.92}{GW200115_combined_highspin}{2.89}{GW200115_NSBH_lowspin}{3.14}}}
\newcommand{\tiltoneninetypercent}[1]{\IfEqCase{#1}{{GW200105_XPHM_lowspin}{2.54}{GW200105_v4PHM_lowspin}{2.51}{GW200105_combined_lowspin}{2.53}{GW200105_XPHM_highspin}{2.53}{GW200105_v4PHM_highspin}{2.57}{GW200105_combined_highspin}{2.55}{GW200105_NSBH_lowspin}{3.14}{GW200115_XPHM_lowspin}{2.77}{GW200115_v4PHM_lowspin}{2.88}{GW200115_combined_lowspin}{2.83}{GW200115_XPHM_highspin}{2.73}{GW200115_v4PHM_highspin}{2.83}{GW200115_combined_highspin}{2.79}{GW200115_NSBH_lowspin}{3.14}}}
\newcommand{\tilttwominus}[1]{\IfEqCase{#1}{{GW200105_XPHM_lowspin}{1.07}{GW200105_v4PHM_lowspin}{1.12}{GW200105_combined_lowspin}{1.09}{GW200105_XPHM_highspin}{1.07}{GW200105_v4PHM_highspin}{0.97}{GW200105_combined_highspin}{1.03}{GW200105_NSBH_lowspin}{0.00}{GW200115_XPHM_lowspin}{1.07}{GW200115_v4PHM_lowspin}{1.13}{GW200115_combined_lowspin}{1.10}{GW200115_XPHM_highspin}{1.11}{GW200115_v4PHM_highspin}{1.34}{GW200115_combined_highspin}{1.24}{GW200115_NSBH_lowspin}{3.14}}}
\newcommand{\tilttwomed}[1]{\IfEqCase{#1}{{GW200105_XPHM_lowspin}{1.54}{GW200105_v4PHM_lowspin}{1.61}{GW200105_combined_lowspin}{1.57}{GW200105_XPHM_highspin}{1.59}{GW200105_v4PHM_highspin}{1.61}{GW200105_combined_highspin}{1.60}{GW200105_NSBH_lowspin}{0.00}{GW200115_XPHM_lowspin}{1.57}{GW200115_v4PHM_lowspin}{1.58}{GW200115_combined_lowspin}{1.58}{GW200115_XPHM_highspin}{1.87}{GW200115_v4PHM_highspin}{1.87}{GW200115_combined_highspin}{1.87}{GW200115_NSBH_lowspin}{3.14}}}
\newcommand{\tilttwoplus}[1]{\IfEqCase{#1}{{GW200105_XPHM_lowspin}{1.11}{GW200105_v4PHM_lowspin}{1.11}{GW200105_combined_lowspin}{1.12}{GW200105_XPHM_highspin}{1.04}{GW200105_v4PHM_highspin}{0.94}{GW200105_combined_highspin}{0.99}{GW200105_NSBH_lowspin}{3.14}{GW200115_XPHM_lowspin}{1.08}{GW200115_v4PHM_lowspin}{1.11}{GW200115_combined_lowspin}{1.09}{GW200115_XPHM_highspin}{0.87}{GW200115_v4PHM_highspin}{0.99}{GW200115_combined_highspin}{0.94}{GW200115_NSBH_lowspin}{0.00}}}
\newcommand{\tilttwoonepercent}[1]{\IfEqCase{#1}{{GW200105_XPHM_lowspin}{0.21}{GW200105_v4PHM_lowspin}{0.21}{GW200105_combined_lowspin}{0.21}{GW200105_XPHM_highspin}{0.24}{GW200105_v4PHM_highspin}{0.27}{GW200105_combined_highspin}{0.25}{GW200105_NSBH_lowspin}{0.00}{GW200115_XPHM_lowspin}{0.24}{GW200115_v4PHM_lowspin}{0.20}{GW200115_combined_lowspin}{0.22}{GW200115_XPHM_highspin}{0.37}{GW200115_v4PHM_highspin}{0.20}{GW200115_combined_highspin}{0.28}{GW200115_NSBH_lowspin}{0.00}}}
\newcommand{\tilttwoninetyninepercent}[1]{\IfEqCase{#1}{{GW200105_XPHM_lowspin}{2.93}{GW200105_v4PHM_lowspin}{2.95}{GW200105_combined_lowspin}{2.94}{GW200105_XPHM_highspin}{2.90}{GW200105_v4PHM_highspin}{2.88}{GW200105_combined_highspin}{2.89}{GW200105_NSBH_lowspin}{3.14}{GW200115_XPHM_lowspin}{2.92}{GW200115_v4PHM_lowspin}{2.94}{GW200115_combined_lowspin}{2.93}{GW200115_XPHM_highspin}{2.96}{GW200115_v4PHM_highspin}{3.02}{GW200115_combined_highspin}{2.99}{GW200115_NSBH_lowspin}{3.14}}}
\newcommand{\tilttwofivepercent}[1]{\IfEqCase{#1}{{GW200105_XPHM_lowspin}{0.47}{GW200105_v4PHM_lowspin}{0.49}{GW200105_combined_lowspin}{0.48}{GW200105_XPHM_highspin}{0.53}{GW200105_v4PHM_highspin}{0.64}{GW200105_combined_highspin}{0.57}{GW200105_NSBH_lowspin}{0.00}{GW200115_XPHM_lowspin}{0.50}{GW200115_v4PHM_lowspin}{0.45}{GW200115_combined_lowspin}{0.48}{GW200115_XPHM_highspin}{0.76}{GW200115_v4PHM_highspin}{0.53}{GW200115_combined_highspin}{0.63}{GW200115_NSBH_lowspin}{0.00}}}
\newcommand{\tilttwoninetyfivepercent}[1]{\IfEqCase{#1}{{GW200105_XPHM_lowspin}{2.65}{GW200105_v4PHM_lowspin}{2.72}{GW200105_combined_lowspin}{2.69}{GW200105_XPHM_highspin}{2.63}{GW200105_v4PHM_highspin}{2.55}{GW200105_combined_highspin}{2.59}{GW200105_NSBH_lowspin}{3.14}{GW200115_XPHM_lowspin}{2.64}{GW200115_v4PHM_lowspin}{2.69}{GW200115_combined_lowspin}{2.67}{GW200115_XPHM_highspin}{2.74}{GW200115_v4PHM_highspin}{2.87}{GW200115_combined_highspin}{2.82}{GW200115_NSBH_lowspin}{3.14}}}
\newcommand{\tilttwoninetypercent}[1]{\IfEqCase{#1}{{GW200105_XPHM_lowspin}{2.45}{GW200105_v4PHM_lowspin}{2.53}{GW200105_combined_lowspin}{2.49}{GW200105_XPHM_highspin}{2.43}{GW200105_v4PHM_highspin}{2.35}{GW200105_combined_highspin}{2.39}{GW200105_NSBH_lowspin}{3.14}{GW200115_XPHM_lowspin}{2.45}{GW200115_v4PHM_lowspin}{2.51}{GW200115_combined_lowspin}{2.48}{GW200115_XPHM_highspin}{2.58}{GW200115_v4PHM_highspin}{2.73}{GW200115_combined_highspin}{2.66}{GW200115_NSBH_lowspin}{3.14}}}
\newcommand{\phionetwominus}[1]{\IfEqCase{#1}{{GW200105_XPHM_lowspin}{2.88}{GW200105_v4PHM_lowspin}{2.89}{GW200105_combined_lowspin}{2.89}{GW200105_XPHM_highspin}{2.81}{GW200105_v4PHM_highspin}{2.79}{GW200105_combined_highspin}{2.80}{GW200105_NSBH_lowspin}{0.00}{GW200115_XPHM_lowspin}{2.87}{GW200115_v4PHM_lowspin}{2.84}{GW200115_combined_lowspin}{2.86}{GW200115_XPHM_highspin}{2.81}{GW200115_v4PHM_highspin}{2.89}{GW200115_combined_highspin}{2.84}{GW200115_NSBH_lowspin}{0.00}}}
\newcommand{\phionetwomed}[1]{\IfEqCase{#1}{{GW200105_XPHM_lowspin}{3.20}{GW200105_v4PHM_lowspin}{3.18}{GW200105_combined_lowspin}{3.19}{GW200105_XPHM_highspin}{3.14}{GW200105_v4PHM_highspin}{3.12}{GW200105_combined_highspin}{3.14}{GW200105_NSBH_lowspin}{0.00}{GW200115_XPHM_lowspin}{3.18}{GW200115_v4PHM_lowspin}{3.16}{GW200115_combined_lowspin}{3.17}{GW200115_XPHM_highspin}{3.13}{GW200115_v4PHM_highspin}{3.21}{GW200115_combined_highspin}{3.16}{GW200115_NSBH_lowspin}{0.00}}}
\newcommand{\phionetwoplus}[1]{\IfEqCase{#1}{{GW200105_XPHM_lowspin}{2.78}{GW200105_v4PHM_lowspin}{2.79}{GW200105_combined_lowspin}{2.78}{GW200105_XPHM_highspin}{2.82}{GW200105_v4PHM_highspin}{2.85}{GW200105_combined_highspin}{2.83}{GW200105_NSBH_lowspin}{0.00}{GW200115_XPHM_lowspin}{2.79}{GW200115_v4PHM_lowspin}{2.82}{GW200115_combined_lowspin}{2.80}{GW200115_XPHM_highspin}{2.85}{GW200115_v4PHM_highspin}{2.75}{GW200115_combined_highspin}{2.80}{GW200115_NSBH_lowspin}{0.00}}}
\newcommand{\phionetwoonepercent}[1]{\IfEqCase{#1}{{GW200105_XPHM_lowspin}{0.06}{GW200105_v4PHM_lowspin}{0.06}{GW200105_combined_lowspin}{0.06}{GW200105_XPHM_highspin}{0.05}{GW200105_v4PHM_highspin}{0.07}{GW200105_combined_highspin}{0.06}{GW200105_NSBH_lowspin}{0.00}{GW200115_XPHM_lowspin}{0.07}{GW200115_v4PHM_lowspin}{0.06}{GW200115_combined_lowspin}{0.06}{GW200115_XPHM_highspin}{0.07}{GW200115_v4PHM_highspin}{0.06}{GW200115_combined_highspin}{0.06}{GW200115_NSBH_lowspin}{0.00}}}
\newcommand{\phionetwoninetyninepercent}[1]{\IfEqCase{#1}{{GW200105_XPHM_lowspin}{6.22}{GW200105_v4PHM_lowspin}{6.23}{GW200105_combined_lowspin}{6.22}{GW200105_XPHM_highspin}{6.22}{GW200105_v4PHM_highspin}{6.21}{GW200105_combined_highspin}{6.22}{GW200105_NSBH_lowspin}{0.00}{GW200115_XPHM_lowspin}{6.21}{GW200115_v4PHM_lowspin}{6.22}{GW200115_combined_lowspin}{6.22}{GW200115_XPHM_highspin}{6.22}{GW200115_v4PHM_highspin}{6.21}{GW200115_combined_highspin}{6.22}{GW200115_NSBH_lowspin}{0.00}}}
\newcommand{\phionetwofivepercent}[1]{\IfEqCase{#1}{{GW200105_XPHM_lowspin}{0.32}{GW200105_v4PHM_lowspin}{0.29}{GW200105_combined_lowspin}{0.31}{GW200105_XPHM_highspin}{0.33}{GW200105_v4PHM_highspin}{0.33}{GW200105_combined_highspin}{0.33}{GW200105_NSBH_lowspin}{0.00}{GW200115_XPHM_lowspin}{0.31}{GW200115_v4PHM_lowspin}{0.32}{GW200115_combined_lowspin}{0.31}{GW200115_XPHM_highspin}{0.32}{GW200115_v4PHM_highspin}{0.32}{GW200115_combined_highspin}{0.32}{GW200115_NSBH_lowspin}{0.00}}}
\newcommand{\phionetwoninetyfivepercent}[1]{\IfEqCase{#1}{{GW200105_XPHM_lowspin}{5.98}{GW200105_v4PHM_lowspin}{5.98}{GW200105_combined_lowspin}{5.98}{GW200105_XPHM_highspin}{5.96}{GW200105_v4PHM_highspin}{5.97}{GW200105_combined_highspin}{5.97}{GW200105_NSBH_lowspin}{0.00}{GW200115_XPHM_lowspin}{5.97}{GW200115_v4PHM_lowspin}{5.98}{GW200115_combined_lowspin}{5.97}{GW200115_XPHM_highspin}{5.98}{GW200115_v4PHM_highspin}{5.96}{GW200115_combined_highspin}{5.97}{GW200115_NSBH_lowspin}{0.00}}}
\newcommand{\phionetwoninetypercent}[1]{\IfEqCase{#1}{{GW200105_XPHM_lowspin}{5.67}{GW200105_v4PHM_lowspin}{5.64}{GW200105_combined_lowspin}{5.66}{GW200105_XPHM_highspin}{5.63}{GW200105_v4PHM_highspin}{5.61}{GW200105_combined_highspin}{5.62}{GW200105_NSBH_lowspin}{0.00}{GW200115_XPHM_lowspin}{5.67}{GW200115_v4PHM_lowspin}{5.65}{GW200115_combined_lowspin}{5.66}{GW200115_XPHM_highspin}{5.65}{GW200115_v4PHM_highspin}{5.62}{GW200115_combined_highspin}{5.64}{GW200115_NSBH_lowspin}{0.00}}}
\newcommand{\phijlminus}[1]{\IfEqCase{#1}{{GW200105_XPHM_lowspin}{2.49}{GW200105_v4PHM_lowspin}{2.87}{GW200105_combined_lowspin}{2.70}{GW200105_XPHM_highspin}{2.79}{GW200105_v4PHM_highspin}{2.81}{GW200105_combined_highspin}{2.80}{GW200105_NSBH_lowspin}{0.00}{GW200115_XPHM_lowspin}{2.78}{GW200115_v4PHM_lowspin}{2.56}{GW200115_combined_lowspin}{2.68}{GW200115_XPHM_highspin}{2.84}{GW200115_v4PHM_highspin}{2.60}{GW200115_combined_highspin}{2.73}{GW200115_NSBH_lowspin}{0.00}}}
\newcommand{\phijlmed}[1]{\IfEqCase{#1}{{GW200105_XPHM_lowspin}{2.88}{GW200105_v4PHM_lowspin}{3.19}{GW200105_combined_lowspin}{3.04}{GW200105_XPHM_highspin}{3.12}{GW200105_v4PHM_highspin}{3.13}{GW200105_combined_highspin}{3.13}{GW200105_NSBH_lowspin}{0.00}{GW200115_XPHM_lowspin}{3.16}{GW200115_v4PHM_lowspin}{2.89}{GW200115_combined_lowspin}{3.03}{GW200115_XPHM_highspin}{3.16}{GW200115_v4PHM_highspin}{2.95}{GW200115_combined_highspin}{3.06}{GW200115_NSBH_lowspin}{0.00}}}
\newcommand{\phijlplus}[1]{\IfEqCase{#1}{{GW200105_XPHM_lowspin}{3.01}{GW200105_v4PHM_lowspin}{2.78}{GW200105_combined_lowspin}{2.89}{GW200105_XPHM_highspin}{2.82}{GW200105_v4PHM_highspin}{2.85}{GW200105_combined_highspin}{2.84}{GW200105_NSBH_lowspin}{0.00}{GW200115_XPHM_lowspin}{2.76}{GW200115_v4PHM_lowspin}{3.05}{GW200115_combined_lowspin}{2.89}{GW200115_XPHM_highspin}{2.80}{GW200115_v4PHM_highspin}{2.97}{GW200115_combined_highspin}{2.88}{GW200115_NSBH_lowspin}{0.00}}}
\newcommand{\phijlonepercent}[1]{\IfEqCase{#1}{{GW200105_XPHM_lowspin}{0.07}{GW200105_v4PHM_lowspin}{0.07}{GW200105_combined_lowspin}{0.07}{GW200105_XPHM_highspin}{0.07}{GW200105_v4PHM_highspin}{0.06}{GW200105_combined_highspin}{0.07}{GW200105_NSBH_lowspin}{0.00}{GW200115_XPHM_lowspin}{0.08}{GW200115_v4PHM_lowspin}{0.07}{GW200115_combined_lowspin}{0.07}{GW200115_XPHM_highspin}{0.06}{GW200115_v4PHM_highspin}{0.07}{GW200115_combined_highspin}{0.06}{GW200115_NSBH_lowspin}{0.00}}}
\newcommand{\phijlninetyninepercent}[1]{\IfEqCase{#1}{{GW200105_XPHM_lowspin}{6.22}{GW200105_v4PHM_lowspin}{6.22}{GW200105_combined_lowspin}{6.22}{GW200105_XPHM_highspin}{6.22}{GW200105_v4PHM_highspin}{6.22}{GW200105_combined_highspin}{6.22}{GW200105_NSBH_lowspin}{0.00}{GW200115_XPHM_lowspin}{6.20}{GW200115_v4PHM_lowspin}{6.21}{GW200115_combined_lowspin}{6.20}{GW200115_XPHM_highspin}{6.21}{GW200115_v4PHM_highspin}{6.21}{GW200115_combined_highspin}{6.21}{GW200115_NSBH_lowspin}{0.00}}}
\newcommand{\phijlfivepercent}[1]{\IfEqCase{#1}{{GW200105_XPHM_lowspin}{0.39}{GW200105_v4PHM_lowspin}{0.32}{GW200105_combined_lowspin}{0.35}{GW200105_XPHM_highspin}{0.33}{GW200105_v4PHM_highspin}{0.33}{GW200105_combined_highspin}{0.33}{GW200105_NSBH_lowspin}{0.00}{GW200115_XPHM_lowspin}{0.38}{GW200115_v4PHM_lowspin}{0.33}{GW200115_combined_lowspin}{0.35}{GW200115_XPHM_highspin}{0.32}{GW200115_v4PHM_highspin}{0.35}{GW200115_combined_highspin}{0.33}{GW200115_NSBH_lowspin}{0.00}}}
\newcommand{\phijlninetyfivepercent}[1]{\IfEqCase{#1}{{GW200105_XPHM_lowspin}{5.90}{GW200105_v4PHM_lowspin}{5.97}{GW200105_combined_lowspin}{5.93}{GW200105_XPHM_highspin}{5.95}{GW200105_v4PHM_highspin}{5.98}{GW200105_combined_highspin}{5.97}{GW200105_NSBH_lowspin}{0.00}{GW200115_XPHM_lowspin}{5.92}{GW200115_v4PHM_lowspin}{5.93}{GW200115_combined_lowspin}{5.92}{GW200115_XPHM_highspin}{5.96}{GW200115_v4PHM_highspin}{5.93}{GW200115_combined_highspin}{5.95}{GW200115_NSBH_lowspin}{0.00}}}
\newcommand{\phijlninetypercent}[1]{\IfEqCase{#1}{{GW200105_XPHM_lowspin}{5.56}{GW200105_v4PHM_lowspin}{5.68}{GW200105_combined_lowspin}{5.62}{GW200105_XPHM_highspin}{5.65}{GW200105_v4PHM_highspin}{5.70}{GW200105_combined_highspin}{5.68}{GW200105_NSBH_lowspin}{0.00}{GW200115_XPHM_lowspin}{5.58}{GW200115_v4PHM_lowspin}{5.57}{GW200115_combined_lowspin}{5.57}{GW200115_XPHM_highspin}{5.65}{GW200115_v4PHM_highspin}{5.56}{GW200115_combined_highspin}{5.61}{GW200115_NSBH_lowspin}{0.00}}}
\newcommand{\costhetajnminus}[1]{\IfEqCase{#1}{{GW200105_XPHM_lowspin}{0.43}{GW200105_v4PHM_lowspin}{0.73}{GW200105_combined_lowspin}{0.56}{GW200105_XPHM_highspin}{0.67}{GW200105_v4PHM_highspin}{0.77}{GW200105_combined_highspin}{0.73}{GW200105_NSBH_lowspin}{0.93}{GW200115_XPHM_lowspin}{1.56}{GW200115_v4PHM_lowspin}{1.67}{GW200115_combined_lowspin}{1.70}{GW200115_XPHM_highspin}{1.38}{GW200115_v4PHM_highspin}{1.66}{GW200115_combined_highspin}{1.70}{GW200115_NSBH_lowspin}{1.64}}}
\newcommand{\costhetajnmed}[1]{\IfEqCase{#1}{{GW200105_XPHM_lowspin}{-0.51}{GW200105_v4PHM_lowspin}{-0.22}{GW200105_combined_lowspin}{-0.39}{GW200105_XPHM_highspin}{-0.28}{GW200105_v4PHM_highspin}{-0.19}{GW200105_combined_highspin}{-0.22}{GW200105_NSBH_lowspin}{-0.04}{GW200115_XPHM_lowspin}{0.85}{GW200115_v4PHM_lowspin}{0.75}{GW200115_combined_lowspin}{0.82}{GW200115_XPHM_highspin}{0.86}{GW200115_v4PHM_highspin}{0.74}{GW200115_combined_highspin}{0.82}{GW200115_NSBH_lowspin}{0.75}}}
\newcommand{\costhetajnplus}[1]{\IfEqCase{#1}{{GW200105_XPHM_lowspin}{1.45}{GW200105_v4PHM_lowspin}{1.17}{GW200105_combined_lowspin}{1.33}{GW200105_XPHM_highspin}{1.22}{GW200105_v4PHM_highspin}{1.14}{GW200105_combined_highspin}{1.17}{GW200105_NSBH_lowspin}{1.01}{GW200115_XPHM_lowspin}{0.13}{GW200115_v4PHM_lowspin}{0.23}{GW200115_combined_lowspin}{0.17}{GW200115_XPHM_highspin}{0.12}{GW200115_v4PHM_highspin}{0.24}{GW200115_combined_highspin}{0.16}{GW200115_NSBH_lowspin}{0.23}}}
\newcommand{\costhetajnonepercent}[1]{\IfEqCase{#1}{{GW200105_XPHM_lowspin}{-0.99}{GW200105_v4PHM_lowspin}{-0.99}{GW200105_combined_lowspin}{-0.99}{GW200105_XPHM_highspin}{-0.99}{GW200105_v4PHM_highspin}{-0.99}{GW200105_combined_highspin}{-0.99}{GW200105_NSBH_lowspin}{-0.99}{GW200115_XPHM_lowspin}{-0.95}{GW200115_v4PHM_lowspin}{-0.99}{GW200115_combined_lowspin}{-0.98}{GW200115_XPHM_highspin}{-0.94}{GW200115_v4PHM_highspin}{-0.99}{GW200115_combined_highspin}{-0.98}{GW200115_NSBH_lowspin}{-0.98}}}
\newcommand{\costhetajnninetyninepercent}[1]{\IfEqCase{#1}{{GW200105_XPHM_lowspin}{0.98}{GW200105_v4PHM_lowspin}{0.99}{GW200105_combined_lowspin}{0.99}{GW200105_XPHM_highspin}{0.99}{GW200105_v4PHM_highspin}{0.99}{GW200105_combined_highspin}{0.99}{GW200105_NSBH_lowspin}{0.99}{GW200115_XPHM_lowspin}{1.00}{GW200115_v4PHM_lowspin}{1.00}{GW200115_combined_lowspin}{1.00}{GW200115_XPHM_highspin}{1.00}{GW200115_v4PHM_highspin}{1.00}{GW200115_combined_highspin}{1.00}{GW200115_NSBH_lowspin}{1.00}}}
\newcommand{\costhetajnfivepercent}[1]{\IfEqCase{#1}{{GW200105_XPHM_lowspin}{-0.95}{GW200105_v4PHM_lowspin}{-0.96}{GW200105_combined_lowspin}{-0.95}{GW200105_XPHM_highspin}{-0.95}{GW200105_v4PHM_highspin}{-0.96}{GW200105_combined_highspin}{-0.95}{GW200105_NSBH_lowspin}{-0.97}{GW200115_XPHM_lowspin}{-0.71}{GW200115_v4PHM_lowspin}{-0.92}{GW200115_combined_lowspin}{-0.88}{GW200115_XPHM_highspin}{-0.52}{GW200115_v4PHM_highspin}{-0.92}{GW200115_combined_highspin}{-0.88}{GW200115_NSBH_lowspin}{-0.89}}}
\newcommand{\costhetajnninetyfivepercent}[1]{\IfEqCase{#1}{{GW200105_XPHM_lowspin}{0.94}{GW200105_v4PHM_lowspin}{0.95}{GW200105_combined_lowspin}{0.94}{GW200105_XPHM_highspin}{0.94}{GW200105_v4PHM_highspin}{0.95}{GW200105_combined_highspin}{0.95}{GW200105_NSBH_lowspin}{0.97}{GW200115_XPHM_lowspin}{0.98}{GW200115_v4PHM_lowspin}{0.98}{GW200115_combined_lowspin}{0.98}{GW200115_XPHM_highspin}{0.98}{GW200115_v4PHM_highspin}{0.98}{GW200115_combined_highspin}{0.98}{GW200115_NSBH_lowspin}{0.98}}}
\newcommand{\costhetajnninetypercent}[1]{\IfEqCase{#1}{{GW200105_XPHM_lowspin}{0.89}{GW200105_v4PHM_lowspin}{0.90}{GW200105_combined_lowspin}{0.89}{GW200105_XPHM_highspin}{0.90}{GW200105_v4PHM_highspin}{0.90}{GW200105_combined_highspin}{0.90}{GW200105_NSBH_lowspin}{0.93}{GW200115_XPHM_lowspin}{0.97}{GW200115_v4PHM_lowspin}{0.97}{GW200115_combined_lowspin}{0.97}{GW200115_XPHM_highspin}{0.97}{GW200115_v4PHM_highspin}{0.96}{GW200115_combined_highspin}{0.97}{GW200115_NSBH_lowspin}{0.96}}}
\newcommand{\psiminus}[1]{\IfEqCase{#1}{{GW200105_XPHM_lowspin}{1.21}{GW200105_v4PHM_lowspin}{2.96}{GW200105_combined_lowspin}{2.12}{GW200105_XPHM_highspin}{1.61}{GW200105_v4PHM_highspin}{2.97}{GW200105_combined_highspin}{2.26}{GW200105_NSBH_lowspin}{1.41}{GW200115_XPHM_lowspin}{1.73}{GW200115_v4PHM_lowspin}{2.85}{GW200115_combined_lowspin}{2.13}{GW200115_XPHM_highspin}{1.74}{GW200115_v4PHM_highspin}{2.85}{GW200115_combined_highspin}{2.11}{GW200115_NSBH_lowspin}{1.40}}}
\newcommand{\psimed}[1]{\IfEqCase{#1}{{GW200105_XPHM_lowspin}{1.30}{GW200105_v4PHM_lowspin}{3.13}{GW200105_combined_lowspin}{2.25}{GW200105_XPHM_highspin}{1.70}{GW200105_v4PHM_highspin}{3.15}{GW200105_combined_highspin}{2.39}{GW200105_NSBH_lowspin}{1.56}{GW200115_XPHM_lowspin}{1.88}{GW200115_v4PHM_lowspin}{3.14}{GW200115_combined_lowspin}{2.32}{GW200115_XPHM_highspin}{1.88}{GW200115_v4PHM_highspin}{3.13}{GW200115_combined_highspin}{2.31}{GW200115_NSBH_lowspin}{1.57}}}
\newcommand{\psiplus}[1]{\IfEqCase{#1}{{GW200105_XPHM_lowspin}{1.75}{GW200105_v4PHM_lowspin}{2.95}{GW200105_combined_lowspin}{3.59}{GW200105_XPHM_highspin}{1.34}{GW200105_v4PHM_highspin}{2.94}{GW200105_combined_highspin}{3.46}{GW200105_NSBH_lowspin}{1.42}{GW200115_XPHM_lowspin}{1.13}{GW200115_v4PHM_lowspin}{2.85}{GW200115_combined_lowspin}{3.42}{GW200115_XPHM_highspin}{1.13}{GW200115_v4PHM_highspin}{2.87}{GW200115_combined_highspin}{3.42}{GW200115_NSBH_lowspin}{1.40}}}
\newcommand{\psionepercent}[1]{\IfEqCase{#1}{{GW200105_XPHM_lowspin}{0.02}{GW200105_v4PHM_lowspin}{0.04}{GW200105_combined_lowspin}{0.02}{GW200105_XPHM_highspin}{0.02}{GW200105_v4PHM_highspin}{0.04}{GW200105_combined_highspin}{0.02}{GW200105_NSBH_lowspin}{0.03}{GW200115_XPHM_lowspin}{0.03}{GW200115_v4PHM_lowspin}{0.06}{GW200115_combined_lowspin}{0.04}{GW200115_XPHM_highspin}{0.02}{GW200115_v4PHM_highspin}{0.05}{GW200115_combined_highspin}{0.03}{GW200115_NSBH_lowspin}{0.03}}}
\newcommand{\psininetyninepercent}[1]{\IfEqCase{#1}{{GW200105_XPHM_lowspin}{3.12}{GW200105_v4PHM_lowspin}{6.25}{GW200105_combined_lowspin}{6.20}{GW200105_XPHM_highspin}{3.12}{GW200105_v4PHM_highspin}{6.25}{GW200105_combined_highspin}{6.21}{GW200105_NSBH_lowspin}{3.11}{GW200115_XPHM_lowspin}{3.11}{GW200115_v4PHM_lowspin}{6.22}{GW200115_combined_lowspin}{6.16}{GW200115_XPHM_highspin}{3.11}{GW200115_v4PHM_highspin}{6.22}{GW200115_combined_highspin}{6.17}{GW200115_NSBH_lowspin}{3.11}}}
\newcommand{\psifivepercent}[1]{\IfEqCase{#1}{{GW200105_XPHM_lowspin}{0.09}{GW200105_v4PHM_lowspin}{0.17}{GW200105_combined_lowspin}{0.12}{GW200105_XPHM_highspin}{0.10}{GW200105_v4PHM_highspin}{0.18}{GW200105_combined_highspin}{0.13}{GW200105_NSBH_lowspin}{0.15}{GW200115_XPHM_lowspin}{0.15}{GW200115_v4PHM_lowspin}{0.29}{GW200115_combined_lowspin}{0.19}{GW200115_XPHM_highspin}{0.14}{GW200115_v4PHM_highspin}{0.29}{GW200115_combined_highspin}{0.19}{GW200115_NSBH_lowspin}{0.17}}}
\newcommand{\psininetyfivepercent}[1]{\IfEqCase{#1}{{GW200105_XPHM_lowspin}{3.05}{GW200105_v4PHM_lowspin}{6.08}{GW200105_combined_lowspin}{5.83}{GW200105_XPHM_highspin}{3.05}{GW200105_v4PHM_highspin}{6.09}{GW200105_combined_highspin}{5.85}{GW200105_NSBH_lowspin}{2.99}{GW200115_XPHM_lowspin}{3.01}{GW200115_v4PHM_lowspin}{6.00}{GW200115_combined_lowspin}{5.74}{GW200115_XPHM_highspin}{3.01}{GW200115_v4PHM_highspin}{6.00}{GW200115_combined_highspin}{5.73}{GW200115_NSBH_lowspin}{2.98}}}
\newcommand{\psininetypercent}[1]{\IfEqCase{#1}{{GW200105_XPHM_lowspin}{2.95}{GW200105_v4PHM_lowspin}{5.83}{GW200105_combined_lowspin}{4.82}{GW200105_XPHM_highspin}{2.96}{GW200105_v4PHM_highspin}{5.85}{GW200105_combined_highspin}{4.89}{GW200105_NSBH_lowspin}{2.83}{GW200115_XPHM_lowspin}{2.88}{GW200115_v4PHM_lowspin}{5.74}{GW200115_combined_lowspin}{5.16}{GW200115_XPHM_highspin}{2.88}{GW200115_v4PHM_highspin}{5.73}{GW200115_combined_highspin}{5.13}{GW200115_NSBH_lowspin}{2.82}}}
\newcommand{\phaseminus}[1]{\IfEqCase{#1}{{GW200105_XPHM_lowspin}{1.12}{GW200105_v4PHM_lowspin}{2.44}{GW200105_combined_lowspin}{3.22}{GW200105_XPHM_highspin}{1.06}{GW200105_v4PHM_highspin}{2.49}{GW200105_combined_highspin}{3.25}{GW200105_NSBH_lowspin}{2.77}{GW200115_XPHM_lowspin}{1.57}{GW200115_v4PHM_lowspin}{2.93}{GW200115_combined_lowspin}{2.09}{GW200115_XPHM_highspin}{1.49}{GW200115_v4PHM_highspin}{2.94}{GW200115_combined_highspin}{2.06}{GW200115_NSBH_lowspin}{2.84}}}
\newcommand{\phasemed}[1]{\IfEqCase{#1}{{GW200105_XPHM_lowspin}{4.72}{GW200105_v4PHM_lowspin}{3.18}{GW200105_combined_lowspin}{4.33}{GW200105_XPHM_highspin}{4.73}{GW200105_v4PHM_highspin}{3.21}{GW200105_combined_highspin}{4.34}{GW200105_NSBH_lowspin}{3.10}{GW200115_XPHM_lowspin}{1.79}{GW200115_v4PHM_lowspin}{3.24}{GW200115_combined_lowspin}{2.34}{GW200115_XPHM_highspin}{1.71}{GW200115_v4PHM_highspin}{3.24}{GW200115_combined_highspin}{2.31}{GW200115_NSBH_lowspin}{3.16}}}
\newcommand{\phaseplus}[1]{\IfEqCase{#1}{{GW200105_XPHM_lowspin}{0.93}{GW200105_v4PHM_lowspin}{2.39}{GW200105_combined_lowspin}{1.29}{GW200105_XPHM_highspin}{0.94}{GW200105_v4PHM_highspin}{2.37}{GW200105_combined_highspin}{1.30}{GW200105_NSBH_lowspin}{2.83}{GW200115_XPHM_lowspin}{4.22}{GW200115_v4PHM_lowspin}{2.76}{GW200115_combined_lowspin}{3.66}{GW200115_XPHM_highspin}{4.31}{GW200115_v4PHM_highspin}{2.74}{GW200115_combined_highspin}{3.68}{GW200115_NSBH_lowspin}{2.81}}}
\newcommand{\phaseonepercent}[1]{\IfEqCase{#1}{{GW200105_XPHM_lowspin}{0.41}{GW200105_v4PHM_lowspin}{0.16}{GW200105_combined_lowspin}{0.22}{GW200105_XPHM_highspin}{0.44}{GW200105_v4PHM_highspin}{0.15}{GW200105_combined_highspin}{0.21}{GW200105_NSBH_lowspin}{0.06}{GW200115_XPHM_lowspin}{0.05}{GW200115_v4PHM_lowspin}{0.06}{GW200115_combined_lowspin}{0.05}{GW200115_XPHM_highspin}{0.05}{GW200115_v4PHM_highspin}{0.06}{GW200115_combined_highspin}{0.05}{GW200115_NSBH_lowspin}{0.06}}}
\newcommand{\phaseninetyninepercent}[1]{\IfEqCase{#1}{{GW200105_XPHM_lowspin}{6.06}{GW200105_v4PHM_lowspin}{6.11}{GW200105_combined_lowspin}{6.09}{GW200105_XPHM_highspin}{6.09}{GW200105_v4PHM_highspin}{6.13}{GW200105_combined_highspin}{6.11}{GW200105_NSBH_lowspin}{6.22}{GW200115_XPHM_lowspin}{6.23}{GW200115_v4PHM_lowspin}{6.23}{GW200115_combined_lowspin}{6.23}{GW200115_XPHM_highspin}{6.23}{GW200115_v4PHM_highspin}{6.22}{GW200115_combined_highspin}{6.23}{GW200115_NSBH_lowspin}{6.22}}}
\newcommand{\phasefivepercent}[1]{\IfEqCase{#1}{{GW200105_XPHM_lowspin}{3.59}{GW200105_v4PHM_lowspin}{0.74}{GW200105_combined_lowspin}{1.10}{GW200105_XPHM_highspin}{3.66}{GW200105_v4PHM_highspin}{0.72}{GW200105_combined_highspin}{1.09}{GW200105_NSBH_lowspin}{0.34}{GW200115_XPHM_lowspin}{0.22}{GW200115_v4PHM_lowspin}{0.30}{GW200115_combined_lowspin}{0.26}{GW200115_XPHM_highspin}{0.22}{GW200115_v4PHM_highspin}{0.30}{GW200115_combined_highspin}{0.25}{GW200115_NSBH_lowspin}{0.31}}}
\newcommand{\phaseninetyfivepercent}[1]{\IfEqCase{#1}{{GW200105_XPHM_lowspin}{5.64}{GW200105_v4PHM_lowspin}{5.57}{GW200105_combined_lowspin}{5.61}{GW200105_XPHM_highspin}{5.67}{GW200105_v4PHM_highspin}{5.58}{GW200105_combined_highspin}{5.64}{GW200105_NSBH_lowspin}{5.94}{GW200115_XPHM_lowspin}{6.01}{GW200115_v4PHM_lowspin}{6.00}{GW200115_combined_lowspin}{6.00}{GW200115_XPHM_highspin}{6.02}{GW200115_v4PHM_highspin}{5.98}{GW200115_combined_highspin}{6.00}{GW200115_NSBH_lowspin}{5.97}}}
\newcommand{\phaseninetypercent}[1]{\IfEqCase{#1}{{GW200105_XPHM_lowspin}{5.43}{GW200105_v4PHM_lowspin}{5.06}{GW200105_combined_lowspin}{5.32}{GW200105_XPHM_highspin}{5.45}{GW200105_v4PHM_highspin}{5.09}{GW200105_combined_highspin}{5.36}{GW200105_NSBH_lowspin}{5.59}{GW200115_XPHM_lowspin}{5.68}{GW200115_v4PHM_lowspin}{5.71}{GW200115_combined_lowspin}{5.70}{GW200115_XPHM_highspin}{5.63}{GW200115_v4PHM_highspin}{5.67}{GW200115_combined_highspin}{5.65}{GW200115_NSBH_lowspin}{5.65}}}
\newcommand{\loglikelihoodminus}[1]{\IfEqCase{#1}{{GW200105_XPHM_lowspin}{4.9}{GW200105_v4PHM_lowspin}{5.4}{GW200105_combined_lowspin}{6.7}{GW200105_XPHM_highspin}{4.7}{GW200105_v4PHM_highspin}{5.8}{GW200105_combined_highspin}{6.9}{GW200105_NSBH_lowspin}{4.2}{GW200115_XPHM_lowspin}{4.9}{GW200115_v4PHM_lowspin}{6.9}{GW200115_combined_lowspin}{8.8}{GW200115_XPHM_highspin}{5.2}{GW200115_v4PHM_highspin}{8.0}{GW200115_combined_highspin}{9.7}{GW200115_NSBH_lowspin}{4.8}}}
\newcommand{\loglikelihoodmed}[1]{\IfEqCase{#1}{{GW200105_XPHM_lowspin}{87.7}{GW200105_v4PHM_lowspin}{82.0}{GW200105_combined_lowspin}{84.5}{GW200105_XPHM_highspin}{87.5}{GW200105_v4PHM_highspin}{81.8}{GW200105_combined_highspin}{84.5}{GW200105_NSBH_lowspin}{84.8}{GW200115_XPHM_lowspin}{59.4}{GW200115_v4PHM_lowspin}{51.5}{GW200115_combined_lowspin}{55.3}{GW200115_XPHM_highspin}{59.7}{GW200115_v4PHM_highspin}{51.4}{GW200115_combined_highspin}{55.4}{GW200115_NSBH_lowspin}{56.8}}}
\newcommand{\loglikelihoodplus}[1]{\IfEqCase{#1}{{GW200105_XPHM_lowspin}{3.1}{GW200105_v4PHM_lowspin}{3.4}{GW200105_combined_lowspin}{5.8}{GW200105_XPHM_highspin}{3.2}{GW200105_v4PHM_highspin}{3.6}{GW200105_combined_highspin}{5.7}{GW200105_NSBH_lowspin}{2.9}{GW200115_XPHM_lowspin}{3.1}{GW200115_v4PHM_lowspin}{4.2}{GW200115_combined_lowspin}{6.6}{GW200115_XPHM_highspin}{3.2}{GW200115_v4PHM_highspin}{4.5}{GW200115_combined_highspin}{6.9}{GW200115_NSBH_lowspin}{3.0}}}
\newcommand{\loglikelihoodonepercent}[1]{\IfEqCase{#1}{{GW200105_XPHM_lowspin}{80.3}{GW200105_v4PHM_lowspin}{73.5}{GW200105_combined_lowspin}{74.8}{GW200105_XPHM_highspin}{80.2}{GW200105_v4PHM_highspin}{72.6}{GW200105_combined_highspin}{74.1}{GW200105_NSBH_lowspin}{78.3}{GW200115_XPHM_lowspin}{51.8}{GW200115_v4PHM_lowspin}{39.2}{GW200115_combined_lowspin}{42.0}{GW200115_XPHM_highspin}{51.9}{GW200115_v4PHM_highspin}{37.3}{GW200115_combined_highspin}{40.1}{GW200115_NSBH_lowspin}{49.5}}}
\newcommand{\loglikelihoodninetyninepercent}[1]{\IfEqCase{#1}{{GW200105_XPHM_lowspin}{91.6}{GW200105_v4PHM_lowspin}{86.4}{GW200105_combined_lowspin}{91.4}{GW200105_XPHM_highspin}{91.7}{GW200105_v4PHM_highspin}{86.4}{GW200105_combined_highspin}{91.3}{GW200105_NSBH_lowspin}{88.6}{GW200115_XPHM_lowspin}{63.4}{GW200115_v4PHM_lowspin}{57.1}{GW200115_combined_lowspin}{63.0}{GW200115_XPHM_highspin}{63.9}{GW200115_v4PHM_highspin}{57.3}{GW200115_combined_highspin}{63.5}{GW200115_NSBH_lowspin}{60.5}}}
\newcommand{\loglikelihoodfivepercent}[1]{\IfEqCase{#1}{{GW200105_XPHM_lowspin}{82.8}{GW200105_v4PHM_lowspin}{76.6}{GW200105_combined_lowspin}{77.9}{GW200105_XPHM_highspin}{82.8}{GW200105_v4PHM_highspin}{76.1}{GW200105_combined_highspin}{77.6}{GW200105_NSBH_lowspin}{80.6}{GW200115_XPHM_lowspin}{54.5}{GW200115_v4PHM_lowspin}{44.6}{GW200115_combined_lowspin}{46.5}{GW200115_XPHM_highspin}{54.5}{GW200115_v4PHM_highspin}{43.3}{GW200115_combined_highspin}{45.7}{GW200115_NSBH_lowspin}{52.0}}}
\newcommand{\loglikelihoodninetyfivepercent}[1]{\IfEqCase{#1}{{GW200105_XPHM_lowspin}{90.8}{GW200105_v4PHM_lowspin}{85.4}{GW200105_combined_lowspin}{90.3}{GW200105_XPHM_highspin}{90.7}{GW200105_v4PHM_highspin}{85.4}{GW200105_combined_highspin}{90.1}{GW200105_NSBH_lowspin}{87.7}{GW200115_XPHM_lowspin}{62.5}{GW200115_v4PHM_lowspin}{55.8}{GW200115_combined_lowspin}{61.9}{GW200115_XPHM_highspin}{62.9}{GW200115_v4PHM_highspin}{55.9}{GW200115_combined_highspin}{62.3}{GW200115_NSBH_lowspin}{59.8}}}
\newcommand{\loglikelihoodninetypercent}[1]{\IfEqCase{#1}{{GW200105_XPHM_lowspin}{90.3}{GW200105_v4PHM_lowspin}{84.8}{GW200105_combined_lowspin}{89.5}{GW200105_XPHM_highspin}{90.1}{GW200105_v4PHM_highspin}{84.7}{GW200105_combined_highspin}{89.3}{GW200105_NSBH_lowspin}{87.2}{GW200115_XPHM_lowspin}{61.9}{GW200115_v4PHM_lowspin}{55.0}{GW200115_combined_lowspin}{61.1}{GW200115_XPHM_highspin}{62.3}{GW200115_v4PHM_highspin}{55.1}{GW200115_combined_highspin}{61.5}{GW200115_NSBH_lowspin}{59.2}}}
\newcommand{\raminus}[1]{\IfEqCase{#1}{{GW200105_XPHM_lowspin}{1.21092}{GW200105_v4PHM_lowspin}{1.09521}{GW200105_combined_lowspin}{1.15845}{GW200105_XPHM_highspin}{2.86443}{GW200105_v4PHM_highspin}{1.07868}{GW200105_combined_highspin}{1.24593}{GW200105_NSBH_lowspin}{1.20706}{GW200115_XPHM_lowspin}{0.10160}{GW200115_v4PHM_lowspin}{0.13758}{GW200115_combined_lowspin}{0.11404}{GW200115_XPHM_highspin}{0.09755}{GW200115_v4PHM_highspin}{0.13670}{GW200115_combined_highspin}{0.11069}{GW200115_NSBH_lowspin}{0.11916}}}
\newcommand{\ramed}[1]{\IfEqCase{#1}{{GW200105_XPHM_lowspin}{1.86665}{GW200105_v4PHM_lowspin}{1.71114}{GW200105_combined_lowspin}{1.79494}{GW200105_XPHM_highspin}{3.54875}{GW200105_v4PHM_highspin}{1.68779}{GW200105_combined_highspin}{1.88625}{GW200105_NSBH_lowspin}{1.81908}{GW200115_XPHM_lowspin}{0.74736}{GW200115_v4PHM_lowspin}{0.78995}{GW200115_combined_lowspin}{0.76240}{GW200115_XPHM_highspin}{0.74228}{GW200115_v4PHM_highspin}{0.78991}{GW200115_combined_highspin}{0.75883}{GW200115_NSBH_lowspin}{0.76549}}}
\newcommand{\raplus}[1]{\IfEqCase{#1}{{GW200105_XPHM_lowspin}{3.06522}{GW200105_v4PHM_lowspin}{3.25738}{GW200105_combined_lowspin}{3.15734}{GW200105_XPHM_highspin}{1.41352}{GW200105_v4PHM_highspin}{3.27940}{GW200105_combined_highspin}{3.07760}{GW200105_NSBH_lowspin}{3.12611}{GW200115_XPHM_lowspin}{3.92058}{GW200115_v4PHM_lowspin}{3.96480}{GW200115_combined_lowspin}{3.96535}{GW200115_XPHM_highspin}{3.89929}{GW200115_v4PHM_highspin}{3.96771}{GW200115_combined_highspin}{3.96373}{GW200115_NSBH_lowspin}{4.30526}}}
\newcommand{\raonepercent}[1]{\IfEqCase{#1}{{GW200105_XPHM_lowspin}{0.42861}{GW200105_v4PHM_lowspin}{0.41852}{GW200105_combined_lowspin}{0.42077}{GW200105_XPHM_highspin}{0.45937}{GW200105_v4PHM_highspin}{0.41348}{GW200105_combined_highspin}{0.43474}{GW200105_NSBH_lowspin}{0.42350}{GW200115_XPHM_lowspin}{0.61133}{GW200115_v4PHM_lowspin}{0.61166}{GW200115_combined_lowspin}{0.61117}{GW200115_XPHM_highspin}{0.61081}{GW200115_v4PHM_highspin}{0.61248}{GW200115_combined_highspin}{0.61131}{GW200115_NSBH_lowspin}{0.60918}}}
\newcommand{\raninetyninepercent}[1]{\IfEqCase{#1}{{GW200105_XPHM_lowspin}{5.07640}{GW200105_v4PHM_lowspin}{5.10228}{GW200105_combined_lowspin}{5.08950}{GW200105_XPHM_highspin}{5.08126}{GW200105_v4PHM_highspin}{5.10347}{GW200105_combined_highspin}{5.09113}{GW200105_NSBH_lowspin}{5.09911}{GW200115_XPHM_lowspin}{5.21838}{GW200115_v4PHM_lowspin}{5.17516}{GW200115_combined_lowspin}{5.20149}{GW200115_XPHM_highspin}{5.20621}{GW200115_v4PHM_highspin}{5.17892}{GW200115_combined_highspin}{5.19491}{GW200115_NSBH_lowspin}{5.24267}}}
\newcommand{\rafivepercent}[1]{\IfEqCase{#1}{{GW200105_XPHM_lowspin}{0.65573}{GW200105_v4PHM_lowspin}{0.61593}{GW200105_combined_lowspin}{0.63649}{GW200105_XPHM_highspin}{0.68432}{GW200105_v4PHM_highspin}{0.60911}{GW200105_combined_highspin}{0.64032}{GW200105_NSBH_lowspin}{0.61202}{GW200115_XPHM_lowspin}{0.64576}{GW200115_v4PHM_lowspin}{0.65237}{GW200115_combined_lowspin}{0.64836}{GW200115_XPHM_highspin}{0.64473}{GW200115_v4PHM_highspin}{0.65322}{GW200115_combined_highspin}{0.64814}{GW200115_NSBH_lowspin}{0.64633}}}
\newcommand{\raninetyfivepercent}[1]{\IfEqCase{#1}{{GW200105_XPHM_lowspin}{4.93187}{GW200105_v4PHM_lowspin}{4.96853}{GW200105_combined_lowspin}{4.95228}{GW200105_XPHM_highspin}{4.96227}{GW200105_v4PHM_highspin}{4.96719}{GW200105_combined_highspin}{4.96384}{GW200105_NSBH_lowspin}{4.94520}{GW200115_XPHM_lowspin}{4.66794}{GW200115_v4PHM_lowspin}{4.75475}{GW200115_combined_lowspin}{4.72775}{GW200115_XPHM_highspin}{4.64157}{GW200115_v4PHM_highspin}{4.75763}{GW200115_combined_highspin}{4.72256}{GW200115_NSBH_lowspin}{5.07075}}}
\newcommand{\raninetypercent}[1]{\IfEqCase{#1}{{GW200105_XPHM_lowspin}{4.75218}{GW200105_v4PHM_lowspin}{4.79874}{GW200105_combined_lowspin}{4.77375}{GW200105_XPHM_highspin}{4.84330}{GW200105_v4PHM_highspin}{4.79370}{GW200105_combined_highspin}{4.82045}{GW200105_NSBH_lowspin}{4.77790}{GW200115_XPHM_lowspin}{4.52819}{GW200115_v4PHM_lowspin}{4.67517}{GW200115_combined_lowspin}{4.63811}{GW200115_XPHM_highspin}{4.25527}{GW200115_v4PHM_highspin}{4.67627}{GW200115_combined_highspin}{4.63128}{GW200115_NSBH_lowspin}{4.69030}}}
\newcommand{\decminus}[1]{\IfEqCase{#1}{{GW200105_XPHM_lowspin}{0.76028}{GW200105_v4PHM_lowspin}{0.70007}{GW200105_combined_lowspin}{0.72252}{GW200105_XPHM_highspin}{0.89837}{GW200105_v4PHM_highspin}{0.69745}{GW200105_combined_highspin}{0.76267}{GW200105_NSBH_lowspin}{0.77046}{GW200115_XPHM_lowspin}{0.57765}{GW200115_v4PHM_lowspin}{0.61315}{GW200115_combined_lowspin}{0.61369}{GW200115_XPHM_highspin}{0.54824}{GW200115_v4PHM_highspin}{0.61460}{GW200115_combined_highspin}{0.60993}{GW200115_NSBH_lowspin}{0.78568}}}
\newcommand{\decmed}[1]{\IfEqCase{#1}{{GW200105_XPHM_lowspin}{-0.12706}{GW200105_v4PHM_lowspin}{-0.19682}{GW200105_combined_lowspin}{-0.16833}{GW200105_XPHM_highspin}{0.02525}{GW200105_v4PHM_highspin}{-0.19946}{GW200105_combined_highspin}{-0.12002}{GW200105_NSBH_lowspin}{-0.12238}{GW200115_XPHM_lowspin}{0.02067}{GW200115_v4PHM_lowspin}{-0.01291}{GW200115_combined_lowspin}{0.00645}{GW200115_XPHM_highspin}{0.02199}{GW200115_v4PHM_highspin}{-0.01189}{GW200115_combined_highspin}{0.00819}{GW200115_NSBH_lowspin}{-0.01042}}}
\newcommand{\decplus}[1]{\IfEqCase{#1}{{GW200105_XPHM_lowspin}{0.89525}{GW200105_v4PHM_lowspin}{0.86689}{GW200105_combined_lowspin}{0.90275}{GW200105_XPHM_highspin}{0.79393}{GW200105_v4PHM_highspin}{0.86668}{GW200105_combined_highspin}{0.89990}{GW200105_NSBH_lowspin}{0.92535}{GW200115_XPHM_lowspin}{0.61978}{GW200115_v4PHM_lowspin}{0.80234}{GW200115_combined_lowspin}{0.72403}{GW200115_XPHM_highspin}{0.28794}{GW200115_v4PHM_highspin}{0.79793}{GW200115_combined_highspin}{0.69886}{GW200115_NSBH_lowspin}{0.72687}}}
\newcommand{\deconepercent}[1]{\IfEqCase{#1}{{GW200105_XPHM_lowspin}{-1.12308}{GW200105_v4PHM_lowspin}{-1.14252}{GW200105_combined_lowspin}{-1.13320}{GW200105_XPHM_highspin}{-1.12677}{GW200105_v4PHM_highspin}{-1.15321}{GW200105_combined_highspin}{-1.14282}{GW200105_NSBH_lowspin}{-1.14378}{GW200115_XPHM_lowspin}{-0.86426}{GW200115_v4PHM_lowspin}{-0.83238}{GW200115_combined_lowspin}{-0.85271}{GW200115_XPHM_highspin}{-0.85922}{GW200115_v4PHM_highspin}{-0.83478}{GW200115_combined_highspin}{-0.84884}{GW200115_NSBH_lowspin}{-0.90066}}}
\newcommand{\decninetyninepercent}[1]{\IfEqCase{#1}{{GW200105_XPHM_lowspin}{0.89562}{GW200105_v4PHM_lowspin}{0.90795}{GW200105_combined_lowspin}{0.89701}{GW200105_XPHM_highspin}{0.90943}{GW200105_v4PHM_highspin}{0.89342}{GW200105_combined_highspin}{0.90477}{GW200105_NSBH_lowspin}{0.92133}{GW200115_XPHM_lowspin}{0.95871}{GW200115_v4PHM_lowspin}{1.07501}{GW200115_combined_lowspin}{1.02271}{GW200115_XPHM_highspin}{0.93921}{GW200115_v4PHM_highspin}{1.08385}{GW200115_combined_highspin}{1.04054}{GW200115_NSBH_lowspin}{1.08033}}}
\newcommand{\decfivepercent}[1]{\IfEqCase{#1}{{GW200105_XPHM_lowspin}{-0.88734}{GW200105_v4PHM_lowspin}{-0.89690}{GW200105_combined_lowspin}{-0.89085}{GW200105_XPHM_highspin}{-0.87312}{GW200105_v4PHM_highspin}{-0.89691}{GW200105_combined_highspin}{-0.88269}{GW200105_NSBH_lowspin}{-0.89284}{GW200115_XPHM_lowspin}{-0.55698}{GW200115_v4PHM_lowspin}{-0.62607}{GW200115_combined_lowspin}{-0.60725}{GW200115_XPHM_highspin}{-0.52624}{GW200115_v4PHM_highspin}{-0.62649}{GW200115_combined_highspin}{-0.60174}{GW200115_NSBH_lowspin}{-0.79610}}}
\newcommand{\decninetyfivepercent}[1]{\IfEqCase{#1}{{GW200105_XPHM_lowspin}{0.76819}{GW200105_v4PHM_lowspin}{0.67007}{GW200105_combined_lowspin}{0.73443}{GW200105_XPHM_highspin}{0.81918}{GW200105_v4PHM_highspin}{0.66722}{GW200105_combined_highspin}{0.77989}{GW200105_NSBH_lowspin}{0.80297}{GW200115_XPHM_lowspin}{0.64045}{GW200115_v4PHM_lowspin}{0.78943}{GW200115_combined_lowspin}{0.73048}{GW200115_XPHM_highspin}{0.30993}{GW200115_v4PHM_highspin}{0.78604}{GW200115_combined_highspin}{0.70705}{GW200115_NSBH_lowspin}{0.71645}}}
\newcommand{\decninetypercent}[1]{\IfEqCase{#1}{{GW200105_XPHM_lowspin}{0.63543}{GW200105_v4PHM_lowspin}{0.50328}{GW200105_combined_lowspin}{0.58533}{GW200105_XPHM_highspin}{0.71685}{GW200105_v4PHM_highspin}{0.49975}{GW200105_combined_highspin}{0.63792}{GW200105_NSBH_lowspin}{0.68063}{GW200115_XPHM_lowspin}{0.19031}{GW200115_v4PHM_lowspin}{0.50469}{GW200115_combined_lowspin}{0.22262}{GW200115_XPHM_highspin}{0.17334}{GW200115_v4PHM_highspin}{0.50023}{GW200115_combined_highspin}{0.20793}{GW200115_NSBH_lowspin}{0.20229}}}
\newcommand{\geocenttimeminus}[1]{\IfEqCase{#1}{{GW200105_XPHM_lowspin}{0.0}{GW200105_v4PHM_lowspin}{0.0}{GW200105_combined_lowspin}{0.0}{GW200105_XPHM_highspin}{0.0}{GW200105_v4PHM_highspin}{0.0}{GW200105_combined_highspin}{0.0}{GW200105_NSBH_lowspin}{0.0}{GW200115_XPHM_lowspin}{0.0}{GW200115_v4PHM_lowspin}{0.0}{GW200115_combined_lowspin}{0.0}{GW200115_XPHM_highspin}{0.0}{GW200115_v4PHM_highspin}{0.0}{GW200115_combined_highspin}{0.0}{GW200115_NSBH_lowspin}{0.0}}}
\newcommand{\geocenttimemed}[1]{\IfEqCase{#1}{{GW200105_XPHM_lowspin}{1262276684.0}{GW200105_v4PHM_lowspin}{1262276684.0}{GW200105_combined_lowspin}{1262276684.0}{GW200105_XPHM_highspin}{1262276684.1}{GW200105_v4PHM_highspin}{1262276684.0}{GW200105_combined_highspin}{1262276684.0}{GW200105_NSBH_lowspin}{1262276684.0}{GW200115_XPHM_lowspin}{1263097407.8}{GW200115_v4PHM_lowspin}{1263097407.8}{GW200115_combined_lowspin}{1263097407.8}{GW200115_XPHM_highspin}{1263097407.8}{GW200115_v4PHM_highspin}{1263097407.8}{GW200115_combined_highspin}{1263097407.8}{GW200115_NSBH_lowspin}{1263097407.8}}}
\newcommand{\geocenttimeplus}[1]{\IfEqCase{#1}{{GW200105_XPHM_lowspin}{0.0}{GW200105_v4PHM_lowspin}{0.0}{GW200105_combined_lowspin}{0.0}{GW200105_XPHM_highspin}{0.0}{GW200105_v4PHM_highspin}{0.0}{GW200105_combined_highspin}{0.0}{GW200105_NSBH_lowspin}{0.0}{GW200115_XPHM_lowspin}{0.0}{GW200115_v4PHM_lowspin}{0.0}{GW200115_combined_lowspin}{0.0}{GW200115_XPHM_highspin}{0.0}{GW200115_v4PHM_highspin}{0.0}{GW200115_combined_highspin}{0.0}{GW200115_NSBH_lowspin}{0.0}}}
\newcommand{\geocenttimeonepercent}[1]{\IfEqCase{#1}{{GW200105_XPHM_lowspin}{1262276684.0}{GW200105_v4PHM_lowspin}{1262276684.0}{GW200105_combined_lowspin}{1262276684.0}{GW200105_XPHM_highspin}{1262276684.0}{GW200105_v4PHM_highspin}{1262276684.0}{GW200105_combined_highspin}{1262276684.0}{GW200105_NSBH_lowspin}{1262276684.0}{GW200115_XPHM_lowspin}{1263097407.7}{GW200115_v4PHM_lowspin}{1263097407.8}{GW200115_combined_lowspin}{1263097407.7}{GW200115_XPHM_highspin}{1263097407.7}{GW200115_v4PHM_highspin}{1263097407.8}{GW200115_combined_highspin}{1263097407.7}{GW200115_NSBH_lowspin}{1263097407.7}}}
\newcommand{\geocenttimeninetyninepercent}[1]{\IfEqCase{#1}{{GW200105_XPHM_lowspin}{1262276684.1}{GW200105_v4PHM_lowspin}{1262276684.0}{GW200105_combined_lowspin}{1262276684.1}{GW200105_XPHM_highspin}{1262276684.1}{GW200105_v4PHM_highspin}{1262276684.0}{GW200105_combined_highspin}{1262276684.1}{GW200105_NSBH_lowspin}{1262276684.1}{GW200115_XPHM_lowspin}{1263097407.8}{GW200115_v4PHM_lowspin}{1263097407.8}{GW200115_combined_lowspin}{1263097407.8}{GW200115_XPHM_highspin}{1263097407.8}{GW200115_v4PHM_highspin}{1263097407.8}{GW200115_combined_highspin}{1263097407.8}{GW200115_NSBH_lowspin}{1263097407.8}}}
\newcommand{\geocenttimefivepercent}[1]{\IfEqCase{#1}{{GW200105_XPHM_lowspin}{1262276684.0}{GW200105_v4PHM_lowspin}{1262276684.0}{GW200105_combined_lowspin}{1262276684.0}{GW200105_XPHM_highspin}{1262276684.0}{GW200105_v4PHM_highspin}{1262276684.0}{GW200105_combined_highspin}{1262276684.0}{GW200105_NSBH_lowspin}{1262276684.0}{GW200115_XPHM_lowspin}{1263097407.7}{GW200115_v4PHM_lowspin}{1263097407.8}{GW200115_combined_lowspin}{1263097407.7}{GW200115_XPHM_highspin}{1263097407.7}{GW200115_v4PHM_highspin}{1263097407.8}{GW200115_combined_highspin}{1263097407.7}{GW200115_NSBH_lowspin}{1263097407.7}}}
\newcommand{\geocenttimeninetyfivepercent}[1]{\IfEqCase{#1}{{GW200105_XPHM_lowspin}{1262276684.1}{GW200105_v4PHM_lowspin}{1262276684.0}{GW200105_combined_lowspin}{1262276684.1}{GW200105_XPHM_highspin}{1262276684.1}{GW200105_v4PHM_highspin}{1262276684.0}{GW200105_combined_highspin}{1262276684.1}{GW200105_NSBH_lowspin}{1262276684.1}{GW200115_XPHM_lowspin}{1263097407.8}{GW200115_v4PHM_lowspin}{1263097407.8}{GW200115_combined_lowspin}{1263097407.8}{GW200115_XPHM_highspin}{1263097407.8}{GW200115_v4PHM_highspin}{1263097407.8}{GW200115_combined_highspin}{1263097407.8}{GW200115_NSBH_lowspin}{1263097407.8}}}
\newcommand{\geocenttimeninetypercent}[1]{\IfEqCase{#1}{{GW200105_XPHM_lowspin}{1262276684.1}{GW200105_v4PHM_lowspin}{1262276684.0}{GW200105_combined_lowspin}{1262276684.1}{GW200105_XPHM_highspin}{1262276684.1}{GW200105_v4PHM_highspin}{1262276684.0}{GW200105_combined_highspin}{1262276684.1}{GW200105_NSBH_lowspin}{1262276684.1}{GW200115_XPHM_lowspin}{1263097407.8}{GW200115_v4PHM_lowspin}{1263097407.8}{GW200115_combined_lowspin}{1263097407.8}{GW200115_XPHM_highspin}{1263097407.8}{GW200115_v4PHM_highspin}{1263097407.8}{GW200115_combined_highspin}{1263097407.8}{GW200115_NSBH_lowspin}{1263097407.8}}}
\newcommand{\totalmassdetminus}[1]{\IfEqCase{#1}{{GW200105_XPHM_lowspin}{1.3}{GW200105_v4PHM_lowspin}{0.9}{GW200105_combined_lowspin}{1.1}{GW200105_XPHM_highspin}{1.4}{GW200105_v4PHM_highspin}{1.0}{GW200105_combined_highspin}{1.2}{GW200105_NSBH_lowspin}{1.8}{GW200115_XPHM_lowspin}{1.2}{GW200115_v4PHM_lowspin}{2.0}{GW200115_combined_lowspin}{1.6}{GW200115_XPHM_highspin}{1.3}{GW200115_v4PHM_highspin}{1.6}{GW200115_combined_highspin}{1.5}{GW200115_NSBH_lowspin}{1.5}}}
\newcommand{\totalmassdetmed}[1]{\IfEqCase{#1}{{GW200105_XPHM_lowspin}{11.5}{GW200105_v4PHM_lowspin}{11.5}{GW200105_combined_lowspin}{11.5}{GW200105_XPHM_highspin}{11.5}{GW200105_v4PHM_highspin}{11.5}{GW200105_combined_highspin}{11.5}{GW200105_NSBH_lowspin}{11.4}{GW200115_XPHM_lowspin}{7.5}{GW200115_v4PHM_lowspin}{8.2}{GW200115_combined_lowspin}{7.8}{GW200115_XPHM_highspin}{7.5}{GW200115_v4PHM_highspin}{7.7}{GW200115_combined_highspin}{7.6}{GW200115_NSBH_lowspin}{8.4}}}
\newcommand{\totalmassdetplus}[1]{\IfEqCase{#1}{{GW200105_XPHM_lowspin}{1.1}{GW200105_v4PHM_lowspin}{0.8}{GW200105_combined_lowspin}{1.0}{GW200105_XPHM_highspin}{1.3}{GW200105_v4PHM_highspin}{0.7}{GW200105_combined_highspin}{1.1}{GW200105_NSBH_lowspin}{2.9}{GW200115_XPHM_lowspin}{1.6}{GW200115_v4PHM_lowspin}{0.9}{GW200115_combined_lowspin}{1.3}{GW200115_XPHM_highspin}{1.4}{GW200115_v4PHM_highspin}{1.7}{GW200115_combined_highspin}{1.6}{GW200115_NSBH_lowspin}{1.6}}}
\newcommand{\totalmassdetonepercent}[1]{\IfEqCase{#1}{{GW200105_XPHM_lowspin}{9.3}{GW200105_v4PHM_lowspin}{10.2}{GW200105_combined_lowspin}{9.6}{GW200105_XPHM_highspin}{9.1}{GW200105_v4PHM_highspin}{9.0}{GW200105_combined_highspin}{9.1}{GW200105_NSBH_lowspin}{9.1}{GW200115_XPHM_lowspin}{6.1}{GW200115_v4PHM_lowspin}{6.1}{GW200115_combined_lowspin}{6.1}{GW200115_XPHM_highspin}{6.0}{GW200115_v4PHM_highspin}{5.9}{GW200115_combined_highspin}{6.0}{GW200115_NSBH_lowspin}{6.5}}}
\newcommand{\totalmassdetninetyninepercent}[1]{\IfEqCase{#1}{{GW200105_XPHM_lowspin}{13.4}{GW200105_v4PHM_lowspin}{13.5}{GW200105_combined_lowspin}{13.5}{GW200105_XPHM_highspin}{13.8}{GW200105_v4PHM_highspin}{13.4}{GW200105_combined_highspin}{13.7}{GW200105_NSBH_lowspin}{16.3}{GW200115_XPHM_lowspin}{10.3}{GW200115_v4PHM_lowspin}{9.8}{GW200115_combined_lowspin}{10.1}{GW200115_XPHM_highspin}{9.7}{GW200115_v4PHM_highspin}{10.0}{GW200115_combined_highspin}{10.0}{GW200115_NSBH_lowspin}{11.0}}}
\newcommand{\totalmassdetfivepercent}[1]{\IfEqCase{#1}{{GW200105_XPHM_lowspin}{10.2}{GW200105_v4PHM_lowspin}{10.6}{GW200105_combined_lowspin}{10.4}{GW200105_XPHM_highspin}{10.1}{GW200105_v4PHM_highspin}{10.5}{GW200105_combined_highspin}{10.2}{GW200105_NSBH_lowspin}{9.6}{GW200115_XPHM_lowspin}{6.3}{GW200115_v4PHM_lowspin}{6.2}{GW200115_combined_lowspin}{6.2}{GW200115_XPHM_highspin}{6.2}{GW200115_v4PHM_highspin}{6.1}{GW200115_combined_highspin}{6.1}{GW200115_NSBH_lowspin}{6.9}}}
\newcommand{\totalmassdetninetyfivepercent}[1]{\IfEqCase{#1}{{GW200105_XPHM_lowspin}{12.6}{GW200105_v4PHM_lowspin}{12.4}{GW200105_combined_lowspin}{12.5}{GW200105_XPHM_highspin}{12.8}{GW200105_v4PHM_highspin}{12.2}{GW200105_combined_highspin}{12.6}{GW200105_NSBH_lowspin}{14.3}{GW200115_XPHM_lowspin}{9.1}{GW200115_v4PHM_lowspin}{9.1}{GW200115_combined_lowspin}{9.1}{GW200115_XPHM_highspin}{9.0}{GW200115_v4PHM_highspin}{9.4}{GW200115_combined_highspin}{9.2}{GW200115_NSBH_lowspin}{10.1}}}
\newcommand{\totalmassdetninetypercent}[1]{\IfEqCase{#1}{{GW200105_XPHM_lowspin}{12.2}{GW200105_v4PHM_lowspin}{12.1}{GW200105_combined_lowspin}{12.2}{GW200105_XPHM_highspin}{12.4}{GW200105_v4PHM_highspin}{11.9}{GW200105_combined_highspin}{12.2}{GW200105_NSBH_lowspin}{13.3}{GW200115_XPHM_lowspin}{8.8}{GW200115_v4PHM_lowspin}{8.9}{GW200115_combined_lowspin}{8.9}{GW200115_XPHM_highspin}{8.7}{GW200115_v4PHM_highspin}{9.1}{GW200115_combined_highspin}{8.9}{GW200115_NSBH_lowspin}{9.6}}}
\newcommand{\massonedetminus}[1]{\IfEqCase{#1}{{GW200105_XPHM_lowspin}{1.6}{GW200105_v4PHM_lowspin}{1.1}{GW200105_combined_lowspin}{1.3}{GW200105_XPHM_highspin}{1.8}{GW200105_v4PHM_highspin}{1.2}{GW200105_combined_highspin}{1.5}{GW200105_NSBH_lowspin}{2.4}{GW200115_XPHM_lowspin}{1.8}{GW200115_v4PHM_lowspin}{2.8}{GW200115_combined_lowspin}{2.3}{GW200115_XPHM_highspin}{2.0}{GW200115_v4PHM_highspin}{2.5}{GW200115_combined_highspin}{2.2}{GW200115_NSBH_lowspin}{1.9}}}
\newcommand{\massonedetmed}[1]{\IfEqCase{#1}{{GW200105_XPHM_lowspin}{9.4}{GW200105_v4PHM_lowspin}{9.5}{GW200105_combined_lowspin}{9.5}{GW200105_XPHM_highspin}{9.5}{GW200105_v4PHM_highspin}{9.5}{GW200105_combined_highspin}{9.5}{GW200105_NSBH_lowspin}{9.3}{GW200115_XPHM_lowspin}{5.9}{GW200115_v4PHM_lowspin}{6.7}{GW200115_combined_lowspin}{6.3}{GW200115_XPHM_highspin}{5.9}{GW200115_v4PHM_highspin}{6.2}{GW200115_combined_highspin}{6.0}{GW200115_NSBH_lowspin}{7.0}}}
\newcommand{\massonedetplus}[1]{\IfEqCase{#1}{{GW200105_XPHM_lowspin}{1.3}{GW200105_v4PHM_lowspin}{1.0}{GW200105_combined_lowspin}{1.1}{GW200105_XPHM_highspin}{1.5}{GW200105_v4PHM_highspin}{0.8}{GW200105_combined_highspin}{1.3}{GW200105_NSBH_lowspin}{3.3}{GW200115_XPHM_lowspin}{1.9}{GW200115_v4PHM_lowspin}{1.0}{GW200115_combined_lowspin}{1.5}{GW200115_XPHM_highspin}{1.7}{GW200115_v4PHM_highspin}{1.9}{GW200115_combined_highspin}{1.9}{GW200115_NSBH_lowspin}{1.9}}}
\newcommand{\massonedetonepercent}[1]{\IfEqCase{#1}{{GW200105_XPHM_lowspin}{6.6}{GW200105_v4PHM_lowspin}{7.9}{GW200105_combined_lowspin}{7.1}{GW200105_XPHM_highspin}{6.3}{GW200105_v4PHM_highspin}{6.1}{GW200105_combined_highspin}{6.2}{GW200105_NSBH_lowspin}{6.3}{GW200115_XPHM_lowspin}{3.7}{GW200115_v4PHM_lowspin}{3.6}{GW200115_combined_lowspin}{3.7}{GW200115_XPHM_highspin}{3.5}{GW200115_v4PHM_highspin}{3.2}{GW200115_combined_highspin}{3.3}{GW200115_NSBH_lowspin}{4.6}}}
\newcommand{\massonedetninetyninepercent}[1]{\IfEqCase{#1}{{GW200105_XPHM_lowspin}{11.7}{GW200105_v4PHM_lowspin}{11.8}{GW200105_combined_lowspin}{11.8}{GW200105_XPHM_highspin}{12.1}{GW200105_v4PHM_highspin}{11.6}{GW200105_combined_highspin}{12.0}{GW200105_NSBH_lowspin}{14.8}{GW200115_XPHM_lowspin}{9.2}{GW200115_v4PHM_lowspin}{8.6}{GW200115_combined_lowspin}{8.9}{GW200115_XPHM_highspin}{8.5}{GW200115_v4PHM_highspin}{8.9}{GW200115_combined_highspin}{8.8}{GW200115_NSBH_lowspin}{9.9}}}
\newcommand{\massonedetfivepercent}[1]{\IfEqCase{#1}{{GW200105_XPHM_lowspin}{7.8}{GW200105_v4PHM_lowspin}{8.4}{GW200105_combined_lowspin}{8.1}{GW200105_XPHM_highspin}{7.7}{GW200105_v4PHM_highspin}{8.3}{GW200105_combined_highspin}{7.9}{GW200105_NSBH_lowspin}{7.0}{GW200115_XPHM_lowspin}{4.1}{GW200115_v4PHM_lowspin}{3.9}{GW200115_combined_lowspin}{4.0}{GW200115_XPHM_highspin}{4.0}{GW200115_v4PHM_highspin}{3.7}{GW200115_combined_highspin}{3.8}{GW200115_NSBH_lowspin}{5.1}}}
\newcommand{\massonedetninetyfivepercent}[1]{\IfEqCase{#1}{{GW200105_XPHM_lowspin}{10.7}{GW200105_v4PHM_lowspin}{10.5}{GW200105_combined_lowspin}{10.6}{GW200105_XPHM_highspin}{11.0}{GW200105_v4PHM_highspin}{10.2}{GW200105_combined_highspin}{10.8}{GW200105_NSBH_lowspin}{12.6}{GW200115_XPHM_lowspin}{7.8}{GW200115_v4PHM_lowspin}{7.7}{GW200115_combined_lowspin}{7.8}{GW200115_XPHM_highspin}{7.7}{GW200115_v4PHM_highspin}{8.1}{GW200115_combined_highspin}{8.0}{GW200115_NSBH_lowspin}{8.9}}}
\newcommand{\massonedetninetypercent}[1]{\IfEqCase{#1}{{GW200105_XPHM_lowspin}{10.3}{GW200105_v4PHM_lowspin}{10.2}{GW200105_combined_lowspin}{10.2}{GW200105_XPHM_highspin}{10.6}{GW200105_v4PHM_highspin}{10.0}{GW200105_combined_highspin}{10.3}{GW200105_NSBH_lowspin}{11.5}{GW200115_XPHM_lowspin}{7.4}{GW200115_v4PHM_lowspin}{7.6}{GW200115_combined_lowspin}{7.6}{GW200115_XPHM_highspin}{7.4}{GW200115_v4PHM_highspin}{7.8}{GW200115_combined_highspin}{7.6}{GW200115_NSBH_lowspin}{8.3}}}
\newcommand{\masstwodetminus}[1]{\IfEqCase{#1}{{GW200105_XPHM_lowspin}{0.2}{GW200105_v4PHM_lowspin}{0.1}{GW200105_combined_lowspin}{0.2}{GW200105_XPHM_highspin}{0.2}{GW200105_v4PHM_highspin}{0.1}{GW200105_combined_highspin}{0.2}{GW200105_NSBH_lowspin}{0.4}{GW200115_XPHM_lowspin}{0.3}{GW200115_v4PHM_lowspin}{0.1}{GW200115_combined_lowspin}{0.2}{GW200115_XPHM_highspin}{0.3}{GW200115_v4PHM_highspin}{0.3}{GW200115_combined_highspin}{0.3}{GW200115_NSBH_lowspin}{0.2}}}
\newcommand{\masstwodetmed}[1]{\IfEqCase{#1}{{GW200105_XPHM_lowspin}{2.0}{GW200105_v4PHM_lowspin}{2.0}{GW200105_combined_lowspin}{2.0}{GW200105_XPHM_highspin}{2.0}{GW200105_v4PHM_highspin}{2.0}{GW200105_combined_highspin}{2.0}{GW200105_NSBH_lowspin}{2.1}{GW200115_XPHM_lowspin}{1.6}{GW200115_v4PHM_lowspin}{1.5}{GW200115_combined_lowspin}{1.5}{GW200115_XPHM_highspin}{1.6}{GW200115_v4PHM_highspin}{1.6}{GW200115_combined_highspin}{1.6}{GW200115_NSBH_lowspin}{1.4}}}
\newcommand{\masstwodetplus}[1]{\IfEqCase{#1}{{GW200105_XPHM_lowspin}{0.3}{GW200105_v4PHM_lowspin}{0.2}{GW200105_combined_lowspin}{0.2}{GW200105_XPHM_highspin}{0.3}{GW200105_v4PHM_highspin}{0.2}{GW200105_combined_highspin}{0.3}{GW200105_NSBH_lowspin}{0.5}{GW200115_XPHM_lowspin}{0.5}{GW200115_v4PHM_lowspin}{0.8}{GW200115_combined_lowspin}{0.7}{GW200115_XPHM_highspin}{0.6}{GW200115_v4PHM_highspin}{0.8}{GW200115_combined_highspin}{0.8}{GW200115_NSBH_lowspin}{0.4}}}
\newcommand{\masstwodetonepercent}[1]{\IfEqCase{#1}{{GW200105_XPHM_lowspin}{1.7}{GW200105_v4PHM_lowspin}{1.7}{GW200105_combined_lowspin}{1.7}{GW200105_XPHM_highspin}{1.7}{GW200105_v4PHM_highspin}{1.8}{GW200105_combined_highspin}{1.7}{GW200105_NSBH_lowspin}{1.5}{GW200115_XPHM_lowspin}{1.2}{GW200115_v4PHM_lowspin}{1.2}{GW200115_combined_lowspin}{1.2}{GW200115_XPHM_highspin}{1.2}{GW200115_v4PHM_highspin}{1.2}{GW200115_combined_highspin}{1.2}{GW200115_NSBH_lowspin}{1.1}}}
\newcommand{\masstwodetninetyninepercent}[1]{\IfEqCase{#1}{{GW200105_XPHM_lowspin}{2.7}{GW200105_v4PHM_lowspin}{2.3}{GW200105_combined_lowspin}{2.5}{GW200105_XPHM_highspin}{2.8}{GW200105_v4PHM_highspin}{2.9}{GW200105_combined_highspin}{2.8}{GW200105_NSBH_lowspin}{2.8}{GW200115_XPHM_lowspin}{2.4}{GW200115_v4PHM_lowspin}{2.4}{GW200115_combined_lowspin}{2.4}{GW200115_XPHM_highspin}{2.5}{GW200115_v4PHM_highspin}{2.7}{GW200115_combined_highspin}{2.6}{GW200115_NSBH_lowspin}{2.0}}}
\newcommand{\masstwodetfivepercent}[1]{\IfEqCase{#1}{{GW200105_XPHM_lowspin}{1.9}{GW200105_v4PHM_lowspin}{1.9}{GW200105_combined_lowspin}{1.9}{GW200105_XPHM_highspin}{1.8}{GW200105_v4PHM_highspin}{1.9}{GW200105_combined_highspin}{1.8}{GW200105_NSBH_lowspin}{1.6}{GW200115_XPHM_lowspin}{1.3}{GW200115_v4PHM_lowspin}{1.3}{GW200115_combined_lowspin}{1.3}{GW200115_XPHM_highspin}{1.3}{GW200115_v4PHM_highspin}{1.3}{GW200115_combined_highspin}{1.3}{GW200115_NSBH_lowspin}{1.2}}}
\newcommand{\masstwodetninetyfivepercent}[1]{\IfEqCase{#1}{{GW200105_XPHM_lowspin}{2.4}{GW200105_v4PHM_lowspin}{2.2}{GW200105_combined_lowspin}{2.3}{GW200105_XPHM_highspin}{2.4}{GW200105_v4PHM_highspin}{2.3}{GW200105_combined_highspin}{2.3}{GW200105_NSBH_lowspin}{2.6}{GW200115_XPHM_lowspin}{2.2}{GW200115_v4PHM_lowspin}{2.3}{GW200115_combined_lowspin}{2.2}{GW200115_XPHM_highspin}{2.2}{GW200115_v4PHM_highspin}{2.4}{GW200115_combined_highspin}{2.3}{GW200115_NSBH_lowspin}{1.8}}}
\newcommand{\masstwodetninetypercent}[1]{\IfEqCase{#1}{{GW200105_XPHM_lowspin}{2.2}{GW200105_v4PHM_lowspin}{2.2}{GW200105_combined_lowspin}{2.2}{GW200105_XPHM_highspin}{2.3}{GW200105_v4PHM_highspin}{2.1}{GW200105_combined_highspin}{2.2}{GW200105_NSBH_lowspin}{2.5}{GW200115_XPHM_lowspin}{2.0}{GW200115_v4PHM_lowspin}{2.1}{GW200115_combined_lowspin}{2.0}{GW200115_XPHM_highspin}{2.0}{GW200115_v4PHM_highspin}{2.3}{GW200115_combined_highspin}{2.2}{GW200115_NSBH_lowspin}{1.7}}}
\newcommand{\thetajnminus}[1]{\IfEqCase{#1}{{GW200105_XPHM_lowspin}{1.75}{GW200105_v4PHM_lowspin}{1.47}{GW200105_combined_lowspin}{1.63}{GW200105_XPHM_highspin}{1.51}{GW200105_v4PHM_highspin}{1.44}{GW200105_combined_highspin}{1.46}{GW200105_NSBH_lowspin}{1.35}{GW200115_XPHM_lowspin}{0.37}{GW200115_v4PHM_lowspin}{0.54}{GW200115_combined_lowspin}{0.43}{GW200115_XPHM_highspin}{0.35}{GW200115_v4PHM_highspin}{0.55}{GW200115_combined_highspin}{0.42}{GW200115_NSBH_lowspin}{0.53}}}
\newcommand{\thetajnmed}[1]{\IfEqCase{#1}{{GW200105_XPHM_lowspin}{2.11}{GW200105_v4PHM_lowspin}{1.79}{GW200105_combined_lowspin}{1.97}{GW200105_XPHM_highspin}{1.85}{GW200105_v4PHM_highspin}{1.76}{GW200105_combined_highspin}{1.79}{GW200105_NSBH_lowspin}{1.61}{GW200115_XPHM_lowspin}{0.55}{GW200115_v4PHM_lowspin}{0.72}{GW200115_combined_lowspin}{0.61}{GW200115_XPHM_highspin}{0.53}{GW200115_v4PHM_highspin}{0.74}{GW200115_combined_highspin}{0.61}{GW200115_NSBH_lowspin}{0.72}}}
\newcommand{\thetajnplus}[1]{\IfEqCase{#1}{{GW200105_XPHM_lowspin}{0.71}{GW200105_v4PHM_lowspin}{1.05}{GW200105_combined_lowspin}{0.86}{GW200105_XPHM_highspin}{0.97}{GW200105_v4PHM_highspin}{1.08}{GW200105_combined_highspin}{1.04}{GW200105_NSBH_lowspin}{1.28}{GW200115_XPHM_lowspin}{1.81}{GW200115_v4PHM_lowspin}{2.03}{GW200115_combined_lowspin}{2.03}{GW200115_XPHM_highspin}{1.58}{GW200115_v4PHM_highspin}{2.01}{GW200115_combined_highspin}{2.03}{GW200115_NSBH_lowspin}{1.94}}}
\newcommand{\thetajnonepercent}[1]{\IfEqCase{#1}{{GW200105_XPHM_lowspin}{0.18}{GW200105_v4PHM_lowspin}{0.15}{GW200105_combined_lowspin}{0.16}{GW200105_XPHM_highspin}{0.16}{GW200105_v4PHM_highspin}{0.14}{GW200105_combined_highspin}{0.15}{GW200105_NSBH_lowspin}{0.11}{GW200115_XPHM_lowspin}{0.08}{GW200115_v4PHM_lowspin}{0.08}{GW200115_combined_lowspin}{0.08}{GW200115_XPHM_highspin}{0.09}{GW200115_v4PHM_highspin}{0.09}{GW200115_combined_highspin}{0.08}{GW200115_NSBH_lowspin}{0.08}}}
\newcommand{\thetajnninetyninepercent}[1]{\IfEqCase{#1}{{GW200105_XPHM_lowspin}{2.97}{GW200105_v4PHM_lowspin}{3.01}{GW200105_combined_lowspin}{2.99}{GW200105_XPHM_highspin}{2.98}{GW200105_v4PHM_highspin}{3.01}{GW200105_combined_highspin}{3.00}{GW200105_NSBH_lowspin}{3.02}{GW200115_XPHM_lowspin}{2.82}{GW200115_v4PHM_lowspin}{2.97}{GW200115_combined_lowspin}{2.92}{GW200115_XPHM_highspin}{2.81}{GW200115_v4PHM_highspin}{2.98}{GW200115_combined_highspin}{2.93}{GW200115_NSBH_lowspin}{2.94}}}
\newcommand{\thetajnfivepercent}[1]{\IfEqCase{#1}{{GW200105_XPHM_lowspin}{0.36}{GW200105_v4PHM_lowspin}{0.32}{GW200105_combined_lowspin}{0.34}{GW200105_XPHM_highspin}{0.34}{GW200105_v4PHM_highspin}{0.32}{GW200105_combined_highspin}{0.33}{GW200105_NSBH_lowspin}{0.26}{GW200115_XPHM_lowspin}{0.18}{GW200115_v4PHM_lowspin}{0.18}{GW200115_combined_lowspin}{0.18}{GW200115_XPHM_highspin}{0.18}{GW200115_v4PHM_highspin}{0.19}{GW200115_combined_highspin}{0.18}{GW200115_NSBH_lowspin}{0.19}}}
\newcommand{\thetajnninetyfivepercent}[1]{\IfEqCase{#1}{{GW200105_XPHM_lowspin}{2.82}{GW200105_v4PHM_lowspin}{2.84}{GW200105_combined_lowspin}{2.83}{GW200105_XPHM_highspin}{2.82}{GW200105_v4PHM_highspin}{2.85}{GW200105_combined_highspin}{2.83}{GW200105_NSBH_lowspin}{2.89}{GW200115_XPHM_lowspin}{2.36}{GW200115_v4PHM_lowspin}{2.75}{GW200115_combined_lowspin}{2.65}{GW200115_XPHM_highspin}{2.12}{GW200115_v4PHM_highspin}{2.75}{GW200115_combined_highspin}{2.64}{GW200115_NSBH_lowspin}{2.66}}}
\newcommand{\thetajnninetypercent}[1]{\IfEqCase{#1}{{GW200105_XPHM_lowspin}{2.71}{GW200105_v4PHM_lowspin}{2.72}{GW200105_combined_lowspin}{2.72}{GW200105_XPHM_highspin}{2.71}{GW200105_v4PHM_highspin}{2.72}{GW200105_combined_highspin}{2.72}{GW200105_NSBH_lowspin}{2.78}{GW200115_XPHM_lowspin}{1.06}{GW200115_v4PHM_lowspin}{2.55}{GW200115_combined_lowspin}{2.35}{GW200115_XPHM_highspin}{0.97}{GW200115_v4PHM_highspin}{2.56}{GW200115_combined_highspin}{2.33}{GW200115_NSBH_lowspin}{2.37}}}
\newcommand{\symmetricmassratiominus}[1]{\IfEqCase{#1}{{GW200105_XPHM_lowspin}{0.02}{GW200105_v4PHM_lowspin}{0.02}{GW200105_combined_lowspin}{0.02}{GW200105_XPHM_highspin}{0.02}{GW200105_v4PHM_highspin}{0.01}{GW200105_combined_highspin}{0.02}{GW200105_NSBH_lowspin}{0.05}{GW200115_XPHM_lowspin}{0.05}{GW200115_v4PHM_lowspin}{0.02}{GW200115_combined_lowspin}{0.03}{GW200115_XPHM_highspin}{0.04}{GW200115_v4PHM_highspin}{0.04}{GW200115_combined_highspin}{0.04}{GW200115_NSBH_lowspin}{0.04}}}
\newcommand{\symmetricmassratiomed}[1]{\IfEqCase{#1}{{GW200105_XPHM_lowspin}{0.15}{GW200105_v4PHM_lowspin}{0.15}{GW200105_combined_lowspin}{0.15}{GW200105_XPHM_highspin}{0.15}{GW200105_v4PHM_highspin}{0.15}{GW200105_combined_highspin}{0.15}{GW200105_NSBH_lowspin}{0.15}{GW200115_XPHM_lowspin}{0.17}{GW200115_v4PHM_lowspin}{0.15}{GW200115_combined_lowspin}{0.16}{GW200115_XPHM_highspin}{0.17}{GW200115_v4PHM_highspin}{0.16}{GW200115_combined_highspin}{0.16}{GW200115_NSBH_lowspin}{0.14}}}
\newcommand{\symmetricmassratioplus}[1]{\IfEqCase{#1}{{GW200105_XPHM_lowspin}{0.03}{GW200105_v4PHM_lowspin}{0.02}{GW200105_combined_lowspin}{0.03}{GW200105_XPHM_highspin}{0.03}{GW200105_v4PHM_highspin}{0.02}{GW200105_combined_highspin}{0.03}{GW200105_NSBH_lowspin}{0.05}{GW200115_XPHM_lowspin}{0.06}{GW200115_v4PHM_lowspin}{0.09}{GW200115_combined_lowspin}{0.07}{GW200115_XPHM_highspin}{0.06}{GW200115_v4PHM_highspin}{0.08}{GW200115_combined_highspin}{0.07}{GW200115_NSBH_lowspin}{0.05}}}
\newcommand{\symmetricmassratioonepercent}[1]{\IfEqCase{#1}{{GW200105_XPHM_lowspin}{0.11}{GW200105_v4PHM_lowspin}{0.11}{GW200105_combined_lowspin}{0.11}{GW200105_XPHM_highspin}{0.11}{GW200105_v4PHM_highspin}{0.11}{GW200105_combined_highspin}{0.11}{GW200105_NSBH_lowspin}{0.08}{GW200115_XPHM_lowspin}{0.10}{GW200115_v4PHM_lowspin}{0.11}{GW200115_combined_lowspin}{0.10}{GW200115_XPHM_highspin}{0.11}{GW200115_v4PHM_highspin}{0.10}{GW200115_combined_highspin}{0.11}{GW200115_NSBH_lowspin}{0.09}}}
\newcommand{\symmetricmassrationinetyninepercent}[1]{\IfEqCase{#1}{{GW200105_XPHM_lowspin}{0.21}{GW200105_v4PHM_lowspin}{0.18}{GW200105_combined_lowspin}{0.19}{GW200105_XPHM_highspin}{0.21}{GW200105_v4PHM_highspin}{0.22}{GW200105_combined_highspin}{0.22}{GW200105_NSBH_lowspin}{0.21}{GW200115_XPHM_lowspin}{0.24}{GW200115_v4PHM_lowspin}{0.24}{GW200115_combined_lowspin}{0.24}{GW200115_XPHM_highspin}{0.24}{GW200115_v4PHM_highspin}{0.25}{GW200115_combined_highspin}{0.25}{GW200115_NSBH_lowspin}{0.21}}}
\newcommand{\symmetricmassratiofivepercent}[1]{\IfEqCase{#1}{{GW200105_XPHM_lowspin}{0.13}{GW200105_v4PHM_lowspin}{0.13}{GW200105_combined_lowspin}{0.13}{GW200105_XPHM_highspin}{0.12}{GW200105_v4PHM_highspin}{0.13}{GW200105_combined_highspin}{0.12}{GW200105_NSBH_lowspin}{0.10}{GW200115_XPHM_lowspin}{0.12}{GW200115_v4PHM_lowspin}{0.12}{GW200115_combined_lowspin}{0.12}{GW200115_XPHM_highspin}{0.13}{GW200115_v4PHM_highspin}{0.12}{GW200115_combined_highspin}{0.12}{GW200115_NSBH_lowspin}{0.10}}}
\newcommand{\symmetricmassrationinetyfivepercent}[1]{\IfEqCase{#1}{{GW200105_XPHM_lowspin}{0.18}{GW200105_v4PHM_lowspin}{0.17}{GW200105_combined_lowspin}{0.17}{GW200105_XPHM_highspin}{0.18}{GW200105_v4PHM_highspin}{0.17}{GW200105_combined_highspin}{0.18}{GW200105_NSBH_lowspin}{0.20}{GW200115_XPHM_lowspin}{0.22}{GW200115_v4PHM_lowspin}{0.23}{GW200115_combined_lowspin}{0.23}{GW200115_XPHM_highspin}{0.23}{GW200115_v4PHM_highspin}{0.24}{GW200115_combined_highspin}{0.24}{GW200115_NSBH_lowspin}{0.19}}}
\newcommand{\symmetricmassrationinetypercent}[1]{\IfEqCase{#1}{{GW200105_XPHM_lowspin}{0.17}{GW200105_v4PHM_lowspin}{0.16}{GW200105_combined_lowspin}{0.16}{GW200105_XPHM_highspin}{0.17}{GW200105_v4PHM_highspin}{0.16}{GW200105_combined_highspin}{0.16}{GW200105_NSBH_lowspin}{0.19}{GW200115_XPHM_lowspin}{0.21}{GW200115_v4PHM_lowspin}{0.22}{GW200115_combined_lowspin}{0.22}{GW200115_XPHM_highspin}{0.22}{GW200115_v4PHM_highspin}{0.23}{GW200115_combined_highspin}{0.22}{GW200115_NSBH_lowspin}{0.18}}}
\newcommand{\phioneminus}[1]{\IfEqCase{#1}{{GW200105_XPHM_lowspin}{2.81}{GW200105_v4PHM_lowspin}{2.81}{GW200105_combined_lowspin}{2.82}{GW200105_XPHM_highspin}{2.93}{GW200105_v4PHM_highspin}{2.72}{GW200105_combined_highspin}{2.82}{GW200105_NSBH_lowspin}{0.00}{GW200115_XPHM_lowspin}{2.95}{GW200115_v4PHM_lowspin}{2.95}{GW200115_combined_lowspin}{2.97}{GW200115_XPHM_highspin}{2.83}{GW200115_v4PHM_highspin}{2.64}{GW200115_combined_highspin}{2.73}{GW200115_NSBH_lowspin}{0.00}}}
\newcommand{\phionemed}[1]{\IfEqCase{#1}{{GW200105_XPHM_lowspin}{3.07}{GW200105_v4PHM_lowspin}{3.11}{GW200105_combined_lowspin}{3.10}{GW200105_XPHM_highspin}{3.21}{GW200105_v4PHM_highspin}{3.06}{GW200105_combined_highspin}{3.13}{GW200105_NSBH_lowspin}{0.00}{GW200115_XPHM_lowspin}{3.26}{GW200115_v4PHM_lowspin}{3.31}{GW200115_combined_lowspin}{3.30}{GW200115_XPHM_highspin}{3.14}{GW200115_v4PHM_highspin}{2.98}{GW200115_combined_highspin}{3.06}{GW200115_NSBH_lowspin}{0.00}}}
\newcommand{\phioneplus}[1]{\IfEqCase{#1}{{GW200105_XPHM_lowspin}{2.94}{GW200105_v4PHM_lowspin}{2.82}{GW200105_combined_lowspin}{2.88}{GW200105_XPHM_highspin}{2.78}{GW200105_v4PHM_highspin}{2.90}{GW200105_combined_highspin}{2.85}{GW200105_NSBH_lowspin}{0.00}{GW200115_XPHM_lowspin}{2.72}{GW200115_v4PHM_lowspin}{2.68}{GW200115_combined_lowspin}{2.69}{GW200115_XPHM_highspin}{2.84}{GW200115_v4PHM_highspin}{3.00}{GW200115_combined_highspin}{2.92}{GW200115_NSBH_lowspin}{0.00}}}
\newcommand{\phioneonepercent}[1]{\IfEqCase{#1}{{GW200105_XPHM_lowspin}{0.05}{GW200105_v4PHM_lowspin}{0.06}{GW200105_combined_lowspin}{0.05}{GW200105_XPHM_highspin}{0.05}{GW200105_v4PHM_highspin}{0.07}{GW200105_combined_highspin}{0.06}{GW200105_NSBH_lowspin}{0.00}{GW200115_XPHM_lowspin}{0.06}{GW200115_v4PHM_lowspin}{0.08}{GW200115_combined_lowspin}{0.07}{GW200115_XPHM_highspin}{0.06}{GW200115_v4PHM_highspin}{0.07}{GW200115_combined_highspin}{0.07}{GW200115_NSBH_lowspin}{0.00}}}
\newcommand{\phioneninetyninepercent}[1]{\IfEqCase{#1}{{GW200105_XPHM_lowspin}{6.22}{GW200105_v4PHM_lowspin}{6.21}{GW200105_combined_lowspin}{6.22}{GW200105_XPHM_highspin}{6.22}{GW200105_v4PHM_highspin}{6.21}{GW200105_combined_highspin}{6.22}{GW200105_NSBH_lowspin}{0.00}{GW200115_XPHM_lowspin}{6.22}{GW200115_v4PHM_lowspin}{6.23}{GW200115_combined_lowspin}{6.23}{GW200115_XPHM_highspin}{6.23}{GW200115_v4PHM_highspin}{6.23}{GW200115_combined_highspin}{6.22}{GW200115_NSBH_lowspin}{0.00}}}
\newcommand{\phionefivepercent}[1]{\IfEqCase{#1}{{GW200105_XPHM_lowspin}{0.25}{GW200105_v4PHM_lowspin}{0.31}{GW200105_combined_lowspin}{0.27}{GW200105_XPHM_highspin}{0.28}{GW200105_v4PHM_highspin}{0.35}{GW200105_combined_highspin}{0.31}{GW200105_NSBH_lowspin}{0.00}{GW200115_XPHM_lowspin}{0.31}{GW200115_v4PHM_lowspin}{0.37}{GW200115_combined_lowspin}{0.33}{GW200115_XPHM_highspin}{0.31}{GW200115_v4PHM_highspin}{0.34}{GW200115_combined_highspin}{0.33}{GW200115_NSBH_lowspin}{0.00}}}
\newcommand{\phioneninetyfivepercent}[1]{\IfEqCase{#1}{{GW200105_XPHM_lowspin}{6.01}{GW200105_v4PHM_lowspin}{5.93}{GW200105_combined_lowspin}{5.97}{GW200105_XPHM_highspin}{5.99}{GW200105_v4PHM_highspin}{5.96}{GW200105_combined_highspin}{5.98}{GW200105_NSBH_lowspin}{0.00}{GW200115_XPHM_lowspin}{5.99}{GW200115_v4PHM_lowspin}{5.99}{GW200115_combined_lowspin}{5.99}{GW200115_XPHM_highspin}{5.98}{GW200115_v4PHM_highspin}{5.98}{GW200115_combined_highspin}{5.98}{GW200115_NSBH_lowspin}{0.00}}}
\newcommand{\phioneninetypercent}[1]{\IfEqCase{#1}{{GW200105_XPHM_lowspin}{5.73}{GW200105_v4PHM_lowspin}{5.64}{GW200105_combined_lowspin}{5.70}{GW200105_XPHM_highspin}{5.69}{GW200105_v4PHM_highspin}{5.63}{GW200105_combined_highspin}{5.66}{GW200105_NSBH_lowspin}{0.00}{GW200115_XPHM_lowspin}{5.70}{GW200115_v4PHM_lowspin}{5.68}{GW200115_combined_lowspin}{5.69}{GW200115_XPHM_highspin}{5.67}{GW200115_v4PHM_highspin}{5.69}{GW200115_combined_highspin}{5.68}{GW200115_NSBH_lowspin}{0.00}}}
\newcommand{\phitwominus}[1]{\IfEqCase{#1}{{GW200105_XPHM_lowspin}{2.90}{GW200105_v4PHM_lowspin}{2.76}{GW200105_combined_lowspin}{2.82}{GW200105_XPHM_highspin}{2.81}{GW200105_v4PHM_highspin}{2.73}{GW200105_combined_highspin}{2.78}{GW200105_NSBH_lowspin}{0.00}{GW200115_XPHM_lowspin}{2.84}{GW200115_v4PHM_lowspin}{2.83}{GW200115_combined_lowspin}{2.83}{GW200115_XPHM_highspin}{2.78}{GW200115_v4PHM_highspin}{2.82}{GW200115_combined_highspin}{2.79}{GW200115_NSBH_lowspin}{0.00}}}
\newcommand{\phitwomed}[1]{\IfEqCase{#1}{{GW200105_XPHM_lowspin}{3.23}{GW200105_v4PHM_lowspin}{3.07}{GW200105_combined_lowspin}{3.14}{GW200105_XPHM_highspin}{3.12}{GW200105_v4PHM_highspin}{2.99}{GW200105_combined_highspin}{3.07}{GW200105_NSBH_lowspin}{0.00}{GW200115_XPHM_lowspin}{3.18}{GW200115_v4PHM_lowspin}{3.13}{GW200115_combined_lowspin}{3.15}{GW200115_XPHM_highspin}{3.10}{GW200115_v4PHM_highspin}{3.12}{GW200115_combined_highspin}{3.10}{GW200115_NSBH_lowspin}{0.00}}}
\newcommand{\phitwoplus}[1]{\IfEqCase{#1}{{GW200105_XPHM_lowspin}{2.76}{GW200105_v4PHM_lowspin}{2.89}{GW200105_combined_lowspin}{2.83}{GW200105_XPHM_highspin}{2.83}{GW200105_v4PHM_highspin}{2.99}{GW200105_combined_highspin}{2.90}{GW200105_NSBH_lowspin}{0.00}{GW200115_XPHM_lowspin}{2.80}{GW200115_v4PHM_lowspin}{2.81}{GW200115_combined_lowspin}{2.81}{GW200115_XPHM_highspin}{2.87}{GW200115_v4PHM_highspin}{2.86}{GW200115_combined_highspin}{2.88}{GW200115_NSBH_lowspin}{0.00}}}
\newcommand{\phitwoonepercent}[1]{\IfEqCase{#1}{{GW200105_XPHM_lowspin}{0.07}{GW200105_v4PHM_lowspin}{0.06}{GW200105_combined_lowspin}{0.07}{GW200105_XPHM_highspin}{0.07}{GW200105_v4PHM_highspin}{0.07}{GW200105_combined_highspin}{0.07}{GW200105_NSBH_lowspin}{0.00}{GW200115_XPHM_lowspin}{0.07}{GW200115_v4PHM_lowspin}{0.06}{GW200115_combined_lowspin}{0.07}{GW200115_XPHM_highspin}{0.06}{GW200115_v4PHM_highspin}{0.06}{GW200115_combined_highspin}{0.06}{GW200115_NSBH_lowspin}{0.00}}}
\newcommand{\phitwoninetyninepercent}[1]{\IfEqCase{#1}{{GW200105_XPHM_lowspin}{6.22}{GW200105_v4PHM_lowspin}{6.23}{GW200105_combined_lowspin}{6.22}{GW200105_XPHM_highspin}{6.21}{GW200105_v4PHM_highspin}{6.22}{GW200105_combined_highspin}{6.22}{GW200105_NSBH_lowspin}{0.00}{GW200115_XPHM_lowspin}{6.23}{GW200115_v4PHM_lowspin}{6.21}{GW200115_combined_lowspin}{6.22}{GW200115_XPHM_highspin}{6.22}{GW200115_v4PHM_highspin}{6.23}{GW200115_combined_highspin}{6.23}{GW200115_NSBH_lowspin}{0.00}}}
\newcommand{\phitwofivepercent}[1]{\IfEqCase{#1}{{GW200105_XPHM_lowspin}{0.33}{GW200105_v4PHM_lowspin}{0.31}{GW200105_combined_lowspin}{0.31}{GW200105_XPHM_highspin}{0.31}{GW200105_v4PHM_highspin}{0.27}{GW200105_combined_highspin}{0.29}{GW200105_NSBH_lowspin}{0.00}{GW200115_XPHM_lowspin}{0.34}{GW200115_v4PHM_lowspin}{0.30}{GW200115_combined_lowspin}{0.32}{GW200115_XPHM_highspin}{0.32}{GW200115_v4PHM_highspin}{0.30}{GW200115_combined_highspin}{0.31}{GW200115_NSBH_lowspin}{0.00}}}
\newcommand{\phitwoninetyfivepercent}[1]{\IfEqCase{#1}{{GW200105_XPHM_lowspin}{5.98}{GW200105_v4PHM_lowspin}{5.96}{GW200105_combined_lowspin}{5.97}{GW200105_XPHM_highspin}{5.95}{GW200105_v4PHM_highspin}{5.98}{GW200105_combined_highspin}{5.97}{GW200105_NSBH_lowspin}{0.00}{GW200115_XPHM_lowspin}{5.97}{GW200115_v4PHM_lowspin}{5.94}{GW200115_combined_lowspin}{5.96}{GW200115_XPHM_highspin}{5.98}{GW200115_v4PHM_highspin}{5.98}{GW200115_combined_highspin}{5.98}{GW200115_NSBH_lowspin}{0.00}}}
\newcommand{\phitwoninetypercent}[1]{\IfEqCase{#1}{{GW200105_XPHM_lowspin}{5.67}{GW200105_v4PHM_lowspin}{5.62}{GW200105_combined_lowspin}{5.64}{GW200105_XPHM_highspin}{5.64}{GW200105_v4PHM_highspin}{5.66}{GW200105_combined_highspin}{5.66}{GW200105_NSBH_lowspin}{0.00}{GW200115_XPHM_lowspin}{5.67}{GW200115_v4PHM_lowspin}{5.62}{GW200115_combined_lowspin}{5.64}{GW200115_XPHM_highspin}{5.65}{GW200115_v4PHM_highspin}{5.69}{GW200115_combined_highspin}{5.68}{GW200115_NSBH_lowspin}{0.00}}}
\newcommand{\chieffminus}[1]{\IfEqCase{#1}{{GW200105_XPHM_lowspin}{0.15}{GW200105_v4PHM_lowspin}{0.10}{GW200105_combined_lowspin}{0.12}{GW200105_XPHM_highspin}{0.18}{GW200105_v4PHM_highspin}{0.11}{GW200105_combined_highspin}{0.15}{GW200105_NSBH_lowspin}{0.22}{GW200115_XPHM_lowspin}{0.27}{GW200115_v4PHM_lowspin}{0.42}{GW200115_combined_lowspin}{0.34}{GW200115_XPHM_highspin}{0.30}{GW200115_v4PHM_highspin}{0.40}{GW200115_combined_highspin}{0.35}{GW200115_NSBH_lowspin}{0.24}}}
\newcommand{\chieffmed}[1]{\IfEqCase{#1}{{GW200105_XPHM_lowspin}{-0.01}{GW200105_v4PHM_lowspin}{-0.00}{GW200105_combined_lowspin}{-0.01}{GW200105_XPHM_highspin}{-0.01}{GW200105_v4PHM_highspin}{-0.01}{GW200105_combined_highspin}{-0.01}{GW200105_NSBH_lowspin}{-0.01}{GW200115_XPHM_lowspin}{-0.19}{GW200115_v4PHM_lowspin}{-0.09}{GW200115_combined_lowspin}{-0.14}{GW200115_XPHM_highspin}{-0.20}{GW200115_v4PHM_highspin}{-0.17}{GW200115_combined_highspin}{-0.19}{GW200115_NSBH_lowspin}{-0.04}}}
\newcommand{\chieffplus}[1]{\IfEqCase{#1}{{GW200105_XPHM_lowspin}{0.10}{GW200105_v4PHM_lowspin}{0.08}{GW200105_combined_lowspin}{0.08}{GW200105_XPHM_highspin}{0.13}{GW200105_v4PHM_highspin}{0.06}{GW200105_combined_highspin}{0.11}{GW200105_NSBH_lowspin}{0.22}{GW200115_XPHM_lowspin}{0.22}{GW200115_v4PHM_lowspin}{0.11}{GW200115_combined_lowspin}{0.17}{GW200115_XPHM_highspin}{0.22}{GW200115_v4PHM_highspin}{0.23}{GW200115_combined_highspin}{0.23}{GW200115_NSBH_lowspin}{0.18}}}
\newcommand{\chieffonepercent}[1]{\IfEqCase{#1}{{GW200105_XPHM_lowspin}{-0.29}{GW200105_v4PHM_lowspin}{-0.15}{GW200105_combined_lowspin}{-0.24}{GW200105_XPHM_highspin}{-0.34}{GW200105_v4PHM_highspin}{-0.37}{GW200105_combined_highspin}{-0.34}{GW200105_NSBH_lowspin}{-0.29}{GW200115_XPHM_lowspin}{-0.53}{GW200115_v4PHM_lowspin}{-0.54}{GW200115_combined_lowspin}{-0.54}{GW200115_XPHM_highspin}{-0.58}{GW200115_v4PHM_highspin}{-0.60}{GW200115_combined_highspin}{-0.60}{GW200115_NSBH_lowspin}{-0.38}}}
\newcommand{\chieffninetyninepercent}[1]{\IfEqCase{#1}{{GW200105_XPHM_lowspin}{0.15}{GW200105_v4PHM_lowspin}{0.16}{GW200105_combined_lowspin}{0.16}{GW200105_XPHM_highspin}{0.18}{GW200105_v4PHM_highspin}{0.15}{GW200105_combined_highspin}{0.17}{GW200105_NSBH_lowspin}{0.31}{GW200115_XPHM_lowspin}{0.14}{GW200115_v4PHM_lowspin}{0.11}{GW200115_combined_lowspin}{0.13}{GW200115_XPHM_highspin}{0.09}{GW200115_v4PHM_highspin}{0.13}{GW200115_combined_highspin}{0.12}{GW200115_NSBH_lowspin}{0.21}}}
\newcommand{\chiefffivepercent}[1]{\IfEqCase{#1}{{GW200105_XPHM_lowspin}{-0.16}{GW200105_v4PHM_lowspin}{-0.10}{GW200105_combined_lowspin}{-0.13}{GW200105_XPHM_highspin}{-0.19}{GW200105_v4PHM_highspin}{-0.12}{GW200105_combined_highspin}{-0.16}{GW200105_NSBH_lowspin}{-0.23}{GW200115_XPHM_lowspin}{-0.46}{GW200115_v4PHM_lowspin}{-0.51}{GW200115_combined_lowspin}{-0.49}{GW200115_XPHM_highspin}{-0.50}{GW200115_v4PHM_highspin}{-0.57}{GW200115_combined_highspin}{-0.54}{GW200115_NSBH_lowspin}{-0.29}}}
\newcommand{\chieffninetyfivepercent}[1]{\IfEqCase{#1}{{GW200105_XPHM_lowspin}{0.09}{GW200105_v4PHM_lowspin}{0.07}{GW200105_combined_lowspin}{0.08}{GW200105_XPHM_highspin}{0.12}{GW200105_v4PHM_highspin}{0.06}{GW200105_combined_highspin}{0.10}{GW200105_NSBH_lowspin}{0.21}{GW200115_XPHM_lowspin}{0.03}{GW200115_v4PHM_lowspin}{0.02}{GW200115_combined_lowspin}{0.03}{GW200115_XPHM_highspin}{0.02}{GW200115_v4PHM_highspin}{0.06}{GW200115_combined_highspin}{0.04}{GW200115_NSBH_lowspin}{0.13}}}
\newcommand{\chieffninetypercent}[1]{\IfEqCase{#1}{{GW200105_XPHM_lowspin}{0.06}{GW200105_v4PHM_lowspin}{0.05}{GW200105_combined_lowspin}{0.05}{GW200105_XPHM_highspin}{0.09}{GW200105_v4PHM_highspin}{0.04}{GW200105_combined_highspin}{0.07}{GW200105_NSBH_lowspin}{0.15}{GW200115_XPHM_lowspin}{-0.01}{GW200115_v4PHM_lowspin}{0.01}{GW200115_combined_lowspin}{0.01}{GW200115_XPHM_highspin}{-0.02}{GW200115_v4PHM_highspin}{0.03}{GW200115_combined_highspin}{0.01}{GW200115_NSBH_lowspin}{0.08}}}
\newcommand{\chipminus}[1]{\IfEqCase{#1}{{GW200105_XPHM_lowspin}{0.08}{GW200105_v4PHM_lowspin}{0.05}{GW200105_combined_lowspin}{0.06}{GW200105_XPHM_highspin}{0.08}{GW200105_v4PHM_highspin}{0.05}{GW200105_combined_highspin}{0.07}{GW200105_NSBH_lowspin}{0.00}{GW200115_XPHM_lowspin}{0.20}{GW200115_v4PHM_lowspin}{0.12}{GW200115_combined_lowspin}{0.17}{GW200115_XPHM_highspin}{0.18}{GW200115_v4PHM_highspin}{0.14}{GW200115_combined_highspin}{0.17}{GW200115_NSBH_lowspin}{0.00}}}
\newcommand{\chipmed}[1]{\IfEqCase{#1}{{GW200105_XPHM_lowspin}{0.09}{GW200105_v4PHM_lowspin}{0.06}{GW200105_combined_lowspin}{0.07}{GW200105_XPHM_highspin}{0.11}{GW200105_v4PHM_highspin}{0.07}{GW200105_combined_highspin}{0.09}{GW200105_NSBH_lowspin}{0.00}{GW200115_XPHM_lowspin}{0.25}{GW200115_v4PHM_lowspin}{0.13}{GW200115_combined_lowspin}{0.19}{GW200115_XPHM_highspin}{0.25}{GW200115_v4PHM_highspin}{0.17}{GW200115_combined_highspin}{0.21}{GW200115_NSBH_lowspin}{0.00}}}
\newcommand{\chipplus}[1]{\IfEqCase{#1}{{GW200105_XPHM_lowspin}{0.16}{GW200105_v4PHM_lowspin}{0.13}{GW200105_combined_lowspin}{0.15}{GW200105_XPHM_highspin}{0.15}{GW200105_v4PHM_highspin}{0.10}{GW200105_combined_highspin}{0.14}{GW200105_NSBH_lowspin}{0.00}{GW200115_XPHM_lowspin}{0.27}{GW200115_v4PHM_lowspin}{0.27}{GW200115_combined_lowspin}{0.28}{GW200115_XPHM_highspin}{0.30}{GW200115_v4PHM_highspin}{0.28}{GW200115_combined_highspin}{0.30}{GW200115_NSBH_lowspin}{0.00}}}
\newcommand{\chiponepercent}[1]{\IfEqCase{#1}{{GW200105_XPHM_lowspin}{0.00}{GW200105_v4PHM_lowspin}{0.00}{GW200105_combined_lowspin}{0.00}{GW200105_XPHM_highspin}{0.01}{GW200105_v4PHM_highspin}{0.01}{GW200105_combined_highspin}{0.01}{GW200105_NSBH_lowspin}{0.00}{GW200115_XPHM_lowspin}{0.01}{GW200115_v4PHM_lowspin}{0.00}{GW200115_combined_lowspin}{0.00}{GW200115_XPHM_highspin}{0.03}{GW200115_v4PHM_highspin}{0.02}{GW200115_combined_highspin}{0.02}{GW200115_NSBH_lowspin}{0.00}}}
\newcommand{\chipninetyninepercent}[1]{\IfEqCase{#1}{{GW200105_XPHM_lowspin}{0.37}{GW200105_v4PHM_lowspin}{0.24}{GW200105_combined_lowspin}{0.31}{GW200105_XPHM_highspin}{0.40}{GW200105_v4PHM_highspin}{0.24}{GW200105_combined_highspin}{0.34}{GW200105_NSBH_lowspin}{0.00}{GW200115_XPHM_lowspin}{0.62}{GW200115_v4PHM_lowspin}{0.49}{GW200115_combined_lowspin}{0.58}{GW200115_XPHM_highspin}{0.67}{GW200115_v4PHM_highspin}{0.54}{GW200115_combined_highspin}{0.64}{GW200115_NSBH_lowspin}{0.00}}}
\newcommand{\chipfivepercent}[1]{\IfEqCase{#1}{{GW200105_XPHM_lowspin}{0.01}{GW200105_v4PHM_lowspin}{0.01}{GW200105_combined_lowspin}{0.01}{GW200105_XPHM_highspin}{0.03}{GW200105_v4PHM_highspin}{0.02}{GW200105_combined_highspin}{0.02}{GW200105_NSBH_lowspin}{0.00}{GW200115_XPHM_lowspin}{0.05}{GW200115_v4PHM_lowspin}{0.01}{GW200115_combined_lowspin}{0.02}{GW200115_XPHM_highspin}{0.07}{GW200115_v4PHM_highspin}{0.04}{GW200115_combined_highspin}{0.05}{GW200115_NSBH_lowspin}{0.00}}}
\newcommand{\chipninetyfivepercent}[1]{\IfEqCase{#1}{{GW200105_XPHM_lowspin}{0.25}{GW200105_v4PHM_lowspin}{0.19}{GW200105_combined_lowspin}{0.22}{GW200105_XPHM_highspin}{0.26}{GW200105_v4PHM_highspin}{0.17}{GW200105_combined_highspin}{0.22}{GW200105_NSBH_lowspin}{0.00}{GW200115_XPHM_lowspin}{0.52}{GW200115_v4PHM_lowspin}{0.40}{GW200115_combined_lowspin}{0.47}{GW200115_XPHM_highspin}{0.55}{GW200115_v4PHM_highspin}{0.45}{GW200115_combined_highspin}{0.51}{GW200115_NSBH_lowspin}{0.00}}}
\newcommand{\chipninetypercent}[1]{\IfEqCase{#1}{{GW200105_XPHM_lowspin}{0.20}{GW200105_v4PHM_lowspin}{0.15}{GW200105_combined_lowspin}{0.18}{GW200105_XPHM_highspin}{0.21}{GW200105_v4PHM_highspin}{0.14}{GW200105_combined_highspin}{0.18}{GW200105_NSBH_lowspin}{0.00}{GW200115_XPHM_lowspin}{0.46}{GW200115_v4PHM_lowspin}{0.34}{GW200115_combined_lowspin}{0.41}{GW200115_XPHM_highspin}{0.48}{GW200115_v4PHM_highspin}{0.39}{GW200115_combined_highspin}{0.44}{GW200115_NSBH_lowspin}{0.00}}}
\newcommand{\costiltoneminus}[1]{\IfEqCase{#1}{{GW200105_XPHM_lowspin}{0.71}{GW200105_v4PHM_lowspin}{0.83}{GW200105_combined_lowspin}{0.77}{GW200105_XPHM_highspin}{0.81}{GW200105_v4PHM_highspin}{0.85}{GW200105_combined_highspin}{0.83}{GW200105_NSBH_lowspin}{0.00}{GW200115_XPHM_lowspin}{0.29}{GW200115_v4PHM_lowspin}{0.34}{GW200115_combined_lowspin}{0.31}{GW200115_XPHM_highspin}{0.32}{GW200115_v4PHM_highspin}{0.27}{GW200115_combined_highspin}{0.30}{GW200115_NSBH_lowspin}{0.00}}}
\newcommand{\costiltonemed}[1]{\IfEqCase{#1}{{GW200105_XPHM_lowspin}{-0.20}{GW200105_v4PHM_lowspin}{-0.08}{GW200105_combined_lowspin}{-0.14}{GW200105_XPHM_highspin}{-0.09}{GW200105_v4PHM_highspin}{-0.09}{GW200105_combined_highspin}{-0.09}{GW200105_NSBH_lowspin}{-1.00}{GW200115_XPHM_lowspin}{-0.68}{GW200115_v4PHM_lowspin}{-0.65}{GW200115_combined_lowspin}{-0.66}{GW200115_XPHM_highspin}{-0.63}{GW200115_v4PHM_highspin}{-0.71}{GW200115_combined_highspin}{-0.66}{GW200115_NSBH_lowspin}{-1.00}}}
\newcommand{\costiltoneplus}[1]{\IfEqCase{#1}{{GW200105_XPHM_lowspin}{1.00}{GW200105_v4PHM_lowspin}{0.96}{GW200105_combined_lowspin}{0.98}{GW200105_XPHM_highspin}{0.96}{GW200105_v4PHM_highspin}{0.96}{GW200105_combined_highspin}{0.96}{GW200105_NSBH_lowspin}{2.00}{GW200115_XPHM_lowspin}{0.86}{GW200115_v4PHM_lowspin}{1.22}{GW200115_combined_lowspin}{1.09}{GW200115_XPHM_highspin}{0.78}{GW200115_v4PHM_highspin}{1.33}{GW200115_combined_highspin}{1.10}{GW200115_NSBH_lowspin}{2.00}}}
\newcommand{\costiltoneonepercent}[1]{\IfEqCase{#1}{{GW200105_XPHM_lowspin}{-0.98}{GW200105_v4PHM_lowspin}{-0.98}{GW200105_combined_lowspin}{-0.98}{GW200105_XPHM_highspin}{-0.98}{GW200105_v4PHM_highspin}{-0.99}{GW200105_combined_highspin}{-0.99}{GW200105_NSBH_lowspin}{-1.00}{GW200115_XPHM_lowspin}{-0.99}{GW200115_v4PHM_lowspin}{-1.00}{GW200115_combined_lowspin}{-1.00}{GW200115_XPHM_highspin}{-0.99}{GW200115_v4PHM_highspin}{-0.99}{GW200115_combined_highspin}{-0.99}{GW200115_NSBH_lowspin}{-1.00}}}
\newcommand{\costiltoneninetyninepercent}[1]{\IfEqCase{#1}{{GW200105_XPHM_lowspin}{0.95}{GW200105_v4PHM_lowspin}{0.98}{GW200105_combined_lowspin}{0.97}{GW200105_XPHM_highspin}{0.97}{GW200105_v4PHM_highspin}{0.98}{GW200105_combined_highspin}{0.97}{GW200105_NSBH_lowspin}{1.00}{GW200115_XPHM_lowspin}{0.67}{GW200115_v4PHM_lowspin}{0.90}{GW200115_combined_lowspin}{0.83}{GW200115_XPHM_highspin}{0.59}{GW200115_v4PHM_highspin}{0.91}{GW200115_combined_highspin}{0.84}{GW200115_NSBH_lowspin}{1.00}}}
\newcommand{\costiltonefivepercent}[1]{\IfEqCase{#1}{{GW200105_XPHM_lowspin}{-0.91}{GW200105_v4PHM_lowspin}{-0.91}{GW200105_combined_lowspin}{-0.91}{GW200105_XPHM_highspin}{-0.91}{GW200105_v4PHM_highspin}{-0.94}{GW200105_combined_highspin}{-0.92}{GW200105_NSBH_lowspin}{-1.00}{GW200115_XPHM_lowspin}{-0.97}{GW200115_v4PHM_lowspin}{-0.98}{GW200115_combined_lowspin}{-0.98}{GW200115_XPHM_highspin}{-0.96}{GW200115_v4PHM_highspin}{-0.98}{GW200115_combined_highspin}{-0.97}{GW200115_NSBH_lowspin}{-1.00}}}
\newcommand{\costiltoneninetyfivepercent}[1]{\IfEqCase{#1}{{GW200105_XPHM_lowspin}{0.80}{GW200105_v4PHM_lowspin}{0.88}{GW200105_combined_lowspin}{0.84}{GW200105_XPHM_highspin}{0.87}{GW200105_v4PHM_highspin}{0.87}{GW200105_combined_highspin}{0.87}{GW200105_NSBH_lowspin}{1.00}{GW200115_XPHM_lowspin}{0.18}{GW200115_v4PHM_lowspin}{0.58}{GW200115_combined_lowspin}{0.42}{GW200115_XPHM_highspin}{0.14}{GW200115_v4PHM_highspin}{0.62}{GW200115_combined_highspin}{0.44}{GW200115_NSBH_lowspin}{1.00}}}
\newcommand{\costiltoneninetypercent}[1]{\IfEqCase{#1}{{GW200105_XPHM_lowspin}{0.63}{GW200105_v4PHM_lowspin}{0.75}{GW200105_combined_lowspin}{0.69}{GW200105_XPHM_highspin}{0.75}{GW200105_v4PHM_highspin}{0.73}{GW200105_combined_highspin}{0.74}{GW200105_NSBH_lowspin}{1.00}{GW200115_XPHM_lowspin}{-0.08}{GW200115_v4PHM_lowspin}{0.27}{GW200115_combined_lowspin}{0.11}{GW200115_XPHM_highspin}{-0.07}{GW200115_v4PHM_highspin}{0.35}{GW200115_combined_highspin}{0.12}{GW200115_NSBH_lowspin}{1.00}}}
\newcommand{\costilttwominus}[1]{\IfEqCase{#1}{{GW200105_XPHM_lowspin}{0.91}{GW200105_v4PHM_lowspin}{0.87}{GW200105_combined_lowspin}{0.90}{GW200105_XPHM_highspin}{0.85}{GW200105_v4PHM_highspin}{0.79}{GW200105_combined_highspin}{0.82}{GW200105_NSBH_lowspin}{2.00}{GW200115_XPHM_lowspin}{0.88}{GW200115_v4PHM_lowspin}{0.89}{GW200115_combined_lowspin}{0.89}{GW200115_XPHM_highspin}{0.62}{GW200115_v4PHM_highspin}{0.66}{GW200115_combined_highspin}{0.65}{GW200115_NSBH_lowspin}{0.00}}}
\newcommand{\costilttwomed}[1]{\IfEqCase{#1}{{GW200105_XPHM_lowspin}{0.03}{GW200105_v4PHM_lowspin}{-0.04}{GW200105_combined_lowspin}{-0.00}{GW200105_XPHM_highspin}{-0.02}{GW200105_v4PHM_highspin}{-0.04}{GW200105_combined_highspin}{-0.03}{GW200105_NSBH_lowspin}{1.00}{GW200115_XPHM_lowspin}{0.00}{GW200115_v4PHM_lowspin}{-0.01}{GW200115_combined_lowspin}{-0.00}{GW200115_XPHM_highspin}{-0.30}{GW200115_v4PHM_highspin}{-0.30}{GW200115_combined_highspin}{-0.30}{GW200115_NSBH_lowspin}{-1.00}}}
\newcommand{\costilttwoplus}[1]{\IfEqCase{#1}{{GW200105_XPHM_lowspin}{0.86}{GW200105_v4PHM_lowspin}{0.92}{GW200105_combined_lowspin}{0.89}{GW200105_XPHM_highspin}{0.89}{GW200105_v4PHM_highspin}{0.84}{GW200105_combined_highspin}{0.88}{GW200105_NSBH_lowspin}{0.00}{GW200115_XPHM_lowspin}{0.88}{GW200115_v4PHM_lowspin}{0.91}{GW200115_combined_lowspin}{0.89}{GW200115_XPHM_highspin}{1.02}{GW200115_v4PHM_highspin}{1.16}{GW200115_combined_highspin}{1.11}{GW200115_NSBH_lowspin}{2.00}}}
\newcommand{\costilttwoonepercent}[1]{\IfEqCase{#1}{{GW200105_XPHM_lowspin}{-0.98}{GW200105_v4PHM_lowspin}{-0.98}{GW200105_combined_lowspin}{-0.98}{GW200105_XPHM_highspin}{-0.97}{GW200105_v4PHM_highspin}{-0.97}{GW200105_combined_highspin}{-0.97}{GW200105_NSBH_lowspin}{-1.00}{GW200115_XPHM_lowspin}{-0.98}{GW200115_v4PHM_lowspin}{-0.98}{GW200115_combined_lowspin}{-0.98}{GW200115_XPHM_highspin}{-0.98}{GW200115_v4PHM_highspin}{-0.99}{GW200115_combined_highspin}{-0.99}{GW200115_NSBH_lowspin}{-1.00}}}
\newcommand{\costilttwoninetyninepercent}[1]{\IfEqCase{#1}{{GW200105_XPHM_lowspin}{0.98}{GW200105_v4PHM_lowspin}{0.98}{GW200105_combined_lowspin}{0.98}{GW200105_XPHM_highspin}{0.97}{GW200105_v4PHM_highspin}{0.96}{GW200105_combined_highspin}{0.97}{GW200105_NSBH_lowspin}{1.00}{GW200115_XPHM_lowspin}{0.97}{GW200115_v4PHM_lowspin}{0.98}{GW200115_combined_lowspin}{0.98}{GW200115_XPHM_highspin}{0.93}{GW200115_v4PHM_highspin}{0.98}{GW200115_combined_highspin}{0.96}{GW200115_NSBH_lowspin}{1.00}}}
\newcommand{\costilttwofivepercent}[1]{\IfEqCase{#1}{{GW200105_XPHM_lowspin}{-0.88}{GW200105_v4PHM_lowspin}{-0.91}{GW200105_combined_lowspin}{-0.90}{GW200105_XPHM_highspin}{-0.87}{GW200105_v4PHM_highspin}{-0.83}{GW200105_combined_highspin}{-0.85}{GW200105_NSBH_lowspin}{-1.00}{GW200115_XPHM_lowspin}{-0.88}{GW200115_v4PHM_lowspin}{-0.90}{GW200115_combined_lowspin}{-0.89}{GW200115_XPHM_highspin}{-0.92}{GW200115_v4PHM_highspin}{-0.96}{GW200115_combined_highspin}{-0.95}{GW200115_NSBH_lowspin}{-1.00}}}
\newcommand{\costilttwoninetyfivepercent}[1]{\IfEqCase{#1}{{GW200105_XPHM_lowspin}{0.89}{GW200105_v4PHM_lowspin}{0.88}{GW200105_combined_lowspin}{0.89}{GW200105_XPHM_highspin}{0.86}{GW200105_v4PHM_highspin}{0.80}{GW200105_combined_highspin}{0.84}{GW200105_NSBH_lowspin}{1.00}{GW200115_XPHM_lowspin}{0.88}{GW200115_v4PHM_lowspin}{0.90}{GW200115_combined_lowspin}{0.89}{GW200115_XPHM_highspin}{0.73}{GW200115_v4PHM_highspin}{0.86}{GW200115_combined_highspin}{0.81}{GW200115_NSBH_lowspin}{1.00}}}
\newcommand{\costilttwoninetypercent}[1]{\IfEqCase{#1}{{GW200105_XPHM_lowspin}{0.79}{GW200105_v4PHM_lowspin}{0.77}{GW200105_combined_lowspin}{0.78}{GW200105_XPHM_highspin}{0.75}{GW200105_v4PHM_highspin}{0.65}{GW200105_combined_highspin}{0.70}{GW200105_NSBH_lowspin}{1.00}{GW200115_XPHM_lowspin}{0.77}{GW200115_v4PHM_lowspin}{0.80}{GW200115_combined_lowspin}{0.79}{GW200115_XPHM_highspin}{0.53}{GW200115_v4PHM_highspin}{0.73}{GW200115_combined_highspin}{0.64}{GW200115_NSBH_lowspin}{1.00}}}
\newcommand{\comovingdistminus}[1]{\IfEqCase{#1}{{GW200105_XPHM_lowspin}{98}{GW200105_v4PHM_lowspin}{105}{GW200105_combined_lowspin}{103}{GW200105_XPHM_highspin}{98}{GW200105_v4PHM_highspin}{105}{GW200105_combined_highspin}{103}{GW200105_NSBH_lowspin}{112}{GW200115_XPHM_lowspin}{81}{GW200115_v4PHM_lowspin}{113}{GW200115_combined_lowspin}{96}{GW200115_XPHM_highspin}{77}{GW200115_v4PHM_highspin}{112}{GW200115_combined_highspin}{93}{GW200115_NSBH_lowspin}{108}}}
\newcommand{\comovingdistmed}[1]{\IfEqCase{#1}{{GW200105_XPHM_lowspin}{263}{GW200105_v4PHM_lowspin}{258}{GW200105_combined_lowspin}{261}{GW200105_XPHM_highspin}{265}{GW200105_v4PHM_highspin}{257}{GW200105_combined_highspin}{261}{GW200105_NSBH_lowspin}{264}{GW200115_XPHM_lowspin}{282}{GW200115_v4PHM_lowspin}{294}{GW200115_combined_lowspin}{287}{GW200115_XPHM_highspin}{280}{GW200115_v4PHM_highspin}{295}{GW200115_combined_highspin}{286}{GW200115_NSBH_lowspin}{275}}}
\newcommand{\comovingdistplus}[1]{\IfEqCase{#1}{{GW200105_XPHM_lowspin}{95}{GW200105_v4PHM_lowspin}{102}{GW200105_combined_lowspin}{98}{GW200105_XPHM_highspin}{94}{GW200105_v4PHM_highspin}{103}{GW200105_combined_highspin}{98}{GW200105_NSBH_lowspin}{99}{GW200115_XPHM_lowspin}{111}{GW200115_v4PHM_lowspin}{140}{GW200115_combined_lowspin}{130}{GW200115_XPHM_highspin}{108}{GW200115_v4PHM_highspin}{141}{GW200115_combined_highspin}{133}{GW200115_NSBH_lowspin}{136}}}
\newcommand{\comovingdistonepercent}[1]{\IfEqCase{#1}{{GW200105_XPHM_lowspin}{131}{GW200105_v4PHM_lowspin}{121}{GW200105_combined_lowspin}{124}{GW200105_XPHM_highspin}{132}{GW200105_v4PHM_highspin}{119}{GW200105_combined_highspin}{124}{GW200105_NSBH_lowspin}{115}{GW200115_XPHM_lowspin}{165}{GW200115_v4PHM_lowspin}{148}{GW200115_combined_lowspin}{154}{GW200115_XPHM_highspin}{170}{GW200115_v4PHM_highspin}{149}{GW200115_combined_highspin}{156}{GW200115_NSBH_lowspin}{138}}}
\newcommand{\comovingdistninetyninepercent}[1]{\IfEqCase{#1}{{GW200105_XPHM_lowspin}{391}{GW200105_v4PHM_lowspin}{396}{GW200105_combined_lowspin}{393}{GW200105_XPHM_highspin}{393}{GW200105_v4PHM_highspin}{393}{GW200105_combined_highspin}{393}{GW200105_NSBH_lowspin}{396}{GW200115_XPHM_lowspin}{450}{GW200115_v4PHM_lowspin}{489}{GW200115_combined_lowspin}{475}{GW200115_XPHM_highspin}{447}{GW200115_v4PHM_highspin}{494}{GW200115_combined_highspin}{477}{GW200115_NSBH_lowspin}{467}}}
\newcommand{\comovingdistfivepercent}[1]{\IfEqCase{#1}{{GW200105_XPHM_lowspin}{164}{GW200105_v4PHM_lowspin}{153}{GW200105_combined_lowspin}{158}{GW200105_XPHM_highspin}{167}{GW200105_v4PHM_highspin}{152}{GW200105_combined_highspin}{158}{GW200105_NSBH_lowspin}{151}{GW200115_XPHM_lowspin}{201}{GW200115_v4PHM_lowspin}{182}{GW200115_combined_lowspin}{191}{GW200115_XPHM_highspin}{203}{GW200115_v4PHM_highspin}{183}{GW200115_combined_highspin}{193}{GW200115_NSBH_lowspin}{167}}}
\newcommand{\comovingdistninetyfivepercent}[1]{\IfEqCase{#1}{{GW200105_XPHM_lowspin}{357}{GW200105_v4PHM_lowspin}{360}{GW200105_combined_lowspin}{359}{GW200105_XPHM_highspin}{359}{GW200105_v4PHM_highspin}{360}{GW200105_combined_highspin}{359}{GW200105_NSBH_lowspin}{363}{GW200115_XPHM_lowspin}{393}{GW200115_v4PHM_lowspin}{434}{GW200115_combined_lowspin}{418}{GW200115_XPHM_highspin}{388}{GW200115_v4PHM_highspin}{436}{GW200115_combined_highspin}{419}{GW200115_NSBH_lowspin}{412}}}
\newcommand{\comovingdistninetypercent}[1]{\IfEqCase{#1}{{GW200105_XPHM_lowspin}{338}{GW200105_v4PHM_lowspin}{340}{GW200105_combined_lowspin}{339}{GW200105_XPHM_highspin}{340}{GW200105_v4PHM_highspin}{340}{GW200105_combined_highspin}{340}{GW200105_NSBH_lowspin}{345}{GW200115_XPHM_lowspin}{364}{GW200115_v4PHM_lowspin}{403}{GW200115_combined_lowspin}{385}{GW200115_XPHM_highspin}{358}{GW200115_v4PHM_highspin}{406}{GW200115_combined_highspin}{385}{GW200115_NSBH_lowspin}{380}}}
\newcommand{\redshiftminus}[1]{\IfEqCase{#1}{{GW200105_XPHM_lowspin}{0.02}{GW200105_v4PHM_lowspin}{0.02}{GW200105_combined_lowspin}{0.02}{GW200105_XPHM_highspin}{0.02}{GW200105_v4PHM_highspin}{0.02}{GW200105_combined_highspin}{0.02}{GW200105_NSBH_lowspin}{0.03}{GW200115_XPHM_lowspin}{0.02}{GW200115_v4PHM_lowspin}{0.03}{GW200115_combined_lowspin}{0.02}{GW200115_XPHM_highspin}{0.02}{GW200115_v4PHM_highspin}{0.03}{GW200115_combined_highspin}{0.02}{GW200115_NSBH_lowspin}{0.03}}}
\newcommand{\redshiftmed}[1]{\IfEqCase{#1}{{GW200105_XPHM_lowspin}{0.06}{GW200105_v4PHM_lowspin}{0.06}{GW200105_combined_lowspin}{0.06}{GW200105_XPHM_highspin}{0.06}{GW200105_v4PHM_highspin}{0.06}{GW200105_combined_highspin}{0.06}{GW200105_NSBH_lowspin}{0.06}{GW200115_XPHM_lowspin}{0.06}{GW200115_v4PHM_lowspin}{0.07}{GW200115_combined_lowspin}{0.07}{GW200115_XPHM_highspin}{0.06}{GW200115_v4PHM_highspin}{0.07}{GW200115_combined_highspin}{0.07}{GW200115_NSBH_lowspin}{0.06}}}
\newcommand{\redshiftplus}[1]{\IfEqCase{#1}{{GW200105_XPHM_lowspin}{0.02}{GW200105_v4PHM_lowspin}{0.02}{GW200105_combined_lowspin}{0.02}{GW200105_XPHM_highspin}{0.02}{GW200105_v4PHM_highspin}{0.02}{GW200105_combined_highspin}{0.02}{GW200105_NSBH_lowspin}{0.02}{GW200115_XPHM_lowspin}{0.03}{GW200115_v4PHM_lowspin}{0.03}{GW200115_combined_lowspin}{0.03}{GW200115_XPHM_highspin}{0.03}{GW200115_v4PHM_highspin}{0.03}{GW200115_combined_highspin}{0.03}{GW200115_NSBH_lowspin}{0.03}}}
\newcommand{\redshiftonepercent}[1]{\IfEqCase{#1}{{GW200105_XPHM_lowspin}{0.03}{GW200105_v4PHM_lowspin}{0.03}{GW200105_combined_lowspin}{0.03}{GW200105_XPHM_highspin}{0.03}{GW200105_v4PHM_highspin}{0.03}{GW200105_combined_highspin}{0.03}{GW200105_NSBH_lowspin}{0.03}{GW200115_XPHM_lowspin}{0.04}{GW200115_v4PHM_lowspin}{0.03}{GW200115_combined_lowspin}{0.04}{GW200115_XPHM_highspin}{0.04}{GW200115_v4PHM_highspin}{0.03}{GW200115_combined_highspin}{0.04}{GW200115_NSBH_lowspin}{0.03}}}
\newcommand{\redshiftninetyninepercent}[1]{\IfEqCase{#1}{{GW200105_XPHM_lowspin}{0.09}{GW200105_v4PHM_lowspin}{0.09}{GW200105_combined_lowspin}{0.09}{GW200105_XPHM_highspin}{0.09}{GW200105_v4PHM_highspin}{0.09}{GW200105_combined_highspin}{0.09}{GW200105_NSBH_lowspin}{0.09}{GW200115_XPHM_lowspin}{0.10}{GW200115_v4PHM_lowspin}{0.11}{GW200115_combined_lowspin}{0.11}{GW200115_XPHM_highspin}{0.10}{GW200115_v4PHM_highspin}{0.11}{GW200115_combined_highspin}{0.11}{GW200115_NSBH_lowspin}{0.11}}}
\newcommand{\redshiftfivepercent}[1]{\IfEqCase{#1}{{GW200105_XPHM_lowspin}{0.04}{GW200105_v4PHM_lowspin}{0.03}{GW200105_combined_lowspin}{0.04}{GW200105_XPHM_highspin}{0.04}{GW200105_v4PHM_highspin}{0.03}{GW200105_combined_highspin}{0.04}{GW200105_NSBH_lowspin}{0.03}{GW200115_XPHM_lowspin}{0.05}{GW200115_v4PHM_lowspin}{0.04}{GW200115_combined_lowspin}{0.04}{GW200115_XPHM_highspin}{0.05}{GW200115_v4PHM_highspin}{0.04}{GW200115_combined_highspin}{0.04}{GW200115_NSBH_lowspin}{0.04}}}
\newcommand{\redshiftninetyfivepercent}[1]{\IfEqCase{#1}{{GW200105_XPHM_lowspin}{0.08}{GW200105_v4PHM_lowspin}{0.08}{GW200105_combined_lowspin}{0.08}{GW200105_XPHM_highspin}{0.08}{GW200105_v4PHM_highspin}{0.08}{GW200105_combined_highspin}{0.08}{GW200105_NSBH_lowspin}{0.08}{GW200115_XPHM_lowspin}{0.09}{GW200115_v4PHM_lowspin}{0.10}{GW200115_combined_lowspin}{0.10}{GW200115_XPHM_highspin}{0.09}{GW200115_v4PHM_highspin}{0.10}{GW200115_combined_highspin}{0.10}{GW200115_NSBH_lowspin}{0.10}}}
\newcommand{\redshiftninetypercent}[1]{\IfEqCase{#1}{{GW200105_XPHM_lowspin}{0.08}{GW200105_v4PHM_lowspin}{0.08}{GW200105_combined_lowspin}{0.08}{GW200105_XPHM_highspin}{0.08}{GW200105_v4PHM_highspin}{0.08}{GW200105_combined_highspin}{0.08}{GW200105_NSBH_lowspin}{0.08}{GW200115_XPHM_lowspin}{0.08}{GW200115_v4PHM_lowspin}{0.09}{GW200115_combined_lowspin}{0.09}{GW200115_XPHM_highspin}{0.08}{GW200115_v4PHM_highspin}{0.09}{GW200115_combined_highspin}{0.09}{GW200115_NSBH_lowspin}{0.09}}}
\newcommand{\massonesourceminus}[1]{\IfEqCase{#1}{{GW200105_XPHM_lowspin}{1.5}{GW200105_v4PHM_lowspin}{1.0}{GW200105_combined_lowspin}{1.3}{GW200105_XPHM_highspin}{1.7}{GW200105_v4PHM_highspin}{1.1}{GW200105_combined_highspin}{1.5}{GW200105_NSBH_lowspin}{2.2}{GW200115_XPHM_lowspin}{1.7}{GW200115_v4PHM_lowspin}{2.6}{GW200115_combined_lowspin}{2.1}{GW200115_XPHM_highspin}{1.8}{GW200115_v4PHM_highspin}{2.3}{GW200115_combined_highspin}{2.1}{GW200115_NSBH_lowspin}{1.8}}}
\newcommand{\massonesourcemed}[1]{\IfEqCase{#1}{{GW200105_XPHM_lowspin}{8.9}{GW200105_v4PHM_lowspin}{9.0}{GW200105_combined_lowspin}{8.9}{GW200105_XPHM_highspin}{8.9}{GW200105_v4PHM_highspin}{8.9}{GW200105_combined_highspin}{8.9}{GW200105_NSBH_lowspin}{8.8}{GW200115_XPHM_lowspin}{5.5}{GW200115_v4PHM_lowspin}{6.3}{GW200115_combined_lowspin}{5.9}{GW200115_XPHM_highspin}{5.6}{GW200115_v4PHM_highspin}{5.8}{GW200115_combined_highspin}{5.7}{GW200115_NSBH_lowspin}{6.6}}}
\newcommand{\massonesourceplus}[1]{\IfEqCase{#1}{{GW200105_XPHM_lowspin}{1.2}{GW200105_v4PHM_lowspin}{0.9}{GW200105_combined_lowspin}{1.1}{GW200105_XPHM_highspin}{1.4}{GW200105_v4PHM_highspin}{0.8}{GW200105_combined_highspin}{1.2}{GW200105_NSBH_lowspin}{3.1}{GW200115_XPHM_lowspin}{1.8}{GW200115_v4PHM_lowspin}{1.0}{GW200115_combined_lowspin}{1.4}{GW200115_XPHM_highspin}{1.6}{GW200115_v4PHM_highspin}{1.8}{GW200115_combined_highspin}{1.8}{GW200115_NSBH_lowspin}{1.8}}}
\newcommand{\massonesourceonepercent}[1]{\IfEqCase{#1}{{GW200105_XPHM_lowspin}{6.2}{GW200105_v4PHM_lowspin}{7.4}{GW200105_combined_lowspin}{6.7}{GW200105_XPHM_highspin}{5.9}{GW200105_v4PHM_highspin}{5.7}{GW200105_combined_highspin}{5.8}{GW200105_NSBH_lowspin}{6.0}{GW200115_XPHM_lowspin}{3.5}{GW200115_v4PHM_lowspin}{3.4}{GW200115_combined_lowspin}{3.4}{GW200115_XPHM_highspin}{3.3}{GW200115_v4PHM_highspin}{3.0}{GW200115_combined_highspin}{3.1}{GW200115_NSBH_lowspin}{4.3}}}
\newcommand{\massonesourceninetyninepercent}[1]{\IfEqCase{#1}{{GW200105_XPHM_lowspin}{11.0}{GW200105_v4PHM_lowspin}{11.2}{GW200105_combined_lowspin}{11.1}{GW200105_XPHM_highspin}{11.4}{GW200105_v4PHM_highspin}{11.0}{GW200105_combined_highspin}{11.3}{GW200105_NSBH_lowspin}{13.9}{GW200115_XPHM_lowspin}{8.6}{GW200115_v4PHM_lowspin}{8.0}{GW200115_combined_lowspin}{8.3}{GW200115_XPHM_highspin}{8.0}{GW200115_v4PHM_highspin}{8.3}{GW200115_combined_highspin}{8.2}{GW200115_NSBH_lowspin}{9.3}}}
\newcommand{\massonesourcefivepercent}[1]{\IfEqCase{#1}{{GW200105_XPHM_lowspin}{7.4}{GW200105_v4PHM_lowspin}{7.9}{GW200105_combined_lowspin}{7.7}{GW200105_XPHM_highspin}{7.3}{GW200105_v4PHM_highspin}{7.8}{GW200105_combined_highspin}{7.4}{GW200105_NSBH_lowspin}{6.6}{GW200115_XPHM_lowspin}{3.9}{GW200115_v4PHM_lowspin}{3.6}{GW200115_combined_lowspin}{3.8}{GW200115_XPHM_highspin}{3.7}{GW200115_v4PHM_highspin}{3.4}{GW200115_combined_highspin}{3.6}{GW200115_NSBH_lowspin}{4.8}}}
\newcommand{\massonesourceninetyfivepercent}[1]{\IfEqCase{#1}{{GW200105_XPHM_lowspin}{10.1}{GW200105_v4PHM_lowspin}{9.9}{GW200105_combined_lowspin}{10.0}{GW200105_XPHM_highspin}{10.4}{GW200105_v4PHM_highspin}{9.7}{GW200105_combined_highspin}{10.2}{GW200105_NSBH_lowspin}{11.9}{GW200115_XPHM_lowspin}{7.3}{GW200115_v4PHM_lowspin}{7.3}{GW200115_combined_lowspin}{7.3}{GW200115_XPHM_highspin}{7.2}{GW200115_v4PHM_highspin}{7.6}{GW200115_combined_highspin}{7.4}{GW200115_NSBH_lowspin}{8.4}}}
\newcommand{\massonesourceninetypercent}[1]{\IfEqCase{#1}{{GW200105_XPHM_lowspin}{9.7}{GW200105_v4PHM_lowspin}{9.6}{GW200105_combined_lowspin}{9.7}{GW200105_XPHM_highspin}{10.0}{GW200105_v4PHM_highspin}{9.4}{GW200105_combined_highspin}{9.7}{GW200105_NSBH_lowspin}{10.9}{GW200115_XPHM_lowspin}{7.0}{GW200115_v4PHM_lowspin}{7.1}{GW200115_combined_lowspin}{7.1}{GW200115_XPHM_highspin}{6.9}{GW200115_v4PHM_highspin}{7.3}{GW200115_combined_highspin}{7.1}{GW200115_NSBH_lowspin}{7.8}}}
\newcommand{\masstwosourceminus}[1]{\IfEqCase{#1}{{GW200105_XPHM_lowspin}{0.2}{GW200105_v4PHM_lowspin}{0.1}{GW200105_combined_lowspin}{0.2}{GW200105_XPHM_highspin}{0.2}{GW200105_v4PHM_highspin}{0.1}{GW200105_combined_highspin}{0.2}{GW200105_NSBH_lowspin}{0.4}{GW200115_XPHM_lowspin}{0.3}{GW200115_v4PHM_lowspin}{0.2}{GW200115_combined_lowspin}{0.2}{GW200115_XPHM_highspin}{0.3}{GW200115_v4PHM_highspin}{0.3}{GW200115_combined_highspin}{0.3}{GW200115_NSBH_lowspin}{0.2}}}
\newcommand{\masstwosourcemed}[1]{\IfEqCase{#1}{{GW200105_XPHM_lowspin}{1.9}{GW200105_v4PHM_lowspin}{1.9}{GW200105_combined_lowspin}{1.9}{GW200105_XPHM_highspin}{1.9}{GW200105_v4PHM_highspin}{1.9}{GW200105_combined_highspin}{1.9}{GW200105_NSBH_lowspin}{1.9}{GW200115_XPHM_lowspin}{1.5}{GW200115_v4PHM_lowspin}{1.4}{GW200115_combined_lowspin}{1.4}{GW200115_XPHM_highspin}{1.5}{GW200115_v4PHM_highspin}{1.5}{GW200115_combined_highspin}{1.5}{GW200115_NSBH_lowspin}{1.3}}}
\newcommand{\masstwosourceplus}[1]{\IfEqCase{#1}{{GW200105_XPHM_lowspin}{0.3}{GW200105_v4PHM_lowspin}{0.2}{GW200105_combined_lowspin}{0.2}{GW200105_XPHM_highspin}{0.3}{GW200105_v4PHM_highspin}{0.2}{GW200105_combined_highspin}{0.3}{GW200105_NSBH_lowspin}{0.5}{GW200115_XPHM_lowspin}{0.5}{GW200115_v4PHM_lowspin}{0.8}{GW200115_combined_lowspin}{0.6}{GW200115_XPHM_highspin}{0.6}{GW200115_v4PHM_highspin}{0.8}{GW200115_combined_highspin}{0.7}{GW200115_NSBH_lowspin}{0.4}}}
\newcommand{\masstwosourceonepercent}[1]{\IfEqCase{#1}{{GW200105_XPHM_lowspin}{1.6}{GW200105_v4PHM_lowspin}{1.6}{GW200105_combined_lowspin}{1.6}{GW200105_XPHM_highspin}{1.6}{GW200105_v4PHM_highspin}{1.6}{GW200105_combined_highspin}{1.6}{GW200105_NSBH_lowspin}{1.4}{GW200115_XPHM_lowspin}{1.1}{GW200115_v4PHM_lowspin}{1.1}{GW200115_combined_lowspin}{1.1}{GW200115_XPHM_highspin}{1.1}{GW200115_v4PHM_highspin}{1.1}{GW200115_combined_highspin}{1.1}{GW200115_NSBH_lowspin}{1.0}}}
\newcommand{\masstwosourceninetyninepercent}[1]{\IfEqCase{#1}{{GW200105_XPHM_lowspin}{2.5}{GW200105_v4PHM_lowspin}{2.2}{GW200105_combined_lowspin}{2.4}{GW200105_XPHM_highspin}{2.6}{GW200105_v4PHM_highspin}{2.7}{GW200105_combined_highspin}{2.7}{GW200105_NSBH_lowspin}{2.7}{GW200115_XPHM_lowspin}{2.2}{GW200115_v4PHM_lowspin}{2.3}{GW200115_combined_lowspin}{2.3}{GW200115_XPHM_highspin}{2.4}{GW200115_v4PHM_highspin}{2.6}{GW200115_combined_highspin}{2.5}{GW200115_NSBH_lowspin}{1.9}}}
\newcommand{\masstwosourcefivepercent}[1]{\IfEqCase{#1}{{GW200105_XPHM_lowspin}{1.7}{GW200105_v4PHM_lowspin}{1.8}{GW200105_combined_lowspin}{1.8}{GW200105_XPHM_highspin}{1.7}{GW200105_v4PHM_highspin}{1.8}{GW200105_combined_highspin}{1.7}{GW200105_NSBH_lowspin}{1.6}{GW200115_XPHM_lowspin}{1.2}{GW200115_v4PHM_lowspin}{1.2}{GW200115_combined_lowspin}{1.2}{GW200115_XPHM_highspin}{1.2}{GW200115_v4PHM_highspin}{1.2}{GW200115_combined_highspin}{1.2}{GW200115_NSBH_lowspin}{1.1}}}
\newcommand{\masstwosourceninetyfivepercent}[1]{\IfEqCase{#1}{{GW200105_XPHM_lowspin}{2.2}{GW200105_v4PHM_lowspin}{2.1}{GW200105_combined_lowspin}{2.2}{GW200105_XPHM_highspin}{2.2}{GW200105_v4PHM_highspin}{2.1}{GW200105_combined_highspin}{2.2}{GW200105_NSBH_lowspin}{2.5}{GW200115_XPHM_lowspin}{2.0}{GW200115_v4PHM_lowspin}{2.1}{GW200115_combined_lowspin}{2.1}{GW200115_XPHM_highspin}{2.1}{GW200115_v4PHM_highspin}{2.3}{GW200115_combined_highspin}{2.2}{GW200115_NSBH_lowspin}{1.7}}}
\newcommand{\masstwosourceninetypercent}[1]{\IfEqCase{#1}{{GW200105_XPHM_lowspin}{2.1}{GW200105_v4PHM_lowspin}{2.0}{GW200105_combined_lowspin}{2.1}{GW200105_XPHM_highspin}{2.1}{GW200105_v4PHM_highspin}{2.0}{GW200105_combined_highspin}{2.1}{GW200105_NSBH_lowspin}{2.3}{GW200115_XPHM_lowspin}{1.9}{GW200115_v4PHM_lowspin}{2.0}{GW200115_combined_lowspin}{1.9}{GW200115_XPHM_highspin}{1.9}{GW200115_v4PHM_highspin}{2.1}{GW200115_combined_highspin}{2.0}{GW200115_NSBH_lowspin}{1.6}}}
\newcommand{\totalmasssourceminus}[1]{\IfEqCase{#1}{{GW200105_XPHM_lowspin}{1.2}{GW200105_v4PHM_lowspin}{0.9}{GW200105_combined_lowspin}{1.0}{GW200105_XPHM_highspin}{1.4}{GW200105_v4PHM_highspin}{1.0}{GW200105_combined_highspin}{1.2}{GW200105_NSBH_lowspin}{1.7}{GW200115_XPHM_lowspin}{1.1}{GW200115_v4PHM_lowspin}{1.8}{GW200115_combined_lowspin}{1.5}{GW200115_XPHM_highspin}{1.3}{GW200115_v4PHM_highspin}{1.5}{GW200115_combined_highspin}{1.4}{GW200115_NSBH_lowspin}{1.4}}}
\newcommand{\totalmasssourcemed}[1]{\IfEqCase{#1}{{GW200105_XPHM_lowspin}{10.8}{GW200105_v4PHM_lowspin}{10.9}{GW200105_combined_lowspin}{10.8}{GW200105_XPHM_highspin}{10.9}{GW200105_v4PHM_highspin}{10.9}{GW200105_combined_highspin}{10.9}{GW200105_NSBH_lowspin}{10.7}{GW200115_XPHM_lowspin}{7.1}{GW200115_v4PHM_lowspin}{7.6}{GW200115_combined_lowspin}{7.3}{GW200115_XPHM_highspin}{7.1}{GW200115_v4PHM_highspin}{7.2}{GW200115_combined_highspin}{7.1}{GW200115_NSBH_lowspin}{7.9}}}
\newcommand{\totalmasssourceplus}[1]{\IfEqCase{#1}{{GW200105_XPHM_lowspin}{1.0}{GW200105_v4PHM_lowspin}{0.8}{GW200105_combined_lowspin}{0.9}{GW200105_XPHM_highspin}{1.3}{GW200105_v4PHM_highspin}{0.7}{GW200105_combined_highspin}{1.1}{GW200105_NSBH_lowspin}{2.7}{GW200115_XPHM_lowspin}{1.5}{GW200115_v4PHM_lowspin}{0.9}{GW200115_combined_lowspin}{1.2}{GW200115_XPHM_highspin}{1.4}{GW200115_v4PHM_highspin}{1.5}{GW200115_combined_highspin}{1.5}{GW200115_NSBH_lowspin}{1.6}}}
\newcommand{\totalmasssourceonepercent}[1]{\IfEqCase{#1}{{GW200105_XPHM_lowspin}{8.8}{GW200105_v4PHM_lowspin}{9.6}{GW200105_combined_lowspin}{9.0}{GW200105_XPHM_highspin}{8.6}{GW200105_v4PHM_highspin}{8.5}{GW200105_combined_highspin}{8.5}{GW200105_NSBH_lowspin}{8.6}{GW200115_XPHM_lowspin}{5.7}{GW200115_v4PHM_lowspin}{5.6}{GW200115_combined_lowspin}{5.7}{GW200115_XPHM_highspin}{5.6}{GW200115_v4PHM_highspin}{5.5}{GW200115_combined_highspin}{5.6}{GW200115_NSBH_lowspin}{6.2}}}
\newcommand{\totalmasssourceninetyninepercent}[1]{\IfEqCase{#1}{{GW200105_XPHM_lowspin}{12.6}{GW200105_v4PHM_lowspin}{12.8}{GW200105_combined_lowspin}{12.7}{GW200105_XPHM_highspin}{13.0}{GW200105_v4PHM_highspin}{12.7}{GW200105_combined_highspin}{12.9}{GW200105_NSBH_lowspin}{15.3}{GW200115_XPHM_lowspin}{9.7}{GW200115_v4PHM_lowspin}{9.2}{GW200115_combined_lowspin}{9.4}{GW200115_XPHM_highspin}{9.1}{GW200115_v4PHM_highspin}{9.4}{GW200115_combined_highspin}{9.3}{GW200115_NSBH_lowspin}{10.3}}}
\newcommand{\totalmasssourcefivepercent}[1]{\IfEqCase{#1}{{GW200105_XPHM_lowspin}{9.6}{GW200105_v4PHM_lowspin}{10.0}{GW200105_combined_lowspin}{9.8}{GW200105_XPHM_highspin}{9.5}{GW200105_v4PHM_highspin}{9.9}{GW200105_combined_highspin}{9.6}{GW200105_NSBH_lowspin}{9.0}{GW200115_XPHM_lowspin}{5.9}{GW200115_v4PHM_lowspin}{5.8}{GW200115_combined_lowspin}{5.8}{GW200115_XPHM_highspin}{5.8}{GW200115_v4PHM_highspin}{5.7}{GW200115_combined_highspin}{5.7}{GW200115_NSBH_lowspin}{6.5}}}
\newcommand{\totalmasssourceninetyfivepercent}[1]{\IfEqCase{#1}{{GW200105_XPHM_lowspin}{11.9}{GW200105_v4PHM_lowspin}{11.7}{GW200105_combined_lowspin}{11.8}{GW200105_XPHM_highspin}{12.1}{GW200105_v4PHM_highspin}{11.5}{GW200105_combined_highspin}{11.9}{GW200105_NSBH_lowspin}{13.5}{GW200115_XPHM_lowspin}{8.6}{GW200115_v4PHM_lowspin}{8.5}{GW200115_combined_lowspin}{8.5}{GW200115_XPHM_highspin}{8.4}{GW200115_v4PHM_highspin}{8.8}{GW200115_combined_highspin}{8.6}{GW200115_NSBH_lowspin}{9.5}}}
\newcommand{\totalmasssourceninetypercent}[1]{\IfEqCase{#1}{{GW200105_XPHM_lowspin}{11.5}{GW200105_v4PHM_lowspin}{11.4}{GW200105_combined_lowspin}{11.5}{GW200105_XPHM_highspin}{11.8}{GW200105_v4PHM_highspin}{11.3}{GW200105_combined_highspin}{11.5}{GW200105_NSBH_lowspin}{12.6}{GW200115_XPHM_lowspin}{8.3}{GW200115_v4PHM_lowspin}{8.4}{GW200115_combined_lowspin}{8.3}{GW200115_XPHM_highspin}{8.2}{GW200115_v4PHM_highspin}{8.5}{GW200115_combined_highspin}{8.4}{GW200115_NSBH_lowspin}{9.0}}}
\newcommand{\chirpmasssourceminus}[1]{\IfEqCase{#1}{{GW200105_XPHM_lowspin}{0.07}{GW200105_v4PHM_lowspin}{0.08}{GW200105_combined_lowspin}{0.07}{GW200105_XPHM_highspin}{0.07}{GW200105_v4PHM_highspin}{0.08}{GW200105_combined_highspin}{0.07}{GW200105_NSBH_lowspin}{0.07}{GW200115_XPHM_lowspin}{0.06}{GW200115_v4PHM_lowspin}{0.07}{GW200115_combined_lowspin}{0.07}{GW200115_XPHM_highspin}{0.06}{GW200115_v4PHM_highspin}{0.07}{GW200115_combined_highspin}{0.07}{GW200115_NSBH_lowspin}{0.07}}}
\newcommand{\chirpmasssourcemed}[1]{\IfEqCase{#1}{{GW200105_XPHM_lowspin}{3.41}{GW200105_v4PHM_lowspin}{3.42}{GW200105_combined_lowspin}{3.41}{GW200105_XPHM_highspin}{3.41}{GW200105_v4PHM_highspin}{3.42}{GW200105_combined_highspin}{3.41}{GW200105_NSBH_lowspin}{3.41}{GW200115_XPHM_lowspin}{2.42}{GW200115_v4PHM_lowspin}{2.42}{GW200115_combined_lowspin}{2.42}{GW200115_XPHM_highspin}{2.42}{GW200115_v4PHM_highspin}{2.41}{GW200115_combined_highspin}{2.42}{GW200115_NSBH_lowspin}{2.43}}}
\newcommand{\chirpmasssourceplus}[1]{\IfEqCase{#1}{{GW200105_XPHM_lowspin}{0.08}{GW200105_v4PHM_lowspin}{0.08}{GW200105_combined_lowspin}{0.08}{GW200105_XPHM_highspin}{0.07}{GW200105_v4PHM_highspin}{0.08}{GW200105_combined_highspin}{0.08}{GW200105_NSBH_lowspin}{0.09}{GW200115_XPHM_lowspin}{0.04}{GW200115_v4PHM_lowspin}{0.06}{GW200115_combined_lowspin}{0.05}{GW200115_XPHM_highspin}{0.04}{GW200115_v4PHM_highspin}{0.06}{GW200115_combined_highspin}{0.05}{GW200115_NSBH_lowspin}{0.06}}}
\newcommand{\chirpmasssourceonepercent}[1]{\IfEqCase{#1}{{GW200105_XPHM_lowspin}{3.32}{GW200105_v4PHM_lowspin}{3.32}{GW200105_combined_lowspin}{3.32}{GW200105_XPHM_highspin}{3.32}{GW200105_v4PHM_highspin}{3.32}{GW200105_combined_highspin}{3.32}{GW200105_NSBH_lowspin}{3.31}{GW200115_XPHM_lowspin}{2.34}{GW200115_v4PHM_lowspin}{2.32}{GW200115_combined_lowspin}{2.32}{GW200115_XPHM_highspin}{2.34}{GW200115_v4PHM_highspin}{2.31}{GW200115_combined_highspin}{2.32}{GW200115_NSBH_lowspin}{2.33}}}
\newcommand{\chirpmasssourceninetyninepercent}[1]{\IfEqCase{#1}{{GW200105_XPHM_lowspin}{3.51}{GW200105_v4PHM_lowspin}{3.52}{GW200105_combined_lowspin}{3.52}{GW200105_XPHM_highspin}{3.51}{GW200105_v4PHM_highspin}{3.52}{GW200105_combined_highspin}{3.52}{GW200105_NSBH_lowspin}{3.53}{GW200115_XPHM_lowspin}{2.49}{GW200115_v4PHM_lowspin}{2.50}{GW200115_combined_lowspin}{2.49}{GW200115_XPHM_highspin}{2.48}{GW200115_v4PHM_highspin}{2.49}{GW200115_combined_highspin}{2.49}{GW200115_NSBH_lowspin}{2.50}}}
\newcommand{\chirpmasssourcefivepercent}[1]{\IfEqCase{#1}{{GW200105_XPHM_lowspin}{3.34}{GW200105_v4PHM_lowspin}{3.34}{GW200105_combined_lowspin}{3.34}{GW200105_XPHM_highspin}{3.34}{GW200105_v4PHM_highspin}{3.34}{GW200105_combined_highspin}{3.34}{GW200105_NSBH_lowspin}{3.34}{GW200115_XPHM_lowspin}{2.36}{GW200115_v4PHM_lowspin}{2.34}{GW200115_combined_lowspin}{2.35}{GW200115_XPHM_highspin}{2.37}{GW200115_v4PHM_highspin}{2.34}{GW200115_combined_highspin}{2.35}{GW200115_NSBH_lowspin}{2.36}}}
\newcommand{\chirpmasssourceninetyfivepercent}[1]{\IfEqCase{#1}{{GW200105_XPHM_lowspin}{3.49}{GW200105_v4PHM_lowspin}{3.50}{GW200105_combined_lowspin}{3.49}{GW200105_XPHM_highspin}{3.49}{GW200105_v4PHM_highspin}{3.50}{GW200105_combined_highspin}{3.49}{GW200105_NSBH_lowspin}{3.50}{GW200115_XPHM_lowspin}{2.47}{GW200115_v4PHM_lowspin}{2.48}{GW200115_combined_lowspin}{2.47}{GW200115_XPHM_highspin}{2.46}{GW200115_v4PHM_highspin}{2.47}{GW200115_combined_highspin}{2.47}{GW200115_NSBH_lowspin}{2.49}}}
\newcommand{\chirpmasssourceninetypercent}[1]{\IfEqCase{#1}{{GW200105_XPHM_lowspin}{3.47}{GW200105_v4PHM_lowspin}{3.48}{GW200105_combined_lowspin}{3.48}{GW200105_XPHM_highspin}{3.47}{GW200105_v4PHM_highspin}{3.48}{GW200105_combined_highspin}{3.48}{GW200105_NSBH_lowspin}{3.48}{GW200115_XPHM_lowspin}{2.46}{GW200115_v4PHM_lowspin}{2.46}{GW200115_combined_lowspin}{2.46}{GW200115_XPHM_highspin}{2.45}{GW200115_v4PHM_highspin}{2.46}{GW200115_combined_highspin}{2.46}{GW200115_NSBH_lowspin}{2.48}}}
\newcommand{\iotaminus}[1]{\IfEqCase{#1}{{GW200105_XPHM_lowspin}{1.69}{GW200105_v4PHM_lowspin}{1.49}{GW200105_combined_lowspin}{1.61}{GW200105_XPHM_highspin}{1.47}{GW200105_v4PHM_highspin}{1.45}{GW200105_combined_highspin}{1.45}{GW200105_NSBH_lowspin}{1.35}{GW200115_XPHM_lowspin}{0.40}{GW200115_v4PHM_lowspin}{0.53}{GW200115_combined_lowspin}{0.45}{GW200115_XPHM_highspin}{0.39}{GW200115_v4PHM_highspin}{0.54}{GW200115_combined_highspin}{0.44}{GW200115_NSBH_lowspin}{0.53}}}
\newcommand{\iotamed}[1]{\IfEqCase{#1}{{GW200105_XPHM_lowspin}{2.08}{GW200105_v4PHM_lowspin}{1.81}{GW200105_combined_lowspin}{1.97}{GW200105_XPHM_highspin}{1.83}{GW200105_v4PHM_highspin}{1.77}{GW200105_combined_highspin}{1.79}{GW200105_NSBH_lowspin}{1.61}{GW200115_XPHM_lowspin}{0.57}{GW200115_v4PHM_lowspin}{0.72}{GW200115_combined_lowspin}{0.63}{GW200115_XPHM_highspin}{0.56}{GW200115_v4PHM_highspin}{0.73}{GW200115_combined_highspin}{0.62}{GW200115_NSBH_lowspin}{0.72}}}
\newcommand{\iotaplus}[1]{\IfEqCase{#1}{{GW200105_XPHM_lowspin}{0.72}{GW200105_v4PHM_lowspin}{1.04}{GW200105_combined_lowspin}{0.85}{GW200105_XPHM_highspin}{0.98}{GW200105_v4PHM_highspin}{1.08}{GW200105_combined_highspin}{1.04}{GW200105_NSBH_lowspin}{1.28}{GW200115_XPHM_lowspin}{1.78}{GW200115_v4PHM_lowspin}{2.03}{GW200115_combined_lowspin}{2.03}{GW200115_XPHM_highspin}{1.54}{GW200115_v4PHM_highspin}{2.03}{GW200115_combined_highspin}{2.02}{GW200115_NSBH_lowspin}{1.94}}}
\newcommand{\iotaonepercent}[1]{\IfEqCase{#1}{{GW200105_XPHM_lowspin}{0.19}{GW200105_v4PHM_lowspin}{0.15}{GW200105_combined_lowspin}{0.17}{GW200105_XPHM_highspin}{0.18}{GW200105_v4PHM_highspin}{0.14}{GW200105_combined_highspin}{0.16}{GW200105_NSBH_lowspin}{0.11}{GW200115_XPHM_lowspin}{0.08}{GW200115_v4PHM_lowspin}{0.08}{GW200115_combined_lowspin}{0.08}{GW200115_XPHM_highspin}{0.08}{GW200115_v4PHM_highspin}{0.08}{GW200115_combined_highspin}{0.08}{GW200115_NSBH_lowspin}{0.08}}}
\newcommand{\iotaninetyninepercent}[1]{\IfEqCase{#1}{{GW200105_XPHM_lowspin}{2.96}{GW200105_v4PHM_lowspin}{3.01}{GW200105_combined_lowspin}{2.99}{GW200105_XPHM_highspin}{2.97}{GW200105_v4PHM_highspin}{3.01}{GW200105_combined_highspin}{3.00}{GW200105_NSBH_lowspin}{3.02}{GW200115_XPHM_lowspin}{2.84}{GW200115_v4PHM_lowspin}{2.97}{GW200115_combined_lowspin}{2.93}{GW200115_XPHM_highspin}{2.81}{GW200115_v4PHM_highspin}{2.98}{GW200115_combined_highspin}{2.93}{GW200115_NSBH_lowspin}{2.94}}}
\newcommand{\iotafivepercent}[1]{\IfEqCase{#1}{{GW200105_XPHM_lowspin}{0.39}{GW200105_v4PHM_lowspin}{0.32}{GW200105_combined_lowspin}{0.35}{GW200105_XPHM_highspin}{0.36}{GW200105_v4PHM_highspin}{0.32}{GW200105_combined_highspin}{0.34}{GW200105_NSBH_lowspin}{0.26}{GW200115_XPHM_lowspin}{0.17}{GW200115_v4PHM_lowspin}{0.18}{GW200115_combined_lowspin}{0.17}{GW200115_XPHM_highspin}{0.17}{GW200115_v4PHM_highspin}{0.19}{GW200115_combined_highspin}{0.18}{GW200115_NSBH_lowspin}{0.19}}}
\newcommand{\iotaninetyfivepercent}[1]{\IfEqCase{#1}{{GW200105_XPHM_lowspin}{2.79}{GW200105_v4PHM_lowspin}{2.85}{GW200105_combined_lowspin}{2.82}{GW200105_XPHM_highspin}{2.81}{GW200105_v4PHM_highspin}{2.85}{GW200105_combined_highspin}{2.83}{GW200105_NSBH_lowspin}{2.89}{GW200115_XPHM_lowspin}{2.35}{GW200115_v4PHM_lowspin}{2.74}{GW200115_combined_lowspin}{2.65}{GW200115_XPHM_highspin}{2.09}{GW200115_v4PHM_highspin}{2.75}{GW200115_combined_highspin}{2.65}{GW200115_NSBH_lowspin}{2.66}}}
\newcommand{\iotaninetypercent}[1]{\IfEqCase{#1}{{GW200105_XPHM_lowspin}{2.69}{GW200105_v4PHM_lowspin}{2.73}{GW200105_combined_lowspin}{2.70}{GW200105_XPHM_highspin}{2.69}{GW200105_v4PHM_highspin}{2.73}{GW200105_combined_highspin}{2.71}{GW200105_NSBH_lowspin}{2.78}{GW200115_XPHM_lowspin}{1.09}{GW200115_v4PHM_lowspin}{2.55}{GW200115_combined_lowspin}{2.35}{GW200115_XPHM_highspin}{1.00}{GW200115_v4PHM_highspin}{2.57}{GW200115_combined_highspin}{2.34}{GW200115_NSBH_lowspin}{2.37}}}
\newcommand{\spinonexminus}[1]{\IfEqCase{#1}{{GW200105_XPHM_lowspin}{0.16}{GW200105_v4PHM_lowspin}{0.11}{GW200105_combined_lowspin}{0.13}{GW200105_XPHM_highspin}{0.16}{GW200105_v4PHM_highspin}{0.10}{GW200105_combined_highspin}{0.13}{GW200105_NSBH_lowspin}{0.00}{GW200115_XPHM_lowspin}{0.35}{GW200115_v4PHM_lowspin}{0.23}{GW200115_combined_lowspin}{0.30}{GW200115_XPHM_highspin}{0.35}{GW200115_v4PHM_highspin}{0.29}{GW200115_combined_highspin}{0.33}{GW200115_NSBH_lowspin}{0.00}}}
\newcommand{\spinonexmed}[1]{\IfEqCase{#1}{{GW200105_XPHM_lowspin}{0.00}{GW200105_v4PHM_lowspin}{0.00}{GW200105_combined_lowspin}{0.00}{GW200105_XPHM_highspin}{0.00}{GW200105_v4PHM_highspin}{-0.00}{GW200105_combined_highspin}{-0.00}{GW200105_NSBH_lowspin}{0.00}{GW200115_XPHM_lowspin}{0.00}{GW200115_v4PHM_lowspin}{0.00}{GW200115_combined_lowspin}{0.00}{GW200115_XPHM_highspin}{0.00}{GW200115_v4PHM_highspin}{-0.00}{GW200115_combined_highspin}{-0.00}{GW200115_NSBH_lowspin}{0.00}}}
\newcommand{\spinonexplus}[1]{\IfEqCase{#1}{{GW200105_XPHM_lowspin}{0.15}{GW200105_v4PHM_lowspin}{0.11}{GW200105_combined_lowspin}{0.13}{GW200105_XPHM_highspin}{0.15}{GW200105_v4PHM_highspin}{0.09}{GW200105_combined_highspin}{0.12}{GW200105_NSBH_lowspin}{0.00}{GW200115_XPHM_lowspin}{0.36}{GW200115_v4PHM_lowspin}{0.25}{GW200115_combined_lowspin}{0.31}{GW200115_XPHM_highspin}{0.37}{GW200115_v4PHM_highspin}{0.28}{GW200115_combined_highspin}{0.33}{GW200115_NSBH_lowspin}{0.00}}}
\newcommand{\spinonexonepercent}[1]{\IfEqCase{#1}{{GW200105_XPHM_lowspin}{-0.25}{GW200105_v4PHM_lowspin}{-0.18}{GW200105_combined_lowspin}{-0.22}{GW200105_XPHM_highspin}{-0.26}{GW200105_v4PHM_highspin}{-0.17}{GW200105_combined_highspin}{-0.22}{GW200105_NSBH_lowspin}{0.00}{GW200115_XPHM_lowspin}{-0.48}{GW200115_v4PHM_lowspin}{-0.36}{GW200115_combined_lowspin}{-0.44}{GW200115_XPHM_highspin}{-0.50}{GW200115_v4PHM_highspin}{-0.45}{GW200115_combined_highspin}{-0.48}{GW200115_NSBH_lowspin}{0.00}}}
\newcommand{\spinonexninetyninepercent}[1]{\IfEqCase{#1}{{GW200105_XPHM_lowspin}{0.25}{GW200105_v4PHM_lowspin}{0.17}{GW200105_combined_lowspin}{0.21}{GW200105_XPHM_highspin}{0.25}{GW200105_v4PHM_highspin}{0.15}{GW200105_combined_highspin}{0.21}{GW200105_NSBH_lowspin}{0.00}{GW200115_XPHM_lowspin}{0.51}{GW200115_v4PHM_lowspin}{0.37}{GW200115_combined_lowspin}{0.46}{GW200115_XPHM_highspin}{0.54}{GW200115_v4PHM_highspin}{0.39}{GW200115_combined_highspin}{0.48}{GW200115_NSBH_lowspin}{0.00}}}
\newcommand{\spinonexfivepercent}[1]{\IfEqCase{#1}{{GW200105_XPHM_lowspin}{-0.15}{GW200105_v4PHM_lowspin}{-0.11}{GW200105_combined_lowspin}{-0.13}{GW200105_XPHM_highspin}{-0.16}{GW200105_v4PHM_highspin}{-0.10}{GW200105_combined_highspin}{-0.13}{GW200105_NSBH_lowspin}{0.00}{GW200115_XPHM_lowspin}{-0.34}{GW200115_v4PHM_lowspin}{-0.23}{GW200115_combined_lowspin}{-0.30}{GW200115_XPHM_highspin}{-0.35}{GW200115_v4PHM_highspin}{-0.29}{GW200115_combined_highspin}{-0.33}{GW200115_NSBH_lowspin}{0.00}}}
\newcommand{\spinonexninetyfivepercent}[1]{\IfEqCase{#1}{{GW200105_XPHM_lowspin}{0.16}{GW200105_v4PHM_lowspin}{0.11}{GW200105_combined_lowspin}{0.13}{GW200105_XPHM_highspin}{0.15}{GW200105_v4PHM_highspin}{0.09}{GW200105_combined_highspin}{0.12}{GW200105_NSBH_lowspin}{0.00}{GW200115_XPHM_lowspin}{0.36}{GW200115_v4PHM_lowspin}{0.25}{GW200115_combined_lowspin}{0.31}{GW200115_XPHM_highspin}{0.38}{GW200115_v4PHM_highspin}{0.28}{GW200115_combined_highspin}{0.33}{GW200115_NSBH_lowspin}{0.00}}}
\newcommand{\spinonexninetypercent}[1]{\IfEqCase{#1}{{GW200105_XPHM_lowspin}{0.12}{GW200105_v4PHM_lowspin}{0.07}{GW200105_combined_lowspin}{0.10}{GW200105_XPHM_highspin}{0.11}{GW200105_v4PHM_highspin}{0.06}{GW200105_combined_highspin}{0.09}{GW200105_NSBH_lowspin}{0.00}{GW200115_XPHM_lowspin}{0.28}{GW200115_v4PHM_lowspin}{0.18}{GW200115_combined_lowspin}{0.24}{GW200115_XPHM_highspin}{0.29}{GW200115_v4PHM_highspin}{0.21}{GW200115_combined_highspin}{0.25}{GW200115_NSBH_lowspin}{0.00}}}
\newcommand{\spinoneyminus}[1]{\IfEqCase{#1}{{GW200105_XPHM_lowspin}{0.14}{GW200105_v4PHM_lowspin}{0.11}{GW200105_combined_lowspin}{0.13}{GW200105_XPHM_highspin}{0.15}{GW200105_v4PHM_highspin}{0.10}{GW200105_combined_highspin}{0.13}{GW200105_NSBH_lowspin}{0.00}{GW200115_XPHM_lowspin}{0.34}{GW200115_v4PHM_lowspin}{0.26}{GW200115_combined_lowspin}{0.30}{GW200115_XPHM_highspin}{0.36}{GW200115_v4PHM_highspin}{0.29}{GW200115_combined_highspin}{0.33}{GW200115_NSBH_lowspin}{0.00}}}
\newcommand{\spinoneymed}[1]{\IfEqCase{#1}{{GW200105_XPHM_lowspin}{0.00}{GW200105_v4PHM_lowspin}{0.00}{GW200105_combined_lowspin}{0.00}{GW200105_XPHM_highspin}{-0.00}{GW200105_v4PHM_highspin}{0.00}{GW200105_combined_highspin}{0.00}{GW200105_NSBH_lowspin}{0.00}{GW200115_XPHM_lowspin}{-0.01}{GW200115_v4PHM_lowspin}{-0.00}{GW200115_combined_lowspin}{-0.00}{GW200115_XPHM_highspin}{0.00}{GW200115_v4PHM_highspin}{0.00}{GW200115_combined_highspin}{0.00}{GW200115_NSBH_lowspin}{0.00}}}
\newcommand{\spinoneyplus}[1]{\IfEqCase{#1}{{GW200105_XPHM_lowspin}{0.14}{GW200105_v4PHM_lowspin}{0.11}{GW200105_combined_lowspin}{0.13}{GW200105_XPHM_highspin}{0.14}{GW200105_v4PHM_highspin}{0.10}{GW200105_combined_highspin}{0.12}{GW200105_NSBH_lowspin}{0.00}{GW200115_XPHM_lowspin}{0.35}{GW200115_v4PHM_lowspin}{0.26}{GW200115_combined_lowspin}{0.32}{GW200115_XPHM_highspin}{0.37}{GW200115_v4PHM_highspin}{0.29}{GW200115_combined_highspin}{0.34}{GW200115_NSBH_lowspin}{0.00}}}
\newcommand{\spinoneyonepercent}[1]{\IfEqCase{#1}{{GW200105_XPHM_lowspin}{-0.24}{GW200105_v4PHM_lowspin}{-0.18}{GW200105_combined_lowspin}{-0.21}{GW200105_XPHM_highspin}{-0.26}{GW200105_v4PHM_highspin}{-0.18}{GW200105_combined_highspin}{-0.22}{GW200105_NSBH_lowspin}{0.00}{GW200115_XPHM_lowspin}{-0.49}{GW200115_v4PHM_lowspin}{-0.37}{GW200115_combined_lowspin}{-0.44}{GW200115_XPHM_highspin}{-0.51}{GW200115_v4PHM_highspin}{-0.42}{GW200115_combined_highspin}{-0.47}{GW200115_NSBH_lowspin}{0.00}}}
\newcommand{\spinoneyninetyninepercent}[1]{\IfEqCase{#1}{{GW200105_XPHM_lowspin}{0.24}{GW200105_v4PHM_lowspin}{0.19}{GW200105_combined_lowspin}{0.21}{GW200105_XPHM_highspin}{0.24}{GW200105_v4PHM_highspin}{0.16}{GW200105_combined_highspin}{0.20}{GW200105_NSBH_lowspin}{0.00}{GW200115_XPHM_lowspin}{0.48}{GW200115_v4PHM_lowspin}{0.40}{GW200115_combined_lowspin}{0.44}{GW200115_XPHM_highspin}{0.52}{GW200115_v4PHM_highspin}{0.42}{GW200115_combined_highspin}{0.49}{GW200115_NSBH_lowspin}{0.00}}}
\newcommand{\spinoneyfivepercent}[1]{\IfEqCase{#1}{{GW200105_XPHM_lowspin}{-0.14}{GW200105_v4PHM_lowspin}{-0.11}{GW200105_combined_lowspin}{-0.13}{GW200105_XPHM_highspin}{-0.15}{GW200105_v4PHM_highspin}{-0.10}{GW200105_combined_highspin}{-0.13}{GW200105_NSBH_lowspin}{0.00}{GW200115_XPHM_lowspin}{-0.35}{GW200115_v4PHM_lowspin}{-0.26}{GW200115_combined_lowspin}{-0.31}{GW200115_XPHM_highspin}{-0.36}{GW200115_v4PHM_highspin}{-0.28}{GW200115_combined_highspin}{-0.32}{GW200115_NSBH_lowspin}{0.00}}}
\newcommand{\spinoneyninetyfivepercent}[1]{\IfEqCase{#1}{{GW200105_XPHM_lowspin}{0.14}{GW200105_v4PHM_lowspin}{0.11}{GW200105_combined_lowspin}{0.13}{GW200105_XPHM_highspin}{0.14}{GW200105_v4PHM_highspin}{0.10}{GW200105_combined_highspin}{0.12}{GW200105_NSBH_lowspin}{0.00}{GW200115_XPHM_lowspin}{0.35}{GW200115_v4PHM_lowspin}{0.26}{GW200115_combined_lowspin}{0.31}{GW200115_XPHM_highspin}{0.37}{GW200115_v4PHM_highspin}{0.29}{GW200115_combined_highspin}{0.34}{GW200115_NSBH_lowspin}{0.00}}}
\newcommand{\spinoneyninetypercent}[1]{\IfEqCase{#1}{{GW200105_XPHM_lowspin}{0.10}{GW200105_v4PHM_lowspin}{0.08}{GW200105_combined_lowspin}{0.09}{GW200105_XPHM_highspin}{0.10}{GW200105_v4PHM_highspin}{0.07}{GW200105_combined_highspin}{0.09}{GW200105_NSBH_lowspin}{0.00}{GW200115_XPHM_lowspin}{0.27}{GW200115_v4PHM_lowspin}{0.18}{GW200115_combined_lowspin}{0.23}{GW200115_XPHM_highspin}{0.29}{GW200115_v4PHM_highspin}{0.22}{GW200115_combined_highspin}{0.25}{GW200115_NSBH_lowspin}{0.00}}}
\newcommand{\spinonezminus}[1]{\IfEqCase{#1}{{GW200105_XPHM_lowspin}{0.20}{GW200105_v4PHM_lowspin}{0.12}{GW200105_combined_lowspin}{0.16}{GW200105_XPHM_highspin}{0.21}{GW200105_v4PHM_highspin}{0.15}{GW200105_combined_highspin}{0.19}{GW200105_NSBH_lowspin}{0.30}{GW200115_XPHM_lowspin}{0.45}{GW200115_v4PHM_lowspin}{0.70}{GW200115_combined_lowspin}{0.57}{GW200115_XPHM_highspin}{0.46}{GW200115_v4PHM_highspin}{0.54}{GW200115_combined_highspin}{0.50}{GW200115_NSBH_lowspin}{0.33}}}
\newcommand{\spinonezmed}[1]{\IfEqCase{#1}{{GW200105_XPHM_lowspin}{-0.01}{GW200105_v4PHM_lowspin}{-0.00}{GW200105_combined_lowspin}{-0.01}{GW200105_XPHM_highspin}{-0.00}{GW200105_v4PHM_highspin}{-0.00}{GW200105_combined_highspin}{-0.00}{GW200105_NSBH_lowspin}{-0.01}{GW200115_XPHM_lowspin}{-0.25}{GW200115_v4PHM_lowspin}{-0.11}{GW200115_combined_lowspin}{-0.18}{GW200115_XPHM_highspin}{-0.21}{GW200115_v4PHM_highspin}{-0.17}{GW200115_combined_highspin}{-0.19}{GW200115_NSBH_lowspin}{-0.05}}}
\newcommand{\spinonezplus}[1]{\IfEqCase{#1}{{GW200105_XPHM_lowspin}{0.11}{GW200105_v4PHM_lowspin}{0.09}{GW200105_combined_lowspin}{0.10}{GW200105_XPHM_highspin}{0.13}{GW200105_v4PHM_highspin}{0.07}{GW200105_combined_highspin}{0.11}{GW200105_NSBH_lowspin}{0.25}{GW200115_XPHM_lowspin}{0.28}{GW200115_v4PHM_lowspin}{0.14}{GW200115_combined_lowspin}{0.21}{GW200115_XPHM_highspin}{0.23}{GW200115_v4PHM_highspin}{0.23}{GW200115_combined_highspin}{0.24}{GW200115_NSBH_lowspin}{0.20}}}
\newcommand{\spinonezonepercent}[1]{\IfEqCase{#1}{{GW200105_XPHM_lowspin}{-0.41}{GW200105_v4PHM_lowspin}{-0.20}{GW200105_combined_lowspin}{-0.33}{GW200105_XPHM_highspin}{-0.43}{GW200105_v4PHM_highspin}{-0.47}{GW200105_combined_highspin}{-0.44}{GW200105_NSBH_lowspin}{-0.42}{GW200115_XPHM_lowspin}{-0.86}{GW200115_v4PHM_lowspin}{-0.89}{GW200115_combined_lowspin}{-0.88}{GW200115_XPHM_highspin}{-0.82}{GW200115_v4PHM_highspin}{-0.83}{GW200115_combined_highspin}{-0.82}{GW200115_NSBH_lowspin}{-0.55}}}
\newcommand{\spinonezninetyninepercent}[1]{\IfEqCase{#1}{{GW200105_XPHM_lowspin}{0.18}{GW200105_v4PHM_lowspin}{0.18}{GW200105_combined_lowspin}{0.18}{GW200105_XPHM_highspin}{0.20}{GW200105_v4PHM_highspin}{0.18}{GW200105_combined_highspin}{0.19}{GW200105_NSBH_lowspin}{0.34}{GW200115_XPHM_lowspin}{0.16}{GW200115_v4PHM_lowspin}{0.12}{GW200115_combined_lowspin}{0.14}{GW200115_XPHM_highspin}{0.11}{GW200115_v4PHM_highspin}{0.15}{GW200115_combined_highspin}{0.14}{GW200115_NSBH_lowspin}{0.23}}}
\newcommand{\spinonezfivepercent}[1]{\IfEqCase{#1}{{GW200105_XPHM_lowspin}{-0.21}{GW200105_v4PHM_lowspin}{-0.13}{GW200105_combined_lowspin}{-0.17}{GW200105_XPHM_highspin}{-0.21}{GW200105_v4PHM_highspin}{-0.15}{GW200105_combined_highspin}{-0.19}{GW200105_NSBH_lowspin}{-0.31}{GW200115_XPHM_lowspin}{-0.70}{GW200115_v4PHM_lowspin}{-0.81}{GW200115_combined_lowspin}{-0.75}{GW200115_XPHM_highspin}{-0.66}{GW200115_v4PHM_highspin}{-0.71}{GW200115_combined_highspin}{-0.69}{GW200115_NSBH_lowspin}{-0.39}}}
\newcommand{\spinonezninetyfivepercent}[1]{\IfEqCase{#1}{{GW200105_XPHM_lowspin}{0.10}{GW200105_v4PHM_lowspin}{0.09}{GW200105_combined_lowspin}{0.09}{GW200105_XPHM_highspin}{0.13}{GW200105_v4PHM_highspin}{0.07}{GW200105_combined_highspin}{0.11}{GW200105_NSBH_lowspin}{0.24}{GW200115_XPHM_lowspin}{0.03}{GW200115_v4PHM_lowspin}{0.03}{GW200115_combined_lowspin}{0.03}{GW200115_XPHM_highspin}{0.02}{GW200115_v4PHM_highspin}{0.06}{GW200115_combined_highspin}{0.05}{GW200115_NSBH_lowspin}{0.15}}}
\newcommand{\spinonezninetypercent}[1]{\IfEqCase{#1}{{GW200105_XPHM_lowspin}{0.07}{GW200105_v4PHM_lowspin}{0.06}{GW200105_combined_lowspin}{0.06}{GW200105_XPHM_highspin}{0.09}{GW200105_v4PHM_highspin}{0.04}{GW200105_combined_highspin}{0.07}{GW200105_NSBH_lowspin}{0.17}{GW200115_XPHM_lowspin}{-0.01}{GW200115_v4PHM_lowspin}{0.01}{GW200115_combined_lowspin}{0.01}{GW200115_XPHM_highspin}{-0.01}{GW200115_v4PHM_highspin}{0.02}{GW200115_combined_highspin}{0.01}{GW200115_NSBH_lowspin}{0.10}}}
\newcommand{\spintwoxminus}[1]{\IfEqCase{#1}{{GW200105_XPHM_lowspin}{0.03}{GW200105_v4PHM_lowspin}{0.03}{GW200105_combined_lowspin}{0.03}{GW200105_XPHM_highspin}{0.57}{GW200105_v4PHM_highspin}{0.40}{GW200105_combined_highspin}{0.50}{GW200105_NSBH_lowspin}{0.00}{GW200115_XPHM_lowspin}{0.03}{GW200115_v4PHM_lowspin}{0.03}{GW200115_combined_lowspin}{0.03}{GW200115_XPHM_highspin}{0.57}{GW200115_v4PHM_highspin}{0.41}{GW200115_combined_highspin}{0.50}{GW200115_NSBH_lowspin}{0.00}}}
\newcommand{\spintwoxmed}[1]{\IfEqCase{#1}{{GW200105_XPHM_lowspin}{0.00}{GW200105_v4PHM_lowspin}{0.00}{GW200105_combined_lowspin}{0.00}{GW200105_XPHM_highspin}{0.00}{GW200105_v4PHM_highspin}{0.00}{GW200105_combined_highspin}{0.00}{GW200105_NSBH_lowspin}{0.00}{GW200115_XPHM_lowspin}{-0.00}{GW200115_v4PHM_lowspin}{-0.00}{GW200115_combined_lowspin}{-0.00}{GW200115_XPHM_highspin}{0.00}{GW200115_v4PHM_highspin}{0.01}{GW200115_combined_highspin}{0.00}{GW200115_NSBH_lowspin}{0.00}}}
\newcommand{\spintwoxplus}[1]{\IfEqCase{#1}{{GW200105_XPHM_lowspin}{0.03}{GW200105_v4PHM_lowspin}{0.03}{GW200105_combined_lowspin}{0.03}{GW200105_XPHM_highspin}{0.57}{GW200105_v4PHM_highspin}{0.40}{GW200105_combined_highspin}{0.49}{GW200105_NSBH_lowspin}{0.00}{GW200115_XPHM_lowspin}{0.03}{GW200115_v4PHM_lowspin}{0.03}{GW200115_combined_lowspin}{0.03}{GW200115_XPHM_highspin}{0.57}{GW200115_v4PHM_highspin}{0.42}{GW200115_combined_highspin}{0.51}{GW200115_NSBH_lowspin}{0.00}}}
\newcommand{\spintwoxonepercent}[1]{\IfEqCase{#1}{{GW200105_XPHM_lowspin}{-0.04}{GW200105_v4PHM_lowspin}{-0.04}{GW200105_combined_lowspin}{-0.04}{GW200105_XPHM_highspin}{-0.79}{GW200105_v4PHM_highspin}{-0.60}{GW200105_combined_highspin}{-0.74}{GW200105_NSBH_lowspin}{0.00}{GW200115_XPHM_lowspin}{-0.04}{GW200115_v4PHM_lowspin}{-0.04}{GW200115_combined_lowspin}{-0.04}{GW200115_XPHM_highspin}{-0.78}{GW200115_v4PHM_highspin}{-0.63}{GW200115_combined_highspin}{-0.73}{GW200115_NSBH_lowspin}{0.00}}}
\newcommand{\spintwoxninetyninepercent}[1]{\IfEqCase{#1}{{GW200105_XPHM_lowspin}{0.04}{GW200105_v4PHM_lowspin}{0.04}{GW200105_combined_lowspin}{0.04}{GW200105_XPHM_highspin}{0.79}{GW200105_v4PHM_highspin}{0.61}{GW200105_combined_highspin}{0.73}{GW200105_NSBH_lowspin}{0.00}{GW200115_XPHM_lowspin}{0.04}{GW200115_v4PHM_lowspin}{0.04}{GW200115_combined_lowspin}{0.04}{GW200115_XPHM_highspin}{0.78}{GW200115_v4PHM_highspin}{0.70}{GW200115_combined_highspin}{0.76}{GW200115_NSBH_lowspin}{0.00}}}
\newcommand{\spintwoxfivepercent}[1]{\IfEqCase{#1}{{GW200105_XPHM_lowspin}{-0.03}{GW200105_v4PHM_lowspin}{-0.03}{GW200105_combined_lowspin}{-0.03}{GW200105_XPHM_highspin}{-0.57}{GW200105_v4PHM_highspin}{-0.40}{GW200105_combined_highspin}{-0.50}{GW200105_NSBH_lowspin}{0.00}{GW200115_XPHM_lowspin}{-0.03}{GW200115_v4PHM_lowspin}{-0.03}{GW200115_combined_lowspin}{-0.03}{GW200115_XPHM_highspin}{-0.57}{GW200115_v4PHM_highspin}{-0.41}{GW200115_combined_highspin}{-0.50}{GW200115_NSBH_lowspin}{0.00}}}
\newcommand{\spintwoxninetyfivepercent}[1]{\IfEqCase{#1}{{GW200105_XPHM_lowspin}{0.03}{GW200105_v4PHM_lowspin}{0.03}{GW200105_combined_lowspin}{0.03}{GW200105_XPHM_highspin}{0.58}{GW200105_v4PHM_highspin}{0.40}{GW200105_combined_highspin}{0.49}{GW200105_NSBH_lowspin}{0.00}{GW200115_XPHM_lowspin}{0.03}{GW200115_v4PHM_lowspin}{0.03}{GW200115_combined_lowspin}{0.03}{GW200115_XPHM_highspin}{0.57}{GW200115_v4PHM_highspin}{0.43}{GW200115_combined_highspin}{0.51}{GW200115_NSBH_lowspin}{0.00}}}
\newcommand{\spintwoxninetypercent}[1]{\IfEqCase{#1}{{GW200105_XPHM_lowspin}{0.02}{GW200105_v4PHM_lowspin}{0.02}{GW200105_combined_lowspin}{0.02}{GW200105_XPHM_highspin}{0.43}{GW200105_v4PHM_highspin}{0.28}{GW200105_combined_highspin}{0.36}{GW200105_NSBH_lowspin}{0.00}{GW200115_XPHM_lowspin}{0.02}{GW200115_v4PHM_lowspin}{0.02}{GW200115_combined_lowspin}{0.02}{GW200115_XPHM_highspin}{0.43}{GW200115_v4PHM_highspin}{0.32}{GW200115_combined_highspin}{0.38}{GW200115_NSBH_lowspin}{0.00}}}
\newcommand{\spintwoyminus}[1]{\IfEqCase{#1}{{GW200105_XPHM_lowspin}{0.03}{GW200105_v4PHM_lowspin}{0.03}{GW200105_combined_lowspin}{0.03}{GW200105_XPHM_highspin}{0.57}{GW200105_v4PHM_highspin}{0.35}{GW200105_combined_highspin}{0.48}{GW200105_NSBH_lowspin}{0.00}{GW200115_XPHM_lowspin}{0.03}{GW200115_v4PHM_lowspin}{0.03}{GW200115_combined_lowspin}{0.03}{GW200115_XPHM_highspin}{0.58}{GW200115_v4PHM_highspin}{0.44}{GW200115_combined_highspin}{0.52}{GW200115_NSBH_lowspin}{0.00}}}
\newcommand{\spintwoymed}[1]{\IfEqCase{#1}{{GW200105_XPHM_lowspin}{-0.00}{GW200105_v4PHM_lowspin}{0.00}{GW200105_combined_lowspin}{0.00}{GW200105_XPHM_highspin}{0.00}{GW200105_v4PHM_highspin}{0.00}{GW200105_combined_highspin}{0.00}{GW200105_NSBH_lowspin}{0.00}{GW200115_XPHM_lowspin}{-0.00}{GW200115_v4PHM_lowspin}{0.00}{GW200115_combined_lowspin}{-0.00}{GW200115_XPHM_highspin}{0.00}{GW200115_v4PHM_highspin}{0.00}{GW200115_combined_highspin}{0.00}{GW200115_NSBH_lowspin}{0.00}}}
\newcommand{\spintwoyplus}[1]{\IfEqCase{#1}{{GW200105_XPHM_lowspin}{0.03}{GW200105_v4PHM_lowspin}{0.03}{GW200105_combined_lowspin}{0.03}{GW200105_XPHM_highspin}{0.59}{GW200105_v4PHM_highspin}{0.40}{GW200105_combined_highspin}{0.50}{GW200105_NSBH_lowspin}{0.00}{GW200115_XPHM_lowspin}{0.03}{GW200115_v4PHM_lowspin}{0.03}{GW200115_combined_lowspin}{0.03}{GW200115_XPHM_highspin}{0.57}{GW200115_v4PHM_highspin}{0.45}{GW200115_combined_highspin}{0.52}{GW200115_NSBH_lowspin}{0.00}}}
\newcommand{\spintwoyonepercent}[1]{\IfEqCase{#1}{{GW200105_XPHM_lowspin}{-0.04}{GW200105_v4PHM_lowspin}{-0.04}{GW200105_combined_lowspin}{-0.04}{GW200105_XPHM_highspin}{-0.78}{GW200105_v4PHM_highspin}{-0.55}{GW200105_combined_highspin}{-0.71}{GW200105_NSBH_lowspin}{0.00}{GW200115_XPHM_lowspin}{-0.04}{GW200115_v4PHM_lowspin}{-0.04}{GW200115_combined_lowspin}{-0.04}{GW200115_XPHM_highspin}{-0.78}{GW200115_v4PHM_highspin}{-0.70}{GW200115_combined_highspin}{-0.75}{GW200115_NSBH_lowspin}{0.00}}}
\newcommand{\spintwoyninetyninepercent}[1]{\IfEqCase{#1}{{GW200105_XPHM_lowspin}{0.04}{GW200105_v4PHM_lowspin}{0.04}{GW200105_combined_lowspin}{0.04}{GW200105_XPHM_highspin}{0.79}{GW200105_v4PHM_highspin}{0.60}{GW200105_combined_highspin}{0.73}{GW200105_NSBH_lowspin}{0.00}{GW200115_XPHM_lowspin}{0.04}{GW200115_v4PHM_lowspin}{0.04}{GW200115_combined_lowspin}{0.04}{GW200115_XPHM_highspin}{0.78}{GW200115_v4PHM_highspin}{0.71}{GW200115_combined_highspin}{0.75}{GW200115_NSBH_lowspin}{0.00}}}
\newcommand{\spintwoyfivepercent}[1]{\IfEqCase{#1}{{GW200105_XPHM_lowspin}{-0.03}{GW200105_v4PHM_lowspin}{-0.03}{GW200105_combined_lowspin}{-0.03}{GW200105_XPHM_highspin}{-0.57}{GW200105_v4PHM_highspin}{-0.35}{GW200105_combined_highspin}{-0.48}{GW200105_NSBH_lowspin}{0.00}{GW200115_XPHM_lowspin}{-0.03}{GW200115_v4PHM_lowspin}{-0.03}{GW200115_combined_lowspin}{-0.03}{GW200115_XPHM_highspin}{-0.58}{GW200115_v4PHM_highspin}{-0.44}{GW200115_combined_highspin}{-0.51}{GW200115_NSBH_lowspin}{0.00}}}
\newcommand{\spintwoyninetyfivepercent}[1]{\IfEqCase{#1}{{GW200105_XPHM_lowspin}{0.03}{GW200105_v4PHM_lowspin}{0.03}{GW200105_combined_lowspin}{0.03}{GW200105_XPHM_highspin}{0.59}{GW200105_v4PHM_highspin}{0.40}{GW200105_combined_highspin}{0.50}{GW200105_NSBH_lowspin}{0.00}{GW200115_XPHM_lowspin}{0.03}{GW200115_v4PHM_lowspin}{0.03}{GW200115_combined_lowspin}{0.03}{GW200115_XPHM_highspin}{0.58}{GW200115_v4PHM_highspin}{0.45}{GW200115_combined_highspin}{0.52}{GW200115_NSBH_lowspin}{0.00}}}
\newcommand{\spintwoyninetypercent}[1]{\IfEqCase{#1}{{GW200105_XPHM_lowspin}{0.02}{GW200105_v4PHM_lowspin}{0.02}{GW200105_combined_lowspin}{0.02}{GW200105_XPHM_highspin}{0.44}{GW200105_v4PHM_highspin}{0.28}{GW200105_combined_highspin}{0.36}{GW200105_NSBH_lowspin}{0.00}{GW200115_XPHM_lowspin}{0.02}{GW200115_v4PHM_lowspin}{0.02}{GW200115_combined_lowspin}{0.02}{GW200115_XPHM_highspin}{0.44}{GW200115_v4PHM_highspin}{0.33}{GW200115_combined_highspin}{0.39}{GW200115_NSBH_lowspin}{0.00}}}
\newcommand{\spintwozminus}[1]{\IfEqCase{#1}{{GW200105_XPHM_lowspin}{0.03}{GW200105_v4PHM_lowspin}{0.03}{GW200105_combined_lowspin}{0.03}{GW200105_XPHM_highspin}{0.49}{GW200105_v4PHM_highspin}{0.18}{GW200105_combined_highspin}{0.36}{GW200105_NSBH_lowspin}{0.03}{GW200115_XPHM_lowspin}{0.03}{GW200115_v4PHM_lowspin}{0.03}{GW200115_combined_lowspin}{0.03}{GW200115_XPHM_highspin}{0.54}{GW200115_v4PHM_highspin}{0.59}{GW200115_combined_highspin}{0.56}{GW200115_NSBH_lowspin}{0.03}}}
\newcommand{\spintwozmed}[1]{\IfEqCase{#1}{{GW200105_XPHM_lowspin}{0.00}{GW200105_v4PHM_lowspin}{-0.00}{GW200105_combined_lowspin}{-0.00}{GW200105_XPHM_highspin}{-0.00}{GW200105_v4PHM_highspin}{-0.01}{GW200105_combined_highspin}{-0.00}{GW200105_NSBH_lowspin}{0.00}{GW200115_XPHM_lowspin}{0.00}{GW200115_v4PHM_lowspin}{-0.00}{GW200115_combined_lowspin}{-0.00}{GW200115_XPHM_highspin}{-0.10}{GW200115_v4PHM_highspin}{-0.06}{GW200115_combined_highspin}{-0.08}{GW200115_NSBH_lowspin}{-0.00}}}
\newcommand{\spintwozplus}[1]{\IfEqCase{#1}{{GW200105_XPHM_lowspin}{0.03}{GW200105_v4PHM_lowspin}{0.03}{GW200105_combined_lowspin}{0.03}{GW200105_XPHM_highspin}{0.52}{GW200105_v4PHM_highspin}{0.17}{GW200105_combined_highspin}{0.38}{GW200105_NSBH_lowspin}{0.03}{GW200115_XPHM_lowspin}{0.03}{GW200115_v4PHM_lowspin}{0.03}{GW200115_combined_lowspin}{0.03}{GW200115_XPHM_highspin}{0.42}{GW200115_v4PHM_highspin}{0.37}{GW200115_combined_highspin}{0.39}{GW200115_NSBH_lowspin}{0.03}}}
\newcommand{\spintwozonepercent}[1]{\IfEqCase{#1}{{GW200105_XPHM_lowspin}{-0.04}{GW200105_v4PHM_lowspin}{-0.04}{GW200105_combined_lowspin}{-0.04}{GW200105_XPHM_highspin}{-0.74}{GW200105_v4PHM_highspin}{-0.25}{GW200105_combined_highspin}{-0.65}{GW200105_NSBH_lowspin}{-0.04}{GW200115_XPHM_lowspin}{-0.04}{GW200115_v4PHM_lowspin}{-0.04}{GW200115_combined_lowspin}{-0.04}{GW200115_XPHM_highspin}{-0.81}{GW200115_v4PHM_highspin}{-0.79}{GW200115_combined_highspin}{-0.80}{GW200115_NSBH_lowspin}{-0.04}}}
\newcommand{\spintwozninetyninepercent}[1]{\IfEqCase{#1}{{GW200105_XPHM_lowspin}{0.04}{GW200105_v4PHM_lowspin}{0.04}{GW200105_combined_lowspin}{0.04}{GW200105_XPHM_highspin}{0.75}{GW200105_v4PHM_highspin}{0.24}{GW200105_combined_highspin}{0.67}{GW200105_NSBH_lowspin}{0.04}{GW200115_XPHM_lowspin}{0.04}{GW200115_v4PHM_lowspin}{0.04}{GW200115_combined_lowspin}{0.04}{GW200115_XPHM_highspin}{0.57}{GW200115_v4PHM_highspin}{0.50}{GW200115_combined_highspin}{0.54}{GW200115_NSBH_lowspin}{0.04}}}
\newcommand{\spintwozfivepercent}[1]{\IfEqCase{#1}{{GW200105_XPHM_lowspin}{-0.03}{GW200105_v4PHM_lowspin}{-0.03}{GW200105_combined_lowspin}{-0.03}{GW200105_XPHM_highspin}{-0.50}{GW200105_v4PHM_highspin}{-0.19}{GW200105_combined_highspin}{-0.37}{GW200105_NSBH_lowspin}{-0.03}{GW200115_XPHM_lowspin}{-0.03}{GW200115_v4PHM_lowspin}{-0.03}{GW200115_combined_lowspin}{-0.03}{GW200115_XPHM_highspin}{-0.64}{GW200115_v4PHM_highspin}{-0.66}{GW200115_combined_highspin}{-0.64}{GW200115_NSBH_lowspin}{-0.03}}}
\newcommand{\spintwozninetyfivepercent}[1]{\IfEqCase{#1}{{GW200105_XPHM_lowspin}{0.03}{GW200105_v4PHM_lowspin}{0.03}{GW200105_combined_lowspin}{0.03}{GW200105_XPHM_highspin}{0.51}{GW200105_v4PHM_highspin}{0.16}{GW200105_combined_highspin}{0.37}{GW200105_NSBH_lowspin}{0.03}{GW200115_XPHM_lowspin}{0.03}{GW200115_v4PHM_lowspin}{0.03}{GW200115_combined_lowspin}{0.03}{GW200115_XPHM_highspin}{0.32}{GW200115_v4PHM_highspin}{0.31}{GW200115_combined_highspin}{0.31}{GW200115_NSBH_lowspin}{0.03}}}
\newcommand{\spintwozninetypercent}[1]{\IfEqCase{#1}{{GW200105_XPHM_lowspin}{0.02}{GW200105_v4PHM_lowspin}{0.02}{GW200105_combined_lowspin}{0.02}{GW200105_XPHM_highspin}{0.37}{GW200105_v4PHM_highspin}{0.12}{GW200105_combined_highspin}{0.23}{GW200105_NSBH_lowspin}{0.02}{GW200115_XPHM_lowspin}{0.02}{GW200115_v4PHM_lowspin}{0.02}{GW200115_combined_lowspin}{0.02}{GW200115_XPHM_highspin}{0.20}{GW200115_v4PHM_highspin}{0.21}{GW200115_combined_highspin}{0.21}{GW200115_NSBH_lowspin}{0.02}}}
\newcommand{\finalspinminus}[1]{\IfEqCase{#1}{{GW200105_XPHM_lowspin}{0.01}{GW200105_v4PHM_lowspin}{0.01}{GW200105_combined_lowspin}{0.01}{GW200105_XPHM_highspin}{0.03}{GW200105_v4PHM_highspin}{0.02}{GW200105_combined_highspin}{0.02}{GW200105_NSBH_lowspin}{0.03}{GW200115_XPHM_lowspin}{0.03}{GW200115_v4PHM_lowspin}{0.02}{GW200115_combined_lowspin}{0.03}{GW200115_XPHM_highspin}{0.05}{GW200115_v4PHM_highspin}{0.04}{GW200115_combined_highspin}{0.04}{GW200115_NSBH_lowspin}{0.02}}}
\newcommand{\finalspinmed}[1]{\IfEqCase{#1}{{GW200105_XPHM_lowspin}{0.43}{GW200105_v4PHM_lowspin}{0.43}{GW200105_combined_lowspin}{0.43}{GW200105_XPHM_highspin}{0.44}{GW200105_v4PHM_highspin}{0.44}{GW200105_combined_highspin}{0.44}{GW200105_NSBH_lowspin}{0.43}{GW200115_XPHM_lowspin}{0.41}{GW200115_v4PHM_lowspin}{0.40}{GW200115_combined_lowspin}{0.40}{GW200115_XPHM_highspin}{0.42}{GW200115_v4PHM_highspin}{0.41}{GW200115_combined_highspin}{0.41}{GW200115_NSBH_lowspin}{0.38}}}
\newcommand{\finalspinplus}[1]{\IfEqCase{#1}{{GW200105_XPHM_lowspin}{0.03}{GW200105_v4PHM_lowspin}{0.02}{GW200105_combined_lowspin}{0.03}{GW200105_XPHM_highspin}{0.04}{GW200105_v4PHM_highspin}{0.02}{GW200105_combined_highspin}{0.03}{GW200105_NSBH_lowspin}{0.04}{GW200115_XPHM_lowspin}{0.06}{GW200115_v4PHM_lowspin}{0.05}{GW200115_combined_lowspin}{0.06}{GW200115_XPHM_highspin}{0.08}{GW200115_v4PHM_highspin}{0.08}{GW200115_combined_highspin}{0.08}{GW200115_NSBH_lowspin}{0.04}}}
\newcommand{\finalspinonepercent}[1]{\IfEqCase{#1}{{GW200105_XPHM_lowspin}{0.42}{GW200105_v4PHM_lowspin}{0.42}{GW200105_combined_lowspin}{0.42}{GW200105_XPHM_highspin}{0.39}{GW200105_v4PHM_highspin}{0.41}{GW200105_combined_highspin}{0.40}{GW200105_NSBH_lowspin}{0.40}{GW200115_XPHM_lowspin}{0.37}{GW200115_v4PHM_lowspin}{0.37}{GW200115_combined_lowspin}{0.37}{GW200115_XPHM_highspin}{0.36}{GW200115_v4PHM_highspin}{0.35}{GW200115_combined_highspin}{0.35}{GW200115_NSBH_lowspin}{0.35}}}
\newcommand{\finalspinninetyninepercent}[1]{\IfEqCase{#1}{{GW200105_XPHM_lowspin}{0.49}{GW200105_v4PHM_lowspin}{0.48}{GW200105_combined_lowspin}{0.49}{GW200105_XPHM_highspin}{0.51}{GW200105_v4PHM_highspin}{0.48}{GW200105_combined_highspin}{0.50}{GW200105_NSBH_lowspin}{0.51}{GW200115_XPHM_lowspin}{0.51}{GW200115_v4PHM_lowspin}{0.48}{GW200115_combined_lowspin}{0.50}{GW200115_XPHM_highspin}{0.53}{GW200115_v4PHM_highspin}{0.51}{GW200115_combined_highspin}{0.52}{GW200115_NSBH_lowspin}{0.45}}}
\newcommand{\finalspinfivepercent}[1]{\IfEqCase{#1}{{GW200105_XPHM_lowspin}{0.42}{GW200105_v4PHM_lowspin}{0.42}{GW200105_combined_lowspin}{0.42}{GW200105_XPHM_highspin}{0.41}{GW200105_v4PHM_highspin}{0.42}{GW200105_combined_highspin}{0.41}{GW200105_NSBH_lowspin}{0.41}{GW200115_XPHM_lowspin}{0.38}{GW200115_v4PHM_lowspin}{0.37}{GW200115_combined_lowspin}{0.38}{GW200115_XPHM_highspin}{0.37}{GW200115_v4PHM_highspin}{0.37}{GW200115_combined_highspin}{0.37}{GW200115_NSBH_lowspin}{0.36}}}
\newcommand{\finalspinninetyfivepercent}[1]{\IfEqCase{#1}{{GW200105_XPHM_lowspin}{0.46}{GW200105_v4PHM_lowspin}{0.46}{GW200105_combined_lowspin}{0.46}{GW200105_XPHM_highspin}{0.48}{GW200105_v4PHM_highspin}{0.46}{GW200105_combined_highspin}{0.47}{GW200105_NSBH_lowspin}{0.47}{GW200115_XPHM_lowspin}{0.47}{GW200115_v4PHM_lowspin}{0.45}{GW200115_combined_lowspin}{0.46}{GW200115_XPHM_highspin}{0.50}{GW200115_v4PHM_highspin}{0.49}{GW200115_combined_highspin}{0.49}{GW200115_NSBH_lowspin}{0.42}}}
\newcommand{\finalspinninetypercent}[1]{\IfEqCase{#1}{{GW200105_XPHM_lowspin}{0.45}{GW200105_v4PHM_lowspin}{0.45}{GW200105_combined_lowspin}{0.45}{GW200105_XPHM_highspin}{0.47}{GW200105_v4PHM_highspin}{0.45}{GW200105_combined_highspin}{0.46}{GW200105_NSBH_lowspin}{0.46}{GW200115_XPHM_lowspin}{0.46}{GW200115_v4PHM_lowspin}{0.44}{GW200115_combined_lowspin}{0.45}{GW200115_XPHM_highspin}{0.48}{GW200115_v4PHM_highspin}{0.47}{GW200115_combined_highspin}{0.47}{GW200115_NSBH_lowspin}{0.41}}}
\newcommand{\finalmasssourceminus}[1]{\IfEqCase{#1}{{GW200105_XPHM_lowspin}{1.2}{GW200105_v4PHM_lowspin}{0.9}{GW200105_combined_lowspin}{1.0}{GW200105_XPHM_highspin}{1.4}{GW200105_v4PHM_highspin}{1.0}{GW200105_combined_highspin}{1.2}{GW200105_NSBH_lowspin}{2.0}{GW200115_XPHM_lowspin}{1.2}{GW200115_v4PHM_lowspin}{1.9}{GW200115_combined_lowspin}{1.5}{GW200115_XPHM_highspin}{1.3}{GW200115_v4PHM_highspin}{1.6}{GW200115_combined_highspin}{1.4}{GW200115_NSBH_lowspin}{1.6}}}
\newcommand{\finalmasssourcemed}[1]{\IfEqCase{#1}{{GW200105_XPHM_lowspin}{10.6}{GW200105_v4PHM_lowspin}{10.7}{GW200105_combined_lowspin}{10.6}{GW200105_XPHM_highspin}{10.7}{GW200105_v4PHM_highspin}{10.6}{GW200105_combined_highspin}{10.7}{GW200105_NSBH_lowspin}{10.4}{GW200115_XPHM_lowspin}{6.9}{GW200115_v4PHM_lowspin}{7.5}{GW200115_combined_lowspin}{7.2}{GW200115_XPHM_highspin}{6.9}{GW200115_v4PHM_highspin}{7.1}{GW200115_combined_highspin}{7.0}{GW200115_NSBH_lowspin}{7.8}}}
\newcommand{\finalmasssourceplus}[1]{\IfEqCase{#1}{{GW200105_XPHM_lowspin}{1.1}{GW200105_v4PHM_lowspin}{0.8}{GW200105_combined_lowspin}{0.9}{GW200105_XPHM_highspin}{1.3}{GW200105_v4PHM_highspin}{0.7}{GW200105_combined_highspin}{1.1}{GW200105_NSBH_lowspin}{2.7}{GW200115_XPHM_lowspin}{1.5}{GW200115_v4PHM_lowspin}{0.9}{GW200115_combined_lowspin}{1.2}{GW200115_XPHM_highspin}{1.4}{GW200115_v4PHM_highspin}{1.6}{GW200115_combined_highspin}{1.5}{GW200115_NSBH_lowspin}{1.4}}}
\newcommand{\finalmasssourceonepercent}[1]{\IfEqCase{#1}{{GW200105_XPHM_lowspin}{8.5}{GW200105_v4PHM_lowspin}{9.3}{GW200105_combined_lowspin}{8.8}{GW200105_XPHM_highspin}{8.3}{GW200105_v4PHM_highspin}{8.2}{GW200105_combined_highspin}{8.3}{GW200105_NSBH_lowspin}{8.1}{GW200115_XPHM_lowspin}{5.5}{GW200115_v4PHM_lowspin}{5.4}{GW200115_combined_lowspin}{5.5}{GW200115_XPHM_highspin}{5.4}{GW200115_v4PHM_highspin}{5.3}{GW200115_combined_highspin}{5.4}{GW200115_NSBH_lowspin}{5.9}}}
\newcommand{\finalmasssourceninetyninepercent}[1]{\IfEqCase{#1}{{GW200105_XPHM_lowspin}{12.4}{GW200105_v4PHM_lowspin}{12.6}{GW200105_combined_lowspin}{12.5}{GW200105_XPHM_highspin}{12.8}{GW200105_v4PHM_highspin}{12.5}{GW200105_combined_highspin}{12.7}{GW200105_NSBH_lowspin}{14.8}{GW200115_XPHM_lowspin}{9.6}{GW200115_v4PHM_lowspin}{9.1}{GW200115_combined_lowspin}{9.3}{GW200115_XPHM_highspin}{9.0}{GW200115_v4PHM_highspin}{9.3}{GW200115_combined_highspin}{9.2}{GW200115_NSBH_lowspin}{10.1}}}
\newcommand{\finalmasssourcefivepercent}[1]{\IfEqCase{#1}{{GW200105_XPHM_lowspin}{9.4}{GW200105_v4PHM_lowspin}{9.8}{GW200105_combined_lowspin}{9.6}{GW200105_XPHM_highspin}{9.3}{GW200105_v4PHM_highspin}{9.7}{GW200105_combined_highspin}{9.4}{GW200105_NSBH_lowspin}{8.5}{GW200115_XPHM_lowspin}{5.7}{GW200115_v4PHM_lowspin}{5.6}{GW200115_combined_lowspin}{5.7}{GW200115_XPHM_highspin}{5.6}{GW200115_v4PHM_highspin}{5.5}{GW200115_combined_highspin}{5.6}{GW200115_NSBH_lowspin}{6.2}}}
\newcommand{\finalmasssourceninetyfivepercent}[1]{\IfEqCase{#1}{{GW200105_XPHM_lowspin}{11.7}{GW200105_v4PHM_lowspin}{11.5}{GW200105_combined_lowspin}{11.6}{GW200105_XPHM_highspin}{11.9}{GW200105_v4PHM_highspin}{11.3}{GW200105_combined_highspin}{11.7}{GW200105_NSBH_lowspin}{13.1}{GW200115_XPHM_lowspin}{8.4}{GW200115_v4PHM_lowspin}{8.4}{GW200115_combined_lowspin}{8.4}{GW200115_XPHM_highspin}{8.3}{GW200115_v4PHM_highspin}{8.6}{GW200115_combined_highspin}{8.5}{GW200115_NSBH_lowspin}{9.2}}}
\newcommand{\finalmasssourceninetypercent}[1]{\IfEqCase{#1}{{GW200105_XPHM_lowspin}{11.3}{GW200105_v4PHM_lowspin}{11.2}{GW200105_combined_lowspin}{11.3}{GW200105_XPHM_highspin}{11.6}{GW200105_v4PHM_highspin}{11.1}{GW200105_combined_highspin}{11.4}{GW200105_NSBH_lowspin}{12.2}{GW200115_XPHM_lowspin}{8.1}{GW200115_v4PHM_lowspin}{8.3}{GW200115_combined_lowspin}{8.2}{GW200115_XPHM_highspin}{8.1}{GW200115_v4PHM_highspin}{8.4}{GW200115_combined_highspin}{8.2}{GW200115_NSBH_lowspin}{8.8}}}
\newcommand{\radiatedenergyminus}[1]{\IfEqCase{#1}{{GW200105_XPHM_lowspin}{0.0}{GW200105_v4PHM_lowspin}{0.0}{GW200105_combined_lowspin}{0.0}{GW200105_XPHM_highspin}{0.0}{GW200105_v4PHM_highspin}{0.0}{GW200105_combined_highspin}{0.0}{GW200115_XPHM_lowspin}{0.0}{GW200115_v4PHM_lowspin}{0.0}{GW200115_combined_lowspin}{0.0}{GW200115_XPHM_highspin}{0.0}{GW200115_v4PHM_highspin}{0.0}{GW200115_combined_highspin}{0.0}}}
\newcommand{\radiatedenergymed}[1]{\IfEqCase{#1}{{GW200105_XPHM_lowspin}{0.2}{GW200105_v4PHM_lowspin}{0.2}{GW200105_combined_lowspin}{0.2}{GW200105_XPHM_highspin}{0.2}{GW200105_v4PHM_highspin}{0.2}{GW200105_combined_highspin}{0.2}{GW200115_XPHM_lowspin}{0.1}{GW200115_v4PHM_lowspin}{0.1}{GW200115_combined_lowspin}{0.1}{GW200115_XPHM_highspin}{0.1}{GW200115_v4PHM_highspin}{0.1}{GW200115_combined_highspin}{0.1}}}
\newcommand{\radiatedenergyplus}[1]{\IfEqCase{#1}{{GW200105_XPHM_lowspin}{0.0}{GW200105_v4PHM_lowspin}{0.0}{GW200105_combined_lowspin}{0.0}{GW200105_XPHM_highspin}{0.0}{GW200105_v4PHM_highspin}{0.0}{GW200105_combined_highspin}{0.0}{GW200115_XPHM_lowspin}{0.0}{GW200115_v4PHM_lowspin}{0.0}{GW200115_combined_lowspin}{0.0}{GW200115_XPHM_highspin}{0.0}{GW200115_v4PHM_highspin}{0.0}{GW200115_combined_highspin}{0.0}}}
\newcommand{\radiatedenergyonepercent}[1]{\IfEqCase{#1}{{GW200105_XPHM_lowspin}{0.2}{GW200105_v4PHM_lowspin}{0.2}{GW200105_combined_lowspin}{0.2}{GW200105_XPHM_highspin}{0.2}{GW200105_v4PHM_highspin}{0.2}{GW200105_combined_highspin}{0.2}{GW200115_XPHM_lowspin}{0.1}{GW200115_v4PHM_lowspin}{0.1}{GW200115_combined_lowspin}{0.1}{GW200115_XPHM_highspin}{0.1}{GW200115_v4PHM_highspin}{0.1}{GW200115_combined_highspin}{0.1}}}
\newcommand{\radiatedenergyninetyninepercent}[1]{\IfEqCase{#1}{{GW200105_XPHM_lowspin}{0.2}{GW200105_v4PHM_lowspin}{0.2}{GW200105_combined_lowspin}{0.2}{GW200105_XPHM_highspin}{0.3}{GW200105_v4PHM_highspin}{0.3}{GW200105_combined_highspin}{0.3}{GW200115_XPHM_lowspin}{0.2}{GW200115_v4PHM_lowspin}{0.2}{GW200115_combined_lowspin}{0.2}{GW200115_XPHM_highspin}{0.2}{GW200115_v4PHM_highspin}{0.2}{GW200115_combined_highspin}{0.2}}}
\newcommand{\radiatedenergyfivepercent}[1]{\IfEqCase{#1}{{GW200105_XPHM_lowspin}{0.2}{GW200105_v4PHM_lowspin}{0.2}{GW200105_combined_lowspin}{0.2}{GW200105_XPHM_highspin}{0.2}{GW200105_v4PHM_highspin}{0.2}{GW200105_combined_highspin}{0.2}{GW200115_XPHM_lowspin}{0.1}{GW200115_v4PHM_lowspin}{0.1}{GW200115_combined_lowspin}{0.1}{GW200115_XPHM_highspin}{0.1}{GW200115_v4PHM_highspin}{0.1}{GW200115_combined_highspin}{0.1}}}
\newcommand{\radiatedenergyninetyfivepercent}[1]{\IfEqCase{#1}{{GW200105_XPHM_lowspin}{0.2}{GW200105_v4PHM_lowspin}{0.2}{GW200105_combined_lowspin}{0.2}{GW200105_XPHM_highspin}{0.2}{GW200105_v4PHM_highspin}{0.2}{GW200105_combined_highspin}{0.2}{GW200115_XPHM_lowspin}{0.2}{GW200115_v4PHM_lowspin}{0.2}{GW200115_combined_lowspin}{0.2}{GW200115_XPHM_highspin}{0.2}{GW200115_v4PHM_highspin}{0.2}{GW200115_combined_highspin}{0.2}}}
\newcommand{\radiatedenergyninetypercent}[1]{\IfEqCase{#1}{{GW200105_XPHM_lowspin}{0.2}{GW200105_v4PHM_lowspin}{0.2}{GW200105_combined_lowspin}{0.2}{GW200105_XPHM_highspin}{0.2}{GW200105_v4PHM_highspin}{0.2}{GW200105_combined_highspin}{0.2}{GW200115_XPHM_lowspin}{0.2}{GW200115_v4PHM_lowspin}{0.2}{GW200115_combined_lowspin}{0.2}{GW200115_XPHM_highspin}{0.2}{GW200115_v4PHM_highspin}{0.2}{GW200115_combined_highspin}{0.2}}}
\newcommand{\networkoptimalsnrminus}[1]{\IfEqCase{#1}{{GW200105_XPHM_lowspin}{1.7}{GW200105_XPHM_highspin}{1.7}{GW200105_NSBH_lowspin}{1.7}{GW200115_XPHM_lowspin}{1.7}{GW200115_XPHM_highspin}{1.7}{GW200115_NSBH_lowspin}{1.7}}}
\newcommand{\networkoptimalsnrmed}[1]{\IfEqCase{#1}{{GW200105_XPHM_lowspin}{13.2}{GW200105_XPHM_highspin}{13.2}{GW200105_NSBH_lowspin}{13.0}{GW200115_XPHM_lowspin}{10.8}{GW200115_XPHM_highspin}{10.8}{GW200115_NSBH_lowspin}{10.5}}}
\newcommand{\networkoptimalsnrplus}[1]{\IfEqCase{#1}{{GW200105_XPHM_lowspin}{1.7}{GW200105_XPHM_highspin}{1.7}{GW200105_NSBH_lowspin}{1.7}{GW200115_XPHM_lowspin}{1.7}{GW200115_XPHM_highspin}{1.7}{GW200115_NSBH_lowspin}{1.7}}}
\newcommand{\networkoptimalsnronepercent}[1]{\IfEqCase{#1}{{GW200105_XPHM_lowspin}{10.8}{GW200105_XPHM_highspin}{10.8}{GW200105_NSBH_lowspin}{10.6}{GW200115_XPHM_lowspin}{8.3}{GW200115_XPHM_highspin}{8.3}{GW200115_NSBH_lowspin}{8.1}}}
\newcommand{\networkoptimalsnrninetyninepercent}[1]{\IfEqCase{#1}{{GW200105_XPHM_lowspin}{15.6}{GW200105_XPHM_highspin}{15.6}{GW200105_NSBH_lowspin}{15.3}{GW200115_XPHM_lowspin}{13.2}{GW200115_XPHM_highspin}{13.2}{GW200115_NSBH_lowspin}{13.0}}}
\newcommand{\networkoptimalsnrfivepercent}[1]{\IfEqCase{#1}{{GW200105_XPHM_lowspin}{11.5}{GW200105_XPHM_highspin}{11.5}{GW200105_NSBH_lowspin}{11.3}{GW200115_XPHM_lowspin}{9.1}{GW200115_XPHM_highspin}{9.1}{GW200115_NSBH_lowspin}{8.8}}}
\newcommand{\networkoptimalsnrninetyfivepercent}[1]{\IfEqCase{#1}{{GW200105_XPHM_lowspin}{14.9}{GW200105_XPHM_highspin}{14.9}{GW200105_NSBH_lowspin}{14.7}{GW200115_XPHM_lowspin}{12.5}{GW200115_XPHM_highspin}{12.5}{GW200115_NSBH_lowspin}{12.3}}}
\newcommand{\networkoptimalsnrninetypercent}[1]{\IfEqCase{#1}{{GW200105_XPHM_lowspin}{14.5}{GW200105_XPHM_highspin}{14.5}{GW200105_NSBH_lowspin}{14.3}{GW200115_XPHM_lowspin}{12.1}{GW200115_XPHM_highspin}{12.1}{GW200115_NSBH_lowspin}{11.9}}}
\newcommand{\networkmatchedfiltersnrminus}[1]{\IfEqCase{#1}{{GW200105_XPHM_lowspin}{0.4}{GW200105_XPHM_highspin}{0.4}{GW200105_NSBH_lowspin}{0.3}{GW200115_XPHM_lowspin}{0.5}{GW200115_XPHM_highspin}{0.5}{GW200115_NSBH_lowspin}{0.5}}}
\newcommand{\networkmatchedfiltersnrmed}[1]{\IfEqCase{#1}{{GW200105_XPHM_lowspin}{13.5}{GW200105_XPHM_highspin}{13.5}{GW200105_NSBH_lowspin}{13.3}{GW200115_XPHM_lowspin}{11.2}{GW200115_XPHM_highspin}{11.2}{GW200115_NSBH_lowspin}{11.0}}}
\newcommand{\networkmatchedfiltersnrplus}[1]{\IfEqCase{#1}{{GW200105_XPHM_lowspin}{0.2}{GW200105_XPHM_highspin}{0.2}{GW200105_NSBH_lowspin}{0.2}{GW200115_XPHM_lowspin}{0.3}{GW200115_XPHM_highspin}{0.3}{GW200115_NSBH_lowspin}{0.3}}}
\newcommand{\networkmatchedfiltersnronepercent}[1]{\IfEqCase{#1}{{GW200105_XPHM_lowspin}{12.9}{GW200105_XPHM_highspin}{12.9}{GW200105_NSBH_lowspin}{12.8}{GW200115_XPHM_lowspin}{10.4}{GW200115_XPHM_highspin}{10.4}{GW200115_NSBH_lowspin}{10.2}}}
\newcommand{\networkmatchedfiltersnrninetyninepercent}[1]{\IfEqCase{#1}{{GW200105_XPHM_lowspin}{13.8}{GW200105_XPHM_highspin}{13.8}{GW200105_NSBH_lowspin}{13.6}{GW200115_XPHM_lowspin}{11.6}{GW200115_XPHM_highspin}{11.6}{GW200115_NSBH_lowspin}{11.3}}}
\newcommand{\networkmatchedfiltersnrfivepercent}[1]{\IfEqCase{#1}{{GW200105_XPHM_lowspin}{13.1}{GW200105_XPHM_highspin}{13.1}{GW200105_NSBH_lowspin}{13.0}{GW200115_XPHM_lowspin}{10.7}{GW200115_XPHM_highspin}{10.7}{GW200115_NSBH_lowspin}{10.5}}}
\newcommand{\networkmatchedfiltersnrninetyfivepercent}[1]{\IfEqCase{#1}{{GW200105_XPHM_lowspin}{13.8}{GW200105_XPHM_highspin}{13.7}{GW200105_NSBH_lowspin}{13.5}{GW200115_XPHM_lowspin}{11.5}{GW200115_XPHM_highspin}{11.5}{GW200115_NSBH_lowspin}{11.2}}}
\newcommand{\networkmatchedfiltersnrninetypercent}[1]{\IfEqCase{#1}{{GW200105_XPHM_lowspin}{13.7}{GW200105_XPHM_highspin}{13.7}{GW200105_NSBH_lowspin}{13.5}{GW200115_XPHM_lowspin}{11.4}{GW200115_XPHM_highspin}{11.5}{GW200115_NSBH_lowspin}{11.2}}}
\newcommand{\cosiotaminus}[1]{\IfEqCase{#1}{{GW200105_XPHM_lowspin}{0.46}{GW200105_v4PHM_lowspin}{0.72}{GW200105_combined_lowspin}{0.56}{GW200105_XPHM_highspin}{0.69}{GW200105_v4PHM_highspin}{0.76}{GW200105_combined_highspin}{0.73}{GW200105_NSBH_lowspin}{0.93}{GW200115_XPHM_lowspin}{1.54}{GW200115_v4PHM_lowspin}{1.68}{GW200115_combined_lowspin}{1.69}{GW200115_XPHM_highspin}{1.35}{GW200115_v4PHM_highspin}{1.67}{GW200115_combined_highspin}{1.69}{GW200115_NSBH_lowspin}{1.64}}}
\newcommand{\cosiotamed}[1]{\IfEqCase{#1}{{GW200105_XPHM_lowspin}{-0.48}{GW200105_v4PHM_lowspin}{-0.23}{GW200105_combined_lowspin}{-0.39}{GW200105_XPHM_highspin}{-0.26}{GW200105_v4PHM_highspin}{-0.20}{GW200105_combined_highspin}{-0.22}{GW200105_NSBH_lowspin}{-0.04}{GW200115_XPHM_lowspin}{0.84}{GW200115_v4PHM_lowspin}{0.75}{GW200115_combined_lowspin}{0.81}{GW200115_XPHM_highspin}{0.85}{GW200115_v4PHM_highspin}{0.75}{GW200115_combined_highspin}{0.81}{GW200115_NSBH_lowspin}{0.75}}}
\newcommand{\cosiotaplus}[1]{\IfEqCase{#1}{{GW200105_XPHM_lowspin}{1.41}{GW200105_v4PHM_lowspin}{1.18}{GW200105_combined_lowspin}{1.32}{GW200105_XPHM_highspin}{1.19}{GW200105_v4PHM_highspin}{1.15}{GW200105_combined_highspin}{1.16}{GW200105_NSBH_lowspin}{1.01}{GW200115_XPHM_lowspin}{0.14}{GW200115_v4PHM_lowspin}{0.23}{GW200115_combined_lowspin}{0.17}{GW200115_XPHM_highspin}{0.14}{GW200115_v4PHM_highspin}{0.24}{GW200115_combined_highspin}{0.17}{GW200115_NSBH_lowspin}{0.23}}}
\newcommand{\cosiotaonepercent}[1]{\IfEqCase{#1}{{GW200105_XPHM_lowspin}{-0.98}{GW200105_v4PHM_lowspin}{-0.99}{GW200105_combined_lowspin}{-0.99}{GW200105_XPHM_highspin}{-0.99}{GW200105_v4PHM_highspin}{-0.99}{GW200105_combined_highspin}{-0.99}{GW200105_NSBH_lowspin}{-0.99}{GW200115_XPHM_lowspin}{-0.95}{GW200115_v4PHM_lowspin}{-0.98}{GW200115_combined_lowspin}{-0.98}{GW200115_XPHM_highspin}{-0.95}{GW200115_v4PHM_highspin}{-0.99}{GW200115_combined_highspin}{-0.98}{GW200115_NSBH_lowspin}{-0.98}}}
\newcommand{\cosiotaninetyninepercent}[1]{\IfEqCase{#1}{{GW200105_XPHM_lowspin}{0.98}{GW200105_v4PHM_lowspin}{0.99}{GW200105_combined_lowspin}{0.99}{GW200105_XPHM_highspin}{0.98}{GW200105_v4PHM_highspin}{0.99}{GW200105_combined_highspin}{0.99}{GW200105_NSBH_lowspin}{0.99}{GW200115_XPHM_lowspin}{1.00}{GW200115_v4PHM_lowspin}{1.00}{GW200115_combined_lowspin}{1.00}{GW200115_XPHM_highspin}{1.00}{GW200115_v4PHM_highspin}{1.00}{GW200115_combined_highspin}{1.00}{GW200115_NSBH_lowspin}{1.00}}}
\newcommand{\cosiotafivepercent}[1]{\IfEqCase{#1}{{GW200105_XPHM_lowspin}{-0.94}{GW200105_v4PHM_lowspin}{-0.96}{GW200105_combined_lowspin}{-0.95}{GW200105_XPHM_highspin}{-0.95}{GW200105_v4PHM_highspin}{-0.96}{GW200105_combined_highspin}{-0.95}{GW200105_NSBH_lowspin}{-0.97}{GW200115_XPHM_lowspin}{-0.70}{GW200115_v4PHM_lowspin}{-0.92}{GW200115_combined_lowspin}{-0.88}{GW200115_XPHM_highspin}{-0.50}{GW200115_v4PHM_highspin}{-0.93}{GW200115_combined_highspin}{-0.88}{GW200115_NSBH_lowspin}{-0.89}}}
\newcommand{\cosiotaninetyfivepercent}[1]{\IfEqCase{#1}{{GW200105_XPHM_lowspin}{0.93}{GW200105_v4PHM_lowspin}{0.95}{GW200105_combined_lowspin}{0.94}{GW200105_XPHM_highspin}{0.93}{GW200105_v4PHM_highspin}{0.95}{GW200105_combined_highspin}{0.94}{GW200105_NSBH_lowspin}{0.97}{GW200115_XPHM_lowspin}{0.99}{GW200115_v4PHM_lowspin}{0.98}{GW200115_combined_lowspin}{0.98}{GW200115_XPHM_highspin}{0.99}{GW200115_v4PHM_highspin}{0.98}{GW200115_combined_highspin}{0.98}{GW200115_NSBH_lowspin}{0.98}}}
\newcommand{\cosiotaninetypercent}[1]{\IfEqCase{#1}{{GW200105_XPHM_lowspin}{0.87}{GW200105_v4PHM_lowspin}{0.90}{GW200105_combined_lowspin}{0.89}{GW200105_XPHM_highspin}{0.89}{GW200105_v4PHM_highspin}{0.90}{GW200105_combined_highspin}{0.89}{GW200105_NSBH_lowspin}{0.93}{GW200115_XPHM_lowspin}{0.97}{GW200115_v4PHM_lowspin}{0.96}{GW200115_combined_lowspin}{0.97}{GW200115_XPHM_highspin}{0.97}{GW200115_v4PHM_highspin}{0.96}{GW200115_combined_highspin}{0.97}{GW200115_NSBH_lowspin}{0.96}}}
\newcommand{\numberofcycles}[1]{\IfEqCase{#1}{{GW200105_XPHM_lowspin}{800}{GW200105_v4PHM_lowspin}{800}{GW200105_combined_lowspin}{800}{GW200105_XPHM_highspin}{800}{GW200105_v4PHM_highspin}{800}{GW200105_combined_highspin}{800}{GW200105_NSBH_lowspin}{800}{GW200115_XPHM_lowspin}{1400}{GW200115_v4PHM_lowspin}{1400}{GW200115_combined_lowspin}{1500}{GW200115_XPHM_highspin}{1400}{GW200115_v4PHM_highspin}{1400}{GW200115_combined_highspin}{1400}{GW200115_NSBH_lowspin}{1500}}}
\newcommand{\skyareaninetypercent}[1]{\IfEqCase{#1}{{GW200105_XPHM_lowspin}{6600}{GW200105_v4PHM_lowspin}{8000}{GW200105_combined_lowspin}{7300}{GW200105_XPHM_highspin}{6500}{GW200105_v4PHM_highspin}{7600}{GW200105_combined_highspin}{7200}{GW200105_NSBH_lowspin}{7200}{GW200115_XPHM_lowspin}{400}{GW200115_v4PHM_lowspin}{0}{GW200115_combined_lowspin}{600}{GW200115_XPHM_highspin}{0}{GW200115_v4PHM_highspin}{800}{GW200115_combined_highspin}{600}{GW200115_NSBH_lowspin}{0}}}
\newcommand{\PEpercentBNS}[1]{\IfEqCase{#1}{{GW200105_XPHM_lowspin}{0}{GW200105_v4PHM_lowspin}{0}{GW200105_combined_lowspin}{0}{GW200105_XPHM_highspin}{0}{GW200105_v4PHM_highspin}{0}{GW200105_combined_highspin}{0}{GW200105_NSBH_lowspin}{0}{GW200115_XPHM_lowspin}{0}{GW200115_v4PHM_lowspin}{0}{GW200115_combined_lowspin}{0}{GW200115_XPHM_highspin}{0}{GW200115_v4PHM_highspin}{1}{GW200115_combined_highspin}{1}{GW200115_NSBH_lowspin}{0}}}
\newcommand{\PEpercentNSBH}[1]{\IfEqCase{#1}{{GW200105_XPHM_lowspin}{100}{GW200105_v4PHM_lowspin}{100}{GW200105_combined_lowspin}{100}{GW200105_XPHM_highspin}{100}{GW200105_v4PHM_highspin}{100}{GW200105_combined_highspin}{100}{GW200105_NSBH_lowspin}{100}{GW200115_XPHM_lowspin}{69}{GW200115_v4PHM_lowspin}{77}{GW200115_combined_lowspin}{73}{GW200115_XPHM_highspin}{71}{GW200115_v4PHM_highspin}{69}{GW200115_combined_highspin}{70}{GW200115_NSBH_lowspin}{93}}}
\newcommand{\PEpercentBBH}[1]{\IfEqCase{#1}{{GW200105_XPHM_lowspin}{0}{GW200105_v4PHM_lowspin}{0}{GW200105_combined_lowspin}{0}{GW200105_XPHM_highspin}{0}{GW200105_v4PHM_highspin}{0}{GW200105_combined_highspin}{0}{GW200105_NSBH_lowspin}{0}{GW200115_XPHM_lowspin}{0}{GW200115_v4PHM_lowspin}{0}{GW200115_combined_lowspin}{0}{GW200115_XPHM_highspin}{0}{GW200115_v4PHM_highspin}{0}{GW200115_combined_highspin}{0}{GW200115_NSBH_lowspin}{0}}}
\newcommand{\PEpercentMassGap}[1]{\IfEqCase{#1}{{GW200105_XPHM_lowspin}{0}{GW200105_v4PHM_lowspin}{0}{GW200105_combined_lowspin}{0}{GW200105_XPHM_highspin}{0}{GW200105_v4PHM_highspin}{0}{GW200105_combined_highspin}{0}{GW200105_NSBH_lowspin}{0}{GW200115_XPHM_lowspin}{31}{GW200115_v4PHM_lowspin}{23}{GW200115_combined_lowspin}{27}{GW200115_XPHM_highspin}{29}{GW200115_v4PHM_highspin}{30}{GW200115_combined_highspin}{30}{GW200115_NSBH_lowspin}{7}}}
\newcommand{\PEpercentchionenegative}[1]{\IfEqCase{#1}{{GW200105_XPHM_lowspin}{61}{GW200105_v4PHM_lowspin}{54}{GW200105_combined_lowspin}{58}{GW200105_XPHM_highspin}{55}{GW200105_v4PHM_highspin}{55}{GW200105_combined_highspin}{55}{GW200105_NSBH_lowspin}{55}{GW200115_XPHM_lowspin}{92}{GW200115_v4PHM_lowspin}{83}{GW200115_combined_lowspin}{87}{GW200115_XPHM_highspin}{92}{GW200115_v4PHM_highspin}{83}{GW200115_combined_highspin}{88}{GW200115_NSBH_lowspin}{68}}}
\newcommand{\logtenBayesPrecession}[1]{\IfEqCase{#1}{{GW200105_Combined_highspin}{-0.24}{GW200115_Combined_highspin}{-0.12}}}

%% file: nature_secondary_macros.tex



\newcommand{\lecFlatEventOne    }{96} 
\newcommand{\lecSalpeterEventOne}{95} 
\newcommand{\lecBPLallEventOne  }{93} 
\newcommand{\lecBPLwoutEventOne }{90} 
\newcommand{\lecPLPallEventOne  }{96} 
\newcommand{\lecPLPwoutEventOne }{95} 



\newcommand{\lecspinFlatEventOne    }{94} 
\newcommand{\lecspinSalpeterEventOne}{94} 
\newcommand{\lecspinBPLallEventOne  }{91} 
\newcommand{\lecspinBPLwoutEventOne }{84} 
\newcommand{\lecspinPLPallEventOne  }{95} 
\newcommand{\lecspinPLPwoutEventOne }{94} 

\newcommand{\fcFlatEventOne    }{93} 
\newcommand{\fcSalpeterEventOne}{92} 
\newcommand{\fcBPLallEventOne  }{89} 
\newcommand{\fcBPLwoutEventOne }{82} 
\newcommand{\fcPLPallEventOne  }{94} 
\newcommand{\fcPLPwoutEventOne }{93} 



\newcommand{\lecFlatEventTwo    }{98} 
\newcommand{\lecSalpeterEventTwo}{97} 
\newcommand{\lecBPLallEventTwo  }{94} 
\newcommand{\lecBPLwoutEventTwo }{88} 
\newcommand{\lecPLPallEventTwo  }{98} 
\newcommand{\lecPLPwoutEventTwo }{97} 



\newcommand{\lecspinFlatEventTwo    }{95} 
\newcommand{\lecspinSalpeterEventTwo}{95} 
\newcommand{\lecspinBPLallEventTwo  }{88} 
\newcommand{\lecspinBPLwoutEventTwo }{78} 
\newcommand{\lecspinPLPallEventTwo  }{95} 
\newcommand{\lecspinPLPwoutEventTwo }{92} 

\newcommand{\fcFlatEventTwo    }{96} 
\newcommand{\fcSalpeterEventTwo}{95} 
\newcommand{\fcBPLallEventTwo  }{87} 
\newcommand{\fcBPLwoutEventTwo }{77} 
\newcommand{\fcPLPallEventTwo  }{96} 
\newcommand{\fcPLPwoutEventTwo }{93} 




\newcommand{\lowestPnsLowSpinEventOne}{100.0}

\FPmin{\lowestPnsLowSpinEventOne}{\lowestPnsLowSpinEventOne}{\lecFlatEventOne}
\FPmin{\lowestPnsLowSpinEventOne}{\lowestPnsLowSpinEventOne}{\lecSalpeterEventOne}
\FPmin{\lowestPnsLowSpinEventOne}{\lowestPnsLowSpinEventOne}{\lecBPLallEventOne}
\FPmin{\lowestPnsLowSpinEventOne}{\lowestPnsLowSpinEventOne}{\lecPLPallEventOne}

\newcommand{\highestPnsLowSpinEventOne}{0.0}

\FPmax{\highestPnsLowSpinEventOne}{\highestPnsLowSpinEventOne}{\lecFlatEventOne}
\FPmax{\highestPnsLowSpinEventOne}{\highestPnsLowSpinEventOne}{\lecSalpeterEventOne}
\FPmax{\highestPnsLowSpinEventOne}{\highestPnsLowSpinEventOne}{\lecBPLallEventOne}
\FPmax{\highestPnsLowSpinEventOne}{\highestPnsLowSpinEventOne}{\lecPLPallEventOne}


\newcommand{\lowestPnsHighSpinEventOne}{100.0}

\FPmin{\lowestPnsHighSpinEventOne}{\lowestPnsHighSpinEventOne}{\lecspinFlatEventOne}
\FPmin{\lowestPnsHighSpinEventOne}{\lowestPnsHighSpinEventOne}{\lecspinSalpeterEventOne}
\FPmin{\lowestPnsHighSpinEventOne}{\lowestPnsHighSpinEventOne}{\lecspinBPLallEventOne}
\FPmin{\lowestPnsHighSpinEventOne}{\lowestPnsHighSpinEventOne}{\lecspinPLPallEventOne}

\FPmin{\lowestPnsHighSpinEventOne}{\lowestPnsHighSpinEventOne}{\fcFlatEventOne}
\FPmin{\lowestPnsHighSpinEventOne}{\lowestPnsHighSpinEventOne}{\fcSalpeterEventOne}
\FPmin{\lowestPnsHighSpinEventOne}{\lowestPnsHighSpinEventOne}{\fcBPLallEventOne}
\FPmin{\lowestPnsHighSpinEventOne}{\lowestPnsHighSpinEventOne}{\fcPLPallEventOne}

\newcommand{\highestPnsHighSpinEventOne}{0.0}

\FPmax{\highestPnsHighSpinEventOne}{\highestPnsHighSpinEventOne}{\lecspinFlatEventOne}
\FPmax{\highestPnsHighSpinEventOne}{\highestPnsHighSpinEventOne}{\lecspinSalpeterEventOne}
\FPmax{\highestPnsHighSpinEventOne}{\highestPnsHighSpinEventOne}{\lecspinBPLallEventOne}
\FPmax{\highestPnsHighSpinEventOne}{\highestPnsHighSpinEventOne}{\lecspinPLPallEventOne}

\FPmax{\highestPnsHighSpinEventOne}{\highestPnsHighSpinEventOne}{\fcFlatEventOne}
\FPmax{\highestPnsHighSpinEventOne}{\highestPnsHighSpinEventOne}{\fcSalpeterEventOne}
\FPmax{\highestPnsHighSpinEventOne}{\highestPnsHighSpinEventOne}{\fcBPLallEventOne}
\FPmax{\highestPnsHighSpinEventOne}{\highestPnsHighSpinEventOne}{\fcPLPallEventOne}


\FPmin{\lowestPnsEventOne}{\lowestPnsLowSpinEventOne}{\lowestPnsHighSpinEventOne}
\FPmax{\highestPnsEventOne}{\highestPnsLowSpinEventOne}{\highestPnsHighSpinEventOne}



\newcommand{\lowestPnsLowSpinEventTwo}{100.0}

\FPmin{\lowestPnsLowSpinEventTwo}{\lowestPnsLowSpinEventTwo}{\lecFlatEventTwo}
\FPmin{\lowestPnsLowSpinEventTwo}{\lowestPnsLowSpinEventTwo}{\lecSalpeterEventTwo}
\FPmin{\lowestPnsLowSpinEventTwo}{\lowestPnsLowSpinEventTwo}{\lecBPLallEventTwo}
\FPmin{\lowestPnsLowSpinEventTwo}{\lowestPnsLowSpinEventTwo}{\lecPLPallEventTwo}

\newcommand{\highestPnsLowSpinEventTwo}{0.0}

\FPmax{\highestPnsLowSpinEventTwo}{\highestPnsLowSpinEventTwo}{\lecFlatEventTwo}
\FPmax{\highestPnsLowSpinEventTwo}{\highestPnsLowSpinEventTwo}{\lecSalpeterEventTwo}
\FPmax{\highestPnsLowSpinEventTwo}{\highestPnsLowSpinEventTwo}{\lecBPLallEventTwo}
\FPmax{\highestPnsLowSpinEventTwo}{\highestPnsLowSpinEventTwo}{\lecPLPallEventTwo}


\newcommand{\lowestPnsHighSpinEventTwo}{100.0}

\FPmin{\lowestPnsHighSpinEventTwo}{\lowestPnsHighSpinEventTwo}{\lecspinFlatEventTwo}
\FPmin{\lowestPnsHighSpinEventTwo}{\lowestPnsHighSpinEventTwo}{\lecspinSalpeterEventTwo}
\FPmin{\lowestPnsHighSpinEventTwo}{\lowestPnsHighSpinEventTwo}{\lecspinBPLallEventTwo}
\FPmin{\lowestPnsHighSpinEventTwo}{\lowestPnsHighSpinEventTwo}{\lecspinPLPallEventTwo}

\FPmin{\lowestPnsHighSpinEventTwo}{\lowestPnsHighSpinEventTwo}{\fcFlatEventTwo}
\FPmin{\lowestPnsHighSpinEventTwo}{\lowestPnsHighSpinEventTwo}{\fcSalpeterEventTwo}
\FPmin{\lowestPnsHighSpinEventTwo}{\lowestPnsHighSpinEventTwo}{\fcBPLallEventTwo}
\FPmin{\lowestPnsHighSpinEventTwo}{\lowestPnsHighSpinEventTwo}{\fcPLPallEventTwo}

\newcommand{\highestPnsHighSpinEventTwo}{0.0}

\FPmax{\highestPnsHighSpinEventTwo}{\highestPnsHighSpinEventTwo}{\lecspinFlatEventTwo}
\FPmax{\highestPnsHighSpinEventTwo}{\highestPnsHighSpinEventTwo}{\lecspinSalpeterEventTwo}
\FPmax{\highestPnsHighSpinEventTwo}{\highestPnsHighSpinEventTwo}{\lecspinBPLallEventTwo}
\FPmax{\highestPnsHighSpinEventTwo}{\highestPnsHighSpinEventTwo}{\lecspinPLPallEventTwo}

\FPmax{\highestPnsHighSpinEventTwo}{\highestPnsHighSpinEventTwo}{\fcFlatEventTwo}
\FPmax{\highestPnsHighSpinEventTwo}{\highestPnsHighSpinEventTwo}{\fcSalpeterEventTwo}
\FPmax{\highestPnsHighSpinEventTwo}{\highestPnsHighSpinEventTwo}{\fcBPLallEventTwo}
\FPmax{\highestPnsHighSpinEventTwo}{\highestPnsHighSpinEventTwo}{\fcPLPallEventTwo}


\FPmin{\lowestPnsEventTwo}{\lowestPnsLowSpinEventTwo}{\lowestPnsHighSpinEventTwo}
\FPmax{\highestPnsEventTwo}{\highestPnsLowSpinEventTwo}{\highestPnsHighSpinEventTwo}



\FPmin{\lowestPnsLowSpin}{\lowestPnsLowSpinEventOne}{\lowestPnsLowSpinEventTwo}
\FPmax{\highestPnsLowSpin}{\lowestPnsLowSpinEventOne}{\lowestPnsLowSpinEventTwo}


\FPmin{\lowestPnsHighSpin}{\lowestPnsHighSpinEventOne}{\lowestPnsHighSpinEventTwo}
\FPmax{\highestPnsHighSpin}{\highestPnsHighSpinEventOne}{\highestPnsHighSpinEventTwo}


\FPmin{\lowestPns}{\lowestPnsEventOne}{\lowestPnsEventTwo}
\FPmax{\highestPns}{\highestPnsEventOne}{\highestPnsEventTwo}



\newcommand{\spindiffFlat}{0}

\FPsub{\spindiffAFlatEventOne}{\lecspinFlatEventOne}{\lecFlatEventOne}
\FPsub{\spindiffBFlatEventOne}{\fcFlatEventOne}{\lecFlatEventOne}
\FPabs{\spindiffAFlatEventOne}{\spindiffAFlatEventOne}
\FPabs{\spindiffBFlatEventOne}{\spindiffBFlatEventOne}
\FPmax{\spindiffFlatEventOne}{\spindiffAFlatEventOne}{\spindiffBFlatEventOne}


\FPsub{\spindiffAFlatEventTwo}{\lecspinFlatEventTwo}{\lecFlatEventTwo}
\FPsub{\spindiffBFlatEventTwo}{\fcFlatEventTwo}{\lecFlatEventTwo}
\FPabs{\spindiffAFlatEventTwo}{\spindiffAFlatEventTwo}
\FPabs{\spindiffBFlatEventTwo}{\spindiffBFlatEventTwo}
\FPmax{\spindiffFlatEventTwo}{\spindiffAFlatEventTwo}{\spindiffBFlatEventTwo}


\FPmax{\spindiffFlat}{\spindiffFlatEventOne}{\spindiffFlatEventTwo}
\FPeval{\spindiffFlat}{round(\spindiffFlat:0)}



\newcommand{\Pnsdiff}{0}

\FPsub{\PnsdiffEventOne}{\lecFlatEventOne}{\lowestPnsEventOne}
\FPsub{\PnsdiffEventTwo}{\lecFlatEventTwo}{\lowestPnsEventTwo}
\FPmax{\Pnsdiff}{\PnsdiffEventOne}{\PnsdiffEventTwo}
\FPeval{\Pnsdiff}{round(\Pnsdiff:0)}

%% file: authors.tex



\author{R.~Abbott}
\affiliation{LIGO Laboratory, California Institute of Technology, Pasadena, CA 91125, USA}
\author{T.~D.~Abbott}
\affiliation{Louisiana State University, Baton Rouge, LA 70803, USA}
\author{S.~Abraham}
\affiliation{Inter-University Centre for Astronomy and Astrophysics, Pune 411007, India}
\author{F.~Acernese}
\affiliation{Dipartimento di Farmacia, Universit\`a di Salerno, I-84084 Fisciano, Salerno, Italy  }
\affiliation{INFN, Sezione di Napoli, Complesso Universitario di Monte S.Angelo, I-80126 Napoli, Italy  }
\author{K.~Ackley}
\affiliation{OzGrav, School of Physics \& Astronomy, Monash University, Clayton 3800, Victoria, Australia}
\author{A.~Adams}
\affiliation{Christopher Newport University, Newport News, VA 23606, USA}
\author{C.~Adams}
\affiliation{LIGO Livingston Observatory, Livingston, LA 70754, USA}
\author{R.~X.~Adhikari}
\affiliation{LIGO Laboratory, California Institute of Technology, Pasadena, CA 91125, USA}
\author{V.~B.~Adya}
\affiliation{OzGrav, Australian National University, Canberra, Australian Capital Territory 0200, Australia}
\author{C.~Affeldt}
\affiliation{Max Planck Institute for Gravitational Physics (Albert Einstein Institute), D-30167 Hannover, Germany}
\affiliation{Leibniz Universit\"at Hannover, D-30167 Hannover, Germany}
% ...




%% file: intro.tex
In January 2020, the LIGO--Virgo detector network observed
gravitational-wave (GW) signals from two compact binary inspirals
which are consistent with neutron star--black hole (NSBH) binaries.
These represent the  first confident observations to date of NSBH binaries via any observational means.
The two events,
carrying the full designations \eventoneFullName and \eventtwoFullName, and abbreviated henceforth as \eventone and \eventtwo, were detected
on \utcone and \utctwo, respectively.  The coincident detection
of \eventtwo by the three detectors LIGO Hanford, LIGO Livingston, and
Virgo gives it high confidence of being an astrophysical GW event.
During the other event, \eventone, the \LHO detector was not
operational and, owing to the small signal-to-noise ratio $(S/N)$ in \virgo,
it was effectively a single-detector event in LIGO Livingston.  While
quantification of the confidence of single-detector events is subject
to significant uncertainty, \eventone stands clearly apart in the \LLO
data from any other candidate with NSBH-like parameters during the $\sim$11 months of the third observing run (O3).

The component masses inferred
from these binary inspirals provide information
on the nature of their components.  The primaries have masses of
$\massonesource \!=\!
\massonesourcemed{GW200105_combined_highspin}^{+\massonesourceplus{GW200105_combined_highspin}}_{-\massonesourceminus{GW200105_combined_highspin}}\,
\msun$ and $\massonesource\!=\!\massonesourcemed{GW200115_combined_highspin}^{+\massonesourceplus{GW200115_combined_highspin}}_{-\massonesourceminus{GW200115_combined_highspin}}\,
\msun$, for \eventone and \eventtwo, respectively, with uncertainties quoted at the $90\%$ credible level.  These primary masses are well above the maximum mass of a neutron star \citep[NS;][]{MaxNSmass_causality, LVCeos2018, PhysRevD.100.023015,Cromartie:2019kug, FarrChatziioannou2020, Nathanail:2021tay,Fonseca:2021wxt}  and within the mass-range of 
black holes (BHs) observed electromagnetically \citep{Ozel:2010,2011ApJ...741..103F,Kreidberg:2012,Miller:2014aaa} and via GWs \citep{O1BBH,2019PhRvX...9c1040A,Abbott:2020niy}.
The masses of the secondaries are $\masstwosource\!=\!
\masstwosourcemed{GW200105_combined_highspin}^{+\masstwosourceplus{GW200105_combined_highspin}}_{-\masstwosourceminus{GW200105_combined_highspin}}\,
\msun$ and $\masstwosource\!=\!\masstwosourcemed{GW200115_combined_highspin}^{+\masstwosourceplus{GW200115_combined_highspin}}_{-\masstwosourceminus{GW200115_combined_highspin}}\,\msun$, respectively, for \eventone and \eventtwo,  within the mass range of known NSs
\citep{2016arXiv160501665A, Alsing:2017bbc, TheLIGOScientific:2017qsa, Abbott:2020uma}.

Detections of NSBHs have so far remained elusive in both
electromagnetic (EM) and GW surveys. In the past four decades,
surveys have identified 19 binary neutron star (BNS) systems in the
Galaxy \citep{2019ApJ...876...18F, agazie2021green}; however,
the discovery of a pulsar in an NSBH
binary remains a key objective
for current and future radio
observations \citep{2014MNRAS.445.3115L,2020PASA...37....2W}.

Similarly, GW observations of LIGO and Virgo through the first part of the O3 run (O3a) have led to the
identification of 48 binary black hole (BBH) candidates 
\citep{2019PhRvX...9c1040A,Abbott:2020niy}
and two BNS
candidates \citep{TheLIGOScientific:2017qsa,GW190412-discovery}.
Independent analyses of the public detector data identified additional GW candidates  \citep{Venumadhav19a,Venumadhav:2019lyq,Zackay:2019btq,Zackay:2019tzo,magee2019sub,Nitz19,Nitz20,Nitz:2021uxj}.
The absence of NSBH candidates in LIGO and Virgo's first and second
observing runs (O1 and O2, respectively) led to the
upper limit on the local merger rate density of NSBH systems of $\mergerratedensity_{\rm NSBH}\leq 610$~\VolRateUnit at the
90\% credible level \citep{2019PhRvX...9c1040A}.

During  O3a, two events were notable as possible NSBH candidates. First,
\marginal \citep{Abbott:2020niy} was identified as a marginal NSBH candidate with false alarm rate ($1.4/$yr) so high that it
could also plausibly be a detector noise artifact.  Second, GW190814 \citep{GW190814} may have been an NSBH
merger.  Although GW190814's secondary mass of
$m_2=2.59^{+0.08}_{-0.09}\,\msun$ likely exceeds the maximum mass
supported by slowly spinning NSs \citep{Essick:2020ghc, Fattoyev_2020,
Tews:2020ylw, Godzieba_2021}, such as those found in known binaries
that will coalesce within a Hubble time, the secondary could
conceivably be an NS spinning near its breakup
frequency \citep{Essick:2020ghc, Tews:2020ylw,Most:2020bba,
Dexheimer_2021}.

The existence of NSBH systems has long been conjectured.
Observations in the Milky Way reveal high-mass X-ray
binaries composed of a massive star and a compact
object \citep{2008ApJ...679L..37L,Orosz:2009,2009ApJ...701.1076G,Orosz:2011,2014ApJ...790...29G}. Binary evolution models show that X-ray binaries with a BH
component are 
possible progenitors of NSBH
systems \citep{Belczynski2013,Grudzinska2015}.

Major uncertainties regarding massive binary evolution, such as
  mass loss, mass transfer and the impact of supernova explosions
  result in a wide range of  merger rate predictions:
0.1--800~\VolRateUnit{}
\citep{2002ApJ...572..407B,Sipior2002,Belczynski2006,dominik2015,Belczynski2016,eldridge2017,mapelli2018,Kruckow2018,neijssel2019,Drozda:2020qab,2020ApJ...899L...1Z,Broekgaarden:2021iew}.
  Comparable NSBH 
 merger rates are predicted from young star
 clusters \citep{ziosi2014,rastello2020} while NSBH merger rates in
globular and nuclear  clusters are predicted to be orders of magnitude lower \citep{Clausen2013, Ye2020, Fragione:2020wac, Hoang:2020gsi, ArcaSedda2020}.
Measuring the NSBH merger rate and properties such as masses and spins is
 crucial in determining formation channels.

This letter presents the
status of the detectors during times around \eventone and \eventtwo (\S\ref{sec:Detectors}),
and the results of the different searches
leading to the detections (\S\ref{sec:Detections}).  We describe
the main properties of the two events~(\S\ref{sec:SourceProperties}) and discuss the nature of the secondary
components (\S\ref{sec:NatureSecondary}).  Finally, we present
the astrophysical implications, including merger rates of this new class
of GW source (\S\ref{sec:AstroImplications}) and conclude (\S\ref{sec:Conclusions}).
Data products associated with the events reported here, such as calibrated
strain time series and parameter estimation posterior samples, are available
through the Gravitational Wave Open Science Center (GWOSC) at \href{https://gw-openscience.org}{gw-openscience.org}.

%% file: detector_split_sections.tex
\section{Detectors and Data}
\label{sec:Detectors}


The two events reported here were observed in the second part of the third observing run (O3b).
At the time of \eventone, \LLO had been in a stable operational state for over 10 hrs, with
a sensitivity, quantified by the angle-averaged BNS inspiral range~\citep{Allen:2005fk},
of \RangeLLOone. \virgo had been in its nominal state for $\sim$22 hrs, with
a BNS range of \RangeVirgoone, and LIGO Hanford was not operational.
At the time of \eventtwo, all three interferometers had been in a stable operational state for
over 2 hrs.
The BNS ranges for \LHO, \LLO, and \virgo around the  detection
 were \RangeLHOtwo, \RangeLLOtwo, and \RangeVirgotwo, respectively.

\begin{figure*}
\includegraphics[width=0.99\columnwidth, height=10.2cm]{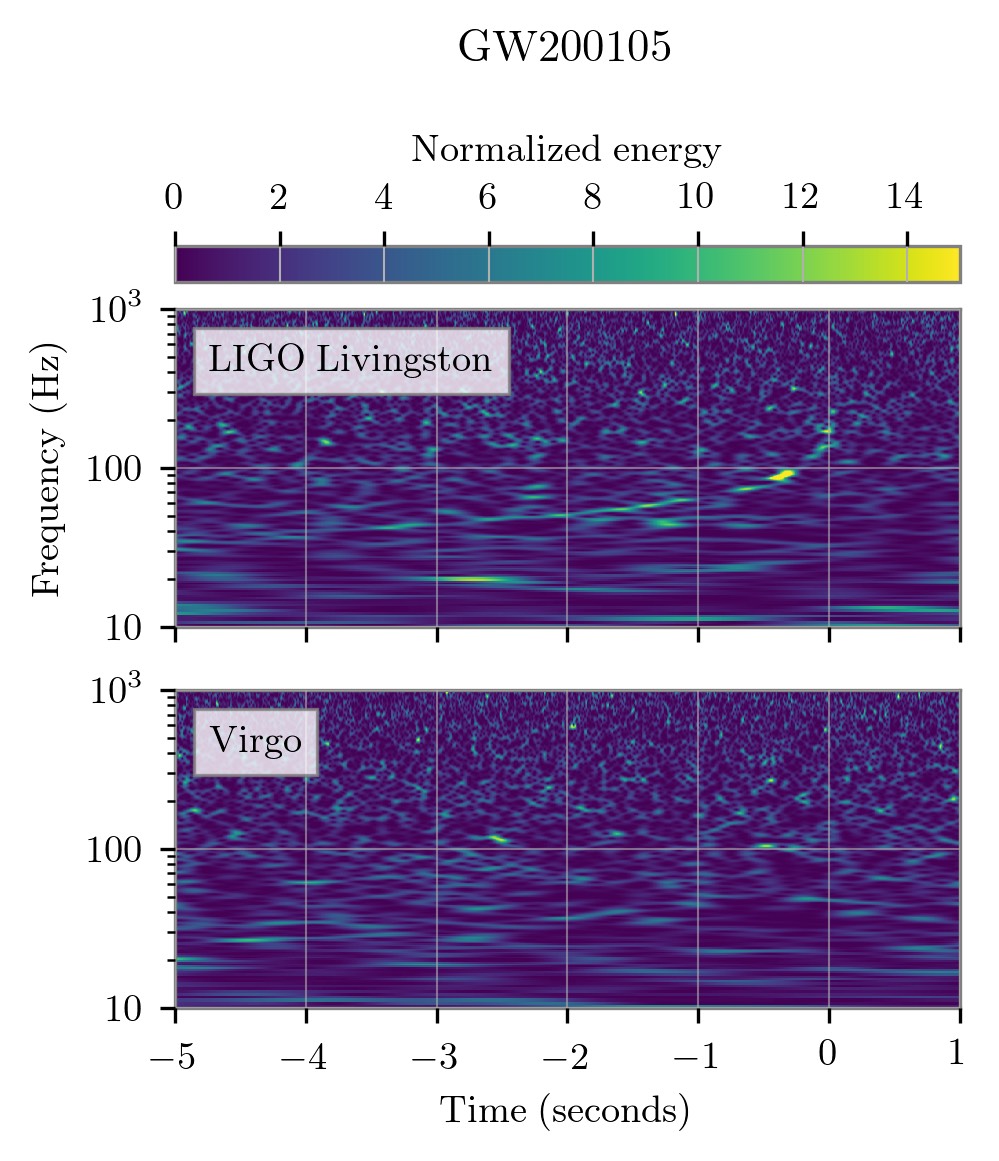}
\includegraphics[width=0.99\columnwidth]{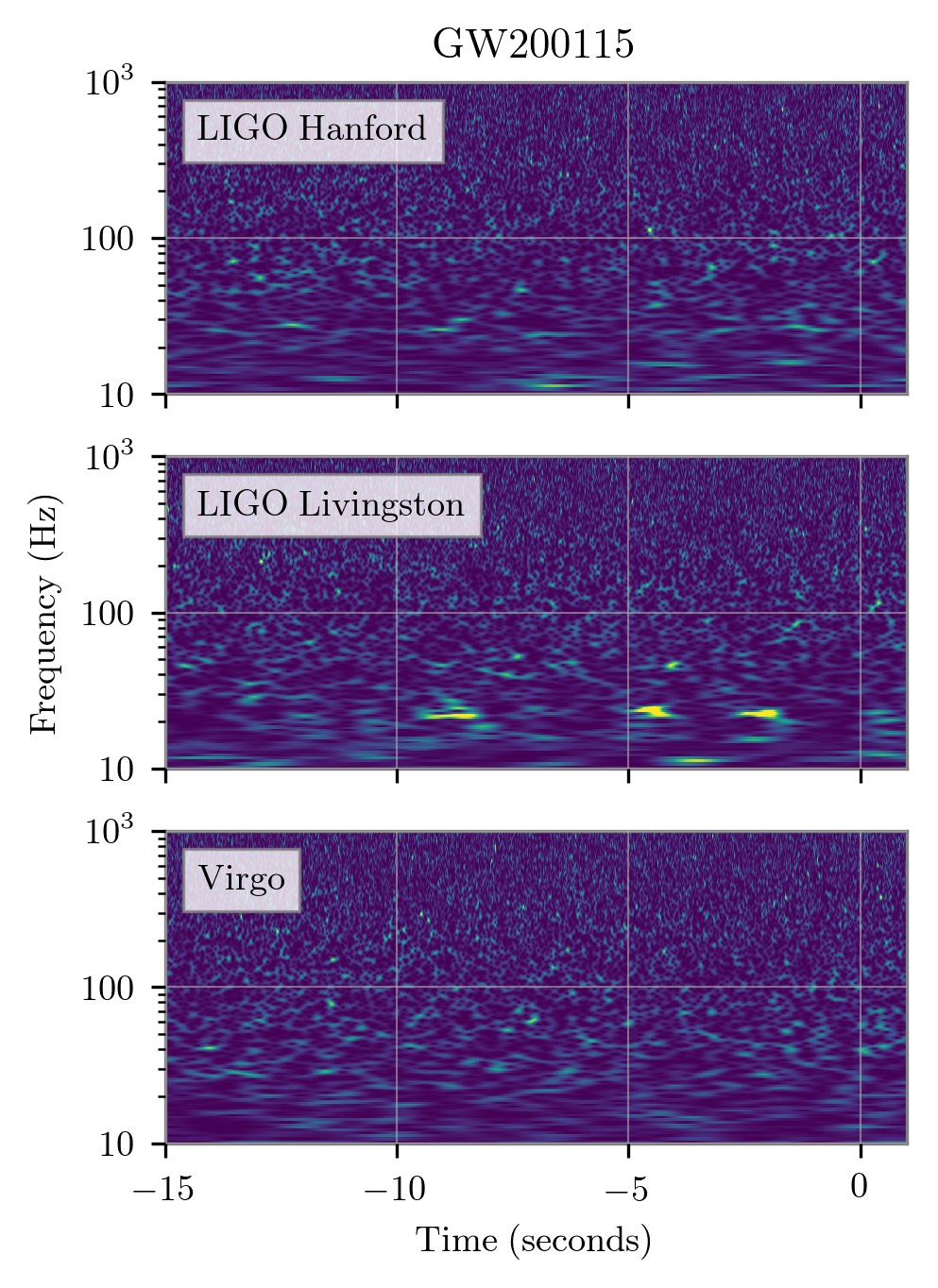}
\caption{\label{fig:omegascans}
  Time--frequency representations
  of the data containing GW200105 (left column) and GW200115 (right column).  Times are shown relative to the signals' merger times, January 05, 2020 at 16:24:26 UTC (left) and January 15, 2020 at 04:23:10 UTC (right).
  The amplitude scale of each time--frequency tile is normalized by the respective detector's noise amplitude spectral density.
The \LLO data of \eventone show
a track of excess power with increasing frequency.
In the other panels, no similar tracks are visible, as the S/N in each of the detectors
is lower and (for \eventtwo) the signal is longer.
  For GW200105, the LIGO Livingston data are shown after glitch subtraction.
  For GW200115, light-scattering noise is visible in \LLO below 25~Hz.
}
\end{figure*}

Figure~\ref{fig:omegascans} shows time--frequency representations~\citep{Chatterji:2004qg}
of the two events.
The \LLO data of \eventone show
a track of excess power with increasing frequency.
For \eventtwo, no similar tracks are visible, as the S/N in each of the detectors
is lower than that of \eventone in \LLO (see Fig.~\ref{fig:105_background}).
Also, light-scattering noise \citep{Soni:2020rbu, Soni:2021cjy}
is visible in Fig.~\ref{fig:omegascans} in LIGO Livingston around 20~Hz.

The LIGO and \virgo GW detectors are calibrated using
radiation pressure from auxiliary lasers at a known frequency and amplitude~\citep{Acernese:2018bfl,Sun:2020wke}.
In LIGO, the calibration pipeline~\citep{Viets:2017yvy} linearly subtracts
the noise from the calibration lines and the harmonics of the power mains
from the data.

The calibration systematic error and associated uncertainty
of the data in the [20--1024]~Hz frequency region
are for the LIGO detectors
no larger than \MaxUncMagLLOone in
amplitude and \MaxUncPhaLLOone in phase for \eventone and
\MaxUncMagLLOtwo in amplitude
and \MaxUncPhaLLOtwo in phase for \eventtwo (68$\%$ credible intervals).  For \virgo,
we use \MaxUncMagVirgoone in
amplitude and \MaxUncPhaVirgoone in phase for both events, except for frequencies 46--51~Hz, where additional calibration systematic error arises from online loops set up to damp
the main power lines at 50~Hz and mechanical resonances of the suspensions close to 48~Hz.
This effect was introduced during detector improvements carried out between O3a and O3b and is not accounted for in the signal reconstruction process.
However, Virgo data in the frequency window $46\text{--}51\,\textrm{Hz}$ are excluded from the parameter estimation analyses in \S\ref{sec:SourceProperties}.

To verify that instrumental noise artifacts do not bias the analysis of source properties of the observed events,
we use data quality validation procedures as in previous events~\citep{TheLIGOScientific:2016zmo, Davis:2021ecd},
employing sensor arrays at LIGO and \virgo to measure environmental disturbances that
could couple into the interferometers~\citep{nguyen2021environmental}.
In \virgo, we find no evidence of excess power from terrestrial sources for both events.
For \eventone, we identify light-scattering noise in \LLO below 25~Hz, 3~s before merger.
As in past detections \citep{Abbott:2020niy}, we
subtract this noise  with the BayesWave algorithm~\citep{Cornish:2020dwh} and use the cleaned data for the source parameter estimation in
\S\ref{sec:SourceProperties}.
The top left panel of Fig.~\ref{fig:omegascans} shows the cleaned data.  A low-energy feature around 20~Hz, not overlapping with the time--frequency track of \eventone, remains after the glitch subtraction.
For \eventtwo, we also identify
light scattering in the \LLO data below 25~Hz.
Due to the increased difficulty of subtracting the long-duration glitching that coincides
with the time--frequency track of \eventtwo, glitch-subtracted data were not available
at the time of analysis. Hence, we exclude \LLO data below 25~Hz in the analysis in
\S\ref{sec:SourceProperties}.

\section{Detections}
\label{sec:Detections}

\eventtwo is a coincident event, and in \S\ref{sec:search115} we describe the established
procedures to identify it and determine its
significance.  Subsequently, we describe the procedures followed for
the single-detector event \eventone.

\subsection{\eventtwo\ -- Multi-detector event}
\label{sec:search115}

\begin{figure}
\centering
\includegraphics[width=\columnwidth,trim=20 0 10 0]{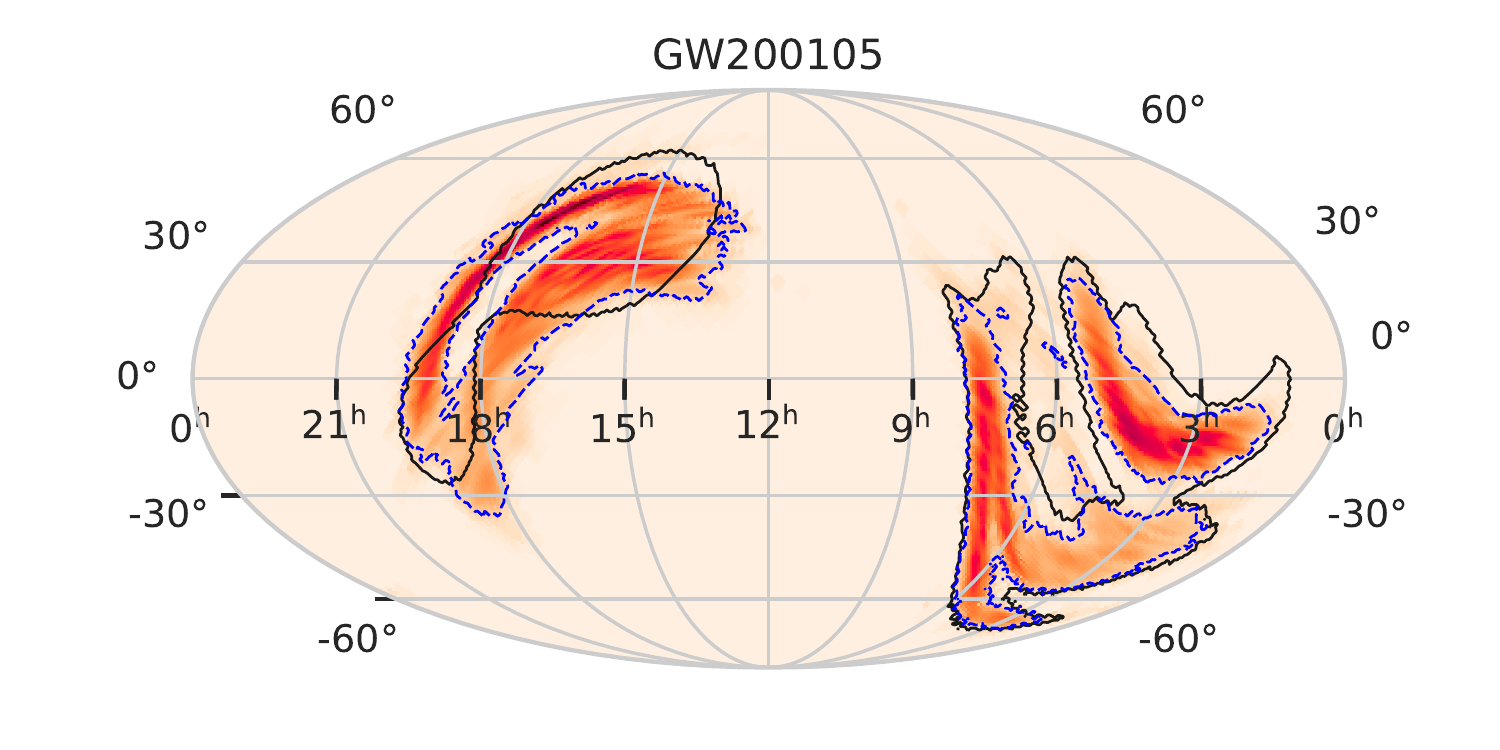}
\includegraphics[width=\columnwidth,trim=20 0 10 0]{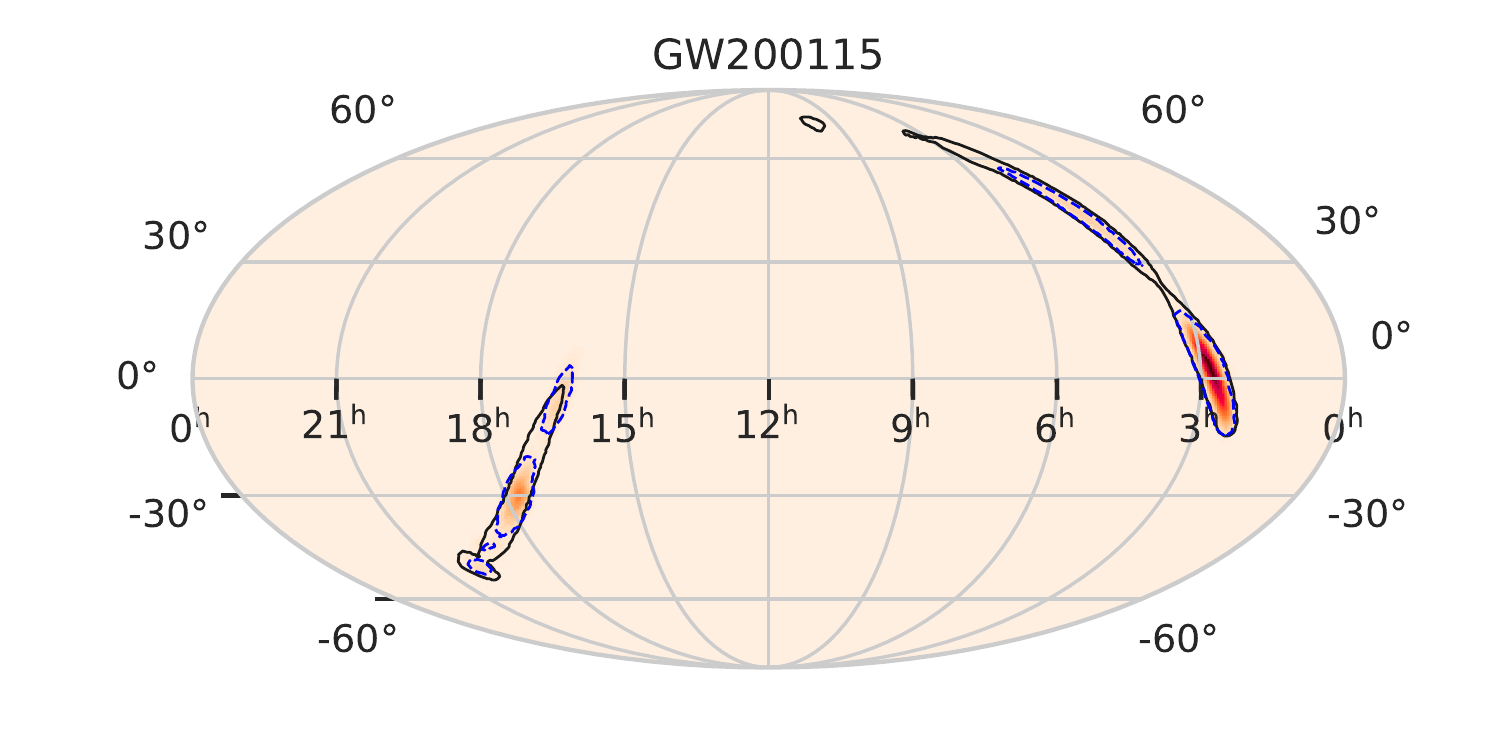}
\caption{Sky localizations for \eventone (top) and for \eventtwo (bottom),
    in terms of right ascension and declination.
The thick, solid contours show the $90\%$ credible regions from the
low-latency sky localization
algorithm \textsc{BAYESTAR} \citep{BAYESTAR}.  The shaded patch is the
sky map obtained from the preferred high-spin analysis of
\S\ref{sec:SourceProperties}, with the $90\%$ credible
regions bounded by the thin dotted contours.
}
\label{fig:skymaps}
\end{figure}

\eventtwo was initially identified by all four low-latency matched-filtering pipelines as a possible compact binary coalescence (CBC)
candidate in \LHO and \LLO:
\gstlal~\citep{Cannon:2011vi, Privitera:2013xza, Messick:2016aqy, Sachdev:2019vvd, Hanna:2019ezx}, \mbtaol~\citep{Aubin:2020goo,2016CQGra..33q5012A},
\pycbclive~\citep{DalCanton:2020vpm, Nitz:2018rgo, pycbc-software, Nitz:2017svb, Usman:2015kfa}, and \spiir~\citep{2012PhRvD..85j2002L, SPIIR2, chu2020spiir,
2018CoPhC.231...62G}.
  As detailed in \cite{Abbott:2020niy}, matched-filtering searches use 
  banks~\citep{
      Owen:1998dk,
      Harry:2009ea,
      Privitera:2013xza, 
      PhysRevD.95.104045,
      PhysRevD.99.024048}
    of modeled gravitational
    waveforms, with the mass range relevant to \eventone and \eventtwo
    covered by the SEOBNRv4\_ROM waveform model \citep{Bohe:2016gbl,Purrer:2015tud}.
The signal in \virgo was not loud enough to further improve the significance of the coincident candidate observed by the LIGO detectors, so
\eventtwo was reported as a two-detector event in the \gdb event database named \gidtwo with a merger time of \utctwo.

A public Preliminary Gamma-ray burst Coordinates Network (GCN) Notice was sent out six minutes after the
event~\citep{2020GCN.26759....1L}. The low-latency \textsc{BAYESTAR}
\citep{BAYESTAR} sky map computed from the original trigger, which was
produced from \mbtaol, indicated a possible discrepancy compared to those
from the other three pipelines. Therefore, the representative trigger was manually
switched to the one from \gstlal \citep{2020GCN.26807....1L}.
The associated sky map is shown in the bottom panel of Fig.~~\ref{fig:skymaps} with a solid black line with 90\% credible area of \OnlineSkymapAreatwo.
The GCN Circular reported a probability $>99\%$ for the lighter compact object
to have a mass below $3 \msun$, derived from a machine-learning analysis \citep{Kapadia:2020, Chatterjee:2019avs}.
The RAVEN pipeline~\citep{2016PhDT.........8U}, which looks for coincident gamma ray bursts (GRBs),
did not report an associated GRB trigger from Fermi-GBM or the Neil Gehrels Swift
Observatory within $-1~\mathrm{s}$ and $+5~\mathrm{s}$ of the GW trigger.
A total of 31 GCN Circulars~\citep{S200115jCircs} reporting
follow-up observations were published for this event. To date, no EM
counterpart has been reported associated with \eventtwo.

\input{search_table.tex}

After the low-latency identification of \eventtwo, the extended
periods of strain data around the event are further analyzed by the
detection pipelines in their offline configurations using improved
calibration and refined data quality information that is not
available in low
latency~\citep{TheLIGOScientific:2017lwt,Abbott:2020niy,Davis:2021ecd}.
In this analysis, we also subtract nonstationary noise due to the
nonlinear coupling of the 60 Hz power mains at the LIGO detectors
using coupling functions derived from machine-learning
techniques~\mbox{\citep{PhysRevD.101.042003}}.

\pycbc uses a template bank constructed with a hybrid geometric-random
algorithm \citep{PhysRevD.99.024048,PhysRevD.95.104045}, while \gstlal and
\mbta use a template bank generated by a stochastic placement
method \citep{Mukherjee:2018yra,Babak:2008rb,Privitera:2013xza}.
 \gstlal identifies \eventtwo with a network S/N of \GstLALOfflineNetSNRtwo and FAR of
\GstLALOfflineFARtwo using the data from November 1, 2019
15:00 UTC to January 22, 2020  18:11 UTC. The \mbta and
\pycbc offline analyses also identified the trigger in data from
January 13, 2020 10:27 UTC to January 22, 2020
 18:11 UTC, yielding consistent network S/Ns as shown in
Table~\ref{table:search_snr}, as well as estimated FARs of
\MBTAOfflineFARtricointwo and \PyCBCOfflineFARtwo, respectively.
Based on two detectors, these FARs robustly indicate the confidence of the detection.

\subsection{\eventone\ -- Single-detector event}
\label{sec:search105}

\eventone was identified by \gstlal running in the low-latency configuration
as a possible CBC candidate in \LLO and \virgo data with network S/N of
\GstLALOnlineNetSNRone  and merger time \utcone.
The candidate identified as \gidone in \gdb was considered a single-detector event
because the S/N in \virgo was below the threshold S/N of \gstlalsinglethreshold.

The low-latency FAR of \GstLALOnlineFARone did not pass the threshold for sending a GCN alert.
Without the information provided by temporal coincidence between
different detectors and consistency in recovered parameters, it is
more difficult to obtain a robust estimate of the event's significance.  In fact, an alternative configuration of \gstlal,
running to test several improvements made in the offline configuration of early O3~\citep{Abbott:2020niy}, found
the trigger at higher significance.  Therefore,
validation procedures and subsequent early-offline analysis were
initiated.
The three other low-latency search pipelines, \mbtaol, \pycbclive, and \spiir,
also generated triggers with consistent S/Ns near the event time of \gidone{} (see Table~\ref{table:search_snr}).
These three pipelines were not configured to assign FARs to single-detector triggers and therefore did not generate automatic alerts for \eventone.

On January 6, 2020 at 19:39:09 UTC, the initial GCN Circular reported
a low-latency BAYESTAR skymap (solid black contours) in the top panel of Fig.~\ref{fig:skymaps} with 90\% credible region of \OnlineSkymapAreaone and
a 12\% chance of formation of
tidally disrupted matter---relevant for a possible EM counterpart---as a result of the coalescence,
as derived from a machine-learning analysis~\citep{Kapadia:2020, Chatterjee:2019avs}.
Using the parameters obtained from low-latency parameter estimation, the chance of formation
of tidally disrupted material was later revised to be negligible (see \citealp{2020GCN.26688....1L} and \S\ref{sec:NatureSecondary} below).

In response to the discovery notice of \eventone, multiple observatories carried out
follow-up and shared their results publicly via 21 GCN circulars~\citep{S200105aeCircs}.
To date, no EM counterpart has been reported associated with \eventone.

After the event's identification in low latency, the strain data are further analyzed by the detection
pipelines in their offline configuration, following procedures for single-detector events discussed in~\citet{Abbott:2020uma}.
\gstlal identifies \eventone with an S/NR of  \GstLALOfflineNetSNRone and FAR of \GstLALOfflineFARone in its offline configuration using data from  November 1, 2019
15:00 UTC up to January 22, 2020 18:11 UTC.
For single-detector events, like \eventone, the FAR estimate involves extrapolation, as explained below.  The
\mbta and \pycbc offline analyses also identify the trigger in the \llo data from
January 4, 2020 17:09 UTC to January 13, 2020 10:27 UTC
yielding the consistent S/Ns shown in Table~\ref{table:search_snr}.
The S/N for \eventone is larger than that for \eventtwo, even though \LHO was not operational.

\begin{figure}
  \includegraphics[width=\columnwidth,trim=5 2 4 0]{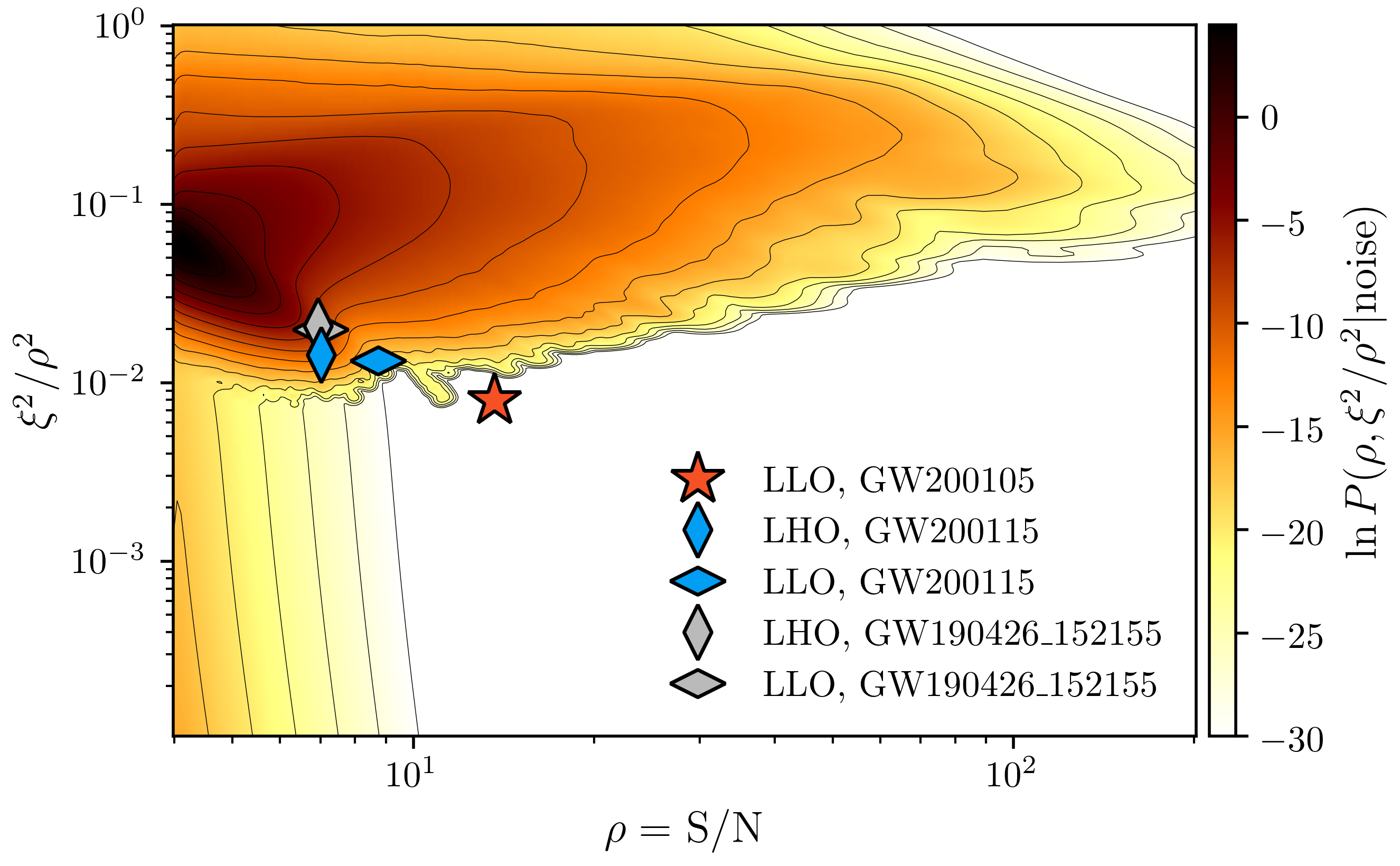}
\caption{\label{fig:105_background} Colored shading shows the joint S/N--$\xi^2$
noise probability density function for \LHO (LHO), \LLO (LLO), and
\virgo, computed from background triggers found by \gstlal in the region of lighter binary systems during the entire O3. The
red star indicates \eventone. At its position, there is no background
present, indicating that it stands above all of the recorded background triggers.
For comparison, the triggers of \eventtwo and the marginal \marginal are
also shown.}
\end{figure}

We cannot rely on only the S/N of the \cbc trigger to estimate its
significance, i.e.\ to quantify how often detector noise mimics a
possible coalescence signal. Because detector noise shows significant
deviation in the tails from the standard Gaussianity assumptions, its
properties have to be estimated
empirically~\citep{TheLIGOScientific:2016qqj}. For multi-detector
triggers like \eventtwo, consistency of the observed
\cbc trigger among two or more detectors forms an integral part of
establishing it as a confident detection with a low FAR.  On the other hand, single-detector
triggers can be assigned only modest FAR values due to limited detector
observational time \citep{Callister:2017urp,2020ApJ...897..169N}.
Nonetheless, alternative methods can be used to estimate confidence in a
single-detector event, as was demonstrated for GW190425
\citep{Abbott:2020uma}.

The signal \eventone measured in \LLO is distinct from all noise
events in the entire O3 observation period.
Specifically, Fig.~\ref{fig:105_background} shows with colored shading the distribution of
background noise triggers identified by \gstlal in the region of binary systems with a chirp mass less than 4 $\msun$.
The colored region indicates
the density of background noise triggers
quantified through the S/N--$\xi^2$ noise probability
density function for the three detectors, where the autocorrelation
$\xi^2$ provides a consistency test similar to
$\chi^2$ \citep{Messick:2016aqy}. The red star
represents \eventone, which lies in a region of this plane without noise
triggers,
indicating that this trigger is unique within the entire O3.

For comparison, Fig.~\ref{fig:105_background} also shows \eventtwo
and the marginal candidate \marginal separately for \LLO
and \LHO.
In general, triggers such as \eventtwo and \marginal in Fig.~\ref{fig:105_background} would not be separable from the noise by
a single detector alone, and it is the coincidence between multiple detectors that raises their
significance. On the other hand, the coincidence was not available for \eventone, yet it clearly stands out from the background in Fig.~\ref{fig:105_background} as strong GW signals typically do.

The uniqueness of \eventone within the data shown leads to a upper bound on
the FAR assignment comparable to the inverse observing time.
A stronger bound, like the FAR  quoted above, must rely on
assumptions made on the properties of the noise triggers if one had collected
the observational strain data for a longer period, a process called extrapolation.  As a consequence
of the assumptions made, the extrapolated FAR estimates have
large uncertainties. Since single-detector FAR estimates rely on the
uniqueness within the dataset, the values may change and errors may reduce as
more data are accumulated.
At the time of the
analysis, neither \mbta nor \pycbc had the capability to assign a significance
to single-detector triggers, although \eventone also stands out as a unique
trigger in their analyses. Based on these considerations, we consider this trigger to be astrophysical and include it in the analysis in the remainder of this
paper.

%% file: search_table.tex
\newcolumntype{a}[1]{D{.}{\cdot}{#1} }

\begin{table}
    \centering
\caption{\label{table:search_snr}
    Network S/N recovered by the search pipelines in
    low-latency and in later offline analysis with improved calibration and refined data-quality information.
An asterisk (*) indicates values where Virgo is not included in the S/N calculation.
}
\begin{tabular}{cccccc} \toprule\toprule
    $\!\!$Event & & $\!\!\!$\gstlal$\!\!$ & $\!\!$\mbta$\!\!\!$ & $\!$\pycbc$\!\!$ &
    $\!\!$\spiir$\!\!\!\!$ \\

        \midrule

            \multirow{2}{*}{$\!\!$\eventone$\!\!$} & $\!\!\!\text{low-latency}\!\!\!$ & $\GstLALOnlineNetSNRone$ & $\MBTAOnlineNetSNRone$ & $\PyCBCLiveNetSNRone^*$ & $\SpiirNetSNRone$ \\

            & $\!\!\text{offline}\!$ & $\GstLALOfflineNetSNRone$ & $\MBTAOfflineNetSNRone$ & $\PyCBCOfflineNetSNRone^*$ & $-$ \\

            \multirow{2}{*}{$\!\!$\eventtwo$\!$\!} & $\!\!\!\text{low-latency}\!\!\!$ & $\GstLALOnlineNetSNRtwo$ & $\MBTAOnlineNetSNRtwo$ & $\PyCBCLiveNetSNRtwo\;\,$ & $\SpiirNetSNRtwo$ \\

            & $\!\!\text{offline}\!$ & $\GstLALOfflineNetSNRtwo$ & $\MBTAOfflineNetSNRtwo$ & $\PyCBCOfflineNetSNRtwo^*$ & $-$ \\
\toprule\toprule
        \end{tabular}
\end{table}

%% file: pe.tex
We infer the physical properties of the two GW events using a coherent
Bayesian analysis following the methodology
described in Appendix B of \cite{2019PhRvX...9c1040A}. For GW200105, data from LIGO Livingston and Virgo are analyzed, whereas for GW200115, data from both LIGO detectors and Virgo are used.

Owing to the different signal durations, we analyze $32\,\textrm{s}$ of data for the higher-mass event GW200105 and $64\,\textrm{s}$ of data for GW200115.
All likelihood evaluations use a low-frequency cutoff of $f_{\textrm{low}}=20\,\textrm{Hz}$, except for LIGO Livingston for \eventtwo, where $f_{\textrm{low}}=25\,\textrm{Hz}$ avoids excess noise localized at low frequencies, as discussed in \S\ref{sec:Detectors}.
The power spectral density used in the likelihood calculations is
the median estimate calculated with BayesLine \citep{Cornish:2014kda}.

The parallel Bilby (\pbilby) inference library, together with the
\textsc{dynesty} nested sampling software \citep{2019ApJS..241...27A, 2020MNRAS.499.3295R,Smith_20, 2019arXiv190402180S} is the primary tool used to sample
the posterior distribution of the sources' parameters and perform hypothesis
testing.  In addition, we use RIFT \citep{2018arXiv180510457L} for the 
most computationally expensive analyses and \lalinference 
\citep{PhysRevD.91.042003} for verification.

\begin{table*}
\caption{\label{table:combined}
Source properties of \eventone and \eventtwo. We report the median values with $90\%$ credible intervals. Parameter estimates are obtained using the Combined PHM samples.}
\input{pe_combined_table.tex}
\end{table*}

We base our main analyses of \eventone and \eventtwo on BBH waveform models that include the effects of spin-induced orbital precession and higher-order multipole GW moments, but do not include tidal effects on the secondary.
Specifically, we use two signal models: IMRPhenomXPHM \citep[Phenom PHM;][]{2020arXiv200406503P} from the phenomenological family and SEOBNRv4PHM
\citep[EOBNR PHM;][]{Ossokine:2020} from the effective one-body numerical relativity family. The acronym PHM stands for Precessing Higher-order multipole Moments. Henceforth, we will use the shortened names for the waveform models.

In order to quantify the impact of neglecting tidal effects, we also
analyze \eventone and \eventtwo using two NSBH waveform models that
include tidal effects and assume that spins are aligned with the
orbital angular momentum: IMRPhenomNSBH \citep[Phenom
NSBH;][]{PhenomNSBH} and SEOBNRv4\_ROM\_NRTidalv2\_NSBH \citep[EOBNR
NSBH;][]{SEOBNRNSBH}.  We restrict the NSBH analyses to the
region of applicability of the NSBH models, i.e.
$\chi_1<0.5, \chi_2<0.05$ for Phenom NSBH and 
$\chi_1<0.9, \chi_2<0.05$ for EOBNR NSBH. We also perform aligned-spin
BBH waveform analyses and find good agreement with the analyses using
NSBH waveform models (see \S\ref{sec:waveform systematics} below),
validating the use of BBH waveform models.  Specifically, we use the
aligned-spin BBH models
IMRPhenomXAS \citep[Phenom;][]{Pratten:2020fqn} and
SEOBNRv4 \citep[EOBNR;][]{Bohe:2016gbl}, which only contain dominant
quadrupole moments, and IMRPhenomXHM \citep[Phenom
HM;][]{Garcia-Quiros:2020qpx} and SEOBNRv4HM \citep[EOBNR
HM;][]{Cotesta:2018fcv,Cotesta:2020qhw}, which contain higher-order
moments.

The secondary objects are probably NSs based on mass estimates, as discussed in detail in \S\ref{sec:NatureSecondary}.  As in earlier GW analyses \citep{TheLIGOScientific:2017qsa, Abbott:2020uma}, we proceed with two different priors on the secondary's spin magnitude:  a \textit{low-spin prior}, $\spintwomag\leq 0.05$, which captures the maximum spin observed in Galactic BNSs that will merge within a Hubble time \citep{Burgay:2003jj}, and a \textit{high-spin prior}, $\spintwomag \leq 0.99$, which is agnostic about the nature of the compact object. The two priors allow us to investigate whether the astrophysically relevant subcase of low NS spin leads to differences in the parameter estimation for the binaries.
All other priors are set as in previous analyses \citep[e.g.,][]{Abbott:2020niy}.
Throughout, we assume a standard flat $\Lambda\text{CDM}$ cosmology with Hubble constant $H_0 =~67.9\, \km\,
\s^{-1}\,\mpc^{-1}$ and matter density parameter $\Omega_{\mathrm{m}}=0.3065$ \citep{Ade:2015xua}.

For each spin prior, we run our main analyses with higher-order multipole moments and precession for both waveform families, EOBNR PHM and Phenom PHM.  The EOBNR PHM model is used in combination with RIFT and the Phenom PHM model with \pbilby.  The parameter estimation results for the individual precessing waveform models yield results in very good agreement; the median values typically differ by $1/10$ of the width of the 90$\%$ credible interval.
Nevertheless, in order to alleviate
potential biases due to different samplers or waveform models, we combine an equal number of samples of each into one data set for each spin prior \citep{TheLIGOScientific:2016wfe, 2020PhRvD.101f4037A, GW190814} and denote these as Combined PHM. The quoted parameter estimates in the  following sections are the Combined PHM high-spin prior analyses. In the figures, we emphasize the high-spin prior results.
The values of the most important parameters of the binaries are summarized in Table~\ref{table:combined}, and we will
present details in the following sections.

\subsection{Masses}
\label{sec:pe:masses}

\begin{figure}
\includegraphics[width=\linewidth,trim=45 60 105 110,clip=true]{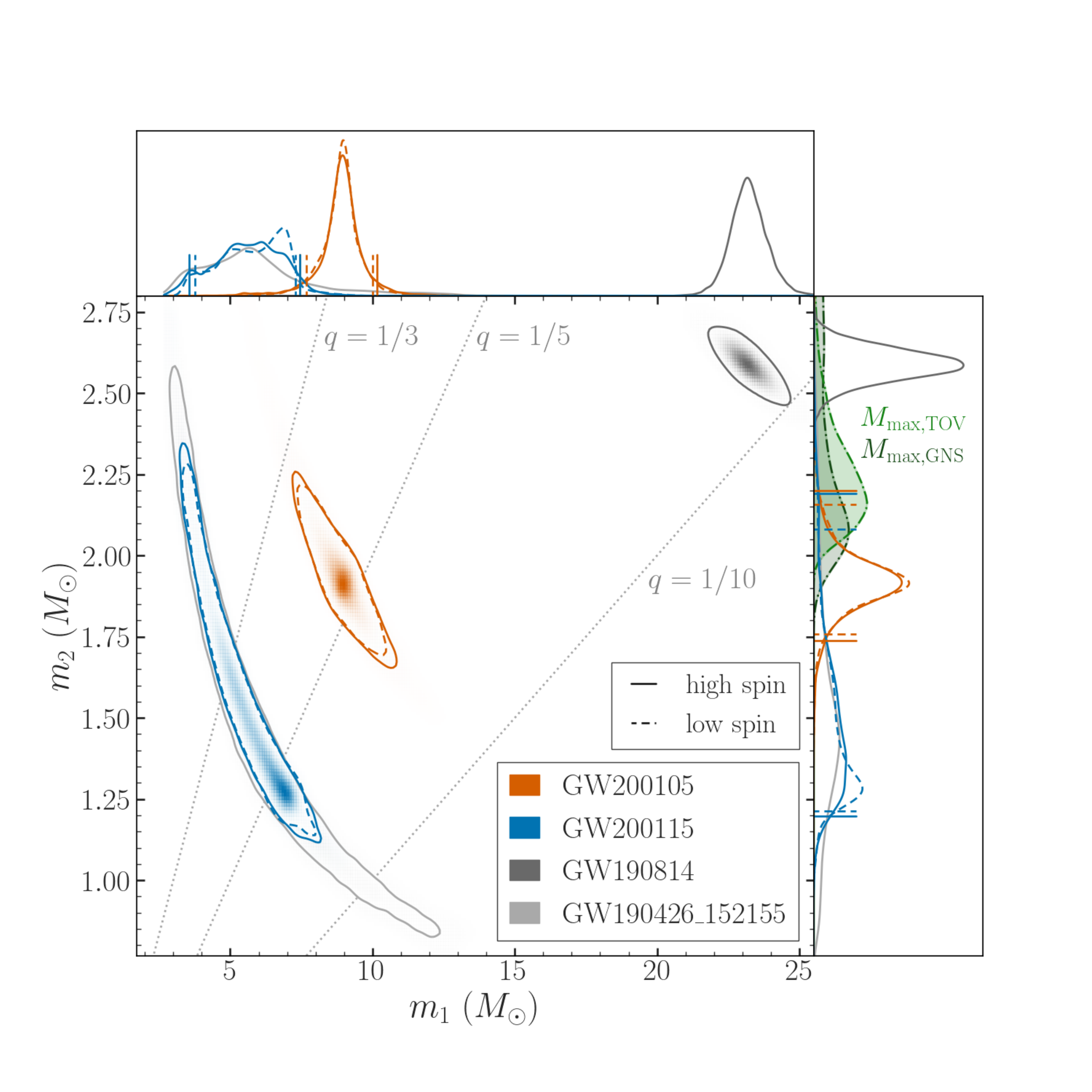}
  \caption{\label{fig:masses_plot_compare}
Component masses of \eventone (red) and \eventtwo (blue), represented by their two- and one-dimensional posterior distributions.
Colored shading and solid curves indicate the high-spin prior, whereas
dashed curves represent the low-spin prior.
The contours in the main panel, as well as the vertical and horizontal lines in the top and right panels, respectively, indicate the 90\% credible intervals.
Also shown in gray are two possible NSBH events, GW190814 and the marginal candidate \marginal, the latter overlapping \eventtwo.
Lines of constant mass ratio are indicated in dashed gray.
The green shaded curves in the right panel represent the one-dimensional probability densities for two estimates of the maximum NS mass,
based on analyses of nonrotating NSs \citep[$M_{\mathrm{max,TOV}}$;][]{LandryEssick2020, landry_philippe_2021_4678703} and Galactic NSs \citep[$M_{\mathrm{max,GNS}}$;][]{FarrChatziioannou2020}.
}
\end{figure}

Figure~\ref{fig:masses_plot_compare} shows the posterior distribution for the component masses of the two binaries. 
Defining the mass parameters such that the heavier mass is the primary object, i.e.\ $\massonesource > \masstwosource$,
 our analysis shows that
\eventone is a binary with a mass ratio of $\massratio =m_2/m_1=
\massratiomed{GW200105_combined_highspin}^{+\massratioplus{GW200105_combined_highspin}}_{-\massratiominus{GW200105_combined_highspin}}$,
with source component masses $\massonesource =
\massonesourcemed{GW200105_combined_highspin}^{+\massonesourceplus{GW200105_combined_highspin}}_{-\massonesourceminus{GW200105_combined_highspin}}\,
\msun$ and $\masstwosource =
\masstwosourcemed{GW200105_combined_highspin}^{+\masstwosourceplus{GW200105_combined_highspin}}_{-\masstwosourceminus{GW200105_combined_highspin}}\,
\msun$. Similarly, \eventtwo is a binary with a mass ratio of $\massratio
=\massratiomed{GW200115_combined_highspin}^{+\massratioplus{GW200115_combined_highspin}}_{-\massratiominus{GW200115_combined_highspin}}$,
with source component masses $\massonesource
=~\massonesourcemed{GW200115_combined_highspin}^{+\massonesourceplus{GW200115_combined_highspin}}_{-\massonesourceminus{GW200115_combined_highspin}}\,
\msun$ and $\masstwosource
=~\masstwosourcemed{GW200115_combined_highspin}^{+\masstwosourceplus{GW200115_combined_highspin}}_{-\masstwosourceminus{GW200115_combined_highspin}}\,
\msun$.

The primary components  of \eventone and \eventtwo are identified as BHs from their mass measurements.
For GW200115, we find that the probability of the primary falling in the lower mass gap \citep[$3\,\msun\! \lesssim\! m_1 \!\lesssim \!5\,\msun$;][]{1998ApJ...499..367B, OroszJain1998} is $\PEpercentMassGap{GW200115_combined_highspin}\%\,$ $(\PEpercentMassGap{GW200115_combined_lowspin}\%)$ for high-spin (low-spin) prior.
For context, Fig.~\ref{fig:masses_plot_compare} also includes two potential NSBH candidates discovered previously;
GW190814 \citep{GW190814} is a high-S/N event with well-measured masses that has a significantly more massive primary and a distinctly more massive secondary than either \eventone or \eventtwo, and the marginal candidate \marginal \citep{Abbott:2020niy}, has (if of astrophysical origin) $m_1$--$m_2$ contours that overlap those of \eventtwo.  The masses of \marginal are less constrained than those of \eventtwo due to its smaller S/N. To highlight how the secondary masses of \eventone and \eventtwo compare to the maximum NS mass, we also show two estimates of the maximum NS mass 
based on an analyses of  non-rotating \citep{LandryEssick2020} and Galactic \citep{FarrChatziioannou2020} NSs.

The secondary masses are consistent with the maximum NS mass, which we quantify in \S\ref{sec:nature:masses}.

\subsection{Sky location, distance, and inclination}
\label{sec:sky_loc}

We localize \eventone's source to a sky area of \skyareaninetypercent{GW200105_combined_highspin} $\,\text{deg}^2$ (90\% credible region).
The large sky area arises due to the absence of data from \LHO.
The luminosity distance of the source is found to be $\luminositydistance = \luminositydistancemed{GW200105_combined_highspin}^{+\luminositydistanceplus{GW200105_combined_highspin}}_{-\luminositydistanceminus{GW200105_combined_highspin}}~\mpc$.
For the second event, \eventtwo, we localize its source to be within $\skyareaninetypercent{GW200115_combined_highspin}\,\text{deg}^2$.
It is better localized than \eventone by an order of magnitude, since \eventtwo was observed with three detectors.
We find the luminosity distance of the source to be $\luminositydistance = \luminositydistancemed{GW200115_combined_highspin}^{+\luminositydistanceplus{GW200115_combined_highspin}}_{-\luminositydistanceminus{GW200115_combined_highspin}}~\mpc$.

\begin{figure}
\centering
\includegraphics[width=\linewidth,trim=45 60 105 110]{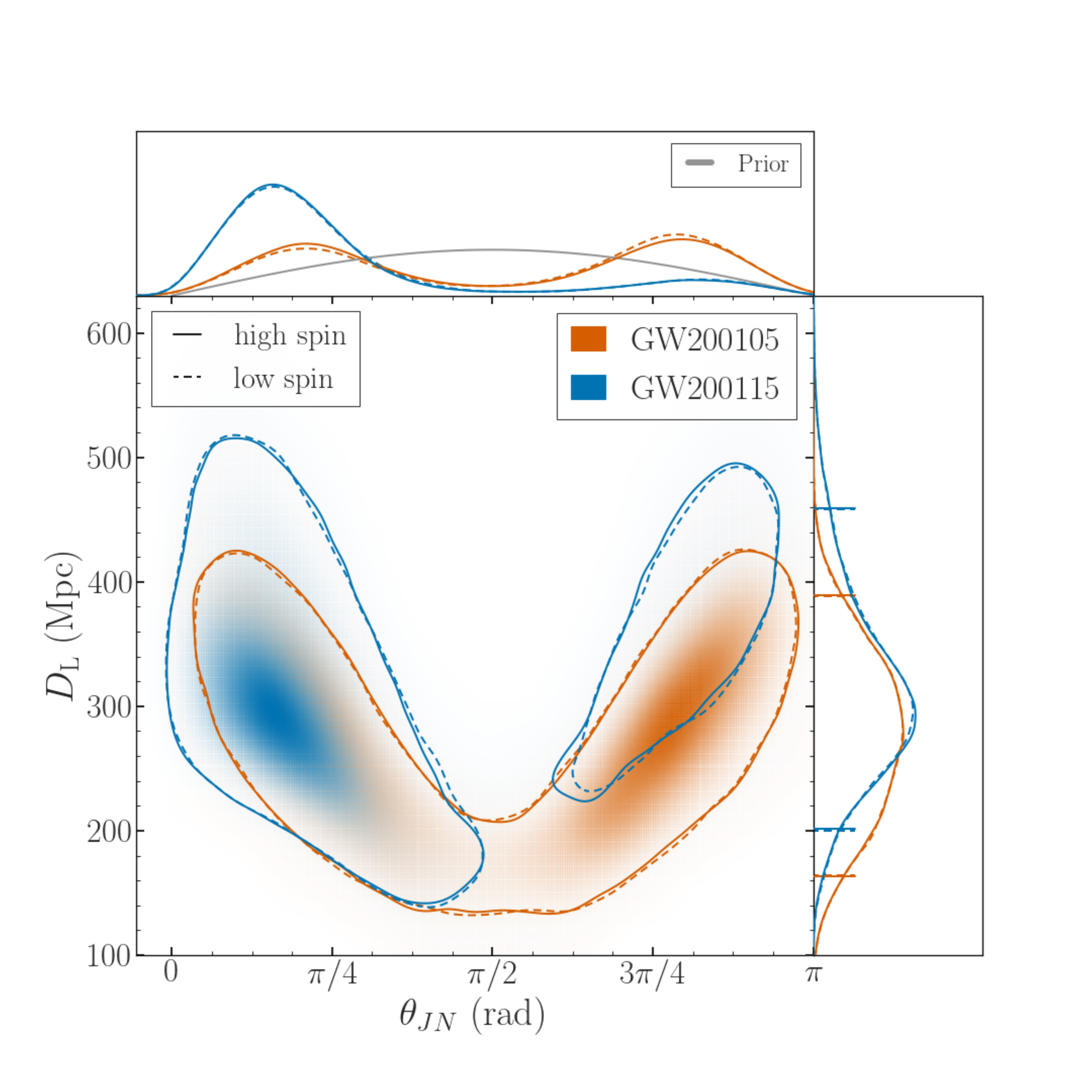}
    \caption{Two- and one-dimensional posterior distributions for distance $\luminositydistance$ and inclinaton $\inclination$.
        The solid (dashed) lines indicate the high-spin (low-spin) prior analysis, and the shading indicates the posterior probability of the high-spin prior analysis.
                  The contours in the main panel and the horizontal lines in the right panel indicate 90\% credible intervals.}
\label{fig:dl_thetajn}
\end{figure}

The luminosity distance is degenerate with the inclination angle $\inclination$ between the line of sight and the binaries' total angular momentum vector \citep{Cutler:1994ys, 2010ApJ...725..496N}. Inclination $\inclination=0$ indicates that the angular momentum vector points toward Earth.
The posterior distribution of the inclination angle is bimodal and strongly correlated with luminosity distance, as shown in Figure~\ref{fig:dl_thetajn}.
The inclination measurement for GW200105 equally favors orbits that are either oriented toward or away from the line of sight. In contrast, GW200115 shows modest preference for an orientation $\inclination \leq \pi/2$.

\subsection{Spins}
\label{sec:spins}

\begin{figure*}
\centering
\includegraphics[width=0.47\linewidth,trim=30 40 10 40]{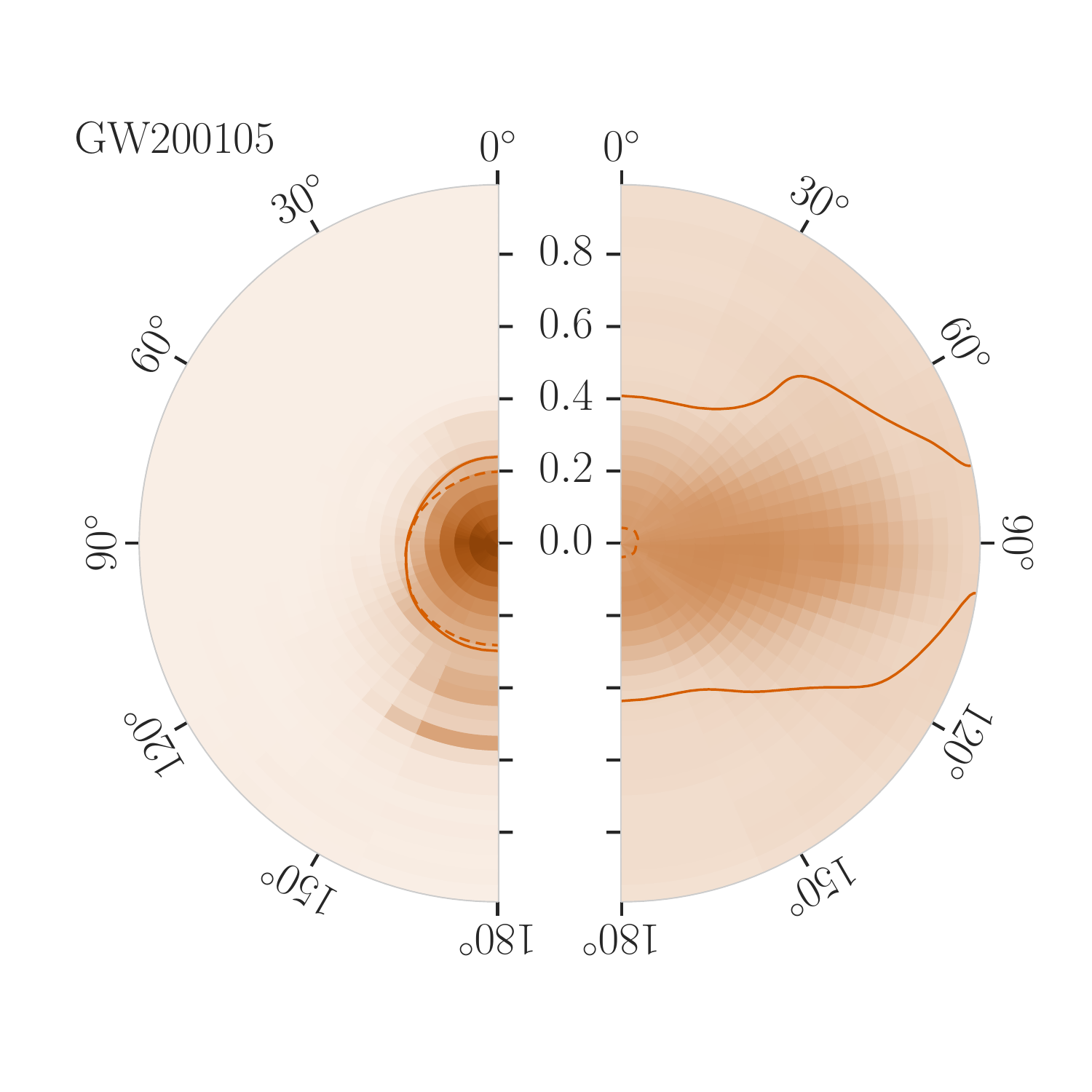}
\hfil
\includegraphics[width=0.47\linewidth,trim=10 40 30 40]{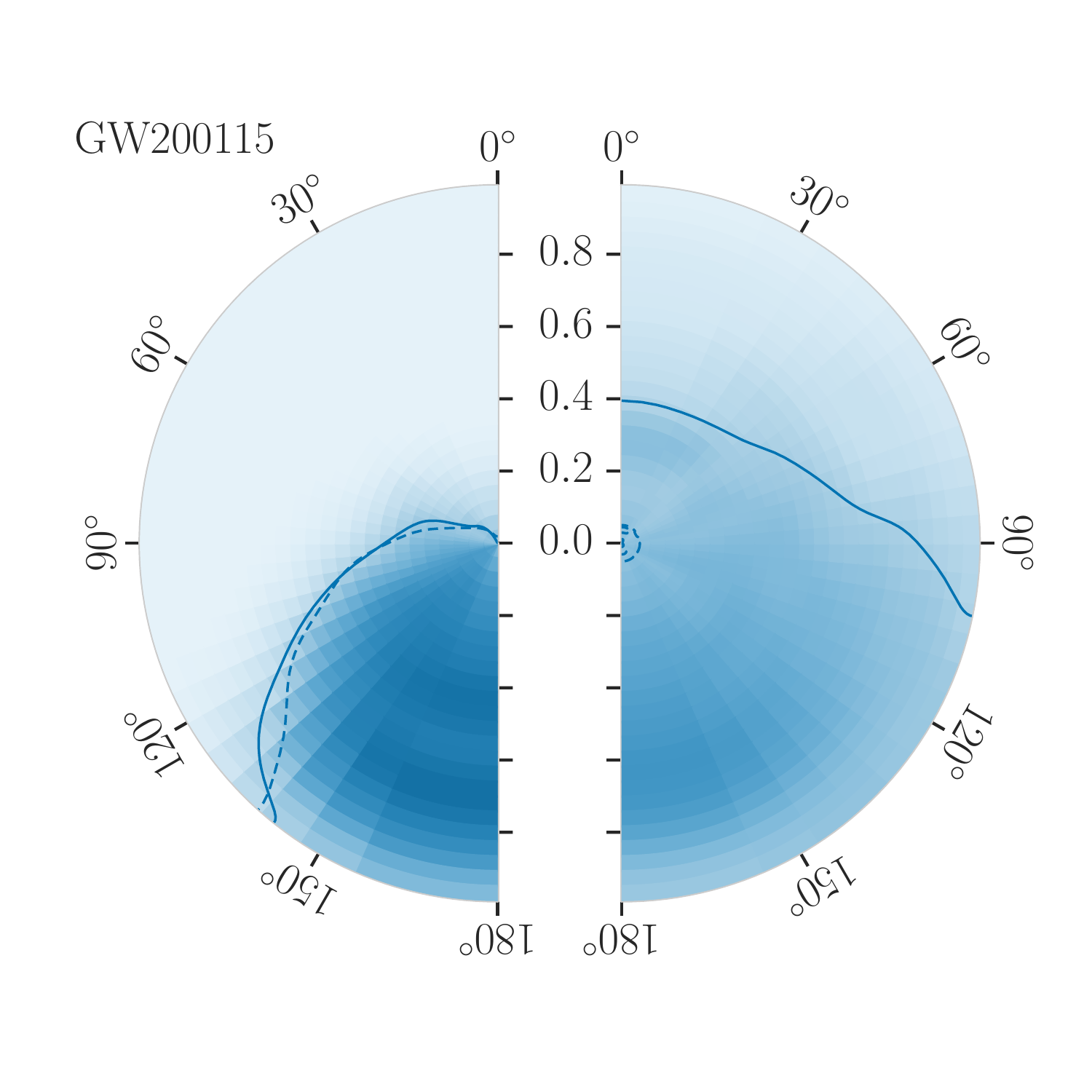}
\caption{
Two-dimensional posterior probability for the spin-tilt angle and spin
magnitude for the primary objects (left hemispheres) and secondary
objects (right hemispheres) for both events.  Spin-tilt angles of
0° (180°) correspond to spins aligned (antialigned) with the orbital
anglular momentum. The color indicates the posterior probability per
pixel of the high-spin prior analysis.  For comparison with the low-spin analysis, the solid (dashed) lines indicate the 90\% credible regions of the high-spin  (low-spin) prior analyses.   The tiles are constructed linearly in spin magnitude and the
cosine of the tilt angles such that each tile contains an identical prior
probability.  The probabilities are marginalized over the azimuthal
angles.\\}
\label{fig:spin_disk}
\end{figure*}

The angular momentum vector $\vec S_i$ of each compact object is related to
its dimensionless spin vector $\vec\chi_i\equiv c\vec S_i/(Gm^2_i)$.  Its magnitude $\chi_i\equiv|\vec\chi_i|$ is bounded by $1$.
For \eventone, we infer
 $\chi_1 = \spinonemed{GW200105_combined_highspin}^{+\spinoneplus{GW200105_combined_highspin}}_{-\spinoneminus{GW200105_combined_highspin}}$, which is consistent
with zero.  For \eventtwo, the spin magnitude is not as tightly constrained, 
$\chi_1 = \spinonemed{GW200115_combined_highspin}^{+\spinoneplus{GW200115_combined_highspin}}_{-\spinoneminus{GW200115_combined_highspin}}$,
 but is also consistent with zero.
The spin of the secondary for both events is unconstrained.

One of the best-constrained spin parameters is the effective
inspiral spin parameter $\spineffective$ \citep{Damour:2001tu, Racine:2008qv,
PhysRevD.82.064016,Ajith:2012mn, Vitale:2016avz}.
It encodes information about the binaries' spin
components parallel to the orbital angular momentum, 
$\chi_{\rm{eff}} = \left(\frac{m_1}{M}\vec{\chi}_{1}+ \frac{m_2}{M}\vec{\chi}_{2}\right)\cdot\hat{L}$,
where $\hat{L}$ is the unit vector along the orbital angular momentum.

For \eventone,  $\spineffective =
\chieffmed{GW200105_combined_highspin}^{+\chieffplus{GW200105_combined_highspin}}_{-\chieffminus{GW200105_combined_highspin}}$ and we find the effective inspiral spin parameter to be strongly peaked about zero, with roughly equal support for being either  positive or negative.
For \eventtwo, we find modest support
for negative effective inspiral spin: $\spineffective =
\chieffmed{GW200115_combined_highspin}^{+\chieffplus{GW200115_combined_highspin}}_{-\chieffminus{GW200115_combined_highspin}}$. Negative values of $\spineffective$ indicate binaries with at least one spin component negatively aligned with respect to the orbital angular momentum, i.e.\  $\chi_{i,z} \equiv \vec \chi_i\cdot \hat L<0$.   We find  $\chi_{1,z} =\spinonezmed{GW200115_combined_highspin}^{+\spinonezplus{GW200115_combined_highspin}}_{-\spinonezminus{GW200115_combined_highspin}}$,  and a probability of $\PEpercentchionenegative{GW200115_combined_highspin}\%$ that $\chi_{1,z} < 0 $.

The joint posterior probability of the dimensionless spin angular
momentum magnitude and tilt angle for both components of both events is shown in
Fig.~\ref{fig:spin_disk}.  The tilt angle with respect to the orbital
angular momentum is defined as
$\arccos\big(\hat{L}.\hat{\chi}_i\big)$.  Deviations from uniform
shading indicate a spin orientation measurement.  The spin orientation
of the primary of \eventone is unconstrained, whereas the orientation of \eventtwo shows support for negatively aligned primary spin.

Orbital precession is caused by a spin component in the
orbital plane of a binary \citep{Apostolatos:1994}, which we parameterize using the
effective precession spin parameter $0\le\spinprecession\le 1$
\citep{2015PhRvD..91b4043S}. We
infer $\spinprecession = \chipmed{GW200105_combined_highspin}^{+\chipplus{GW200105_combined_highspin}}_{-\chipminus{GW200105_combined_highspin}}$ for \eventone and $\spinprecession =\chipmed{GW200115_combined_highspin}^{+\chipplus{GW200115_combined_highspin}}_{-\chipminus{GW200115_combined_highspin}}$ for \eventtwo. To assess the significance of a measurement of precession, we compute a Bayes factor between a precessing and nonprecessing
signal model and the precession
S/N $\rho_{\mathrm{p}}$~\citep{Fairhurst:2019srr, Fairhurst:2019_2harm}.  For \eventone, we
find a log Bayes factor in favor of spin precession of $\logB=\logtenBayesPrecession{GW200105_Combined_highspin}$ and
precession S/N $\rho_{\mathrm{p}} = 0.74^{+1.35}_{-0.61}$. For \eventtwo,
$\logB=\logtenBayesPrecession{GW200115_Combined_highspin}$  and $\rho_{\mathrm{p}} = 0.97^{+1.57}_{-0.79}$. For both events and both diagnostics, this
indicates inconclusive evidence of
precession.  This result is expected given the S/Ns and inferred inclination angles of the binaries \citep{Vitale:2014mka,Green:2020ptm, Pratten:2020igi}.

\begin{figure}
\includegraphics[width=\linewidth, trim=25 5 50 35]{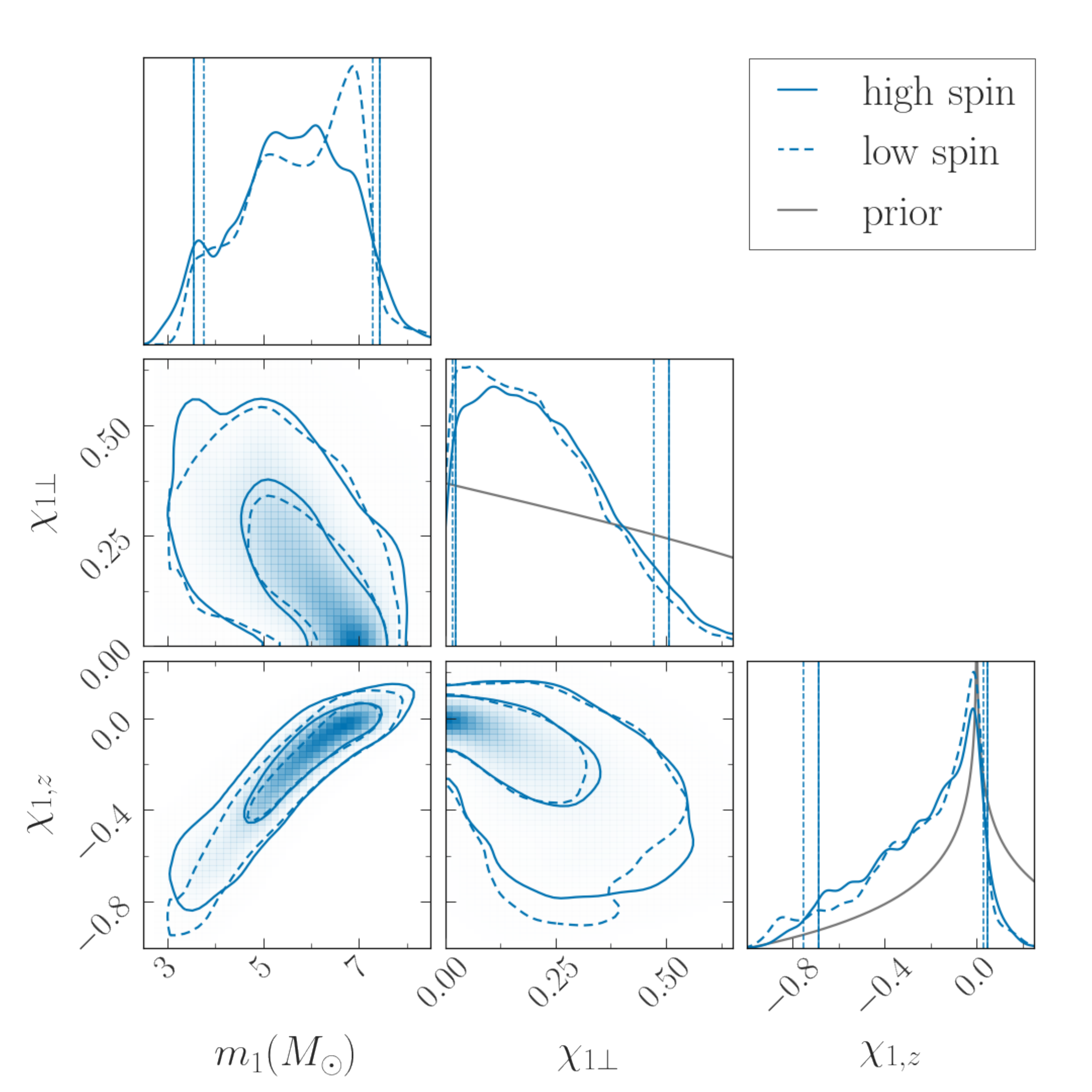}
    \caption{Properties of the primary component of \eventtwo.  The corner plot shows the one-dimensional (diagonal) and two-dimensional (off-diagonal) marginal posterior distributions for the primary's mass and perpendicular and parallel spin components.
The shading indicates the posterior probability of the high-spin prior analysis.  The solid (dashed) lines indicate the 50\% and 90\% credible regions of the high-spin (low-spin) prior analyses.
The vertical lines indicate the 90$\%$ credible intervals for the analyses with high-spin (solid lines) and low-spin (dashed lines) prior.} 
            \label{fig:m1_a1z}
\end{figure}

Low values of the primary mass of \eventtwo ($m_1 \lesssim 5\,\msun$)
are strongly correlated with negative values of the primary
parallel spin component $\chi_{1,z}$, as shown in
Fig.~\ref{fig:m1_a1z}. The astrophysical implications of the mass and
spin correlation are discussed in
\S\ref{sec:AstroImplications}. Figure~\ref{fig:m1_a1z} 
also shows the in-plane spin component $\chi_\perp$, which is peaked
about zero. The lack of conclusive evidence for spin precession
in \eventtwo is consistent with the measurement
of $\chi_\perp$.  Apparent differences between the probability density of the primary spin
in Fig.~\ref{fig:spin_disk} and the posteriors of $\chi_{1\perp}$--$\chi_{1,z}$ in Fig.~\ref{fig:m1_a1z} arise from different choices in visualizing the spin orientation posteriors.

\subsection{Remnant properties}

Under the hypothesis of NSBH coalescence for the two events, estimates
for the final mass and final spin of the remnant BH can be made using
the models of \citet{Zappa:2019ntl}.  We use samples obtained by combining those from Phenom NSBH
and EOB NSBH.
 For \eventone, the remnant mass and spin are
$M_\mathrm{f} = \finalmasssourcemed{GW200105_NSBH_lowspin}^{+\finalmasssourceplus{GW200105_NSBH_lowspin}}_{- \finalmasssourceminus{GW200105_NSBH_lowspin}}$ and $\chi_\mathrm{f} = \finalspinmed{GW200105_NSBH_lowspin}^{+\finalspinplus{GW200105_NSBH_lowspin}}_{- \finalspinminus{GW200105_NSBH_lowspin}}$, while
for \eventtwo,
$M_\mathrm{f} = \finalmasssourcemed{GW200115_NSBH_lowspin}^{+\finalmasssourceplus{GW200115_NSBH_lowspin}}_{- \finalmasssourceminus{GW200115_NSBH_lowspin}}$ and $\chi_\mathrm{f} = \finalspinmed{GW200115_NSBH_lowspin}^{+\finalspinplus{GW200115_NSBH_lowspin}}_{- \finalspinminus{GW200115_NSBH_lowspin}}$.
We do not investigate any post-merger GW signals. The S/Ns
of \eventone and \eventtwo are around a factor of 3 less than
that of GW170817, for which there was no evidence of GWs after the
merger \citep{Abbott:2017dke}.
In the absence of tidal disruption, the postmerger signals of
GW200105 and GW200115 would likely resemble a BH ringdown \citep{2013PhRvD..88f4017F}. The
GW signal associated with such ringdowns would appear well outside of
LIGO's and Virgo’s sensitive bandwidth given the remnant masses and
spins of the systems \citep{2021GReGr..53...59S}.

\subsection{Tests of general relativity and higher-order GW multipole moments}

Results from parameterized tests of general relativity~\citep[GR;][]{
Blanchet:1994ez,PhysRevD.80.122003,PhysRevD.82.064010,PhysRevD.85.082003,Li:2011vx,
Agathos:2013upa,Meidam:2017dgf,LIGOScientific:2018rfd}, show that \eventone and \eventtwo have too low
an S/N to allow for tighter constraints than those already presented
in \citet{LIGOScientific:2020ros}. Within their measurement uncertainties, our
results do not show statistically significant evidence for deviations
from the prediction of GR.

To quantify the evidence for higher-order GW multipole moments, we calculate the orthogonal optimal S/N $\rho_{lm}^{\perp}$ for the subdominant
multipole moments \citep{GW190412-discovery, GW190814,Mills:2020thr}.  We find the $(\ell, |m|)=(3,3)$ to be the loudest subdominant
multipole moment, as expected for binaries with asymmetric masses.
Using the Phenom HM waveform model,
we infer $\rho_{33}^{\perp}= 1.70^{+0.92}_{-1.09}$ ($1.70^{+0.94}_{-1.11}$)
for \eventone and $\rho_{33}^{\perp}=0.91^{+0.93}_{-0.66}$ ($0.86^{+0.90}_{-0.65}$)
for \eventtwo with the low (high) spin prior.
In
Gaussian noise, the median of $\rho_{33}^{\perp}$ is approximately 
chi-distributed with two degrees of freedom, and values greater than $2.1$ indicate significant higher-order multipole content.
The measured
$\rho_{33}^{\perp}$ are therefore consistent with Gaussian noise, as
expected for the majority of NSBHs at these S/Ns, except for those
viewed close to edge-on.

\subsection{Waveform systematics}
\label{sec:waveform systematics}

\begin{figure*}
    \centering
    \includegraphics[width=0.45\linewidth,trim=40 80 90 80]{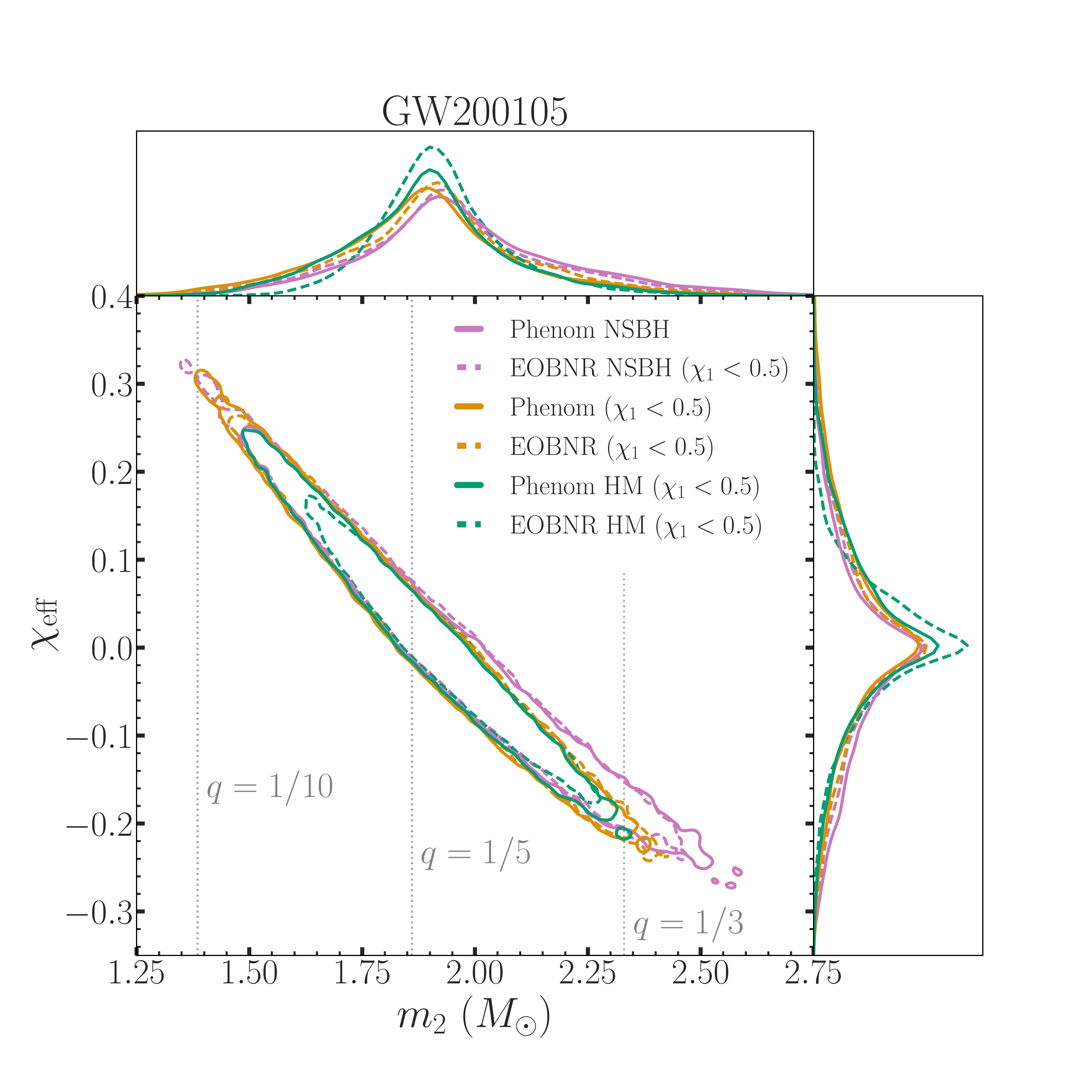}
    \hfil\hfil
   \includegraphics[width=0.45\linewidth,trim=40 80 90 80]{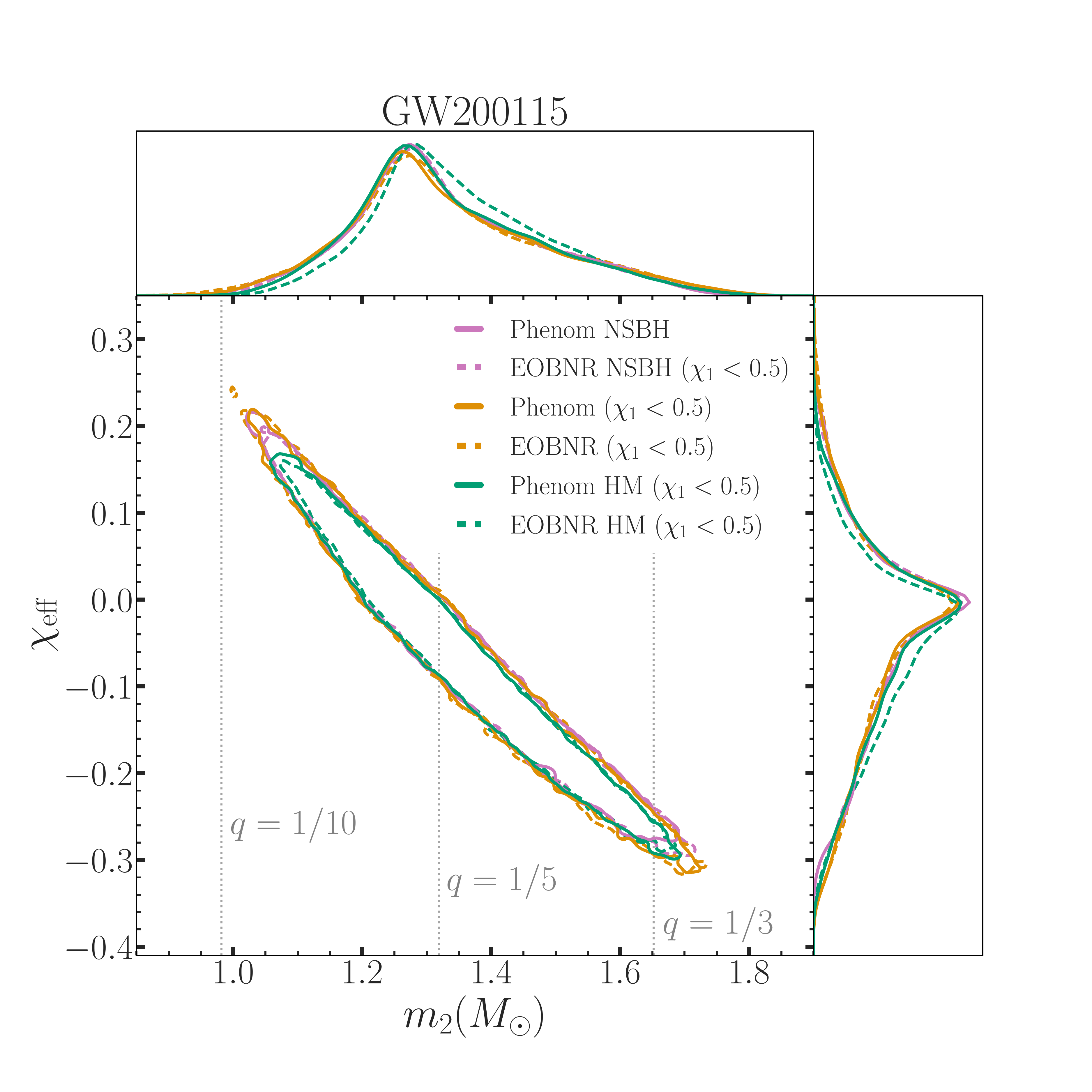}
   \par\medskip \caption{ \label{fig:nsbh_bbh_systematics}
   Comparison of two-dimensional $\masstwosource\text{--}\spineffective$ posteriors
   for the two events reported here, using various NSBH and BBH signal
   models. The vertical dashed lines indicate several mass-ratio
   references mapped to $\masstwosource$ for the median estimate of the chirp masses of \eventone and \eventtwo.\\}
\end{figure*}

Our primary results are obtained using precessing
BBH models with higher-order multipole moments,  Phenom PHM and EOBNR PHM. We now justify this choice by investigating potential systematic uncertainties due to our waveform choice.

First, we investigate the agreement
between independent waveform models that incorporate identical physics.
Figure~\ref{fig:nsbh_bbh_systematics} shows the two-dimensional
$\masstwosource\text{--}\spineffective$ posteriors for both events obtained
using a variety of NSBH and aligned-spin BBH models. Because some NSBH models only cover  $\chi_1<0.5$, we
restrict the prior range of all models to $\chi_1<0.5$ for consistency.

The main panels of Fig.~\ref{fig:nsbh_bbh_systematics} are dominated
by a correlation of the effective inspiral spin
parameter $\spineffective$ with the secondary mass
$\masstwosource$ \citep{Cutler:1994ys,Ng:2018neg}.
Both NSBH models (Phenom NSBH and EOBNR NSBH) give consistent results with each other, as do both BBH models (Phenom and EOBNR), both with and without higher-order multipole moments,
with the most notable difference being that EOBNR HM yields tighter posteriors than Phenom HM.
This demonstrates that waveform models including the same physics give comparable results, but more studies are warranted to improve the understanding of the BBH waveform models in the NSBH region of parameter space.
While not shown in Fig.~\ref{fig:nsbh_bbh_systematics}, we also find good agreement between the primary precessing BBH waveform models; see \S\ref{sec:SourceProperties}.

Second, comparing the NSBH models with the BBH models without higher-order multipole moments (Phenom and EOBNR), the
NSBH models recover similar posterior contours in the
$\masstwosource\text{--}\spineffective$ plane.  This is expected given the asymmetric mass ratio and low S/N of these NSBH observations; see, e.g.~\cite{Huang:2020} for a demonstration that higher S/Ns would be needed to see notable systematic effects. We observe differences
at the extreme ends of the $\masstwosource\text{--}\spineffective$ contours (i.e.\ at the smallest and largest values of
$\masstwosource$).
The construction of the NSBH waveform models used here did not rely on numerical relativity results at mass ratios $q\lesssim 1/8$, nor did they include simulations with $\chi_{1z}<0$ or NS masses $m_2>1.4\msun$ \citep{SEOBNRNSBH,PhenomNSBH}.  Therefore, some differences should be expected, especially for large $m_2$ in \eventone.
Furthermore, for \eventone, the tails of the $m_2$--$\chi_{\rm eff}$ distribution for Phenom NSBH and EOB NSBH at high $m_2$ are also impacted by the inability of the data to constrain the tidal deformability. Hence, the posterior samples include combinations of high $m_2$ with large $\Lambda_2$, despite such combinations being unphysical.  This effect is not apparent for \eventtwo because of its smaller secondary mass.
The isolated islands of probability in the
extreme tails of the distributions are due to sampling noise.

Last, when adding the extra physical content of higher-order multipole moments in BBH models (through Phenom HM and EOBNR HM), the extreme ends of the
$\masstwosource\text{--}\spineffective$ contours are excluded, while the bulk of the distributions are consistent with the posteriors obtained with the NSBH models.  In summary, these comparisons indicate that (i) waveform models including the same physics give comparable results;
(ii) going
from NSBH models to comparable BBH models changes the results only
marginally, i.e.\ any effects of tides are small; and (iii) inclusion of
higher-order multipole moments changes the posterior contours more substantially
than inclusion of tides. We conclude that the the inclusion of precession and higher-order multipole moments afforded by the BBH waveform models is more important than the impact of
tides in the NSBH models.

%% file: pe_combined_table.tex
\newcolumntype{d}[1]{D..{#1}}
\newcommand\mc[1]{\multicolumn{1}{c}{#1}} 
\newcommand\mr[1]{\multicolumn{1}{r}{#1}}
{\centering

\begin{tabular}{ l *{2}{d{4.5}} @{\extracolsep{10pt}} *{1}{d{4.5}} @{\extracolsep{0pt}} *{1}{d{4.5}} }

\toprule\toprule

& \multicolumn{2}{c}{\normalsize GW200105} & \multicolumn{2}{c}{\normalsize GW200115} \\ 

\cmidrule(lr){2-3} \cmidrule(lr){4-5}
& \mc{Low Spin} & \mc{High Spin} & \mc{Low Spin} & \mc{High Spin} \\ 

& \mc{($\chi_2 < 0.05$)} & \mc{($\chi_2 < 0.99$)} &\mc{($\chi_2 < 0.05$)} & \mc{($\chi_2 < 0.99$)} \\ 

\midrule  

$\text{Primary\, mass\, }m_1/M_{\odot}$ & \massonesourcemed{GW200105_combined_lowspin}^{+\massonesourceplus{GW200105_combined_lowspin}}_{-\massonesourceminus{GW200105_combined_lowspin}} & \massonesourcemed{GW200105_combined_highspin}^{+\massonesourceplus{GW200105_combined_highspin}}_{-\massonesourceminus{GW200105_combined_highspin}} & \massonesourcemed{GW200115_combined_lowspin}^{+\massonesourceplus{GW200115_combined_lowspin}}_{-\massonesourceminus{GW200115_combined_lowspin}} & \massonesourcemed{GW200115_combined_highspin}^{+\massonesourceplus{GW200115_combined_highspin}}_{-\massonesourceminus{GW200115_combined_highspin}} \\ 
$\text{Secondary\, mass\, }m_2/M_{\odot}$ & \masstwosourcemed{GW200105_combined_lowspin}^{+\masstwosourceplus{GW200105_combined_lowspin}}_{-\masstwosourceminus{GW200105_combined_lowspin}} & \masstwosourcemed{GW200105_combined_highspin}^{+\masstwosourceplus{GW200105_combined_highspin}}_{-\masstwosourceminus{GW200105_combined_highspin}} & \masstwosourcemed{GW200115_combined_lowspin}^{+\masstwosourceplus{GW200115_combined_lowspin}}_{-\masstwosourceminus{GW200115_combined_lowspin}} & \masstwosourcemed{GW200115_combined_highspin}^{+\masstwosourceplus{GW200115_combined_highspin}}_{-\masstwosourceminus{GW200115_combined_highspin}} \\ 
$\text{Mass\, ratio\, }q$ & \massratiomed{GW200105_combined_lowspin}^{+\massratioplus{GW200105_combined_lowspin}}_{-\massratiominus{GW200105_combined_lowspin}} & \massratiomed{GW200105_combined_highspin}^{+\massratioplus{GW200105_combined_highspin}}_{-\massratiominus{GW200105_combined_highspin}} & \massratiomed{GW200115_combined_lowspin}^{+\massratioplus{GW200115_combined_lowspin}}_{-\massratiominus{GW200115_combined_lowspin}} & \massratiomed{GW200115_combined_highspin}^{+\massratioplus{GW200115_combined_highspin}}_{-\massratiominus{GW200115_combined_highspin}} \\ 
$\text{Total\, mass\, }M/M_{\odot}$ & \totalmasssourcemed{GW200105_combined_lowspin}^{+\totalmasssourceplus{GW200105_combined_lowspin}}_{-\totalmasssourceminus{GW200105_combined_lowspin}} & \totalmasssourcemed{GW200105_combined_highspin}^{+\totalmasssourceplus{GW200105_combined_highspin}}_{-\totalmasssourceminus{GW200105_combined_highspin}} & \totalmasssourcemed{GW200115_combined_lowspin}^{+\totalmasssourceplus{GW200115_combined_lowspin}}_{-\totalmasssourceminus{GW200115_combined_lowspin}} & \totalmasssourcemed{GW200115_combined_highspin}^{+\totalmasssourceplus{GW200115_combined_highspin}}_{-\totalmasssourceminus{GW200115_combined_highspin}} \\ 
$\text{Chirp\, mass\, }\mathcal{M}/M_{\odot}$ & \chirpmasssourcemed{GW200105_combined_lowspin}^{+\chirpmasssourceplus{GW200105_combined_lowspin}}_{-\chirpmasssourceminus{GW200105_combined_lowspin}} & \chirpmasssourcemed{GW200105_combined_highspin}^{+\chirpmasssourceplus{GW200105_combined_highspin}}_{-\chirpmasssourceminus{GW200105_combined_highspin}} & \chirpmasssourcemed{GW200115_combined_lowspin}^{+\chirpmasssourceplus{GW200115_combined_lowspin}}_{-\chirpmasssourceminus{GW200115_combined_lowspin}} & \chirpmasssourcemed{GW200115_combined_highspin}^{+\chirpmasssourceplus{GW200115_combined_highspin}}_{-\chirpmasssourceminus{GW200115_combined_highspin}} \\ 
$\text{Detector-frame chirp\, mass\, } (1+z)\mathcal{M}/M_{\odot}$ & \chirpmassdetmed{GW200105_combined_lowspin}^{+\chirpmassdetplus{GW200105_combined_lowspin}}_{-\chirpmassdetminus{GW200105_combined_lowspin}} & \chirpmassdetmed{GW200105_combined_highspin}^{+\chirpmassdetplus{GW200105_combined_highspin}}_{-\chirpmassdetminus{GW200105_combined_highspin}} & \chirpmassdetmed{GW200115_combined_lowspin}^{+\chirpmassdetplus{GW200115_combined_lowspin}}_{-\chirpmassdetminus{GW200115_combined_lowspin}} & \chirpmassdetmed{GW200115_combined_highspin}^{+\chirpmassdetplus{GW200115_combined_highspin}}_{-\chirpmassdetminus{GW200115_combined_highspin}} \\ 
\addlinespace 
$\text{Primary\, spin\, magnitude\, }\chi_1$ & \spinonemed{GW200105_combined_lowspin}^{+\spinoneplus{GW200105_combined_lowspin}}_{-\spinoneminus{GW200105_combined_lowspin}} & \spinonemed{GW200105_combined_highspin}^{+\spinoneplus{GW200105_combined_highspin}}_{-\spinoneminus{GW200105_combined_highspin}} & \spinonemed{GW200115_combined_lowspin}^{+\spinoneplus{GW200115_combined_lowspin}}_{-\spinoneminus{GW200115_combined_lowspin}} & \spinonemed{GW200115_combined_highspin}^{+\spinoneplus{GW200115_combined_highspin}}_{-\spinoneminus{GW200115_combined_highspin}} \\ 
$\text{Effective\, inspiral\, spin\, parameter\, }\chi_{{\mathrm{eff}}}$ & \chieffmed{GW200105_combined_lowspin}^{+\chieffplus{GW200105_combined_lowspin}}_{-\chieffminus{GW200105_combined_lowspin}} & \chieffmed{GW200105_combined_highspin}^{+\chieffplus{GW200105_combined_highspin}}_{-\chieffminus{GW200105_combined_highspin}} & \chieffmed{GW200115_combined_lowspin}^{+\chieffplus{GW200115_combined_lowspin}}_{-\chieffminus{GW200115_combined_lowspin}} & \chieffmed{GW200115_combined_highspin}^{+\chieffplus{GW200115_combined_highspin}}_{-\chieffminus{GW200115_combined_highspin}} \\ 
$\text{Effective\, precession\, spin\, parameter\, }\chi_p$ & \chipmed{GW200105_combined_lowspin}^{+\chipplus{GW200105_combined_lowspin}}_{-\chipminus{GW200105_combined_lowspin}} & \chipmed{GW200105_combined_highspin}^{+\chipplus{GW200105_combined_highspin}}_{-\chipminus{GW200105_combined_highspin}} & \chipmed{GW200115_combined_lowspin}^{+\chipplus{GW200115_combined_lowspin}}_{-\chipminus{GW200115_combined_lowspin}} & \chipmed{GW200115_combined_highspin}^{+\chipplus{GW200115_combined_highspin}}_{-\chipminus{GW200115_combined_highspin}} \\ 
\addlinespace 
$\text{Luminosity distance\, }D_\mathrm{L}/{\mathrm{Mpc}}$ & \mr{$\luminositydistancemed{GW200105_combined_lowspin}^{+\luminositydistanceplus{GW200105_combined_lowspin}}_{-\luminositydistanceminus{GW200105_combined_lowspin}}$} & \mr{$\luminositydistancemed{GW200105_combined_highspin}^{+\luminositydistanceplus{GW200105_combined_highspin}}_{-\luminositydistanceminus{GW200105_combined_highspin}}$} & \mr{$\luminositydistancemed{GW200115_combined_lowspin}^{+\luminositydistanceplus{GW200115_combined_lowspin}}_{-\luminositydistanceminus{GW200115_combined_lowspin}}$} & \mr{$\luminositydistancemed{GW200115_combined_highspin}^{+\luminositydistanceplus{GW200115_combined_highspin}}_{-\luminositydistanceminus{GW200115_combined_highspin}}$} \\ 
$\text{Source\, redshift\, }z$ & \redshiftmed{GW200105_combined_lowspin}^{+\redshiftplus{GW200105_combined_lowspin}}_{-\redshiftminus{GW200105_combined_lowspin}} & \redshiftmed{GW200105_combined_highspin}^{+\redshiftplus{GW200105_combined_highspin}}_{-\redshiftminus{GW200105_combined_highspin}} & \redshiftmed{GW200115_combined_lowspin}^{+\redshiftplus{GW200115_combined_lowspin}}_{-\redshiftminus{GW200115_combined_lowspin}} & \redshiftmed{GW200115_combined_highspin}^{+\redshiftplus{GW200115_combined_highspin}}_{-\redshiftminus{GW200115_combined_highspin}} \\ 

\toprule\toprule
\end{tabular}
}

%% file: nature_secondary.tex
In this section, we describe the investigations to establish the nature of the secondary objects.
In \S\ref{sec:tidal}, we look for imprints of tidal deformations of the secondaries and conclude that the masses, spins, distances and S/Ns of the detections
make definitive identifications of NSs unlikely, both in GW and
EM measurements.  However, in \S\ref{sec:nature:masses}, we show that the posterior
distributions of the secondary masses agree with those of known NSs.\\


\subsection{Tidal deformability and tidal disruption}
\label{sec:tidal}

The tidal deformability of NSs is imprinted in the GW signal, and is
investigated using the parameter estimation techniques
of \S\ref{sec:SourceProperties}.  In contrast, BHs have zero
tidal deformability \citep{Binnington:2009bb,Damour:2009vw, 2020arXiv201007300C, Charalambous:2021mea, 2021PhRvL.126m1102L}.
We infer the tidal
deformability $\tidaltwo$ of the NSs in \eventone and \eventtwo using
the NSBH waveform models that include tides. We find that the tidal
deformabilities are uninformative relative to a uniform prior in
$\tidaltwo \in [0, 5000]$. This measurement cannot establish the
presence of NSs, which is expected given the mass ratios and the S/N
of the detections \citep{2013PhRvD..88f4017F, Kumar:2016zlj,
Fasano_2020,Huang20}.

Toward the end of the inspiral, the BH may tidally disrupt the NS and form an accretion disk \citep{Pannarale:2012ux,
Foucart:2018rjc}.  This is hypothesized to drive a relativistic jet \citep{Pannarale:2010vs,Paschalidis2015}.
Given the mass ratios for both events and the aligned spins $\chi_{1,z}$ of their primaries (near zero for \eventone, probably negative for \eventtwo),
we do not expect tidal disruption to occur, which would require more equal masses or more positive $\chi_{1,z}$
\citep{2008ApJ...680.1326R,Shibata:2007zm,Etienne2009,Foucart11,PhysRevD.84.064018,Pannarale:2012ux, Foucart:2018rjc}.

To quantitatively confirm this expectation, we use the spectral
representation of equations of state from
\cite{LVCeos2018}, which uses
an SLY low-density crust model \citep{Douchin:2001sv} and parameterizes
the adiabatic index into a polynomial in the logarithm of the
pressure \citep{Lindblom:2010bb,LindblomIndik2012,LindblomIndik2014}.
Following \cite{2021arXiv210301733S}, we marginalize the parameter estimation
samples from the NSBH analyses over these equations of
state. For a fixed equation of state, we compute the maximum
Tolman--Oppenheimer--Volkov (TOV) mass, allowing us to infer the nature (NS or
BH) of the lighter binary compact object, as well as its radius $R_2$,
compactness $C_2 = Gm_2/(R_2c^2)$ and baryon mass. 
Based on these, we
define a total ejecta mass $m_{\rm ej}$  \citep{Fernandez2019} as the sum of
dynamical ejecta \citep{Kruger2020} and $15$\% of
the mass of disk winds \citep{Foucart:2018rjc}. 
For both events, we find that
$m_{\rm ej}<10^{-6}\msun$ for 99\% of the samples.

The absence of ejecta is compatible with the lack of observed
EM counterparts.  However,  given the large distances of the
 mergers ($\simeq 300$ Mpc) and the large uncertainties of their sky
 localization,  EM emission would have been difficult to detect and associate with these GW events.

Estimating the impact of nonlinear $p$--$g$
tidal coupling \citep{PhysRevLett.122.061104}, we find that it
would produce a relative frequency-domain phase shift
for \eventone (\eventtwo) of approximately 134 (38)
times smaller than the equivalent phase shift for GW170817.
This strongly reduced effect is caused by the larger chirp masses, more asymmetric mass ratios, and the presence of only a single NS.
Since $p$--$g$ effects were not detected for GW170817  \citep{PhysRevLett.122.061104}, they will be unobservable within the new NSBH systems.

\subsection{Consistency of component masses with the NS maximum mass}
\label{sec:nature:masses}

\begin{table}
    \caption{\label{table:mmax model selection}  
        Probability that the secondary mass is below the maximum NS mass $M_{\rm max}$ for each event, given different spin assumptions and different choices for the maximum NS mass.
The values shown use a flat prior in $m_2$; alternative, astrophysically motivated mass priors, can cause the estimates to vary by up to \Pnsdiff\% across our chosen models. }
\centering
\begin{tabular}{lccc}
        \toprule\toprule
        &  & \multicolumn{2}{c}{$p(m_2<M_{\rm max})$} \\
        spin prior$\!\!$ & choice of $M_{\rm max}$ & $\!$\eventone$\!$ & $\!$\eventtwo$\!$ \\ \midrule
        $\!\!|\chi_2| < 0.05$ & $M_\mathrm{max,TOV}\!$ & \lecFlatEventOne\% & \lecFlatEventTwo\% \\
        $\!\!|\chi_2| < 0.99$ & $M_\mathrm{max}(\chi_{2})$ & \lecspinFlatEventOne\% & \lecspinFlatEventTwo\% \\
        $\!\!|\chi_2| < 0.99$ & $M_\mathrm{max,GNS}$  & \fcFlatEventOne\% & \fcFlatEventTwo\% \\
        \toprule\toprule
    \end{tabular}
\end{table}

Even  without  definite  identification  of  matter  signatures in the signals, we can compare the observed $m_2$ for \eventone and \eventtwo with
the maximum NS mass, $M_{\rm max}$.
The existence of massive pulsars~\citep{Antoniadis:2013pzd, Cromartie:2019kug,Fonseca:2021wxt} places a lower bound of $M_{\rm max} \gtrsim 2\,\msun$ on the maximum NS mass. 
Studies of GW170817's remnant typically suggest that the maximum mass of a nonrotating NS---the TOV mass---is $M_{\rm max,TOV} \lesssim 2.3\,\msun$~\citep[e.g.,][]{PhysRevD.100.023015, GW170817eosMtov, Nathanail:2021tay}. However, rapid rotation could support a larger $M_{\rm max}$. Given the considerable uncertainty in $M_{\rm max}$, we examine three different scenarios. Following \citet{Essick:2020ghc}, we compute for each scenario the probability $p(m_2<M_{\rm max})$ that the secondary mass is below the maximum NS mass by marginalizing over the uncertainty in $M_{\rm max}$ and the uncertainty of our $m_2$ measurement.

Supposing that the secondaries are slowly spinning, we consider in the first row of Table~\ref{table:mmax model selection} an estimate of $M_{\rm max,TOV}$ from a nonparametric astrophysical inference of the equation of state~\citep{LandryEssick2020}, which predicts $M_{\rm max,TOV} = 2.22^{+0.30}_{-0.20}\,\msun$ and is shown in Fig.~\ref{fig:masses_plot_compare}. 
We then relax the low-spin assumption, estimating in the second row the maximum rotationally supported mass $M_{\rm max}(\chi_2)$, and the breakup spin $\chi_{\rm max}$ with the universal relations from~\citet{10.1093/mnras/stw575}. 
In this scenario, we require $m_2 \leq M_{\rm max}(\chi_2)$ and $\chi_2 \leq \chi_{\rm max}$ for consistency with an NS. 
Finally, in the third row, we consider a parametric fit to the entire distribution of observed Galactic NSs, including rapidly rotating pulsars \citep{FarrChatziioannou2020}, which predicts $M_{\rm max,GNS} = 2.25^{+0.71}_{-0.31}\,\msun$, and is shown in Fig.~\ref{fig:masses_plot_compare}. 
This scenario accounts for the possibility that the maximum mass in the NS population is limited by the astrophysical processes that form compact binaries.
Assuming low spin (first row), we find probabilities of \lecFlatEventOne\% and \lecFlatEventTwo\% that the secondaries in GW200105 and GW200115, respectively, are consistent with an NS assuming a uniform prior in $m_2$. The possibility of large secondary spin reduces these probabilities by up to \spindiffFlat\% (second and third rows).

So far, this analysis has assumed priors that are uniform in component masses.  However, there is considerable uncertainty in the astrophysical mass
priors of such systems, and different prior assumptions can affect the
component mass posteriors for detections with moderate S/N.  To
illustrate the impact of population assumptions, we consider
three alternative priors: one based on Salpeter mass distributions,
$p(m) \sim m^{-2.3}$ \citep{1955ApJ...121..161S}, independently for
each component; one based on an extrapolation of the BBH mass
model \textsc{Broken Power Law} from~\citet{Abbott:2020gyp} down to
$0.5\,\msun$ for both components; and another based on a similar extrapolation of
the \textsc{Power Law + Peak} BBH mass model from the same study.  We
marginalize over the uncertainties in the latter two models, which are
fit to the BBH population from~\citet{Abbott:2020niy}, including the
outlier event GW190814 with a secondary component mass below
$3\,\msun$~\citep{GW190814}. 

These different mass priors change the numbers in Table~\ref{table:mmax model selection} by at most \Pnsdiff\%,
with the smallest values for \eventone and \eventtwo being  $\lowestPnsEventOne\%$ and $\lowestPnsEventTwo\%$, respectively. 
The decrease is due to the three priors assigning more probability density to equal-mass systems, thus favoring larger $m_2$. 
Thus, the secondaries of both systems are consistent with NSs based on our assumptions about the equation of state and the mass distribution.

However, consistency with the maximum NS mass does not exclude the
possibility that the secondaries could be BHs or exotic compact
objects, if such objects also exist within the NS mass range.
For instance, models of primordial BHs predict a peak in the primordial BH mass function at $\sim 1\,\msun$  \citep{Carr:2021}. These models also predict that primordial BHs may form coalescing binaries at mass ratios comparable to those reported here.

%% file: astro.tex
The first confident observations of NSBH binaries enable us to study this novel
type of astrophysical system in entirely new ways.  We pursue three
different avenues in this section. First, we infer the merger rate of
NSBH binaries in the local universe. We
then place the inferred source properties and merger rate in the
context of models of NSBH formation channels and previous EM and GW
observations of BHs and NSs.
Finally, we investigate to what extent the events
reported here can serve to measure the Hubble constant and whether
lensing of GWs may have played a role in the observations.

\subsection{Merger rate density}

\begin{figure}
\centering
\includegraphics[width=0.98\columnwidth]{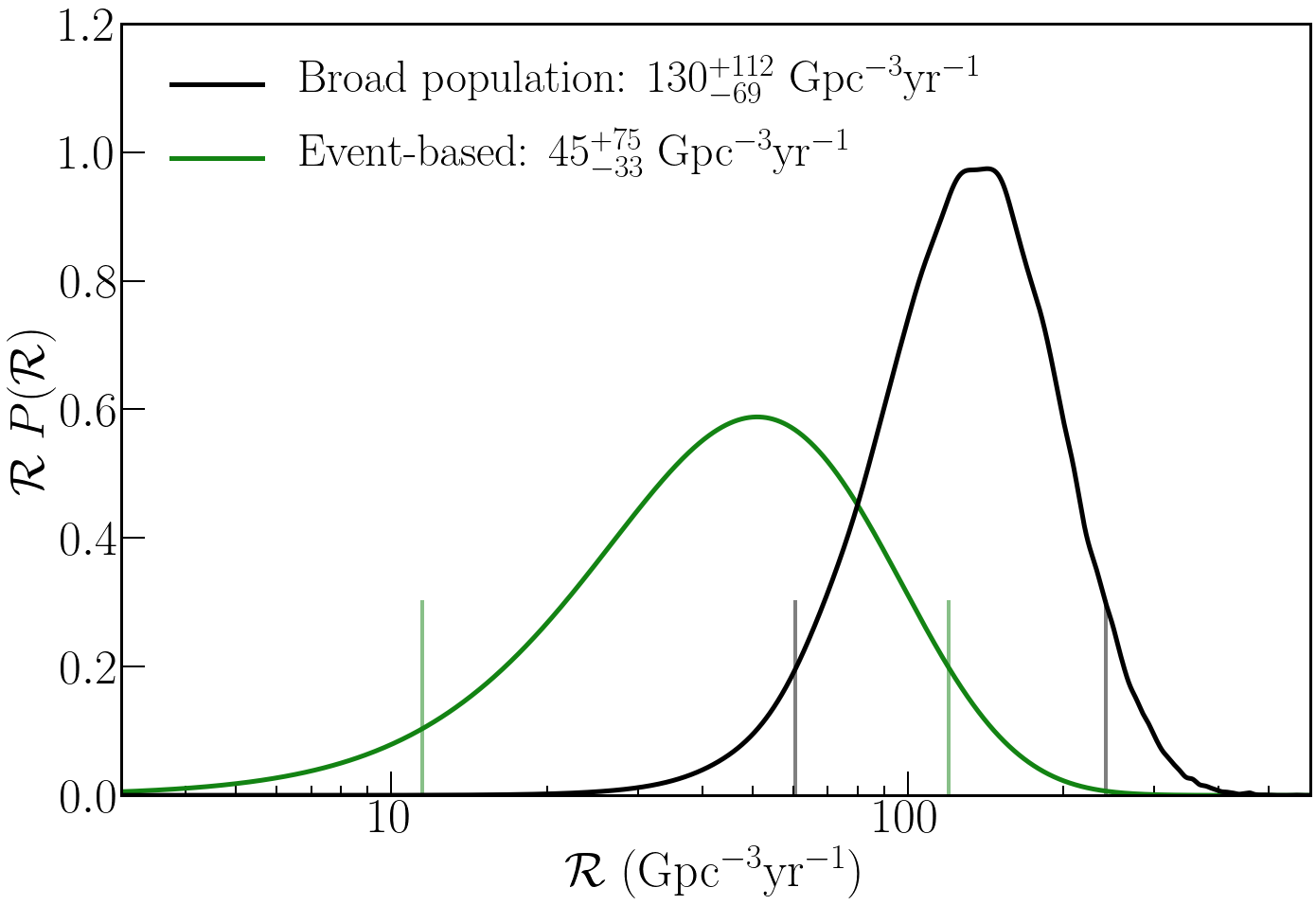}
\caption{\label{fig:rates}
  Inferred probability densities for the NSBH merger rate. Green line:
  rate assuming one count each from an \eventone and \eventtwo-like NSBH population.
  Black line: rate for a broad NSBH population with a low threshold that accounts for 
  marginal triggers. The short vertical lines indicate the 90\% credible intervals.}
\end{figure}

We infer the NSBH merger rate density with our observations using two different approaches.

In the first approach, we
  consider only \eventone and \eventtwo.
Following the method
of \citet{Kim2003} as previously used in, e.g., \citet{GW150914:rates,GW190814},
we calculate an event-based merger rate assuming one Poisson-distributed count each from \eventone- and \eventtwo-like populations.
We calculate semianalytically the search sensitivities across O1, O2, and the first nine months of O3 to
 NSBH populations corresponding to the mass and spin posteriors
 for the two events.  We then calibrate these sensitivities
 to the
 results of \gstlal \citep{Cannon:2011vi, Privitera:2013xza, Messick:2016aqy, Sachdev:2019vvd, Hanna:2019ezx} using a broad NSBH-like population and an FAR threshold of
 \GstLALOfflineFARone. This FAR threshold is chosen to include only
 \eventone and \eventtwo while excluding low-significance triggers
   like \marginal, though a more conservative threshold of 1/(100 yr)
 as used in, e.g., \citet{GW150914:rates,GW190814},
changes the estimated sensitivities by less than 15\%.  Applying the Poisson Jeffreys prior 
proportional to $\mergerratedensity_i^{-1/2}$, we find  
$\mergerratedensity_{200105}=$ \SingleRateOhFiveGSTLAL{}~\VolRateUnit, and 
$\mergerratedensity_{200115}=$ \SingleRateOneFiveGSTLAL{}~\VolRateUnit. 
The \pycbc detection of \eventtwo, using the same method but with an 
independent set of injections for calibration \citep{2018CQGra..35n5009T}, 
yields a consistent $\mergerratedensity_{200115}=\SingleRateOneFivePYCBC{}$~\VolRateUnit.
Combining the likelihoods over $\mergerratedensity_{200105}$ and 
$\mergerratedensity_{200115}$ according to \citet{Kim2003} and applying the Jeffreys prior to the total rate, we find the total event-based NSBH merger rate density $\mergerratedensity_{\rm NSBH}$ = \KKLRateGSTLAL{}~\VolRateUnit, 
plotted in green in Fig.~\ref{fig:rates}.

The second approach to calculating a merger rate takes into account
not only \eventone and \eventtwo, but also less significant
search triggers with masses consistent with the typical range
associated with NSBH binaries.  Specifically, we consider triggers
across O1, O2 and the first nine months of O3 with associated
component masses $m_1\!\in\![2.5,40]\,\msun$ and
$m_2\!\in\![1,3]\,\msun$, i.e.\ broader ranges than the component
mass posteriors for \eventone and \eventtwo obtained
in \S\ref{sec:SourceProperties}.  The cut-off of $m_2 \leq 3 \msun$ is chosen as a robust upper limit on the maximum NS mass \citep{Rhoades:1974fn, Kalogera:1996ci}.
The population is defined to be uniformly distributed in comoving volume, uniform in log component masses in the given ranges, with aligned
spins with spin magnitudes distributed uniformly in $[0,0.95]$.
 We use as
  input \gstlal search triggers above an S/N threshold such that the number of noise triggers exceeds the number of
  astrophysical signals by a factor $\sim$100.  Following
  \citet{2015PhRvD..91b3005F, Kapadia:2020}, we find the resulting
  joint likelihood on Poisson parameters for signals of each
  astrophysical category
($\Lambda_{\rm{BNS}}, \Lambda_{\rm{NSBH}}, \Lambda_{\rm{BBH}}$),
  and for terrestrial noise triggers
$\Lambda_{\rm background}$. Here, the BBH and BNS categories are defined to include triggers where both component masses fall within 5--100 $\msun$ or 1--2.5 $\msun$, respectively.
We apply the Jeffreys prior and recover a merger rate density
 $\mergerratedensity_{\rm{NSBH}} = \Lambda_{\rm{NSBH}}/\langle VT \rangle_{\rm{NSBH}}$, where $\langle VT \rangle_{\rm{NSBH}}$ is the population-averaged
 sensitive time--volume estimated using the \gstlal pipeline and the NSBH population defined above. This yields an estimated NSBH merger rate density of
 $\mergerratedensity_{\rm{NSBH}}=$ \FGMCRate{}~\VolRateUnit, the black line in Fig.~\ref{fig:rates}.

While this rate is higher than our event-based rate, it considers a
wider population that includes additional triggers; for example, the
component masses of GW190814 (although the nature of its secondary is
uncertain) fall within this population.
The calculation is also based
on the component mass values of \gstlal search triggers, adjusted with
an analytical model \citep{Fong:2018elx} 
  of the response of the template bank to signals with a given distribution of true masses.
This procedure is expected to be less precise than
Bayesian parameter estimation, which is impractical for the large
numbers of triggers involved.

The merger rate density measured here is consistent with the
upper bound of 610~\VolRateUnit derived from the absence of NSBH detections in
O1 and O2  \citep{2019PhRvX...9c1040A}.
Revised merger rate estimates for all CBC sources will be provided in a future analysis of the full O3 data set.

\subsection{System origins}
\label{sec:formation_channels}

To understand the origin of \eventone and \eventtwo, we compare their observed masses and spins with theoretical predictions.
Population synthesis studies  modeling the various formation
channels of merging compact object binaries 
distinguish between NSs and BHs with a simple mass cut, typically
between 2 and 3 $\msun$. This is consistent with the secondary
masses of both events being classified as NSs, so for the purposes
of discussing formation channels for these events and predicted rates, 
we will discuss \eventone{} and \eventtwo{} in the context of them being
NSBHs.

\subsubsection{Formation channels} Formation channels of NSBH can be broadly
categorized as either isolated binary evolution or one of several dynamical
formation channels (e.g.,  globular clusters or
  nuclear star clusters). 
 Since isolated binaries form in young star clusters and can be
  influenced by dynamical interactions before the cluster dissolves
  and the binary effectively becomes isolated, rates
  from young star clusters naturally encompass rates from isolated binaries.
Predicted rates of NSBH mergers in the local universe vary by
orders of magnitude across the various formation channels.

Models of the canonical isolated
binary evolution channel---in which stellar progenitors evolve together, shedding orbital angular momentum 
through phases of stable and/or unstable mass transfer
prior to compact object formation---predict NSBH merger rate densities around 0.1--800~\VolRateUnit{}
\citep{2002ApJ...572..407B,Sipior2002,Belczynski2006,Belczynski2016,dominik2015,eldridge2017,Kruckow2018,mapelli2018,neijssel2019,Drozda:2020qab,2020ApJ...899L...1Z,Broekgaarden:2021iew}.
The high
uncertainty is driven by the lack of observed NSBHs and the wide range of model assumptions. Merger rates are
sensitive to the treatment of common envelopes, which may be a 
necessary evolutionary phase for producing
compact binaries in tight orbits capable of merging in a Hubble time \citep{Ivanova:2013b}. They are
also sensitive to prescriptions for supernova kick magnitudes. While moderate kicks
can produce eccentric orbits that merge on short timescales, high supernova kicks
may disrupt the progenitor binaries and suppress the merger
rate \citep{2002ApJ...572..407B,2020ApJ...891..141G,2020MNRAS.493L...6T}.

Models of star formation in the dynamical environments of young star
clusters predict NSBH merger rate densities of 0.1--100~\VolRateUnit{}
\citep{Fragione:2020wac,Hoang:2020gsi,rastello2020, Santoliquido:2020bry}. In this scenario, most systems that form merging NSBH ($\sim$80\%) 
are ejected without undergoing dynamical exchanges,
proceeding to merge in the field.

Models of dynamical formation channels in denser
environments typically predict much lower merger rates. For instance, in globular clusters and nuclear star clusters, BHs segregate and dominate
the core, where the bulk of dynamical interactions
occur \citep{PortegiesZwart2000, Morscher15},  so that encounters between NSs and BHs are
relatively rare, with NSBH merger rate densities on the order of $10^{-2}$~\VolRateUnit
 \citep{Clausen2013, ArcaSedda2020,Ye2020}. In disks of active
 galactic nuclei, the presence of gas could possibly increase
 the NSBH merger rate density up to 300 Gpc$^{-3}$ yr$^{-1}$ \citep{McKernan:2020lgr}.

NSBHs may also merge via
hierarchical triple interactions, where inner NSBH binaries are driven to high
eccentricity by massive tertiary companions and merge on rapid timescales
\citep{antonini2017,silsbee2017}. However, the predicted merger
rates are negligible unless supernova kicks are assumed to be zero
\citep{fragione2019b}.

It is likely that a combination of the above channels contribute to the astrophysical NSBH merger rate.
However, the isolated binary evolution, young star cluster, and active galactic nuclei channels are capable of individually accounting for the NSBH merger rate estimated here.

\subsubsection{Masses}
While there are no
observed NSBHs in the Milky Way, we can place the component masses of \eventone and \eventtwo in the
context of the observed population of BH and NS masses, as well as the predicted populations of NSBHs.
 Observations suggest that the mass distribution of the Galactic population of NSs peaks around
$1.33\,\msun$, with a secondary peak around $1.9\,\msun$
\citep{2016arXiv160501665A, Alsing:2017bbc}.  The
secondary mass observed in \eventtwo and marginal event \marginal are consistent with the population
 peaking at $1.33\,\msun$, while the secondary observed in \eventone ($\simeq 1.9\,\msun$)
and the primary component from BNS merger GW190425 \citep[$m_1=$ 1.60--1.87$\,\msun$; ][]{Abbott:2020uma} are consistent with the high-mass population. However,
a rigorous association of the events with different components of the NS population would require a thorough population analysis.
Radio observations of BNS systems do not
find such massive NSs,  leading to speculation as to the origin of GW190425
\citep{Romero-Shaw:2020aaj, Safarzadeh:2020efa, Galaudage_2021,Mandel21}.
Stellar metallicities in the Milky Way are not representative of all populations 
of GW sources \citep{2008ApJ...675..566O,2010ApJ...716..615O,2010ApJ...715L.138B,eldridge2017,neijssel2019}.

The BH masses observed in
\eventone
and \eventtwo ($\massonesourcemed{GW200105_combined_highspin}^{+\massonesourceplus{GW200105_combined_highspin}}_{-\massonesourceminus{GW200105_combined_highspin}}\,\msun$
and $\massonesourcemed{GW200115_combined_highspin}^{+\massonesourceplus{GW200115_combined_highspin}}_{-\massonesourceminus{GW200115_combined_highspin}}\,\msun$, respectively) are in line with predictions from
population synthesis models for NSBH mergers from
  isolated binary evolution and young star clusters. In NSBHs, current binary evolution
  models do not predict BH masses above $\simeq$ 10 $\msun$ \citep{eldridge2017,Kruckow2018,mapelli2018,Broekgaarden:2021iew}, while Population III NSBHs \citep{Kinugawa_2017} and 
  dynamical interactions in low-metallicity young star clusters allow for higher BH
  masses \citep{ziosi2014,rastello2020}. 

Electromagnetic observations of X-ray binaries have not uncovered BHs between 3 and 5$\,\msun$,
leading to speculation about a mass gap
\citep{Ozel:2010,2011ApJ...741..103F,Kreidberg:2012,Miller:2014aaa}.  Analysis of GWTC-2
 has also found evidence for a gap or dip in the BH mass spectrum between $\sim$2.6 and 4 $\msun$ \citep{2020ApJ...899L...8F}.
For \eventtwo, 
we find nonnegligible support for the primary lying in this mass gap, with 
$p(3 \, \msun\! <\! m_1 \!<\! 5 \, \msun)=\PEpercentMassGap{GW200115_combined_highspin}\%$ (\PEpercentMassGap{GW200115_combined_lowspin}$\%$) 
under the high-spin (low-spin) priors. This low-mass region is correlated 
with negative values of the parallel component of the primary spin; see \S\ref{sec:astro:spins} below.

In summary, the masses inferred for \eventone and \eventtwo are
consistent with expectations for NSBHs; their primary masses are in
agreement with predictions for BH masses in
population synthesis models of the dominant formation
scenarios. Meanwhile, their secondary masses are compatible
with the observed population of Galactic NSs, as well
as the masses inferred from GW
observations of BNS mergers.

\subsubsection{Spins} 
\label{sec:astro:spins}

Spin information encoded in GWs from binaries are a probe of the their evolutionary
history \citep{Farr:2017, Farr:2018,2017MNRAS.471.2801S,Talbot:2017, Vitale:2017, 
  Wysocki:2019}. The BHs in binaries are
expected to exhibit a range of spin magnitudes and orientations, depending on how they formed
\citep{Kalogera_2000, Mandel_2010, Rodriguez_2016, Liu_2017,
    Liu_2018, Antonini_2018,Kruckow2018,
    Bavera:2019fkg,Chattopadhyay:2020lff}.  The highest
dimensionless NS spin implied by pulsar-timing observations of
binaries that merge within a Hubble time is 
$\sim$0.04 \citep{Stovall:2018ouw,Zhu:2018}.

While the secondary spins of both events reported here  are
poorly constrained due to the unequal masses, the primary spins of \eventone and \eventtwo can be placed in the context
of predictions for BH spins from models of stellar and dynamical evolution and EM observations of NSBH progenitors. 
As can be seen in Fig.~\ref{fig:spin_disk} and Fig.~\ref{fig:m1_a1z}, the primary spins of 
\eventone and \eventtwo are consistent with zero 
($\spinonemed{GW200105_combined_highspin}^{+\spinoneplus{GW200105_combined_highspin}}_{-\spinoneminus{GW200105_combined_highspin}}$
and
$\spinonemed{GW200115_combined_highspin}^{+\spinoneplus{GW200115_combined_highspin}}_{-\spinoneminus{GW200115_combined_highspin}}$,
respectively), but moderate values of spin are not ruled out. The primary in \eventtwo
may even have relatively high spin, with a 90\% upper limit of $\spinoneninetypercent{GW200115_combined_highspin}$.
Several
studies of the observed population of high-mass X-ray binaries \citep{2008ApJ...679L..37L, 2009ApJ...701.1076G,2014ApJ...790...29G,Zhao:2021bhj} 
find that the BHs have large spins \citep{Valsecchi:2010, Wong:2011, Qin:2018sxk,Zhao:2021bhj}. Given the short lifetime of the secondary, mass
transfer is argued to be insufficient to generate BHs with
such high spins, implying that the BHs were born with high spins.
\citet{Belczynski:2011bv} found that one such high-mass X-ray binary, Cygnus X-1, is expected to form
an NSBH with a BH that carries near-maximal spin, although it would not
merge within a Hubble time. However, following revised estimates of the component masses of Cygnus X-1 \citep{2021arXiv210209091M},
\citet{Neijssel:2021imj} found that it will most likely form a BBH.
Meanwhile, analyses of GWTC-1 and GWTC-2 have found evidence for BH spin
  \citep{GW151226, 2017PhRvL.119y1103V, Chatziioannou,
Kimball:2020opk,Abbott:2020gyp}, though they do not determine whether those BHs may have been formed with that spin.
Altogether, these EM and GW
  observations of compact binaries and their progenitors suggest a
  range of BH natal spins in NSBH binaries.
   
Along with their magnitudes, the alignments of component spins with the
overall binary orbital angular momentum are of astrophysical interest. In
particular, we find evidence that the primary BH spin in \eventtwo is
negatively aligned with respect to the orbital angular momentum axis, with $p(\chi_{1,z}\!<\!0)=$
\PEpercentchionenegative{GW200115_combined_highspin}\%
(\PEpercentchionenegative{GW200115_combined_lowspin}\%) under the high-spin (low-spin)
prior and with the more negative values of $\chi_{1,z}$ correlated with smaller
$m_1$. This negative alignment is consistent with dynamical formation channels, which typically
form binaries with random spin orientations \citep{Rodriguez_2016}, but the
predicted rates from these channels, discussed in \S\ref{sec:formation_channels}, are small.
Binaries born in isolation are expected to form with only small misalignments
\citep[$\lesssim\,$30°;][]{Kalogera_2000}, though they may become misaligned
by supernova kicks at compact object formation
\citep{Rodriguez_2016,PhysRevD.98.084036,PhysRevD.97.043014}, and possibly during subsequent 
evolution via mass transfer \citep{Stegmann:2020kgb}.
Meanwhile, NSBH
progenitor binaries originating in young star clusters can be perturbed
via close dynamical encounters before being ejected into the field.
Therefore, a misaligned spin in the primary of \eventtwo would not necessarily rule out any of
the plausible NSBH formation channels.

\subsection{Cosmology and lensing}

Gravitational wave sources are standard sirens, providing a direct
measurement of their luminosity distance~\citep{Schutz:1986gp,
Holz:2005df}, and they can be used to measure the Hubble
constant~\citep{2017Natur.551...85A,2019ApJ...871L..13F,Abbott:2019yzh}.
Due to the lack of a confirmed EM counterpart and large numbers of galaxies inside the localization volumes of each of the two events, we do not obtain any informative bounds on $H_0$ from these observations.

The detections of \eventone and \eventtwo are separated by only $\sim$10~days.
  One explanation for the small time-delay could be that the two
  events are created by gravitational lensing by a
  galaxy~\citep{Haris:2018vmn}.   Gravitational lensing is unlikely even a
  priori~\citep{Smith:2018gle,Ng:2017yiu}, and the nonoverlapping mass posteriors (Fig.~\ref{fig:masses_plot_compare})
further exclude it as a possible explanation~\citep{Haris:2018vmn}.
  While \eventtwo and
  \marginal exhibit agreement in their source mass posteriors, their sky localization
  areas do not overlap, and their detector-frame (redshifted) chirp masses show only marginal overlap~\citep{Abbott:2020niy}, ruling out lensing as a possible explanation.

%% file: conclusions.tex
During its third observing run, the LIGO--Virgo GW detector
network observed \eventone and \eventtwo, two GW events
consistent with  NSBH coalescences. 
\eventone is effectively a single-detector event observed in \LLO with an S/N of \GstLALOfflineNetSNRone.  It clearly stands apart from all recorded noise transients, but its statistical confidence is difficult to establish.  \eventtwo was observed in coincidence by the network with an S/N of \GstLALOfflineNetSNRtwo and FAR of \GstLALOfflineFARtwo.

The source component masses of \eventone and \eventtwo make it likely that these events originated from NSBH
coalescences. Their primary masses are found to be
$\massonesource =
\massonesourcemed{GW200105_combined_highspin}^{+\massonesourceplus{GW200105_combined_highspin}}_{-\massonesourceminus{GW200105_combined_highspin}}\,
\msun$ and $\massonesource =\massonesourcemed{GW200115_combined_highspin}^{+\massonesourceplus{GW200115_combined_highspin}}_{-\massonesourceminus{GW200115_combined_highspin}}\,
\msun$, which are consistent with predictions of BH masses in population
synthesis models for NSBHs. Their secondary masses, inferred to be $\masstwosource =
\masstwosourcemed{GW200105_combined_highspin}^{+\masstwosourceplus{GW200105_combined_highspin}}_{-\masstwosourceminus{GW200105_combined_highspin}}\,
\msun$
and $\masstwosource
=~\masstwosourcemed{GW200115_combined_highspin}^{+\masstwosourceplus{GW200115_combined_highspin}}_{-\masstwosourceminus{GW200115_combined_highspin}}\,
\msun$, respectively, are consistent with the observed NS
mass distribution in the Milky Way, as well as population synthesis predictions for
secondary masses in merging NSBHs.

We find no evidence of measurable tides or tidal disruption for either of the two signals, and no EM counterparts to either detection
have been identified. 
As such, there is no direct evidence that the secondaries
are NSs, and our observations are consistent with either event being a BBH merger.
However, the absence of tidal measurements and EM counterparts is to be expected given the properties and distances of the two events. Moreover, the comparisons of the secondary masses to
the maximum allowed NS mass yield a probability $p(m_2\leq
M_{\rm max})$ of $\lowestPnsEventOne\%$--$\highestPnsEventOne\%$ and $\lowestPnsEventTwo\%$--$\highestPnsEventTwo\%$ for the secondaries in \eventone and \eventtwo, respectively, being compatible with NSs (see \S\ref{sec:nature:masses}).

The effective inspiral spin parameter of \eventone is strongly peaked around zero: $\spineffective =
\chieffmed{GW200105_combined_highspin}^{+\chieffplus{GW200105_combined_highspin}}_{-\chieffminus{GW200105_combined_highspin}}$.
For the second event, \eventtwo, the effective inspiral spin parameter is inferred to be
$\spineffective =
\chieffmed{GW200115_combined_highspin}^{+\chieffplus{GW200115_combined_highspin}}_{-\chieffminus{GW200115_combined_highspin}}$.
For \eventtwo, the spin component parallel to the orbital angular
momentum of the primary is $\chi_{1,z}
=\spinonezmed{GW200115_combined_highspin}^{+\spinonezplus{GW200115_combined_highspin}}_{-\spinonezminus{GW200115_combined_highspin}}$,
and we find support for negatively aligned primary spin ($\chi_{1,z} <
0 $) at $\PEpercentchionenegative{GW200115_combined_highspin}\%$
probability.  More negative values of $\chi_{1,z}$ in \eventtwo are
correlated with particularly small primary masses reaching into the
lower mass gap.  We find $p(3 \, \msun\! <\! m_1\! <\!
5 \, \msun)=\PEpercentMassGap{GW200115_combined_highspin}\%$
(\PEpercentMassGap{GW200115_combined_lowspin}$\%$) under the high-spin
(low-spin) parameter estimation priors.  We find no conclusive
evidence for spin-induced orbital precession in either system.

We estimate the merger rate density of NSBH binaries with two approaches.  Assuming that \eventone and \eventtwo are representative of the entire NSBH population, we find $\mergerratedensity_{\rm NSBH} = \KKLRateGSTLAL$~\VolRateUnit.
Conversely, assuming a broader range of allowed primary and secondary masses, and considering all triggers in O3, we find 
$\mergerratedensity_{\rm NSBH} = \FGMCRate$~\VolRateUnit.
These are the first direct measurements of
the NSBH merger rate. Both estimates are broadly
consistent with the rate predicted from NSBH formation in isolated binaries or
young star clusters.  Formation channels in dense star clusters
(globular or nuclear) and in triples predict lower rates than those inferred from the
two events and are unlikely to be the dominant NSBH formation channels.

The observations of
\eventone and \eventtwo are consistent with predictions for merging NSBHs and observations of 
BHs and NSs in the Milky Way. Given their significantly unequal component masses, future observations of NSBH systems will provide new opportunities to study matter under extreme conditions, including tidal disruption, and search for potential deviations from GR.

%% file: LVK_acknowledgements_P2000488_v5.tex
This material is based upon work supported by NSF’s LIGO Laboratory which is a major facility
fully funded by the National Science Foundation.
The authors also gratefully acknowledge the support of
the Science and Technology Facilities Council (STFC) of the
United Kingdom, the Max-Planck-Society (MPS), and the State of
Niedersachsen/Germany for support of the construction of Advanced LIGO 
and construction and operation of the GEO600 detector. 
Additional support for Advanced LIGO was provided by the Australian Research Council.
The authors gratefully acknowledge the Italian Istituto Nazionale di Fisica Nucleare (INFN),  
the French Centre National de la Recherche Scientifique (CNRS) and
the Netherlands Organization for Scientific Research, 
for the construction and operation of the Virgo detector
and the creation and support  of the EGO consortium. 
The authors also gratefully acknowledge research support from these agencies as well as by 
the Council of Scientific and Industrial Research of India, 
the Department of Science and Technology, India,
the Science \& Engineering Research Board (SERB), India,
the Ministry of Human Resource Development, India,
the Spanish Agencia Estatal de Investigaci\'on,
the Vicepresid\`encia i Conselleria d'Innovaci\'o, Recerca i Turisme and the Conselleria d'Educaci\'o i Universitat del Govern de les Illes Balears,
the Conselleria d'Innovaci\'o, Universitats, Ci\`encia i Societat Digital de la Generalitat Valenciana and
the CERCA Programme Generalitat de Catalunya, Spain,
the National Science Centre of Poland and the Foundation for Polish Science (FNP),
the Swiss National Science Foundation (SNSF),
the Russian Foundation for Basic Research, 
the Russian Science Foundation,
the European Commission,
the European Regional Development Funds (ERDF),
the Royal Society, 
the Scottish Funding Council, 
the Scottish Universities Physics Alliance, 
the Hungarian Scientific Research Fund (OTKA),
the French Lyon Institute of Origins (LIO),
the Belgian Fonds de la Recherche Scientifique (FRS-FNRS), 
Actions de Recherche Concertées (ARC) and
Fonds Wetenschappelijk Onderzoek – Vlaanderen (FWO), Belgium,
the Paris \^{I}le-de-France Region, 
the National Research, Development and Innovation Office Hungary (NKFIH), 
the National Research Foundation of Korea,
the Natural Science and Engineering Research Council Canada,
Canadian Foundation for Innovation (CFI),
the Brazilian Ministry of Science, Technology, and Innovations,
the International Center for Theoretical Physics South American Institute for Fundamental Research (ICTP-SAIFR), 
the Research Grants Council of Hong Kong,
the National Natural Science Foundation of China (NSFC),
the Leverhulme Trust, 
the Research Corporation, 
the Ministry of Science and Technology (MOST), Taiwan,
the United States Department of Energy,
and
the Kavli Foundation.
The authors gratefully acknowledge the support of the NSF, STFC, INFN and CNRS for provision of computational resources.

This work was supported by MEXT, the JSPS Leading-edge Research Infrastructure Program, JSPS Grant-in-Aid for Specially Promoted Research 26000005, JSPS Grant-in-Aid for Scientific Research on Innovative Areas 2905: JP17H06358, JP17H06361, and JP17H06364, JSPS Core-to-Core Program A. Advanced Research Networks, JSPS Grant-in-Aid for Scientific Research (S) 17H06133, the joint research program of the Institute for Cosmic Ray Research, University of Tokyo, National Research Foundation (NRF) and Computing Infrastructure Project of KISTI-GSDC in Korea, Academia Sinica (AS), AS Grid Center (ASGC) and the Ministry of Science and Technology (MoST) in Taiwan under grants including AS-CDA-105-M06, Advanced Technology Center (ATC) of NAOJ, and Mechanical Engineering Center of KEK.